\documentclass[%
 reprint,
 nofootinbib,
 amsmath,amssymb,
 aps,
]{revtex4-1}

\usepackage{hyperref}
\hypersetup{
    colorlinks,
    linkcolor={red!50!black},
    citecolor={blue!50!black},
    urlcolor={blue!80!black},
}

\usepackage{aas_macros}
\usepackage{natbib}
\usepackage{color}
\usepackage{mathtools,amssymb} 
\usepackage{graphicx,psfrag,subfigure}
\usepackage{sidecap}
\usepackage{soul,float}
\usepackage{algorithm,algpseudocode}
\usepackage{xcolor}
\usepackage{enumitem}
\usepackage{gensymb}
\usepackage{xspace}
\usepackage{nicefrac}
\usepackage{relsize}
\usepackage[utf8x]{inputenc}
\usepackage{fontawesome}
\usepackage{fancyvrb}

\providecommand{\sorthelp}[1]{}

\DeclareUnicodeCharacter{981}{$\phi$}

\DeclareMathAlphabet{\mathbbgreek}{U}{bbold}{m}{n}

\newcommand{\op}[1]{\mathbb{#1}}
\newcommand{\Aphi}{A_{\phi}}
\newcommand{\Len}[1][]{\op L\ifthenelse{\equal{#1}{}}{}{(#1)}}
\newcommand{\Cflen}[1][]{\op{\widetilde{C}}_{f}\ifthenelse{\equal{#1}{}}{}{(#1)}}
\newcommand{\Cf}[1][]{\op C_{f}\ifthenelse{\equal{#1}{}}{}{(#1)}}
\newcommand{\Cphi}[1][]{\op C_{\phi}\ifthenelse{\equal{#1}{}}{}{(#1)}}
\newcommand{\Cn}{\op C_{n}}
\newcommand{\opG}[1][]{\op G\ifthenelse{\equal{#1}{}}{}{(#1)}}
\newcommand{\D}[1][]{\op D\ifthenelse{\equal{#1}{}}{}{(#1)}}
\newcommand{\I}{\mathbbgreek{1}}
\newcommand{\myvspace}{$\vphantom{\big[^\dagger}$}
\newcommand{\mix}{\prime}

\newcommand{\configA}{\texttt{2PARAM}\xspace}
\newcommand{\configB}{\texttt{MANY}\xspace}
\newcommand{\configC}{\texttt{BIG}\xspace}

\begin{document}

\title{Bayesian delensing delight: sampling-based inference of the primordial CMB and gravitational lensing}

\author{Marius Millea\footnotemark[\value{footnote}]\textsuperscript{\hyperlink{affil}{$\dagger$},}}
\email{mariusmillea@gmail.com}
\affiliation{Berkeley Center for Cosmological Physics and Department of Physics, University of California, Berkeley, CA 94720, USA}

\author{Ethan Anderes}
\altaffiliation{\hypertarget{affil}{Authors contributed equally.}}
\affiliation{Department of Statistics, University of California, Davis, CA 95616, USA}

\author{Benjamin D. Wandelt}
\affiliation{Sorbonne Universit\'e, CNRS, UMR 7095, Institut d'Astrophysique de Paris, 98 bis bd Arago, 75014 Paris, France}
\affiliation{Sorbonne Universit\'e, Institut  Lagrange  de  Paris  (ILP),  98  bis bd Arago, 75014 Paris, France}
\affiliation{Center for Computational Astrophysics, Flatiron Institute, 162 5th Avenue, 10010, New York, NY, USA}

\begin{abstract}
The search for primordial gravitational waves in the Cosmic Microwave Background (CMB) will soon be limited by our ability to remove the lensing contamination to $B$-mode polarization. The often-used quadratic estimator for lensing is known to be suboptimal for surveys that are currently operating and will continue to become less and less efficient as instrumental noise decreases. While foregrounds can in principle be mitigated by observing in more frequency bands, progress in delensing hinges entirely on algorithmic advances. We demonstrate here a new inference method that solves this problem by sampling the exact Bayesian posterior of any desired cosmological parameters, of the gravitational lensing potential, and of the delensed CMB maps, given lensed temperature and polarization data.  We validate the method using simulated CMB data with non-white noise and masking on up to 650\,deg$^2$ patches of sky. A unique strength of this approach is the ability to jointly estimate cosmological parameters which control both the primordial CMB and the lensing potential, which we demonstrate here for the first time by sampling both the tensor-to-scalar ratio, $r$, and the amplitude of the lensing potential, $A_\phi$. The method allows us to perform the most precise check to-date of several important approximations underlying CMB-S4 $r$ forecasting, and we confirm these yield the correct expected uncertainty on $r$ to better than 10\%. 
\end{abstract}

\maketitle

\section{Introduction}
\label{sec:intro}

The gravitational lensing of the Cosmic Microwave Background (CMB) is a key cosmological observable. Current and next generation CMB probes are all targeting significant improvements in sensitivity to the lensing effect \cite{benson2014,anderson2018,henderson2016,suzuki2016,thesimonsobservatorycollaboration2018,abitbol2017,hanany2019,abazajian2019}. These will correspond to large improvements in the precision with which we can reconstruct the gravitational lensing potential, $\phi$, and with which we can ``delense'' the CMB to reveal the unaltered primordial signal. The inferred maps of $\phi$ encode a wealth of information about the late-time structure and geometry of the universe, both by themselves and in cross-correlation with other tracers of matter. Delensing, which can remove the spurious foreground $B$-mode polarization generated by lensing, will be crucial in searching for the hypothesized primordial $B$-mode signal sourced by inflationary gravitational waves. Despite the importance of the lensing effect, however, it is still an open question how in practice to optimally extract cosmological information from the very low-noise observations of the lensed CMB achievable in the near future.

Up until very recently, all CMB lensing analyses have used a quadratic estimator (QE) \cite{hu2002} to produce a point estimate of $\phi$. Obtaining cosmological constraints then proceeds by either 1) taking the auto power spectrum of this reconstructed $\phi$, debiasing the spectrum and computing error bars with a combination of analytic calculations and Monte Carlo simulations, then comparing to model $C_\ell^{\phi\phi}$ power spectra, or 2) cross-correlating $\phi$ with other low-redshift probes of structure, and similarly computing the expected response with various semi-analytic techniques. This is the approach taken in the first detection of the lensing effect in the CMB from cross-correlating WMAP with NVSS galaxies \cite{smith2007}, the first CMB-only detection by the Atacama Cosmology Telescope \cite{das2011a}, the first detection of lensing in the $B$-mode polarization by the South Pole Telescope \cite{hanson2013}, the Planck lensing results \cite{planck2013-p12,planck2013-p13,planck2014-a17,planck2015-XLI,planck2016-l08}, as well as in the large body of other work steadily improving the fidelity of the lensing measurements \cite{vanengelen2012,ade2014,story2015,bicep2collaboration2016a,omori2017,wu2019}. Delensing can be implemented by using the estimate of $\phi$ to undo the lensing deflection in the data maps or by creating a $B$ mode template which can be subtracted. Again, this requires using Monte Carlo simulations to quantify the resulting bias and uncertainties in the power spectra of the delensed maps. The first CMB-only delensing analysis used the QE to estimate $\phi$ maps from {\it Planck} temperature data, and then inverted the lensing deflection \cite{carron2017a}. 

As successful as the QE has been, however, it will soon become obsolete because it becomes statistically suboptimal as instrumental noise levels dip below $\sim5\,\rm \mu K$-arcmin \cite{seljak2004,smith2012,horowitz2017}. This threshold is being crossed with currently available data sets.

Several methods have been proposed to improve upon aspects of the standard QE procedure. \citet{mirmelstein2019} derive a more optimal spatial weighting of the quadratically estimated $\phi$ before taking its powerspectrum, although do not improve the $\phi$ estimate itself. \citet{horowitz2017} and \citet{hadzhiyska2019} work in the small-scale limit ($\ell\gtrsim5000$) where a lower variance $\phi$ estimator can be analytically derived, but which is not optimal on all scales, in particular not on the intermediate and large scales which are relevant for $r$ estimation. \citet{caldeira2018} train a neural network to extract a $\phi$ map from noisy lensed CMB data, finding near-optimality on relevant scales, but it is not straightforward how one would quantify uncertainties on $\phi$ in such an analysis. Finally, there are a class of near-optimal maximum a posteriori (MAP) estimators of $\phi$ generated by maximizing the Bayesian posterior $\mathcal{P}(\phi\,|\,d,\theta)$ where $d$ is the data and $\theta$ represents cosmological parameters or directly the theoretical bandpowers (we will refer to this as the ``marginal posterior'' and the associated ``marginal MAP'' for reasons which will be clear in a moment). \citet{hirata2003,hirata2003b} were the first to explore such an approach and to develop an approximate maximization technique, while \citet{carron2017} recently made the maxmization procedure exact.

A major challenge associated with any new point estimate of $\phi$ is the quantification and propagation of uncertainty when trying to estimate cosmological parameters from the estimated $\phi$ or from data delensed by the estimate. Although Monte Carlo simulations can help, these will generally depend on the same cosmological parameters one is trying to estimate in the first place.  As an example, consider attempting to use the marginal MAP $\phi$ to infer the theoretical $\phi$ bandpowers (in our notation, the case where $\theta\,{\equiv}\,\{C_\ell^{\phi\phi}\}$). Since $\mathcal{P}(\phi\,|\,d,\theta)$ depends on $C_\ell^{\phi\phi}$, the resulting estimate inherits a Wiener-filter-like multiplicative bias which depends explicitly---but not analytically---on $C_\ell^{\phi\phi}$ itself. This circularity seriously complicates any attempt to debias and/or probe properties of $C_\ell^{\phi\phi}$ in this way.

Despite these challenges, some progress has been made using these new $\phi$ estimates. \citet{adachi2019} were recently the first to apply a non-QE method to actual CMB data, demonstrating that delensing data from the POLARBEAR telecope with the algorithm from \cite{carron2017} yielded a 22\% reduction in lensing $B$-modes, compared to only 14\% when delensing with the QE. The circularity problem is partially ameliorated by a procedure they develop termed ``overlapping $B$-mode deprojection,'' wherein for each bandpower which is delensed, a $\phi$ estimate is constructed only from modes outside of that multipole range. This reduces the size of the bias and its dependence on the theoretical spectra themselves, but at the price of a 5\%--35\% reduction in the delensing efficiency depending on the multipole range considered. Skipping ahead slightly, we remark that the new methodology introduced in this paper would fully remove this delensing efficiency penalty, as well allowing inference of other cosmological parameters governing $C_\ell^{\phi\phi}$ or the delensed bandpowers themselves.

In parallel, there have also been attempts to unify near-optimal estimation of $\phi$ with simultaneous inference of cosmological parameters. The main approach has been to extend the marginal posterior from $\mathcal{P}(\phi\,|\,d,\theta)$ to include the $\theta$ as free parameters rather than fixing them, then marginalize out $\phi$ to arrive at constraints on $\theta$ given by $\mathcal{P}(\theta\,|\,d) = \int \mathrm d\phi\, \mathcal{P}(\phi, \theta\,|\,d)$. \citet{hirata2003,hirata2003b} consider the case of $\theta\,{\equiv}\,\{C_\ell^{\phi\phi}\}$, use the Laplace approximation to perform the integral over $\phi$, then compute a maximum likelihood estimator with Gaussian error bars for the resulting $\mathcal{P}(C_\ell^{\phi\phi}\,|\,d)$. \citet{carron2018} developed a similar method for $\theta\equiv r$ which does not assume Gaussian error bars on $r$, but still uses an underlying Laplace approximation. Both are useful forecasting methods, but the former has never been checked in the presence of required analysis complexities such as pixel masking, and the brute-force integration employed by the latter does not scale computationally to these cases.

In this paper, we develop a complete Bayesian solution which unifies optimal inference of $\phi$ along with delensing and cosmological parameter inference. This is achieved by further extending $\mathcal{P}(\phi,\theta\,|\,d)$ to include the unlensed CMB fields, hereafter $f$, rather that analytically marginalizing over them as was implicit in the marginal posterior (hence the name). The resulting ``joint posterior", $\mathcal{P}(f, \phi,\theta\,|\,d)$, theoretically extracts all of the information in $d$ for $(f,\phi,\theta)$ and completely summarizes the uncertainty on all of these quantities. As we will demonstrate, it also allows us to perform parameter inference by using Monte Carlo sampling to compute the integral in $\mathcal{P}(\theta\,|\,d)\,{=}\,\int \mathrm df \mathrm d\phi\, \mathcal{P}(f, \phi, \theta\,|\,d)$. This avoids use of the Laplace approximation, whose accuracy is difficult to check and may be poor due to the non-linearity of the lensing problem. 

The challenge is that this is a very high-dimensional and non-Gaussian posterior, with around ${\sim}\,10^{7}$ dimensions for the cases considered in this work. Previous attempts at sampling in this space have been blocked by the extreme degeneracies generated by parameter expansion---from $\phi$ to $(f,\phi,\theta)$---resulting in more parameter degrees of freedom than data. These non-linear degeneracies render the exploration of the joint posterior surface extremely difficult. To make progress, one has to find a way to condition the posterior into a more manageable form. We do so here by finding a reparametrization of the posterior from variables $(f,\phi,\theta)$ to new variables $(f^\mix,\phi^\mix,\theta)$ which have a posterior distribution which we are then able to  sample efficiently with the combination of a Gibbs block sampler and Hamiltonian Monte Carlo (HMC) \cite{betancourt2017}. The resulting fast-mixing chain yields samples of $(f^\mix,\phi^\mix,\theta)$, which can be easily converted to samples of $(f,\phi,\theta)$ in post-processing. 

The final piece of the procedure is \textsc{LenseFlow}, which is a numerical algorithm for lensing a map \cite{millea2019}. \textsc{LenseFlow} reformulates lensing into solving an ordinary differential equation (ODE), and makes it possible to compute the gradients and determinants that arise in the reparametrization.

We use our method to compute, for the first time, the exact Bayesian posterior, $\mathcal{P}(r\,|\,d)$, in the presence of realistic analysis complexities, notably pixel masking. Doing so, we can check existing forecasts for $r$ similar to those performed for CMB-S4, South Pole Observatory, or Simons Observatory \cite{abazajian2016,abazajian2019b,thesimonsobservatorycollaboration2018,southpoleobservatorycollaboration2020}. These rely on approximations which, among other things, ignore masking \cite{smith2012}. Pixel masking couples modes together and leaks $E$ into $B$ mode polarization exactly like lensing, so it is particularly worrisome that it might impact delensing in some unexpected way. We present these results in Sec.~\ref{sec:forecasting}.

The power of the methodology developed here is not just that it works for forecasting, but that it is ready to be applied to analysis of real data including the many extra complexities which arise. We demonstrate this with simulations which include the effect of beams, non-white noise, and Fourier and pixel masking. We work in the flat-sky approximation and consider patches of sky as large as 512$\times$512 pixels, or ${\sim}\,650{\rm deg}^2$. We focus on the specific problem of delensing and inference on the tensor-to-scalar, $r$, and the amplitude of the lensing potential, $\Aphi$. The procedure is conceptually straightforward to generalize to sampling other cosmological parameters (or to sample bandpowers directly), to the curved sky and larger sky area, and to include foreground components. An accompanying software package, \textsc{CMBLensing.jl} (see Sec.~\ref{sec:code}), is available online \href{https://github.com/marius311/CMBLensing.jl}{\faGithub} \footnote{\url{https://github.com/marius311/CMBLensing.jl}}.

\section{The data model and prior assumptions}
\label{sec:datamodel}

\begin{figure}
\includegraphics[width=\columnwidth]{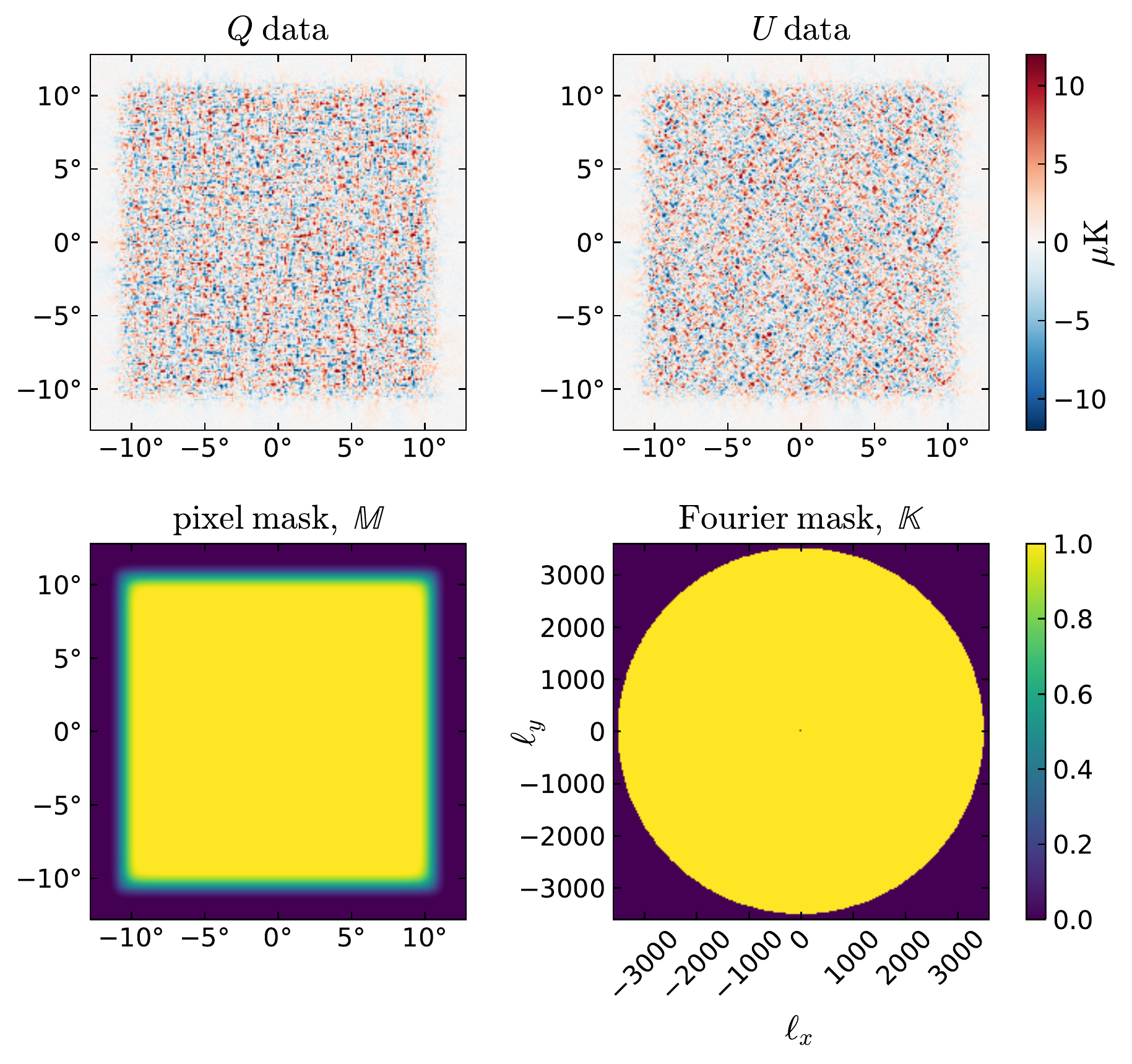}
\caption{Typical simulated data and mask choices used in this work. Specifically, thsese correspond to configuration \configC (see Table~\ref{table:configurations}) with a true value of $r\,{=}\,0.01$. Reconstructed maps from this exact data are shown in Fig.~\ref{fig:configCmaps}. We note that an apodized pixel mask and an isotropic Fourier mask are not algorithm requirements, rather arbitrary choices we made for this example.}
\label{fig:configCdata}
\end{figure}

The Bayesian posterior for the lensing problem is completely specified by a data model and a set of priors. The data model we use, which is flexible enough to handle real experiments, is
\begin{align}
\label{eq:datamodel}
d = \mathbb A\,  \Len[\phi] f + n,
\end{align}
where $d$ is the data, $f$ are the unlensed CMB fields, and $n$ is the instrumental noise. In this paper, we will work with only polarization data since it gives the tightest constraints for low noise levels, although the equations (and our code) are generic to temperature, polarization, or temperature and polarization data. The term $\Len[\phi]$ encodes the lensing displacement operation, which can be written for $f$ in the $T/Q/U$ basis as a function of 2D position on the sky $\boldsymbol x$,
\begin{align}
\big(\Len[\phi] f\big)(\boldsymbol x) = f(\boldsymbol x + \nabla \phi(\boldsymbol x)).
\end{align}
Note that $\Len[\phi]$ is a linear operator acting on $f$, but has non-linear dependence on $\phi$. We use \textsc{LenseFlow} \citep{millea2019} to  implement $\Len[\phi]$ numerically. This is a necessity for our application because no other known numerical approximation allows practical calculation of determinants or of gradients of inverse lensing\footnote{Gradients of forward lensing are simple for many algorithms, but easy gradients of inverse lensing appear unique to \textsc{LenseFlow}.} with respect to $\phi$, both of which are needed by the reparametrization which we will describe in Sec.~\ref{sec:reparam}. Another advantage of \textsc{LenseFlow} is that it  allows us easily to apply the full lensing displacement, rather than, e.g., having to rely on a truncated Taylor approximation. We will assume the lensing Born approximation, although it would be straightforward to include a curl potential to the deflection field to model these effects. We omit a detailed treatment of post-Born effects because their importance in the context of this paper will be marginal for current and upcoming surveys \cite{beck2018,bohm2018,lewis2017,pratten2016}. 

Instrumental transfer functions and user-chosen masking are encoded in the operator
\begin{align}
    \op A \equiv \op K\, \op M\, \op B,
\end{align}
which is the product of a Fourier mask $\op K$, a pixel mask $\op M$, and a beam $\op B$. In general, $\op M$ can be chosen to mask the boundaries of the field and any foreground contaminated areas (such the as the areas around detected discrete sources), and $\op K$ can be chosen to restrict the analysis to only certain modes in the 2D Fourier plane. Typical choices we use for these operators as well as data simulated according to Eqn.~\eqref{eq:datamodel} are shown in Fig.~\ref{fig:configCdata}.

We take Gaussian priors on the fields $f$, $\phi$, and $n$
\begin{align}
\label{eqn:fprior}
f &\sim \mathcal N\big(0, \Cf[r]\big) \\ 
\label{eqn:phiprior}
\phi &\sim \mathcal N\big(0, \Cphi[\Aphi]\big) \\
\label{eqn:nprior}
n &\sim \mathcal N\big(0, \Cn\big),
\end{align}
where $\Cn$, $\Cf[r]$ and $\Cphi[\Aphi]$ denote the covariance operators for the experimental noise, unlensed CMB polarization, and lensing potential. The latter two have explicit dependence on the scalar-to-tenser ratio, $r$, and a lensing spectral amplitude parameter, $\Aphi$, given by
\begin{align}
\label{eq:Cfr}
\Cf[r] &= \op C_{sf}^* + (r/r^*)\,\op C_{tf}^* \\
\Cphi[\Aphi] &= \Aphi \Cphi^*,
\end{align}
where $\op C_{sf}^*$, $\op C_{tf}^*$ and $\op C_{\phi}^*$ are covariance operators for CMB scalar perturbations, tensor perturbations, and the lensing potential field, computed at fiducial $\Lambda$CDM parameters\footnote{Spectra are computed using CAMB (see \texttt{http://camb.info}) with fiducial settings $k^* = 0.002$, $r^* = 0.1$, $A^*_{\phi} = 1$, $\omega_{\rm b} = 0.0224567$, $\omega_{\rm c}=0.118489$, $\tau = 0.055$, $\theta {\rm s} = 0.0104098$, $\text{log}A = 3.043$, $n_{\rm s} = 0.968602$ and $n_{\rm t} = -r^*/8$. Note that for simplicity, in Eqn.~\eqref{eq:Cfr} we are implicitly fixing $n_{\rm t}$ rather than enforcing the single field consistency relation.}

Finally, we chose the following weakly-informative priors for $r$ and $\Aphi$ \cite{gelman2006}, 
\begin{align}
\pi(r)\propto r^{-\nicefrac{1}{2}}, \qquad \pi(\Aphi) \propto \Aphi^{-\nicefrac{1}{2}}.
\end{align}
We find little impact on our sampling algorithm for different priors, and different choices can be importance sampled into the final chains if desired.

With this final ingredient specified, the posterior distribution is now fully defined and given by Eqn.~\eqref{eq:jointposterior},
\begin{widetext}
\begin{align}
\label{eq:jointposterior}
\mathcal{P}(f,\phi,r,\Aphi\,|\,d) &\propto 
\frac{\exp\left\{ -\cfrac{(d-\op A \,\Len[\phi] f)^2}{2 \,\Cn} \right\}}{\det \Cn^{\nicefrac{1}{2}}} \;
\frac{\exp\left\{ -\cfrac{f^2}{2\,\Cf[r]} \right\}}{\det  \Cf[r]^{\nicefrac{1}{2}}} \;
\frac{\exp\left\{ -\cfrac{\phi^2}{2\,\Cphi[\Aphi]} \right\}}{\det  \Cphi[\Aphi]^{\nicefrac{1}{2}}} \;
\frac{1}{(r \Aphi)^{\nicefrac{1}{2}}} \\[20pt]
\label{eq:marginalposterior}
\mathcal{P}(\phi,r,\Aphi\,|\,d) &\propto 
\frac{\exp\left\{ -\cfrac{d^2}{2 \mathbbgreek{\Sigma}_d} \right\}}{\det \mathbbgreek{\Sigma}_d^{\nicefrac{1}{2}}} \;
\frac{\exp\left\{ -\cfrac{\phi^2}{2\,\Cphi[\Aphi]} \right\}}{\det  \Cphi[\Aphi]^{\nicefrac{1}{2}}} \;
\frac{1}{(r \Aphi)^{\nicefrac{1}{2}}}
\end{align}
\begin{multline}
\nonumber \hfill \textrm{where} \;\; \mathbbgreek{\Sigma}_d \equiv \Cn + \op A \, \Len[\phi] \Cf[r] \Len[\phi]^\dagger \op A^\dagger \;\; \textrm{and we use the shorthand} \; x^2/\op N \equiv x^\dagger \op N^{-1} x.
\end{multline}
\end{widetext}

Note that the conditional distribution  $\mathcal{P}(f\,|\,\phi,r,\Aphi,d)$ is Gaussian in $f$ (although all the other conditionals are non-Gaussian). Because of this, it is possible analytically to  marginalize over $f$,
\begin{align}
    \label{eq:marginalizejoint}
    \mathcal{P}(\phi,r,\Aphi\,|\,d) = \int \mathrm df \, \mathcal{P}(f,\phi,r,\Aphi\,|\,d),
\end{align}
to arrive at Eqn.~\eqref{eq:marginalposterior}, which, as previously mentioned, we refer to as the marginal posterior.

As discussed in \cite{millea2019}, the joint and marginal posteriors have a crucial distinction. All of the operators whose determinants and inverses appear in the joint posterior are sparse in simple bases, e.g., $\Cn$ is sparse in pixel space for typical instrumental noise, and $\Cf$ and $\Cphi$ are diagonal (and even isotropic) in Fourier space. The action of these operators can thus be evaluated in $\mathcal{O}(N_{\rm pix}\log N_{\rm pix})$ where $N_{\rm pix}$ is the number of pixels in the maps, as the limiting step is an FFT to transform into the sparse bases. However, $\mathbbgreek{\Sigma}_d$, which is introduced in the marginal posterior, is not sparse in any simple basis. 

This would limit us in several ways if we were attempting to use the marginal posterior for sampling. Evaluating gradients of $\det \mathbbgreek{\Sigma}_d$ with respect to $\phi$, which would be needed by the HMC sampler (see Sec.~\ref{section: gibbs chain}), would now have to be done through a costly Monte Carlo procedure \cite{carron2017}. This procedure involves solving $N_{\rm MC}\sim500$ conjugate gradient problems, each of which require $N_{\rm CG}\sim100$ conjugate gradient iterations, with each iteration having similar computation cost as a single joint posterior gradient. Hence, marginal posterior gradients are slower than joint posterior gradients by a factor of order $N_{\rm MC} N_{\rm CG}$, which can in practice be a very large number. Even if this were overcome (if the total CPU cost was not prohibitive, the $N_{\rm MC}$ steps can at least be done in parallel), there is another even more serious limitation. No algorithm we are aware of can robustly evaluate $\det \mathbbgreek{\Sigma}_d$ itself faster than $\mathcal{O}(N_{\rm pix}^3)$, which in practice makes this impossible for maps larger than about $32\times32$ pixels. Without an ability to evaluate this determinant and hence the value of our log-posterior, the accept/reject step of the HMC is impossible. For these reasons, we find that sampling the joint posterior is the more promising path, and the one which we take.

In summary, we choose to work with the higher dimensional joint posterior because it has a structure that allows the use of powerful Markov Chain Monte Carlo (MCMC) sampling techniques such as HMC. This approach is typical for the implementation of high-dimensional Bayesian Hierarchical Models, starting with their first application in cosmology \cite{WandeltLarsonLakshminarayanan2004} which applied Gibbs sampling to CMB power spectrum inference, or the more recent application to non-linear large scale structure reconstruction and inference in the Bayesian Origin Reconstruction from Galaxies (BORG) sampler \cite{JascheWandelt2013,LavauxJascheLeclercq2019,DoogeshEtAl2019}.

\section{Reparametrizing the posterior}
\label{sec:reparam}

\begin{figure*}
    \includegraphics[width=\textwidth]{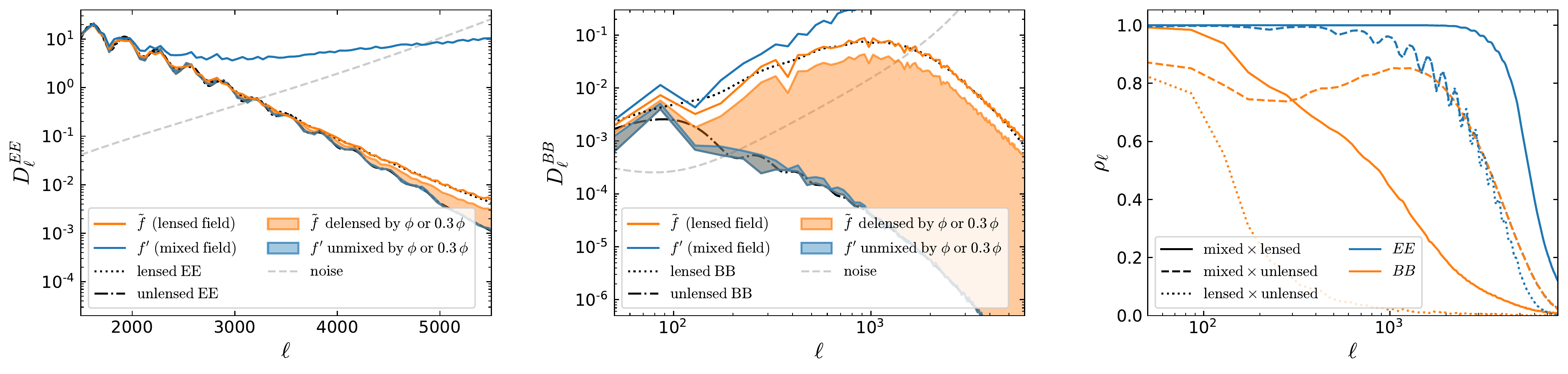}
    \caption{Three figures which are helpful in understanding the benefit of the reparametrization (described in Sec.~\ref{sec:reparam}) which makes sampling possible. Data configuration \configA (see Table~\ref{table:configurations}) is assumed for these figures. {\it (Left two panels)} The reparamaterization includes switching from sampling the unlensed CMB fields, $f$, to sampling the ``mixed'' fields, $f^\mix$. The left two panels show that the power variation in a typical $f^\mix$ unmixed by various $\phi$ is very small, an indication that large moves are allowed in Gibbs samples from the conditional $\mathcal{P}(\phi^\mix\,|\,f^\mix,d)$. For comparison, the much larger variation in a typical $\tilde f$ when delensed by various $\phi$ is shown, indicating that the lensed parametrization performs very poorly for polarization as is the case here. {\it (Right panel)} Empirically, one finds that mixed $E$ is mostly lensed $E$ at all scales, while mixed $B$ is lensed $B$ at large scales but unlensed $B$ at small scales. We demonstrate this here by cross-correlating the mixed with the lensed or unlensed fields. This qualitatively conforms to the expectations of what should give a parametrization which is minimally degenerate (see discussion in Sec.~\ref{sec:reparam}).}
    \label{fig:mixing_spectra}
\end{figure*}

The joint posterior, parametrized as in Eqn.~\eqref{eq:jointposterior} by the unlensed CMB fields and the lensing potential, is nearly unusable in practice due to the presence of large non-Gaussianities and degeneracies. These issues already appeared in a milder form in the temperature-only CMB lensing posterior \cite{anderes2015} where the solution was to change from the unlensed to the lensed (or from a ``sufficient'' to an ``ancillary'') parametrization. The situation is more challenging for the polarized CMB lensing/de-lensing problem we treat in this paper and the solution in \cite{anderes2015} is not powerful enough. In the context of polarization, \citet{millea2019} encountered the same underlying problem when maximizing $\mathcal{P}(f,\phi\,|\,d,\theta)$, but the ``cooling scheme'' solution presented there does not have an obvious analog for sampling. Additionally, here we have the  complexity of degeneracies in the full $(f,\phi,\theta)$ space which must be dealt with.

A key aspect of this work is that we develop a physically motivated reparametrization which works for polarization and yields a posterior which is significantly less degenerate and more Gaussian than the original $\mathcal{P}(f,\phi,r,\Aphi\,|\,d)$. The reparametrization is fully invertible, and consequently does not introduce any approximations to the inference, it only serves to increase the efficiency of sampling or maximization. We first describe the reparametrization (which we also refer to as ``mixing,'' since it mixes the various parameters) and afterwards explain the motivation behind it. 

We perform a change of variables from $(f,\phi)$ to new variables which we call $(f^\mix,\phi^\mix)$, which are defined by
\begin{align}
\label{eq: def phimix} \phi^\mix &\equiv \opG[\Aphi]\, \phi \\
\label{eq: def fmix} f^\mix      &\equiv \Len[\phi] \, \D[r]\, f .
\end{align}
The operator $\D[r]$ is defined to be diagonal in the $E,B$ Fourier domain, and $\opG[\Aphi]$ is diagonal in the Fourier domain, with
\begin{align}
\label{eqn: Dr}
\D[r] &\equiv \left[ \frac{\Cflen[r] + 2 \, \op{N}_f}{\Cflen[r]} \right]^{\nicefrac{1}{2}}  \left[ \frac{\Cflen[r]}{\Cf[r]} \right]^{\nicefrac{1}{2}}\\
\label{eqn: GA}
\opG[\Aphi] &\equiv \left[ \frac{\Cphi[\Aphi] + 2 \, \op{N}_\phi}{\Cphi[\Aphi]} \right]^{\nicefrac{1}{2}}
\end{align}
where ${\widetilde {\op C}}_{f}(r) = \Cf[r] + \op N_{len}$ and $\op N_{len}$ denotes the effective power contribution of lensing to the CMB polarization, which we set equal to $5\,\mu$K-arcmin white noise (this seems to work better than using the actual lensing contribution which rolls off at higher $\ell$). The operators $\op{N}_f$ and $\op{N}_\phi$ are taken to be diagonal in the Fourier domain and are intended to represent the effective noise for $f$ and $\phi$ in the data. Even if the noise covariance is not actually diagonal in Fourier space, the requirement is only that it needs to be approximated sufficiently well by a Fourier diagonal approximation.  Since we explicitly take the instrumental noise in our simulations to be diagonal in Fourier space, we use directly $\op{N}_f\,{=}\,\Cn$. For $\op{N}_\phi$, we compute an iterated ``$N_0$'' noise as described in \citet{smith2012}.

The reparametrized posterior needs the determinant of the Jacobian of the transformation, where the Jacobian is
\begin{align}
    \label{eq:mixjacobian}
    \frac{\partial(f^\mix, \phi^\mix)}{\partial(f, \phi)} = 
    \begin{bmatrix}
    \Len[\phi] \op{D}(r) & \tfrac{\partial}{\partial \phi} \Len[\phi] \D[r]f \\
    0  & \opG[\Aphi]
    \end{bmatrix}.
\end{align}
We have intentionally chosen the reparametrization such that the Jacobian is upper triangular, since in this case the determinant does not involve the complicated off-diagonal term. Additionally, because we model $\Len[\phi]$ with \textsc{LenseFlow}, we have $\det \Len[\phi]\,{=}\,1$, independent of $\phi$ \cite{millea2019}. Also, since $\op{D}(r)$ and $\opG[\Aphi]$ are diagonal in Fourier space, their determinants are easy to compute. This gives a final tractable reparametrized posterior which is given by
\begin{multline}
    \log\mathcal{P}(f^\mix, \phi^\mix, r,\Aphi \,|\, d) = \\ 
    \log\mathcal{P}(f(f^\mix,\phi^\mix,r,\Aphi), \phi(\phi^\mix,\Aphi), r,\Aphi \,|\, d) \\ - \log\det \opG[\Aphi] - \log\det \op{D}(r).
\end{multline}
Note that, by design, the new determinant terms are indepent of $f$ and $\phi$. This means that the best-fit $(f,\phi)$ at fixed $(r,\Aphi)$ can be computed by running the maximization in the mixed parametrization, then taking the best-fit $(f^\mix,\phi^\mix)$ and unmixing them. The maximization is much easier in the mixed parametrization, and can be done with coordinate descent similarly as in \cite{millea2019}, but with the cooling scheme no longer needed.

Gradients of the mixed posterior can be computed from gradients of the original posterior with an application of the chain rule using the Jacobian in Eqn.~\eqref{eq:mixjacobian}. Both evaluating the value of and gradients of the reparametrized posterior are only about twice the computational cost of the original posterior, stemming from the presence of a second lensing operation $\Len[\phi]$ which appears in Eqn.~\eqref{eq: def fmix}.

The choice of $\D[r]$ and $\opG[\Aphi]$ can be motivated as follows. Consider the toy statistical problem of obtaining constraints on a scalar parameter, $\theta$, given data, $d$, where
\[
d = s + n,\quad s \sim\mathcal{N}(0, \op{S}(\theta)), \quad n\sim\mathcal{N}(0, \op{N}).
\]
The field $n$ represents noise and $s$ the signal field, with covariance operator $\op{S}(\theta)$ depending on the unknown parameter. The goal in this toy example is to find an invertible reparametrization $s \rightarrow s^\mix$ of the form $s^\mix = \op{G}(\theta)\,s$ which minimizes the dependence between $\theta$ and $s^\mix$ given $d$. In the ideal case, such a choice of $\op{G}(\theta)$ would have the property that $\mathcal{P}(\theta\,|\,s^\mix,d)\approx \mathcal{P}(\theta\,|\,d)$, meaning $s^\mix$ provides minimal additional information for $\theta$ beyond what is already contained in $d$. Such a property would imply that a single iteration of a Gibbs sampling algorithm for $(\theta,s^\mix)$ would return an approximate marginal draw from $\mathcal{P}(\theta\,|\,d)$. 

Another way of phrasing this goal is to choose $\op{G}(\theta)$ such that the information content in $(d,s^\mix)$ for $\theta$ is minimized. Note that the marginal information in $d$ for $\theta$ is fixed regardless of $\op{G}(\theta)$ since we are simply considering reparametrization of the same data model. So by minimizing the joint information in $(d,s^\mix)$ for $\theta$ we are implicitly minimizing the additional information in $s^\mix$ for $\theta$ beyond that given by $d$.

A way to describe this mathematically is to start by letting $\mathcal{F}(\theta; \op{G})$ denote the Fisher information for $\theta$ given $(d,s^\mix)$ ; in particular,
\begin{equation}
\mathcal{F}(\theta; \op{G}) = \left\langle -\frac{\partial^2}{\partial \theta^2} \log\mathcal{P}(d, s^\mix\,|\,\theta)\right\rangle_{d,s^\mix \,\sim\, \mathcal{P}(d,s^\mix\,|\,\theta)}
\end{equation}
where the dependence on $\op{G}(\theta)$ is implicit in the reparametrized density $\mathcal{P}(d, s^\mix\,|\,\theta)$. Then, $\mathcal{F}(\theta; \op{G})$ can be explicitly computed using standard matrix algebra/calculus to arrive at
\begin{multline}
\mathcal{F}(\theta; \op{G}) = {\rm tr} \Big[\op{S}\,(\op{G}^{-1}\dot{\op{G}})^\dagger\big(\op{N}^{-1}+\op{S}^{-1}\big)\big(\op{G}^{-1}\dot{\op{G}}\big) \\ + \big(\op{G}^{-1}\dot{\op{G}}\big)^2
+ 2 \, \dot{\op{S}}\,\op{S}^{-1} \big(\op{G}^{-1}\dot{\op{G}}\big) + \tfrac{1}{2}\big(\op{S}^{-1} \dot{\op{S}}\big)^2\Big]
\end{multline}
where the overdots refer to derivative with respect to the scalar $\theta$. 
Finally, we seek to minimize the Fisher information, and rather than doing so at any fixed $\theta$, we integrate over the prior for $\theta$, which can be any arbitrary probability function, $\mathcal{P(\theta)}$. Thus, we seek $\op{G}(\theta)$ which is a minimizer in
\begin{align}
\label{eq: min G}
 \text{arg}\min_{\op{G}} \int \mathrm{d}\theta \, \mathcal{F}(\theta;\op{G}) \, \mathcal{P}(\theta).
\end{align}
In the case that $\op{G}(\theta)$, $\op{S}(\theta)$, and $\op{N}$ are diagonal with positive entries we can define $\op{H}(\theta)\,{=}\,\log\op{G}(\theta)$ such that $\dot{\op{H}}\,{=}\,\op{G}^{-1}\dot{\op{G}}$. Now  $\mathcal{F}(\theta;\op{G})\mathcal{P(\theta)} = \mathcal{L}(\theta,\dot{\op{H}}(\theta))$ for a Lagrangian $\mathcal{L}$ which yields $N$ Euler-Lagrange equations (corresponding to $N$ diagonal entires of $\op{G}$) that characterize the stationary points of \eqref{eq: min G} given by
\begin{align}
    \label{eq:EL}
    \big(\op{N}^{-1}\op{S} + 2\op{I}\big) \frac{d}{d\theta} \log\op{G}(\theta)  + \frac{d}{d\theta}\log\op{S}(\theta) = 0,
\end{align}
where we have applied a boundary condition such that Eqn.~\eqref{eq:EL} is invariant to the choice of prior. An explicit solution is then given by
\begin{align}
    \label{eq:Gmixtoy}
    \op{G}(\theta) \propto \left[\frac{\op{S}(\theta) + 2 \op{N}}{\op{S}(\theta)}\right]^{\nicefrac{1}{2}},
\end{align}
which is additionally invariant to multiplication by any diagonal matrix which does not depend on $\theta$.

Notice that for coordinates which are noise dominated, we have $\op{G}(\theta)\,{\propto}\,\op{S}(\theta)^{-1/2}$. It is simple to see analytically that in this limit, the posterior becomes exactly separable given this choice of $\op{G}(\theta)$. This also conforms to the expectation from \citet{jewell2009}, who derived the same result in this limit. For coordinates which are signal dominated, we instead have $\op{G}(\theta)\propto\I$, which also matches intuition since in this limit the data determine $s$ perfectly. Eqn.~\eqref{eq:Gmixtoy} is thus in some sense an optimal way to connect these two limits. One can additionally regard this result as an extension of \citet{racine2016}, who derived a modified Gibbs proposal step which also works in both limits. The advantage of our result is that it is generic and not limited to Gibbs sampling, and that it does not affect the detailed balance of the Monte Carlo chains or force us to include any extra efficiency-reducing accept/reject steps. 

This toy example directly explains the mixing matrix $\opG[\Aphi]$ given in Eqn.~\eqref{eqn: GA}; it is just Eqn.~\eqref{eq:Gmixtoy} with $\op{S}=\Cphi$ and $\op{N}$ chosen as previously described. Although the $(\phi,\Aphi)$ block of the lensing posterior is not exactly the same as $(s,\theta)$ in the toy example, in particular the $\mathcal{P}(s\,|\,\theta,d)$ conditional is Gaussian in the toy example whereas as the corresponding $\phi$ conditional is not, the problems are sufficiently similar that this works very well. 

The mixing matrix $\D[r]$ given in Eqn.~\eqref{eq: def fmix} is also similar; the first term indeed is just Eqn.~\eqref{eq:Gmixtoy} with $\op{S}\,{=}\,\Cflen$ and $\op{N}\,{=}\,\Cn$. There is, however, an additional prefactor of $(\Cflen / \Cf[])^{\nicefrac{1}{2}}$ present, and also a lensing operation in Eqn.~\eqref{eq: def fmix} before arriving at the final mixed field, $f^\mix$. The motivation for this can be understood by applying a similar argument as in our toy example. Suppose we wish to make $f^\mix$ and $\phi^\mix$ more independent, and increase the width of the conditional $\mathcal{P}(\phi^\mix\,|\,f^\mix,\theta,d)$ so that it is on the order of the marginal distribution $\mathcal{P}(\phi^\mix\,|\,\theta,d)$. We have argued that a way to do this is to decrease the information content for $\phi^\mix$ in $(d,f^\mix)$. One way to do so is to prevent the power in $f^\mix$ from being informative about $\phi^\mix$. The prefactor in Eqn.~\eqref{eq: def fmix} serves exactly this purpose, since it boosts power in $f^\mix$ to look like lensed power, independent of whether $\phi^\mix$ causes a large or small lensing. This shifts the information in $\mathcal{P}(\phi^\mix\,|\,f^\mix,\theta,d)$ from the lensed $B$-mode power to the less informative lensed $B$-mode phase coupling. Typical power spectra of $f^\mix$ are shown in the left two panels of Fig.~\ref{fig:mixing_spectra}, as well as an illustration of how the power in $f^\mix$ is less informative than, e.g. the power in the lensed field, $\tilde f$, explaining why $\tilde f$ does not work well as a parameter when considering polarization data.

Another way to understand why the mixed parametrization works well is to ask what choice of variables render the posterior distribution in Eqn.~\eqref{eq:jointposterior} explicitly independent between $f^\mix$ and $\phi^\mix$. In the limit of low signal-to-noise where only the prior terms matter, an independent choice of variables is trivially $(f,\phi)$ since the prior is explicitly separable between them. As we move away from this limit, the data likelihood begins to couple $f$ and $\phi$, so it is clear the right choice will be some combination of them. The mixing indeed has exactly this behavior in these limits, as demonstrated in the right panel of Fig.~\ref{fig:mixing_spectra}. Here, we plot the cross-correlation coefficient at different scales between the mixed maps and either lensed or unlensed ones. For scales where signal-to-noise is low (like medium and small scales in $B$), the mixed field $f^\mix$ looks like the unlensed field. In the high signal-to-noise limit (such as in $E$, or at very large scales in $B$), $f$ becomes a mixture of the two, in particular, we find it tracks the lensed field.

The end result of all of this is a dramatically better conditioned posterior, resulting in large Gibbs moves and much faster chain mixing for the sampling procedure we describe in the next section. The improvement is not limited to our particular Gibbs sampler, however, and we expect that any sampling algorithm applied to this problem would benefit drastically from this reparametrization. Finally, we note that although the reparametrization in our toy example is optimal in the sense that it can be rigorously and analytically derived, the full mixing in Eqns.~\eqref{eq: def phimix} and \eqref{eq: def fmix} is almost certainly not optimal. Instead, it is based on physical intuition and simple analogy to the covariance estimation problem, and it would be worthwhile to investigate even better choices.

\begin{algorithm}[H]
\fontsize{9}{14}\selectfont
\vspace{0.1cm}
\caption{$\mathcal{P}\big(f^\mix,\, \phi^\mix,\, r,\, \Aphi \,|\, d\big)$ sampler}\label{alg:gibbs-hmc}
\begin{algorithmic}[1]
\State Initialize $A_{\phi,0}$ and $r_0$ anywhere within the prior range.
\State Initialize fields $f^\mix_0$ and $\phi^\mix_0$ with quasi-samples.  \label{alg: initialization of f^o and phi^o}
\For{$i=1\,...\,n$  }
\State $f^\mix_i \sim \mathcal{P}( f^\mix \,|\, \phi^\mix_{i-1},\, A_{\phi,i-1},\, r_{i-1}, \, d)$ \Comment{CG}\label{alg: sampling f} 
\State\label{alg: sampling phi} 
$\phi^\mix_i \sim \mathcal{P}(\phi^\mix \,|\, A_{\phi,i-1},\, r_{i-1},\, f^\mix_i,\, d)$ \Comment{HMC}
\State\label{alg: sampling A}  
$A_{\phi,i} \sim \mathcal{P}( \Aphi \,|\, r_{i-1}, \, f^\mix_i,\, \phi^\mix_i,  d)$ \Comment{Slice} 
\State\label{alg: sampling r}  
$r_i \sim \mathcal{P}( r \,|\,  f^\mix_i,\, \phi^\mix_i,\, A_{\phi,i},\,  d)$ \Comment{Slice}
\EndFor \myvspace
\end{algorithmic}
\end{algorithm}

\section{The Gibbs chain}
\label{section: gibbs chain}

{
\renewcommand{\arraystretch}{1.2}
\setlength{\tabcolsep}{8pt}
\begin{table*}
 \centering
\begin{tabular}{lll}
    \hline
    \\[-0.3cm]
    {$\mathbb N_f$} 
    & Effective noise used in $\D[r]$ & see Sec.~\ref{sec:reparam}
    \\
    {$\mathbb N_\phi$} 
    & Effective noise used in $\opG[\Aphi]$ & see Sec.~\ref{sec:reparam}
    \\
    {$\mathbbgreek{\tilde\Lambda}_f(r)$, $n_\text{\tiny  cg}$} 
    & Parameters for conjugate gradient sample of $f^\mix$ & see Sec.~\ref{sec:fpass}
    \\
    {$\epsilon_\text{\tiny h}$, $n_\text{\tiny h}$, $\mathbbgreek{\Lambda}_{\phi^\mix}(\Aphi)$}
    & HMC leap-frog and momentum parameters for $\phi^\mix$ & see Sec.~\ref{sec:phipass}
    \\
    {$K$} 
    & Number of over-relaxation samples for $r$ and $\Aphi$ & see Sec.~\ref{sec:thetapass}
    \\[0.1cm]\hline
\end{tabular}
\caption{List of tuning parameters used in Gibbs Algorithm \ref{alg:gibbs-hmc}.}
\label{table:tunning parameters}
\end{table*}
}

Next, we outline the details of our Gibbs chain for sampling $\mathcal{P}(f^\mix,\, \phi^\mix,\, r,\, \Aphi \,|\, d)$. The procedure itself is summarized in Algorithm \ref{alg:gibbs-hmc}, and is a standard block Gibbs sampler with each of $f^\mix$, $\phi^\mix$, $r$, and $\Aphi$ sampled on separate passes. A list of all of the tuning parameters which will be needed are also summarized in Table~\ref{table:tunning parameters}.

There is a fair amount of freedom in setting up the sampler; our motivation comes from two considerations. First, the conditional distribution of $f^\mix$ is Gaussian, hence it is advantageous to split this piece off into its own Gibbs pass and use a sampling technique specifically tailored for this situation. Second, the $r$ and $\Aphi$ slices are qualitatively quite different from the other parameters since they are ``global'' parameters that are correlated at a small level with everything else, making it more difficult to simply include them in a joint HMC pass. We therefore split these off as well, and since they are one-dimensional, it is easy to use slice sampling. This also has the advantage of letting us build up a Blackwell-Rao posterior for these parameters. 

We now describe the different passes in more detail.

\subsection{Initializing $f^\mix_0$ and $\phi^\mix_0$ with quasi-samples}

The choice of initialization can shorten the ``burn-in time,'' that is, the number of samples required for the Markov chain to equilibrate. Although initialization is less critical for our case since our reparametrization results in good mixing properties of the chains, the method described here is so simple it is worth utilizing. First, we note that while we do have easy access to the best-fit of the distribution, which would seem like reasonable starting point, in very high-dimensional spaces, the best-fit is extremely far from the bulk of the posterior mass. Instead, we use the following cheap way to generate a point which more closely resembles a true sample and should reside closer to the bulk of the posterior.

First, we randomly sample $A_{\phi,0}$ and $ r_0$ from their priors to generate their starting values in the chain. We then initialize $f^\mix_0$ and $\phi^\mix_0$ to zero and iterate the following two steps
\begin{align}
f^\mix_0 &\sim \mathcal{P}( f^\mix \,|\, \phi^\mix_0,\, A_{\phi,0},\, r_0, \, d) \label{eq:quasisample_f} \\ 
\phi^\mix_0 &= \phi^\mix_0 + \alpha \, \mathbbgreek{\Lambda}^{-1}_{\phi^\mix} \nabla^{}_{\phi^\mix} \log\mathcal{P}(\phi^\mix\,|\,f^\mix_0, A_{\phi,0},\, r_0,d)\big\rvert_{\phi^\mix_0}
\label{eq:quasisample_phi}
\end{align}
The first step (Eqn.~\ref{eq:quasisample_f}) is a draw from the conditional distribution of $f^\mix$, which, as we will describe below, can be done with one run of a conjugate gradient solver. The second step (Eqn.~\ref{eq:quasisample_phi}) is a quasi Newton-Raphson iteration where $\alpha$ is a step-size which we compute via line-search to maximize the resulting $\log\mathcal{P}$ at each iteration, and $\mathbbgreek{\Lambda}_{\phi^\mix}$ is an approximate negative Hessian of $\log\mathcal{P}$ with respect to $\phi^\mix$, which we take as
\begin{align}
\label{eq:hessphi}
\mathbbgreek{\Lambda}_{\phi^\mix}(\Aphi) = \opG[\Aphi]^{-2} \Big[\,\mathbb N_\phi^{-1} +  \Cphi[\Aphi]^{-1}  \Big]
\end{align}
where ${\mathbb N}_{\phi}$ is the same approximate noise covariance appearing in Eqn.~\eqref{eqn: GA}. 

Note that if we replaced Eqn.~\eqref{eq:quasisample_phi} with a conditional sample of $\phi^\mix$, we would recover exactly our sampling algorithm given in Algorithm~\ref{alg:gibbs-hmc} with fixed $A_{\phi,0}$ and $r_0$. Hence we call the point generated by this procedure a ``quasi-sample,'' since it involves sampling in the $f^\mix$ direction but maximization in the $\phi^\mix$ direction. In practice, an important aspect of quasi-samples is that they do not contain the mean-field feature which would otherwise exist in the joint best-fit, $\hat\phi_J$ \cite{millea2019}, and which would slow the initial convergence of our chains. We find 20 iterations of Eqns.~(\ref{eq:quasisample_f}-\ref{eq:quasisample_phi}) are sufficient.

\subsection{The $f^\mix$ Gibbs pass}
\label{sec:fpass}

The first step of each full chain iteration is to draw a conditional sample of $f^\mix$. We can do so by solving one conjugate gradient problem \cite{WandeltLarsonLakshminarayanan2004}. This is because the conditional $f$ posterior is Gaussian,
\begin{multline}
\label{eq: unlensed conditional}
\mathcal{P}(f \, | \, \phi,\, \Aphi,\, r, \, d) = \\ \mathcal N\big( \mathbbgreek{\Lambda}_f(r, \phi)^{-1} \Len[\phi]^\dagger \op{A}^\dagger \Cn^{-1} \, d\,,\,\, \mathbbgreek{\Lambda}_f(r, \phi)^{-1}\big ),
\end{multline}
where the inverse covariance $\mathbbgreek{\Lambda}_f(r, \phi)$ is given by
\begin{align}
\mathbbgreek{\Lambda}_f(r, \phi) &= {\Len[\phi]}^\dagger \op{A}^\dagger \Cn^{-1} \op{A} \, \Len[\phi] + {\Cf[r]}^{-1}.
\end{align}
A sample, $f_i$, is then drawn by computing
\begin{multline}
\label{eq:gibbs_f_sample}
f_i = \mathbbgreek{\Lambda}_f(r, \phi)^{-1} \times \Big[ \Len[\phi]^\dagger \op{A}^\dagger \Cn^{-1} \, d \\ + \Len[\phi]^\dagger \op{A}^\dagger \Cn^{-\nicefrac{1}{2}} \xi_1 + \Cf[r]^{-\nicefrac{1}{2}} \xi_2 \Big]
\end{multline}
where $\xi_1$ and $\xi_2$ are independent unit normal random fields, resampled at each iteration, and the inversion of $\mathbbgreek{\Lambda}_f(r, \phi)$ is done via conjugate gradient. Finally, because the mixing is a linear function of $f$, a sample of the mixed field, $f^\mix_i$, is simply given by $f^\mix_i = \Len[\phi_i]\,\D[r_i]\,f_i$. Note that conjugate gradient is, by design, tailored to exploit the positive-definiteness of $\mathbbgreek{\Lambda}_f$, or equivalently, the convexity of the $f$ conditional. This is why it is advantageous to split $f^\mix$ into its own Gibbs step, rather than, e.g. including it in a larger HMC pass which would not be exploiting the convexity and hence be much less efficient.

For the conjugate gradient solver, we use a simple diagonal pre-conditioner, $\mathbbgreek{\tilde \Lambda}_f(r)$, given by
\begin{equation}
\label{def: Lambda_cg}
\mathbbgreek{\tilde \Lambda}_f(r) = {\op B}^\dagger {\op K}^\dagger \Cn^{-1} \op K \, \op B  + \Cf[r]^{-1}.
\end{equation}
Although we find this is sufficient for the simulated data considered here, this step does account for roughly half of the total run time of the entire sampling algorithm, and is thus worth improving further. A promising avenue we expect to try in the future is to use the neural network-based Wiener filter given by \citet{munchmeyer2019}; this assumes $\phi\,{=}\,0$, but could potentially be a powerful preconditioner. Other techniques developed for Wiener filtering without pre-conditioner could be adapted to the lensing problem, possibly in combination with a neural preconditioner \cite{elsner2013efficient,2019MNRAS.490..947K}. We also note that one could absorb the final mixing step into the quantity in brackets in Eqn.~\eqref{eq:gibbs_f_sample}, although in practice we do not do so and instead solve Eqn.~\eqref{eq:gibbs_f_sample} exactly as written, which we find to be more numerically stable.

\subsection{The $\phi^\mix$ Gibbs pass}
\label{sec:phipass}

The next step of the sampling algorithm is to draw a conditional sample of $\phi^\mix$. Because this conditional distribution is not Gaussian, no specialized tricks like in the previous subsection exist, and we instead use a single HMC pass \cite{betancourt2017} to draw a sample. 

There are only two tunable inputs to the HMC algorithm: 1) a mass matrix, which should approximate the Hessian of the distribution to give the most efficient sampling, and 2) a prescription for the length of each Hamiltonian trajectory. For the mass matrix, we again use the Hessian approximation, $\mathbbgreek{\Lambda}_{\phi^\prime}$, given in Eqn.~\eqref{eq:hessphi}. For the trajectories, we perform a leap-frog symplectic integration with $n_h\,{=}\,25$ steps of size $\epsilon_h=0.02$. This choice is hand-tuned to work well for a range of configurations similar to the main ones we consider in this work, but may need to be re-tuned for sufficiently different analyses.

Fortunately, it is fairly straightforward to perform this tuning. To begin with, the choice of $\epsilon_h$ is set uniquely by the need to limit symplectic integration error. This error comes from two sources: 1) errors in the posterior gradient itself, and 2) errors due to the finite step-size, $\epsilon_h$. Before choosing $\epsilon_h$, we first make sure the contribution from (1) is sub-dominant. For this, the number of \textsc{LenseFlow} ODE steps is relevant because we compute gradients of the lensing operator by running a separate ODE for the gradient, rather than by backpropagating a gradient through the original ODE \citep[see Sect.~4 of][]{millea2019}. The gradient generated by the gradient ODE will differ from the true gradient due to ODE integration error. In practice, we find we need a $4^{\rm th}$ order Runge-Kutta integration with 10 steps before the \textsc{LenseFlow} gradient error is a sub-dominant contribution to the symplectic integration error. Another source of error in the posterior gradient is floating point truncation. We find the dominant source comes from the sums involved in the inner products in the posterior in Eqn.~\eqref{eq:jointposterior}, and that these errors can be significantly reduced with Kahan summation \cite{kahan1965}. With this, we are able to run the entire analysis with 32-bit instead of 64-bit floating point numbers, which doubles performance on most CPUs, and gives potentially much more drastic speed improvements on GPUs, depending on hardware (fast 64-bit support on GPUs is limited to high-end models). Once this and the number of \textsc{LenseFlow} ODE steps are set, $\epsilon_h$ is then simply tuned to give small enough integration errors such that the HMC acceptance is near 80\%.

Given $\epsilon_h$, the choice of $n_h$ comes from integrating long enough to meet the ``No U-Turn Criteria'' \cite{hoffman2014}. We have checked the integration length on representative data configurations and multiple random starting points, and find $n_h=25$ is adequate. We note that we do not adaptively change either $n_h$ or $\epsilon_h$ throughout our chains (the full ``No U-Turn Sampler'' of \cite{hoffman2014} usually refers to an algorithm where the integration length is adaptively chosen at each step). We do this for simplicity, and since we have not found very obvious regions of parameter space which appear to need significantly different values. The reparametrization of Sec.~\ref{sec:reparam} in particular helps us avoid the ``funnel problem'' \cite{neal2003} which might otherwise cause such a need. Nevertheless, it is worth exploring more sophisticated HMC sampling techniques in the future, since, as we will discuss in Sec.~\ref{sec:convergence}, our chains have auto-correlation lengths which could be even further improved. 

{
\renewcommand{\arraystretch}{1.2}
\setlength{\tabcolsep}{8pt}
\begin{table*}
    \begin{tabular}{l|c|c|c}
        \toprule
        & Configuration \configA         & Configuration \configB     & Configuration \configC  \\
        \colrule
        Map size                               & 256$\times$256          & 256$\times$256      & 512$\times$512          \\
        Pixel width                            & 2\,arcmin               & 3\,arcmin           & 3\,arcmin               \\
        Total area                             & 73\,deg$^2$             & 160\,deg$^2$        & 650\,deg$^2$            \\
        White noise level in $P$               & $1\;\mu$K-arcmin        & $1\;\mu$K-arcmin    & $1\;\mu$K-arcmin \\
        $(\ell_{\rm knee}, \alpha_{\rm knee})$ & (100,3)                 & (100,3)             & (100,3)                 \\
        Beam FWHM                              & 2\,arcmin               & 3\,arcmin           & 3\,arcmin               \\
        Fourier masking $(\op{K})$             & $2<\ell<5000$           & $2<\ell<3500$       & $2<\ell<3500$           \\
        Pixel masking $(\op{M})$               & 0.4\degree\;border\,+\,0.6\degree\;apod       & 0.6\degree\;border\,+\,0.9\degree\;apod    & 1.2\degree\;border\,+\,1.8\degree\;apod \\
        Sampled parameters ($\theta$)          & $r,\, \Aphi$            & $r$                 & $r$                     \\
        Fiducial $r$                           & $r=0.04$                & $r=\{0.04,0.02,0\}$ & $r=\{0.02,0.01,0\}$                  \\
        Chain iterations                       & 10000 & 5000 & 4000 \\
        Auto-correlation length for $\theta$   & 22 & 5--33 & 12 \\
        Wall-time (one GPU)                    & 48 hours & 19 hours & 50 hours \\
        \colrule
    \end{tabular}
    \caption{Parameters for the different configurations of simulated data used in this work.}
    \label{table:configurations}
\end{table*}
}

\subsection{The $A_{\phi}$ and $r$ Gibbs passes}
\label{sec:thetapass}

Finally, we sample the conditional distribution of each of $A_{\phi}$ and $r$ on separate Gibbs passes. Because these are one-dimensional probability distributions, we can directly probe these functions on a grid and use inverse transform sampling (often called ``slice sampling'') to draw a sample. Moreover, we find that the log conditional densities are typically quite smooth and close to quadratic, so we can compute a very accurate interpolation of the log probability. For the simulations given in this paper we use $200$ grid points over the intervals $\Aphi\in [0.75, 1.25]$ and $r\in [10^{-6}, 0.1]$, respectively, with the $r$ grid points quadratically spaced to ensure sufficient resolution near $r=0$.

There are two additional tricks, which come at no extra computational cost, which we utilize to reduce the number of samples required for convergence. First, we use MCMC over-relaxation \citep{neal1995}; instead of drawing a single sample from the discretized density, $K$ samples are drawn independently, one of which is chosen depending on the rank (among the $K$ draws) of the parameter value from the previous Gibbs iteration parameter. In the simulations below we set $K\,{=}\,15$, and we find that this can sometimes reduce the chain auto-correlation time by 10-20\%. Second, we save the interpolated conditional densities at each step, and use these to construct Rao-Blackwell estimates of the marginal posterior densities,
\begin{align}
\mathcal{P}( r \,|\,d) &\approx \frac{1}{n}\sum_{i=1}^{n} \mathcal{P}( r \,|\,  f^\mix_i,\, \phi^\mix_i,\, A_{\phi,i},\,  d) \\
\mathcal{P}( \Aphi \,|\,d) &\approx \frac{1}{n}\sum_{i=1}^{n} \mathcal{P}( \Aphi \,|\, r_{i-1}, \, f^\mix_i,\, \phi^\mix_i,  d)
\end{align}
This helps reduce the variance of the estimated posteriors slightly faster than just building up a histogram of the Monte Carlo samples, particularly deep in the tails of distribution.

\section{Simulation results}
\label{section: simulation}

\subsection{Description of runs}

\begin{figure*}
\includegraphics[width=\textwidth]{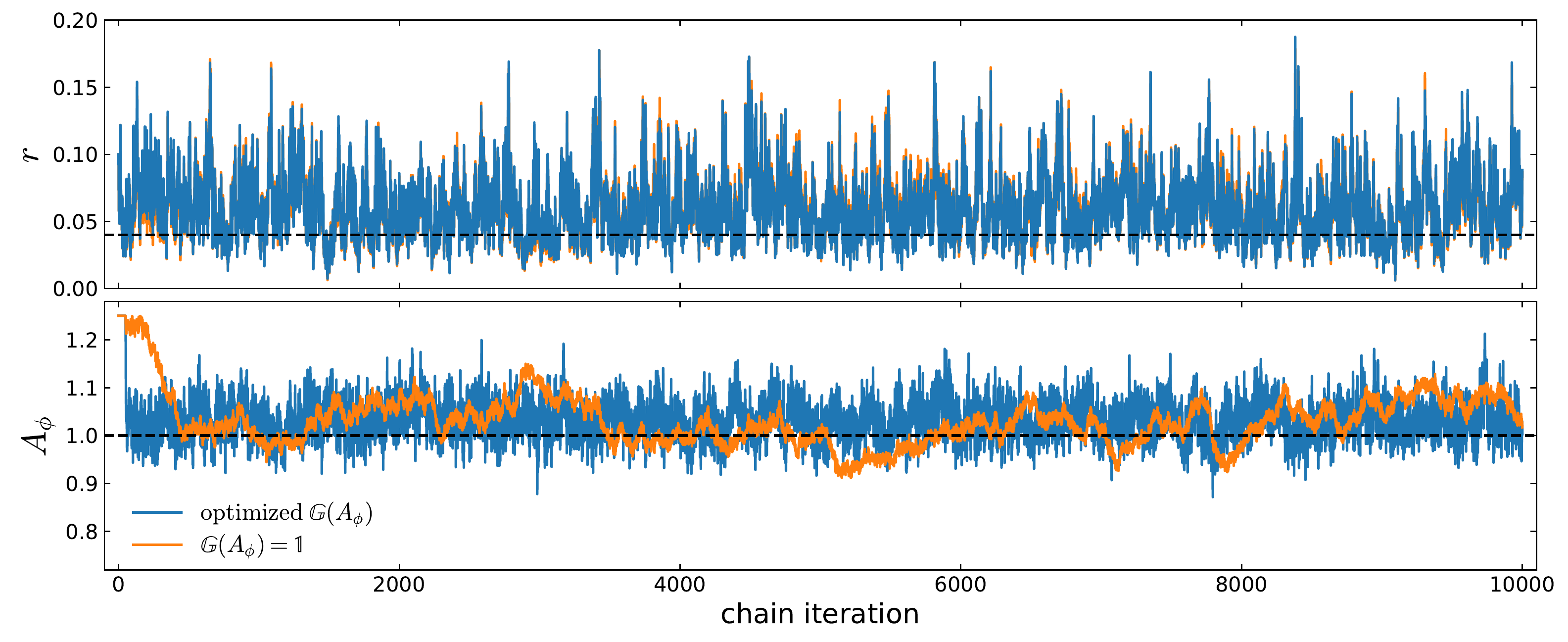}
\caption{Samples of $r$ and $\Aphi$ at each iteration for two chains with the same data and starting random seed, but different choices for parametrizing the posterior (see Sec.~\ref{sec:reparam}). The blue line corresponds to using our optimized $\opG[\Aphi]$ reparametrization, whereas the orange line shows the highly suboptimal choice of $\opG[\Aphi]\,{=}\,\mathbbgreek{1}$. No burn-in is removed in either case. The simulated data here is generated according to configuration \configA (see Table~\ref{table:configurations}). The run-time for a chain of this length is 48 hours on one GPU.}
\label{fig:traceplot}
\end{figure*}

\begin{figure}[h!]
\begin{centering}
\includegraphics[height=3.3in]{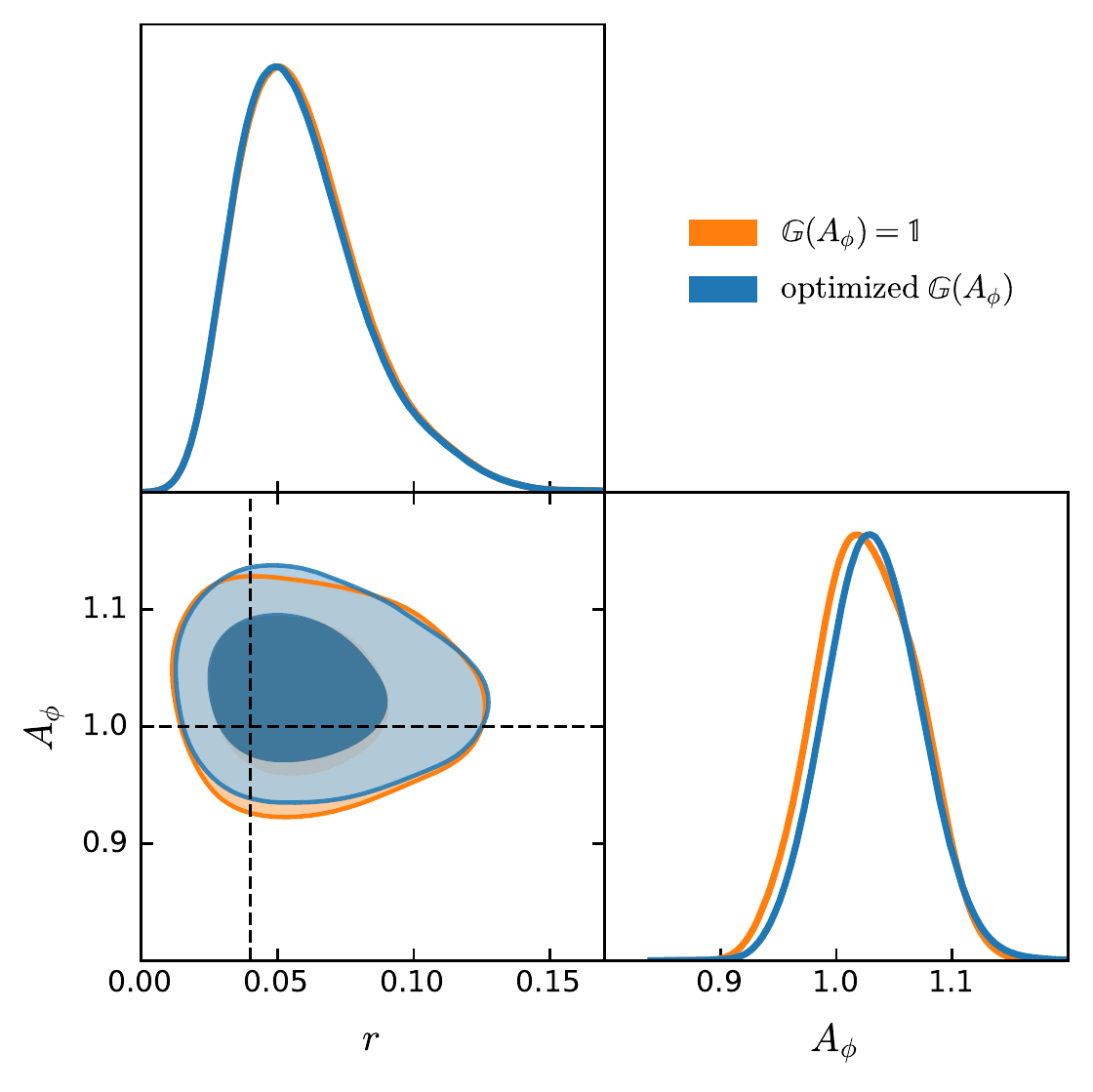}
\end{centering}
\caption{Posterior distribution for $r$ and $\Aphi$ from a chain on simulated data in configuration \configA (see Table~\ref{table:configurations}). The samples that comprise this plot are shown in Fig.~\ref{fig:traceplot}. For demonstration, here we use the \texttt{getdist} \cite{lewis2019} plotting package rather than our Blackwell-Rao posterior density estimate. The ability to  examine joint constraints on these parameters while performing optimal delensing for a realistic data set  with masking is a unique strength of our approach. Here, we find these two parameters are highly uncorrelated, providing evidence that $\Aphi$ can be fixed without impacting $r$ estimation. The orange curve shows a suboptimal choice of the $\op{G}$ matrix, which causes that chain to converge more slowly.}
\label{fig: rao-blackwell densities for simulation 1}
\end{figure}

With the details of our posterior and the sampling algorithm specified, we now turn to actually running chains and interpreting results. We have picked three different configurations of simulated data, the details summarized in Table.~\ref{table:configurations}, which are meant to resemble possible CMB-S4 resolutions and noise levels, but slightly smaller sky area. We will describe these runs first, then come back to a more quantitative discussion of chain convergence as well as some scientific conclusions that can be extracted from these results.

All of our configurations include a Gaussian beam with a 2--3 arcmin full-width-half-max (FWHM). We take isotropic Gaussian 1\,$\mu$K-arcmin polarization noise with a power-spectrum which includes a contribution from a $1/f$ knee, modeled via $\ell_{\rm knee}$ and $\alpha_{\rm knee}$ parameters \cite{barron2017}. The runs are all in the flat-sky approximation, and include a border mask, $\op{M}$, of various widths. Although the runs we have chosen here use an apodized border mask, we find that unapodized masks work just as well. This is helpful if, for example, there are so many point sources that apodizing them all would discard too much data. In addition to a pixel mask, we also apply an isotropic low-pass mask in Fourier space, $\op{K}$, generally near the Nyquist frequency. Although we do not do so here, it is completely straightforward to use an anisotropic Fourier mask instead, which can be useful in limiting systematics by masking scan-parallel and scan-perpendicular directions differently. Finally, we use grid sizes between 256$\times$256 and 512$\times$512 pixels. The latter is around the limit of what is currently computationally possible on performance hardware and covers about 650\,deg$^2$, with an effective unmasked region of around 450\,deg$^2$. This is about a third to a fifth of the planned CMB-S4 deep field where our procedure is most applicable, with several years remaining to scale up to the full patch or beyond.

The first run we describe uses data simulated in configuration \configA. In this configuration, we sample both $r$ and $\Aphi$. We show a trace of the sampled values for these two parameters in Fig.~\ref{fig:traceplot}. We will asses convergence and correctness of the chains in the next subsection, but for now one can at least see by eye the stationarity of the samples and that they cover the true input values, as expected. For this case, we have also run an identical copy of the chain, including identical starting random seed, but which uses $\opG[\Aphi]=\mathbbgreek{1}$ instead of the fiducial choice which we described in Sec.~\ref{sec:reparam}. The impact of not using the fiducial $\opG(\Aphi)$ is shown in orange. There is a  dramatic reduction in the convergence of the $\Aphi$ samples (the auto correlation length is $\sim$\,25 times larger), highlighting the importance of our reparametrization. We do not show a case where we set the other mixing matrix, $\D[r]$, to the identity matrix; in that case, the impact would be so drastic that it would be impossible to even run a chain at all.

In Fig.~\ref{fig: rao-blackwell densities for simulation 1}, we show the posterior distribution for $r$ and $\Aphi$ computed from these samples, for demonstration plotted using the \texttt{getdist} \cite{lewis2019} package instead of our Blackwell-Rao estimate. This ability to compute joint constraints on parameters which control both the unlensed CMB fields and lensing potential, with the Bayesian procedure having implicitly performed an optimal lensing reconstruction and delensing, is a unique strength of our procedure and a key result of this work. Note the very small correlation between $r$ and $\Aphi$ ($\rho\,{=}\,0.10$); this is evidence that estimates of $r$ are not strongly limited by knowledge of the theoretical lensing spectrum, or conversely that lensing reconstruction and hence delensing efficiency is not strongly limited by the true value of $r$. This was expected from the intuition that the lensing reconstruction is mostly dominated by small scales whereas $r$ is mainly estimated from large scales, but our precedure allows us to quantify this explicitly.

\begin{figure}
\begin{centering}
\includegraphics[width=\columnwidth]{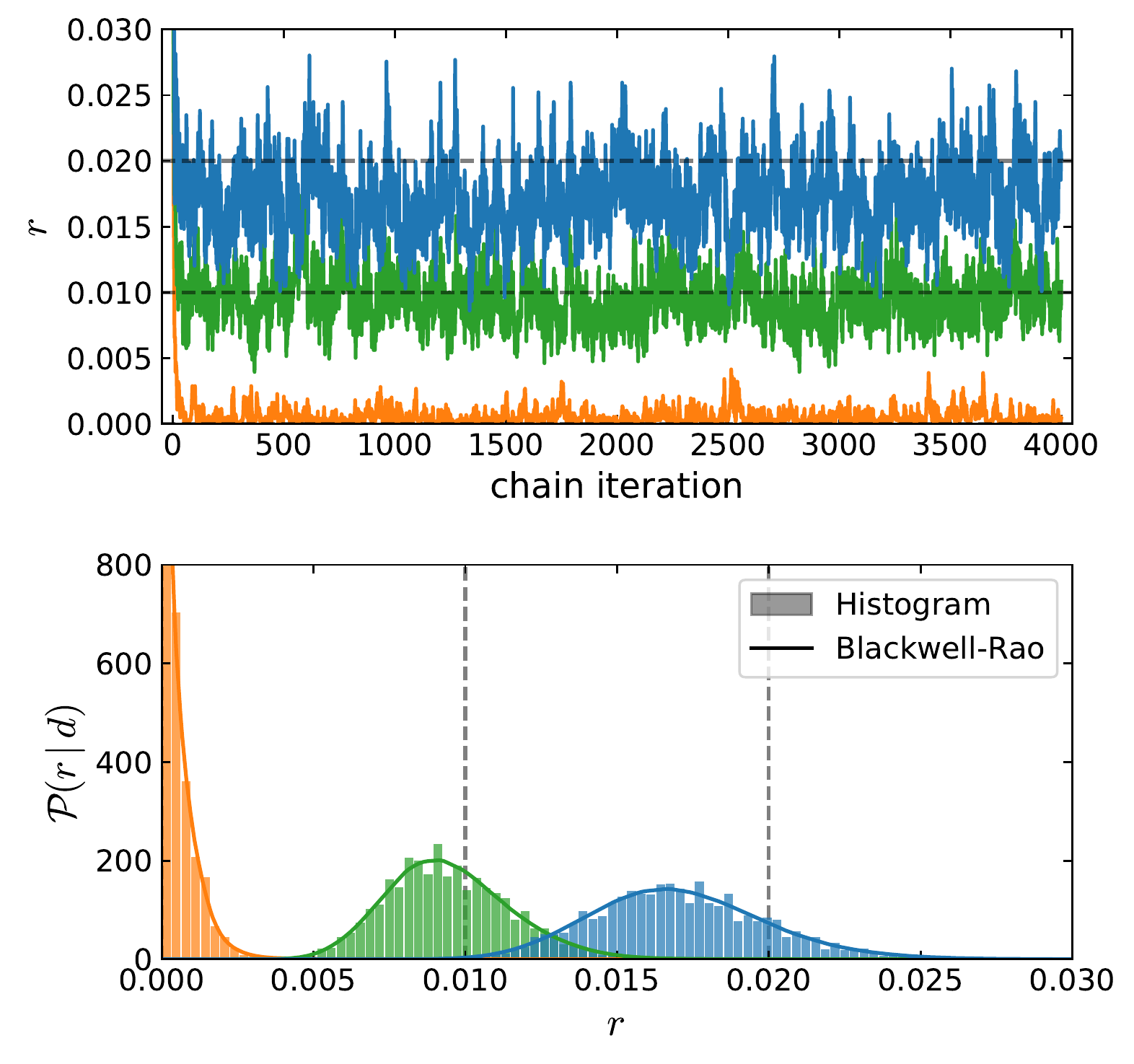}
\caption{(Top panel) The trace of $r$ samples from chains in configuration \configC (see Table~\ref{table:configurations}). Three different fiducial values of $r$ are explored, with the true value given by the black dashed line and each chain in a different color. No burn-in is removed. (Bottom panel) The same samples binned into histograms, as well Blackwell-Rao estimates of the posterior density, as described in Sec.~\ref{sec:thetapass}. These estimates recover very smooth distributions, even in the case where the true $r$ is zero and the constraint is just an upper bound.}
\label{fig:configCposteriors}
\end{centering}
\end{figure}

\begin{figure*}
\includegraphics[width=\textwidth]{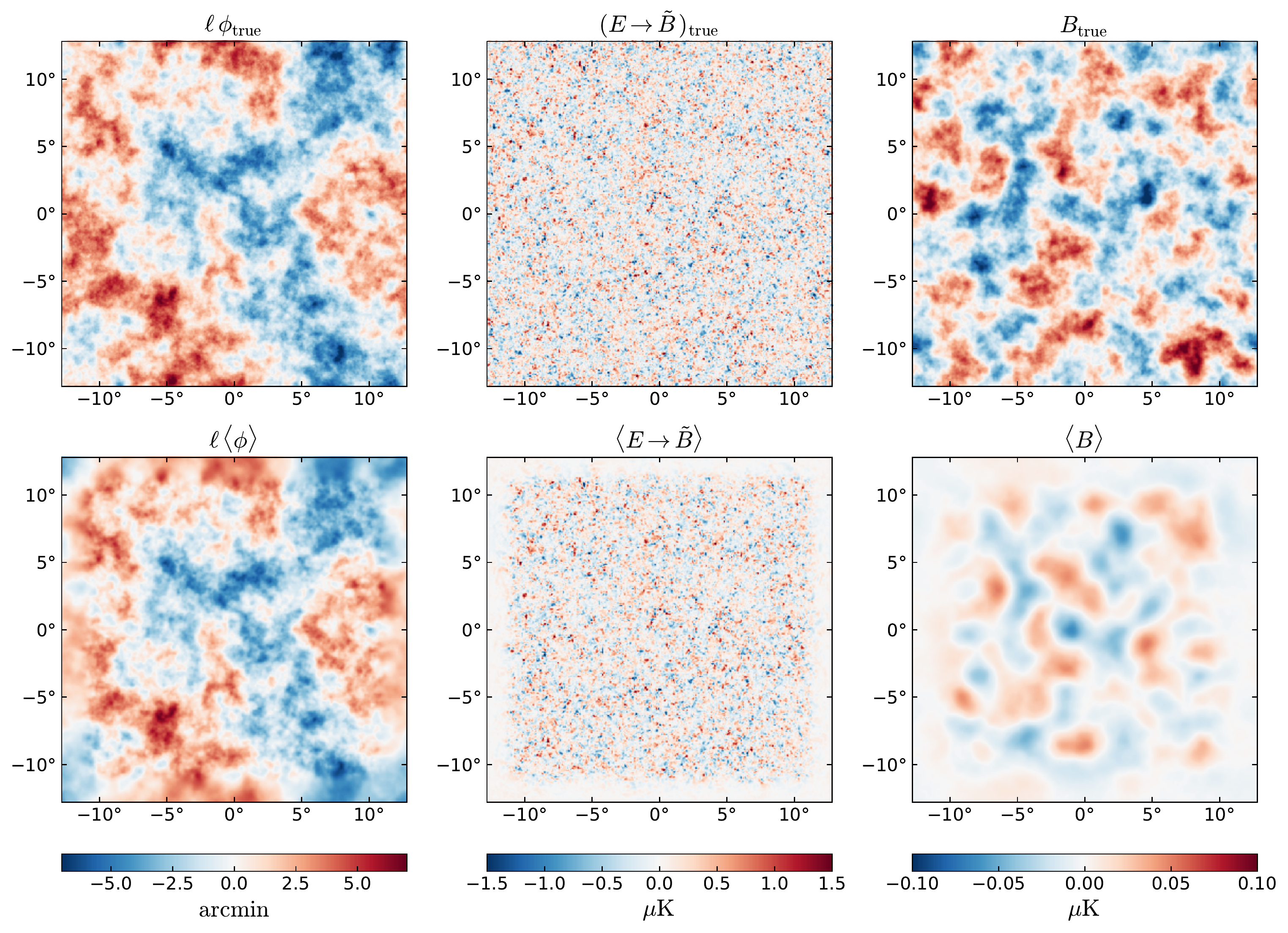}
\caption{True input maps (top row) as compared to posterior mean maps (bottom row) computed by averaging over chain samples. This chain uses configuration \configC (see Table~\ref{table:configurations}) with a true value of $r\,{=}\,0.01$. The data for this chain is shown in Fig.~\ref{fig:configCdata}. The first column shows the $\phi$ map multiplied in Fourier space by $\ell$ to visually enhance smaller scales, the middle column shows $E$ modes which have been lensed into $B$, and the final column shows the reconstructed primordial $B$ modes.}
\label{fig:configCmaps}
\end{figure*}

\begin{figure*}
\begin{centering}
\includegraphics[width=\textwidth]{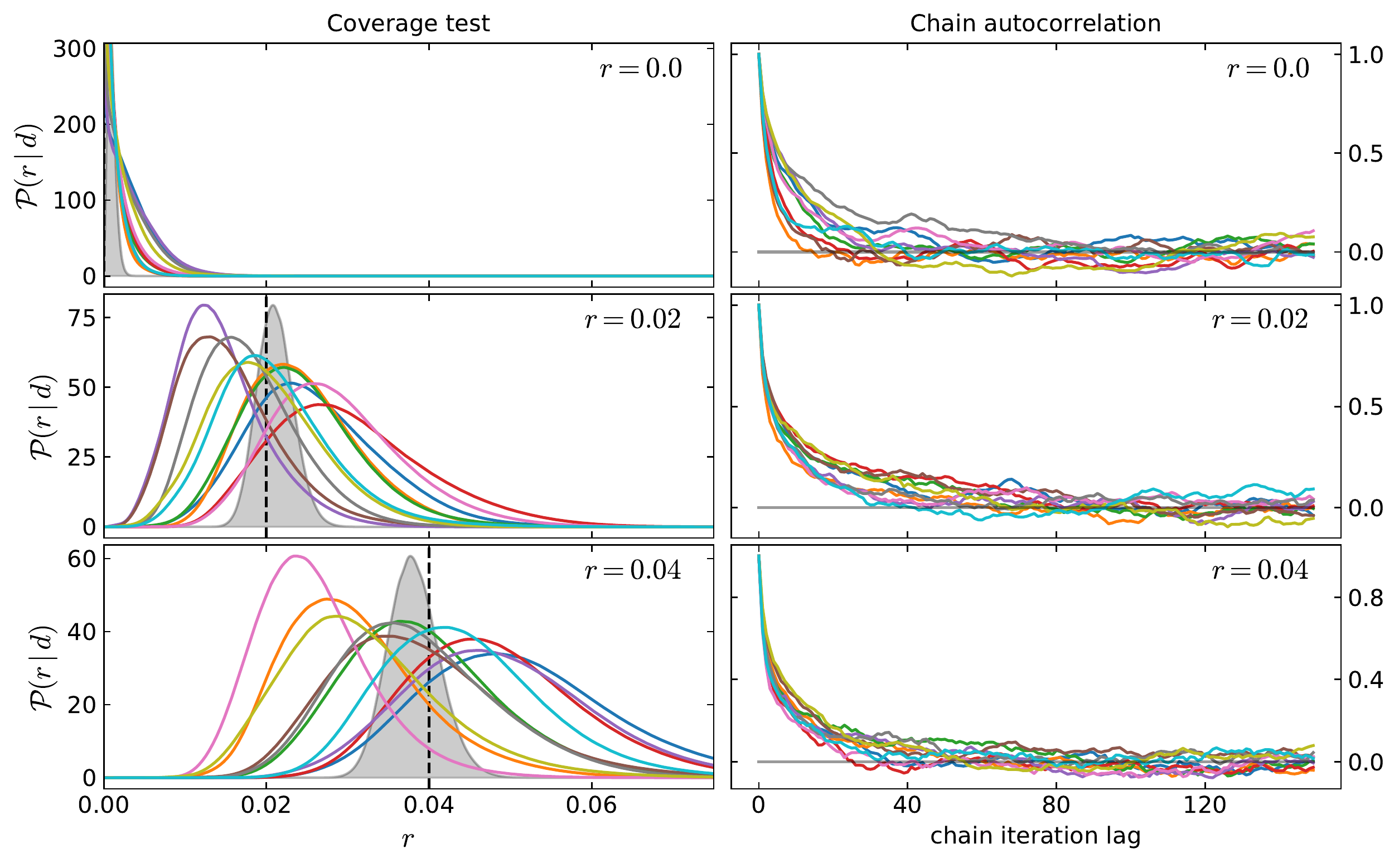}
\caption{(Left column) Blackwell-Rao posteriors from each of 10 chains on different simulated data for three different true values of the tensor-to-scalar ratio indicated in each row. The gray band is the product of the posteriors for each case, with the prior on $r$ importance sampled to be uniform (and with arbitrary normalization constant so as to fit on these axes). We expect that the gray band covers the fiducial value of $r$ to within its own width, as is indeed the case. This is a test of the coverage of our $\mathcal{P}(r\,|\,d)$ posteriors and hence a test of the correctness of our procedure (Right column) The chain auto-correlation function for each of the chains in the left panel. The integrated auto-correlation time for these chains ranges from 5--33.}
\label{fig: coverage and correlation}
\end{centering}
\end{figure*}

\begin{figure}
\begin{centering}
\includegraphics[width=\columnwidth]{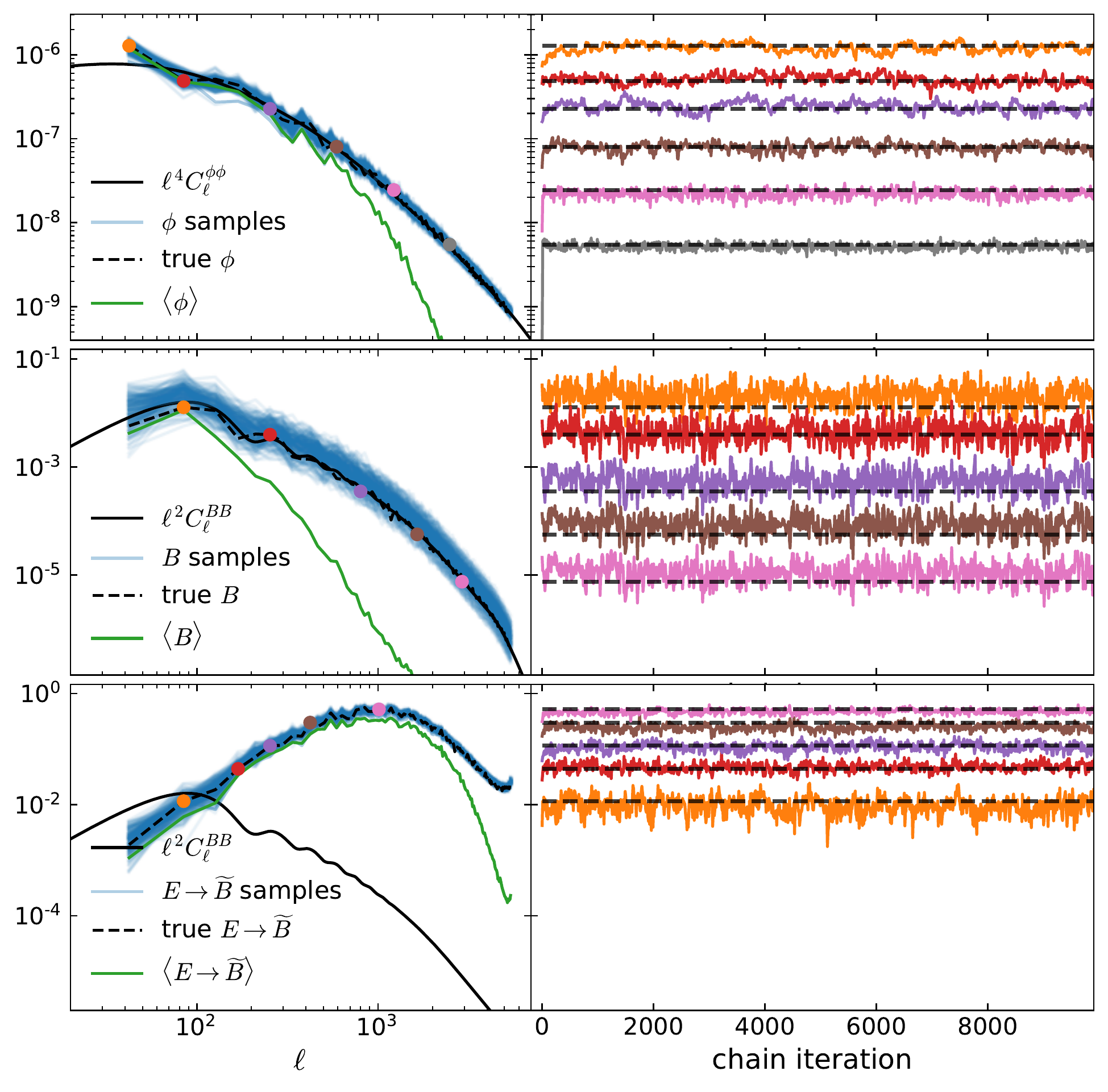}
\end{centering}
\caption{(Left column) In blue, we overlay the power spectra of chain samples of $\phi$ and of two quantities derived from $f$. The black dashed line gives the power spectrum of the truth, and the green line is the power spectrum of the posterior mean map. The three rows correspond to $\phi$, unlensed $B$, and $E$-lensed-into-$B$. The posterior mean maps exhibit Wiener-filter like suppression, as expected, while the samples scatter around the true spectrum and quantify uncertainty. (Right column) The same power spectra which are overlayed on the left, but picking some specific multipoles and plotting the trace of their value throughout the chain. Visually one can see the good convergence of the power spectrum samples. This is the same chain in configuration \configA (see Table~\ref{table:configurations}) used in Figs.~\ref{fig:traceplot} and \ref{fig: rao-blackwell densities for simulation 1}.}
\label{fig: bandpower_cloud_and_trace}
\end{figure}

\begin{figure}
    \includegraphics[width=\columnwidth]{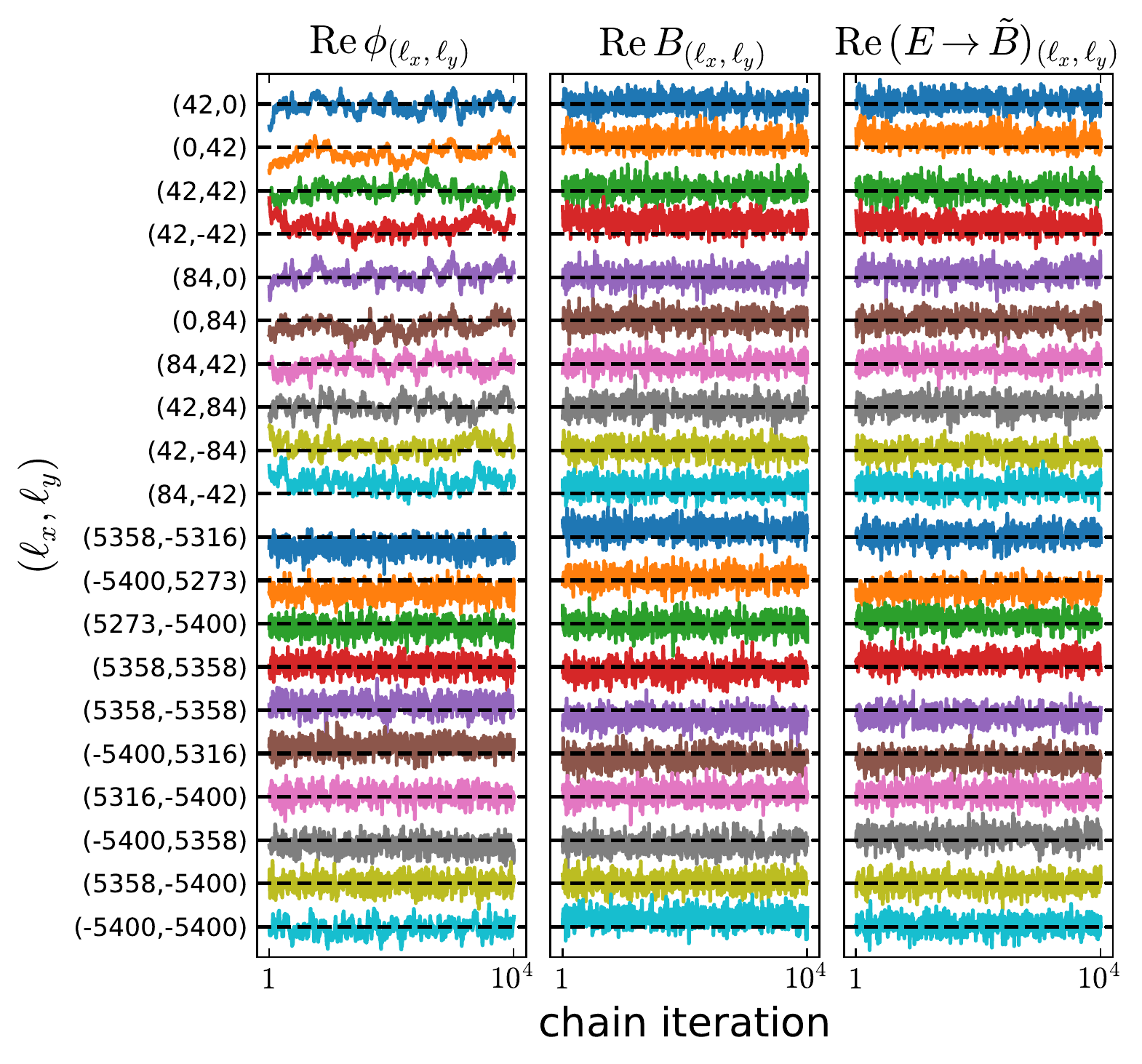}
    \caption{Chain samples of the real part of the ten largest-scale and ten smallest-scale Fourier modes of the posterior $\phi$, $B$, and $E$-lensed-into-$B$ maps. Each set of samples is normalized to unit variance, but the relative distance to the truth (shown in the black dashed line) is preserved. This chain uses configuration \configA (see Table~\ref{table:configurations}). Visually, we achieve great convergence even at the individual mode level. Out of the $\sim$\,200,000 modes which are sampled, the only exceptions are perhaps the two largest scale $\phi$ modes, which could benefit from a slightly longer chain. However, these two modes are not informative for $\Aphi$, which remains very well converged (Fig.~\ref{fig:traceplot}).}
    \label{fig:configAmodetrace}
\end{figure}

Next, we describe a set of simulations in configuration \configC. Since we have ascertained that there is little dependence on $\Aphi$ for $r$ estimation, in these runs, we fix $\Aphi\,{=}\,1$. We also increase the grid size to 512$\times$512 and the pixel size to 3\,arcmin pixels, giving a total sky area of ${\sim}\,650{\rm deg}^2$, which is the largest sky area we analyze in this work. We note that although 3\,arcmin pixels may seem large compared to ${\sim}\,1\,{\rm arcmin}$ typical lensing deflections, \textsc{LenseFlow} is able to  lense maps accurately  up to scales very close to the Nyquist frequency \cite{millea2019}, which here is $\ell=3400$ and contains nearly all of the available information given our choice of beam for this configuration. For these runs, we use simulated data with three different fiducial values for the tensor-to-scalar ratio, $r=\{0, 0.01, 0.02\}$. Posterior distributions for $r$ are shown in Fig.~\ref{fig:configCposteriors}, this time, using the Blackwell-Rao estimate. We can see that each case covers the truth, and that in the $r\,{=}\,0$ case, the chain samples of $r$ oscillate against zero, as expected.

Of course, the chains contain not just samples of the parameters $\theta$, but also samples of $f$ and $\phi$ at each iteration. In Fig.~\ref{fig:configCmaps}, we compare the posterior mean of $\phi$ and some quantities derived from $f$ against the simulation truth for configuration \configC. In the first column, we show the posterior mean reconstructed $\phi$, multiplied in Fourier space by the wavenumber $\ell$ to make smaller scale structure more easily visible. The posterior mean can be regarded as the ``optimal'' point estimate of $\phi$ in the sense that it minimizes the posterior expected squared error against the truth. This estimate is slightly lower variance than the marginal MAP estimate given by \cite{carron2017} (the two differ only due to the non-Gaussianity of $\mathcal{P}(\phi\,|\,d)$), although we leave to a future work determining whether there is a meaningful difference. The remaining two columns of Fig.~\ref{fig:configCmaps} show the posterior mean ``$E$-lensed-into-$B$'' maps (the average over all chain samples of unlensed $E$ and zero $B$, lensed by $\phi$), as well as the posterior mean of the unlensed $B$ map. These latter two quantities are useful data products from the chains, as we will describe in the next section. 

\subsection{What can the $f$ and $\phi$ samples be used for?}

Despite the seemingly valuable information contained in the samples of full maps or their associated posterior mean, it is worth asking ``what explicitly can these actually be used for?" In terms of a principled statistical analysis for parameter inference within a standard cosmological sky model with Gaussian initial conditions,  the answer is actually ``not much"; the map samples are just a byproduct of the Monte Carlo marginalization which we used to obtain constraints on the cosmological quantities which we were really after, here $r$ and $\Aphi$. Indeed, we cannot readily use the samples of $f$ and $\phi$ to estimate any other cosmological parameters which were not jointly sampled in the first place.

The real situation is somewhat less pessimistic, however. For example, if we have a physical reason to believe that having jointly sampled extra parameters would not actually impact the lensing reconstruction and delensing, then it may be still be a valid approximation to derive further constraints from the samples. One such case is the search for primordial scalar non-Gaussianity, where constraints on local-type non-Gaussianity become limited at small scales by lensing-induced variance and could be significantly improved by delensing \cite{coulton2019}. Although part of the locally non-Gaussian primordial signal would affect the reconstruction, \citet{coulton2019} demonstrated this effect is small and quantifiable, meaning our posterior mean delensed maps would be excellent candidates to be used in these searches. Furthermore, our posterior delensed $B$ maps could be used in the search for primordial tensor non-Gaussianity as well \cite{meerburg2016}, with near-optimal results as long as any potential non-Gaussianity is perturbatively small.

The outlook on samples is even better when we consider what can be done in cross-correlation with other probes. Take, for example, the posterior unlensed $B$ map. We could cross-correlate this map with some tracer of foreground $B$ contamination from the Milky Way; if any correlation were detected, it would indicate that whatever foreground cleaning had been performed was insufficient and we would deduce that our corresponding $r$ samples could be biased. Similarly, the sampled maps, their mean, or even the mean power spectrum of the maps, could be inspected for anything that correlates with an instrumental effect as a way to search for systematics.

From searching for contaminants, it is only a small step to using our posterior samples to check all aspects of the data model (containing the cosmological model, the lensed sky signal, noise, etc.) itself. It is worth recalling the well-known quote by George Box that  ``all models are wrong but some are useful" \cite{BOX1979201}. This quote applies to CMB data just as much as to any other data set. One way to check if the standard model of lensed CMB data is useful is to use it to simulate data starting from the posterior samples and then to check whether this replicated data reproduces the salient features of the actual data. This  technique for model evaluation is called ``posterior predictive checks'' (PPCs) and was introduced in a Bayesian context in \cite{rubin1984}; see \cite{2019PhRvL.122f1105F} for a recent application in cosmology. In the literature, PPCs are typically based on the parameters $\theta$; using the samples of the latent fields $f$, and $\phi$ would allow defining much more fine-grained PPCs of the model. 

The samples can also be used in other more quantitative ways. Consider, for example, a cross-correlation analysis between the CMB and another low-redshift probe of matter fluctuations. One can generally write down the likelihood, $\mathcal{L}(d_{\rm low\text{-}z}\,|\,\phi,\theta)$, where $d_{\rm low\text{-}z}$ is the  low-redshift data. The full posterior given both datasets is
\begin{align}
    \mathcal{P}(f,\phi,\theta\,|\,d,d_{\rm low\text{-}z}) = \mathcal{P}(f,\phi,\theta\,|\,d) \mathcal{L}(d_{\rm low\text{-}z}\,|\,\phi,\theta).
\end{align}
If the low-redshift data is sufficiently less constraining on $\phi$ than the CMB data, then importance sampling the CMB chain is an easy and efficient way of obtaining a Monte Carlo representation of the new posterior for both datasets. 

Another analysis which could use the samples would be to split delensing into two steps: 1) obtain $E$-lensed-into-$B$ samples from small scale CMB data, then 2) use these samples to delense large-scale CMB data and search for non-zero $r$. Delensing via the samples rather than via a single point estimate of $\phi$ is a convenient way to propagate the (fully non-Gaussian) delensing uncertainty into the large-scale analysis. A practical reason for doing such a split analysis instead of simply jointly estimating $r$ from the entire CMB dataset might be that large-scale foregrounds and systematics are easier to deal with outside of the Bayesian framework. 

We leave further development of any of these ideas to future work. Regardless of how these samples may be used, the key point is that they are a useful way to capture the entire information content in the CMB data that generate them, and they fully represent the uncertainty in the reconstruction due to noise, modeled systematics, and incomplete knowledge of the cosmological parameters.

\subsection{Convergence diagnostics}
\label{sec:convergence}

Having described some of the results from the chains, we now turn to more quantitatively assessing chain convergence. We begin using a final set of chains with data simulated from configuration \configB. These chains only sample $r$ and have been reduced to 256$\times$256 pixels, however we run 10 chains on different simulated data for each of three fiducial values, $r=\{0, 0.02, 0.04\}$.

The posteriors from each of these chains are shown in Fig.~\ref{fig: coverage and correlation}.  It is worth noting the scatter in the mean and width of the different data realizations (here $\sigma_r$ can vary by almost a factor of two) as a reminder that any one experiment can be lucky or unlucky depending on the particular patch of sky observed. It would be interesting to determine how much of the contribution to this scatter comes from the non-Gaussian uncertainty in the lensing reconstruction as opposed to Gaussian sample variance, although that is beyond our scope here.

One way to check the correctness and convergence of these chains is to multiply the 10 posteriors together. We expect that the resulting distribution should tighten around the true of $r$, with scatter such that roughly ${\sim}~68\%$ of the time the truth will be covered by the $1\,\sigma$ contours. This is indeed what we see in Fig.~\ref{fig: coverage and correlation} for all values of $r$. Formally, with only 10 chains, we can only check for the presence of biases in our posteriors at the $\sigma/\sqrt{10}\approx30\%\,\sigma$ level, however in the absence of a coding error, there is no reason to believe these contours would not continue to shrink further around the truth.

Another way to check the convergence of our chains is by computing the integrated auto-correlation time and the accompanying effective sample size \cite{goodman2010}. The right hand panel of Fig.~\ref{fig: coverage and correlation} shows the auto-correlation function for the $r$ samples from each of these chains. In all cases, it takes about ${\sim}\,40$ iterations of our sampler before the auto-correlation drops to near-zero and we obtain an independent sample. More exactly, the integrated auto-correlation time is in the range of 5\,--\,33, corresponding to an effective sample size of 150\,--\,1000 given the 5000 total iterations in each chain (auto-correlation lengths for all configurations are listed in Table~\ref{table:configurations}). In turn, this means we should expect a Monte Carlo error on the posterior mean of $r$ on the order of 3\%\,--\,10\% of $\sigma_r$. 

This is consistent with another estimate of the error which we can get by splitting our chains into multiple pieces or running multiple chains, and computing the mean from each. We have performed this test for the chain in configuration \configA by splitting the 10000 samples into two halves and checking the difference in the resulting posterior mean for both $\Aphi$ and for $r$. We find that the mean agrees to within 5\% of $\sigma_{\Aphi}$ and 8\% of $\sigma_r$, respectively. 

The posterior distribution of any quantity derived from $(f,\phi,\theta)$ can be explored by post-processing the Monte Carlo chain, and its convergence can be tested. Bandpowers are one such quantity, and these have a very direct relation to the convergence of $r$ and $\Aphi$. In particular, only the bandpowers of $f$ and $\phi$ enter the $\mathcal{P}(r,\Aphi\,|\,f,\phi,d)$ conditional distribution. In Fig.~\ref{fig: bandpower_cloud_and_trace}, we show the trace of various bandpowers of $\phi$, $B$, and $E$-lensed-into-$B$. Visually, we see these samples are still consistent with being drawn from a stationary distribution. 

Delving deeper into the ${\sim}\,200,\!000$ parameters which are sampled in this configuration, we plot in Fig.~\ref{fig:configAmodetrace} the trace of the real part of individual Fourier modes of $\phi$, $B$, and $E$-lensed-into-$B$. The choice of plotting the real part is arbitrary as it has identical statistical properties to the imaginary part under the assumption of isotropy (nevertheless, we have checked that the imaginary part does behave similarly). Even here, we mostly see very good convergence of the samples. For an internal CMB analysis, the convergence of these individual modes is not particularly important, since, as previously stated, what really matters is the convergence of the $\theta$ parameters. However, for a cross-correlation analysis such as the ones described in the previous subsection, the individual modes (and hence the full maps themselves) must be adequately converged. Fig.~\ref{fig:configAmodetrace} is evidence that this is indeed the case. 

We do note that $\phi$ modes at the largest scales converge slightly slower than others, as can be seen in Figs.~\ref{fig: bandpower_cloud_and_trace} and \ref{fig:configAmodetrace}. We believe this is related to the mean-field which also arises in both quadratic or MAP estimation \cite{carron2017}. At these large scales where the mean-field is very big, frequentist analyses require a large number of Monte Carlo simulations to estimate the mean-field precisely enough so that the error on the mean-field determination is sub-dominant to sample variance. In our Bayesian analysis, this challenge is not solved ``for free'', rather it manifests as a need for longer chains to overcome the larger correlation length at these same scales. Evidence that this is the case comes from the fact that removing the mask and hence reducing the mean-field yields more rapid relative convergence at these large scales. We do stress, however, that because the majority of information on $\Aphi$ is not sourced by these handful of largest scale modes, their slower convergence does not significantly impact the very good convergence of $\Aphi$ that we see in Fig.~\ref{fig:traceplot}.

The results in this section demonstrate that the $\theta$, the bandpowers, and even individual Fourier modes are well converged in these chains. However, it is not implausible that one could find pathological combinations of parameters for which this is not the case. We caution users of these chains to first verify convergence of arbitrary derived quantities which they may need. This can be done using tests similar to the ones described in this section.

\subsection{Fisher information on $r$ and S4 forecasting}
\label{sec:forecasting}

The chains give us the ability to check existing forecasts for e.g. CMB-S4, South Pole Observatory, or Simons Observatory to a precision which has not been possible before. We will refer to these as CMB-S4-like forecasts since the methodology we are testing is the same between all of them. The approach is to use chains on simulated data to compute exactly (up to Monte Carlo errors) the Fisher information on $r$ contained in lensed CMB data. This can be done even in the presence of real instrumental complexities such as the pixel masking we apply here. We will use chains in configuration \configB for this test. Although this is a smaller patch of sky than the planned CMB-S4 observations, the noise levels are similar and this lets us validate the forecasting procedure itself.

\begin{figure}
\begin{centering}
\includegraphics[width=\columnwidth]{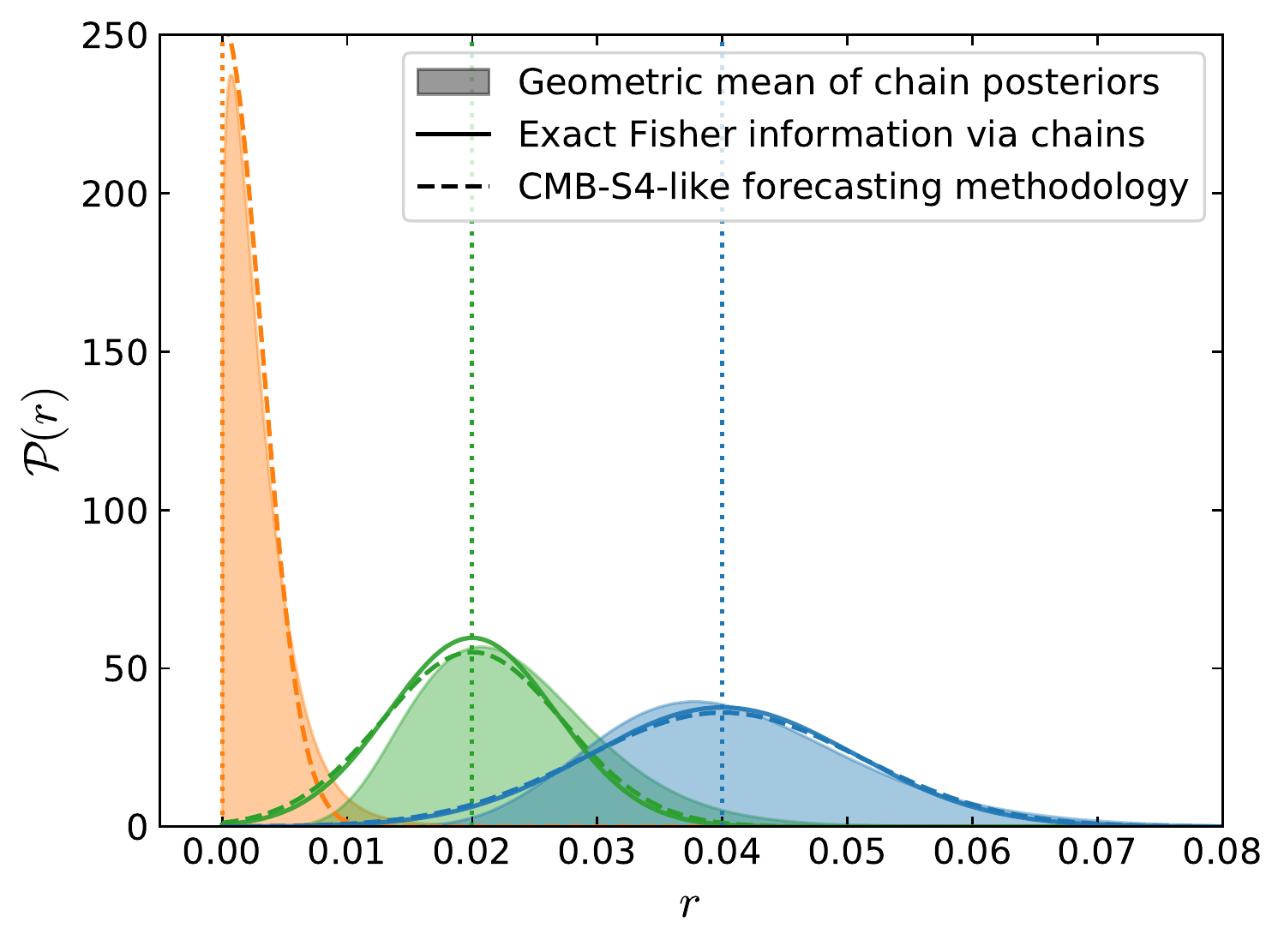}
\caption{A comparison of different methods for forecasting constraints on $r$, assuming configuration \configB (see Table~\ref{table:configurations}). Three different possible true values of $r$ are explored, indicated by vertical dotted lines. The dashed lines show expected Gaussian constraints forecasted with a method very similar to that used for CMB-S4. In solid lines, we show Gaussian distributions with standard deviation computed from the Fisher information on $r$. This work is the first time the Fisher information on $r$ from lensed CMB data has been calculated without approximation. In filled contours, we show the geometric mean of the posteriors from several chains. The excellent agreement between all of these is an important validation of the CMB-S4 $r$ forecasting methodology even in the presence of instrumental effects and masking, as is considered here.}
\label{fig:fisher_information}
\end{centering}
\end{figure}

To begin, consider the Fisher information, 
\begin{align}
    \label{eq:fisher}
    \mathcal{F}_{rr}(r_{\rm fid}) = -\left\langle \left.\frac{d^2}{dr^2} \log \mathcal{L}(d\,|\,r) \right|_{r_{\rm fid}}\right\rangle_{d\sim\mathcal{L}(d\,|\,r_{\rm fid})}.
\end{align}
It is an average over data, $d$, of the Hessian of the log-likelihood of $r$ for each of these data, evaluated at $r\,{=}\,r_{\rm fid}$, and where the data are themselves simulated given $r\,{=}\,r_{\rm fid}$. If we run our chains with a uniform prior on $r$ (or importance sample it to be uniform after the fact), then we have $\mathcal{L}(d\,|\,r)=\mathcal{P}(r\,|\,d)$. Thus we can take the log of the posterior $\mathcal{P}(r\,|\,d)$ estimated from the chains, numerically compute the second derivative, and explicitly perform the average in Eqn.~\eqref{eq:fisher} over several chains with different simulated data. Alternatively, we can swap the order of the derivative and expectation value in Eqn.~\eqref{eq:fisher} and take the geometric mean of the chain posteriors first. The second derivative of the log of this function at $r_{\rm fid}$ is then again the Fisher information, but instead of looking just at one value, we can simply plot the entire function. Loosely speaking, this maps out something like the ``typical posterior'' that one might expect given possible data, which is also a useful forecasting quantity, particularly for $r_{\rm fid}\,{=}\,0$ where Monte Carlo noise prevents us from computing a stable numerical derivative. For configuration \configB, these functions, as well as Gaussians with standard deviations given by $1/\sqrt{\mathcal{F}_{rr}}$ are shown in Fig.~\ref{fig:fisher_information}. 

We would like to compare against CMB-S4-like forecasts. These types of forecasts are broken up into two steps: 1) first, a post-delensing residual lensed $B$ power is computed, then 2) this is treated as Gaussian noise in a second step to estimate $r$. For the forecasts in \cite{abazajian2016}, the first step has been based on the method given in \citet{smith2012}. This method follows the heuristic idea that to perform optimal delensing, one iterates computing the $EB$ quadratic estimate for $\phi$, delenses the data by this $\phi$, then recomputes the $\phi$ estimate, which should now be lower variance because part of the contribution to the error of this estimate, namely the lensed $B$ modes, have been reduced. We note that this computation works only to first order in $\phi$, ignores $\ell$-to-$\ell$ correlations and non-Gaussianities in both the $\phi$ noise and the residual lensed $B$ modes, and ignores pixel masking. So that information is not double-counted, only modes at $\ell\,{\gtrsim}\,150$ are used in step (1) and only modes at $\ell\,{\lesssim}\,150$ are used in step (2). Although conceptually the procedure is very reasonable, \citet{smith2012} do not explicitly check these simplifications, but rather validate the entire approximation by comparing their residual lensed $B$ amplitude against a more exact computation given for several configurations in Table I of \citet{seljak2004} and finding agreement at the $\approx10\%$ level. The numbers computed in \citet{seljak2004} in turn come from computing an approximate marginal MAP estimate of $\phi$ and using this for delensing, with error bars on the delensed $B$ power computed via Monte Carlo. \citet{carron2017} further sharpen up this result by performing the same test with their exact maximization procedure rather than an approximate one, finding good agreement. Once the residual lensed $B$ mode power spectrum is computed, the residual modes are approximated as isotropic and Gaussian, and a traditional power spectrum Fisher forecast is computed for $r$ \cite{smith2012}, or a more sophisticated simulated power spectrum analysis is performed \cite{abazajian2016}.

Our procedure allows us to validate the CMB-S4 forecasting procedure in a much more direct and straightforward way than the long chain of validation steps above, by simply comparing against the Fisher information on $r$ that we derive. This also tests a few remaining assumptions in the CMB-S4-like method, mainly that the residual $B$ modes are Gaussian, that minimal information is lost by the $\ell\,{\lesssim}\,150$ filter, and that the impact of masking is only a reduction in the number of modes which can be captured by an $f_{\rm sky}$ factor. This latter assumption has never been checked but is particularly worrisome, since masking couples modes across $\ell$ and will leak $E$ into $B$, mimicking lensing.

For configuration \configB, we have computed forecasts using the CMB-S4-like procedure described above, accounting for all experimental details listed in Table~\ref{table:configurations} except for the mask, which is instead treated with an $f_{\rm sky}$ factor. Our results are summarized in Fig.~\ref{fig:fisher_information}. One can see the excellent visual agreement between the results from our chains and those from the CMB-S4-like forecast for all values of $r_{\rm fid}$ tested. For $r_{\rm fid}\,{=}\,[0.02,\,0.04]$ where we can compute accurate numerical derivatives, our exact Fisher calculation gives $\sigma_r=1/\sqrt{\mathcal{F}_{rr}}=[0.0067,\,0.0106]$ as compared to the CMB-S4-like forecasts which give $\sigma_r=1/\sqrt{\mathcal{F}_{rr}}=[0.0072,\,0.0111]$, or a difference of only 4\% and 8\%, respectively. This excellent agreement is further proof of the fidelity of existing $r$ forecasts for CMB-S4 \cite{abazajian2016}  and of other current and future forecasts using this same method. We note, though, that this does not necessarily imply that implementing a real analysis pipeline following the heuristic CMB-S4-like treatment would yield an unbiased estimate of $r$, only that this gives very accurate error bars as a forecasting procedure. Our chains, however, could be used to check this in the future.

\section{Concluding Remarks}

\subsection{The CMBLensing.jl package}
\label{sec:code}

Throughout the development of our sampling algorithm, we have used two branches of code in parallel. The first code was initially used to produce the chains presented in the previous sections. The second code, \textsc{CMBLensing.jl}\,\href{https://github.com/marius311/CMBLensing.jl}{\faGithub}, was developed for wider-spread use and is now faster, and is what we recommend for anyone wishing to use, reproduce, or extend our results. The two have been checked for agreement. 

The design of \textsc{CMBLensing.jl} was motivated by the desire for: 1) the ability to transparently run the code on CPUs or GPUs, 2) access to automatic differentiation so that gradients of our posterior or of any future modifications do not need to be hand-coded, and 3) no sacrifice on performance. To our knowledge, only two truly practical avenues exist to achieve this: either describing the posterior as a neural network-like graph in a machine learning library such as \textsc{TensorFlow}, or writing our code in \textsc{Julia} \cite{bezanson2017}. We have chosen the latter as it allows writing normal high-level code and avoids the additional complexity involved in translating our algorithm into the language of computational graphs.

\begin{figure}
\begin{centering}
\includegraphics[width=\columnwidth]{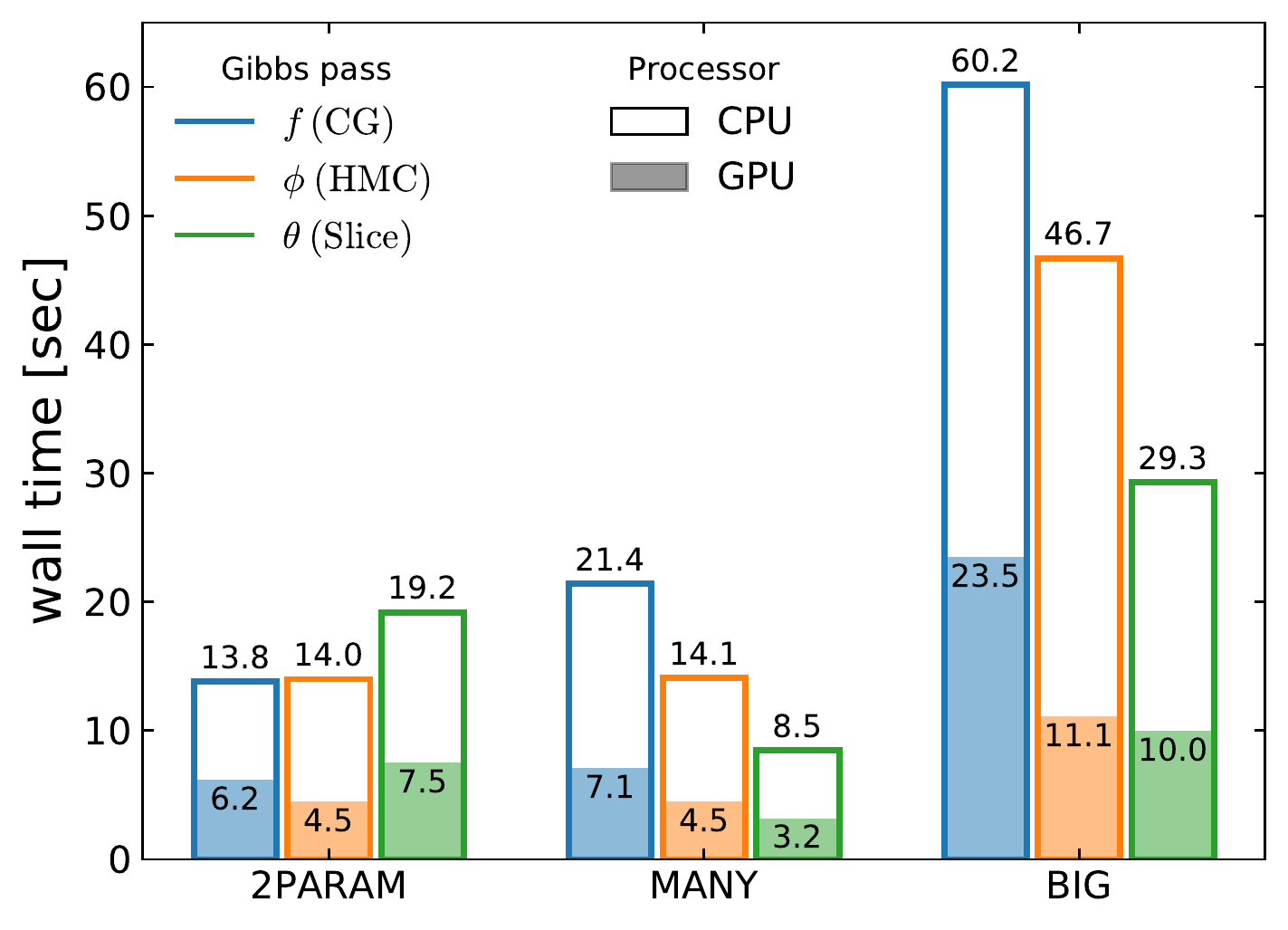}
\caption{The wall-time in seconds for each Gibbs pass, for each configuration (see Table~\ref{table:configurations}), and for running on CPU vs. GPU. The CPU benchmarks utilize a full NERSC Cori Haswell node (Intel Xeon Processor E5-2698 v3), and the GPU benchmarks a single NVIDIA GTX 1080Ti GPU. Although our CPU code is highly optimized, our GPU code likely has room for significant improvement, despite already being faster.}
\label{fig:timing}
\end{centering}
\end{figure}

As a simple example of the ease of this approach, consider the first order Taylor series expansion for lensing, i.e. $f(x+\nabla\phi) \approx f + \nabla\phi \cdot \nabla f.$ This can be written succinctly and true to the underlying mathematical expression in \textsc{CMBLensing.jl} as
\begin{center}
\includegraphics[height=0.45cm]{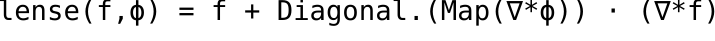}
\end{center}
\vspace{-0.3cm}
and the resulting function is no slower than having written out the necessary \texttt{FFT}s and array multiplications by hand. The arguments of this function are \textsc{CMBLensing.jl} field objects which are just thin wrappers around arrays storing the maps or Fourier coefficients for the fields. Depending on a user setting, these arrays can reside on CPU or NVIDIA GPU, and the above code works transparently in either case. \textsc{Julia} GPU integration is such that only $30$ lines of GPU-specific code are needed in the entire codebase. Fig.~\ref{fig:timing} summarizes the timing for each step in our Gibbs sampler and compares the CPU and GPU performance. We reach improvements in performance of factors of several when running on GPUs\footnote{There is a large dependence on GPU hardware; for example, our experience is that laptop-grade GPUs offer little to no improvement, in contrast to the more performant GPU used in Fig.~\ref{fig:timing}.}, and, encouragingly, the relative improvement grows as we go to larger maps. Additionally, the GPU code is not particularly optimized yet so we expect room for significant improvement, despite it already outperforming the highly optimized CPU code.

Once a function like \texttt{lense} is defined, source-to-source reverse-mode automatic differentiation can be used to compute gradients for most functions on $\mathbb{R}^n\rightarrow\mathbb{R}^1$ which use \texttt{lense} anywhere within their evaluation \cite{innes2018}. Here is a very simple example which takes a gradient with respect to $\phi$, evaluated at $\phi\,{=}\,0$:
\begin{center}
\includegraphics[height=0.5cm]{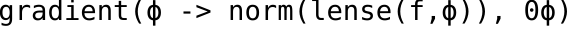}
\end{center}
\vspace{-0.35cm}
Both above code snippets are unmodified from what could be run in a real \textsc{Julia} session. The flexibility afforded by this system is invaluable to the type of quick exploration which was necessary in arriving at the results in this paper and which will be necessary for applying these methods to increasingly complex datasets moving forward. This package should serve as a useful tool for the CMB lensing community in the future, or as a ``black-box" target function (and gradient) for a broader audience wishing to try other inference methods on the CMB lensing problem.

\subsection{Brief summary of main results}
In this work, we have developed a method for joint inference of cosmological parameters, unlensed CMB fields, and the gravitational lensing potential, from CMB temperature and polarization data. By working with the Bayesian posterior, we are guaranteed to have extracted all available information from the data, hence our a, and (very) loosely corresponds to what is sometimes referred to as ``iterative delensing.'' Although several methods exist which can produce point estimates of the lensing potential which are lower-variance than the current state-of-the-art quadratic estimate (see Sec.~\ref{sec:intro}), our method is unique in making it completely straightforward how to actually extract cosmological information including uncertainty estimates from the lensing potential or from the delensed fields.

We have demonstrated this ability by jointly estimating $r$ and $\Aphi$ from simulated data. The analysis hinges on three key pieces, and without any one of them our results would not be possible. These are 1) reparametrizing the posterior to a new set of variables whose posterior distribution is more Gaussian and less degenerate 2) tuning our Monte Carlo sampler, in particular making use of HMC to sample the very high dimensional posterior which remains mildly non-Gaussian even after reparametrizing and 3) numerically implementing the lensing operation with \textsc{LenseFlow} which gives us the needed gradients through the inverse lensing operation, and allows to us to avoid an otherwise prohibitive determinant calculation.

We have used this method to arrive at two useful scientific results. First, we have explicitly demonstrated that the correlation between between $r$ and $\Aphi$ is small ($\rho\,{=}\,0.10$), showing that $r$ inference is not strongly limited by knowledge of the true lensing power spectrum amplitude. Second, we have given the first-ever exact computation of the Fisher information on $r$ in the context of delensing, even including several real instrumental effects, notably pixel masking. Using this, we have validated the $r$ forecasting procedure used for experiments such as CMB-S4, which has never been checked in the presence of pixel masking. Encouragingly, we find that the standard procedure yields results very close (within 8\% in terms of the uncertainty on $r$) to our exact Fisher calculation, giving further evidence that CMB-S4 delensing will work as expected.

\subsection{Future work and new possibilities}

The algorithm presented in this work is ready to be applied to current generation CMB data targeting deep observations over patches of sky of several hundreds of square degrees. There is ongoing work to apply these methods to South Pole Telescope data, and, as mentioned previously, they could also be applied to POLARBEAR data where it would be expected that the delensing effeciency achieved in \citet{adachi2019} could be even further improved.

The Bayesian sampling solution still has some challenges which need to be overcome before analyzing a dataset of the complexity expected from CMB-S4. One main future challenge is simply scaling up the number of pixels and moving beyond the flat-sky approximation to deal with sky curvature. Conceptually, it is completely straightforward to include sky curvature in our method. In terms of performance, the chains presented here run in 24-48 hours on one GPU, and scaling up to nearly full-sky observations will likely require improving this by a few factors of ten. Part of this can be trivially gained by running more chains in parallel, which we have not done here but should work well given that we do not find very significant chain burn-in time is necessary. It seems very possible that the remaining improvements could come from some combination of optimizing the GPU code, discovering even better reparametrizations, accelerating Wiener filtering, and going beyond the very basic HMC sampling algorithm we have used. 

Another challenge which must be tackled is the inclusion of foregrounds. A simple solution which may work well is simply to run our procedure on component separated maps. A more ambitious approach would be to compute a full forward model for the foregrounds and jointly infer them. This sounds difficult, but at least in the medium to small scale regime in polarization (which will be almost solely responsible for lensing reconstruction in the future), expected foregrounds are surprisingly small and simple. The only component expected to be significantly present is shot noise from radio galaxies \cite{crites2015}, which may be quite simple to forward model. We note that forward modeling the foregrounds may put an even bigger requirement on us to work with the joint posterior, because the analytic marginalization in Eqn.~\eqref{eq:marginalizejoint} is likely impossible in the presence of other non-Gaussian components.

Finally, we note that sampling is not the unique way to explore a Bayesian posterior, and many other methods exist which could potentially be accurate enough while being cheaper computationally. Some examples (but by no means an exhaustive list) include ``variational inference'' methods \cite{blei2017,seljak2019,knollmuller2019}, Laplace or higher-order approximations \cite{seljak2017}, or fall under the category of ``likelihood-free inference'' \cite{marin2011,price2018}. Many or all of these methods, however, rely on approximations which are extremely difficult to check in the context of the very high dimensional and non-Gaussian CMB lensing problem. By having explored and built intuition about the lensing posterior, and by having developed a sampling method which can be used on realistic-sized datasets to compute an compute an approximation-free answer, these other methods can, for the first time, be explicitly validated for lensing. If they prove to be sufficiently accurate, then perhaps they offer an advantageous way to perform this analysis in the future.

\begin{acknowledgments}
EA acknowledges support from NSF grants DMS-1252795, DMS-1812199 and a CARMIN research fellowship at IHES and IHP. BDW acknoledges support  from  the  BIG4 project, grant ANR-16-CE23-0002 of the French Agence Nationale de la Recherche (ANR). The Center for Computational Astrophysics is supported by the Simons Foundation. MM thanks Uros Seljak and Bill Holzapfel for useful discussions. 
\end{acknowledgments}

\onecolumngrid
\appendix

\bibliography{notes,marius,ben,Planck_bib}

\begin{thebibliography}{76}%
\makeatletter
\providecommand \@ifxundefined [1]{%
 \@ifx{#1\undefined}
}%
\providecommand \@ifnum [1]{%
 \ifnum #1\expandafter \@firstoftwo
 \else \expandafter \@secondoftwo
 \fi
}%
\providecommand \@ifx [1]{%
 \ifx #1\expandafter \@firstoftwo
 \else \expandafter \@secondoftwo
 \fi
}%
\providecommand \natexlab [1]{#1}%
\providecommand \enquote  [1]{``#1''}%
\providecommand \bibnamefont  [1]{#1}%
\providecommand \bibfnamefont [1]{#1}%
\providecommand \citenamefont [1]{#1}%
\providecommand \href@noop [0]{\@secondoftwo}%
\providecommand \href [0]{\begingroup \@sanitize@url \@href}%
\providecommand \@href[1]{\@@startlink{#1}\@@href}%
\providecommand \@@href[1]{\endgroup#1\@@endlink}%
\providecommand \@sanitize@url [0]{\catcode `\\12\catcode `\$12\catcode
  `\&12\catcode `\#12\catcode `\^12\catcode `\_12\catcode `\%12\relax}%
\providecommand \@@startlink[1]{}%
\providecommand \@@endlink[0]{}%
\providecommand \url  [0]{\begingroup\@sanitize@url \@url }%
\providecommand \@url [1]{\endgroup\@href {#1}{\urlprefix }}%
\providecommand \urlprefix  [0]{URL }%
\providecommand \Eprint [0]{\href }%
\providecommand \doibase [0]{http://dx.doi.org/}%
\providecommand \selectlanguage [0]{\@gobble}%
\providecommand \bibinfo  [0]{\@secondoftwo}%
\providecommand \bibfield  [0]{\@secondoftwo}%
\providecommand \translation [1]{[#1]}%
\providecommand \BibitemOpen [0]{}%
\providecommand \bibitemStop [0]{}%
\providecommand \bibitemNoStop [0]{.\EOS\space}%
\providecommand \EOS [0]{\spacefactor3000\relax}%
\providecommand \BibitemShut  [1]{\csname bibitem#1\endcsname}%
\let\auto@bib@innerbib\@empty
\bibitem [{\citenamefont {Benson}\ \emph {et~al.}(2014)\citenamefont {Benson},
  \citenamefont {Ade}, \citenamefont {Ahmed}, \citenamefont {Allen},
  \citenamefont {Arnold}, \citenamefont {Austermann}, \citenamefont {Bender},
  \citenamefont {Bleem}, \citenamefont {Carlstrom}, \citenamefont {Chang},
  \citenamefont {Cho}, \citenamefont {Ciocys}, \citenamefont {Cliche},
  \citenamefont {Crawford}, \citenamefont {Cukierman}, \citenamefont {{de
  Haan}}, \citenamefont {Dobbs}, \citenamefont {Dutcher}, \citenamefont
  {Everett}, \citenamefont {Gilbert}, \citenamefont {Halverson}, \citenamefont
  {Hanson}, \citenamefont {Harrington}, \citenamefont {Hattori}, \citenamefont
  {Henning}, \citenamefont {Hilton}, \citenamefont {Holder}, \citenamefont
  {Holzapfel}, \citenamefont {Irwin}, \citenamefont {Keisler}, \citenamefont
  {Knox}, \citenamefont {Kubik}, \citenamefont {Kuo}, \citenamefont {Lee},
  \citenamefont {Leitch}, \citenamefont {Li}, \citenamefont {McDonald},
  \citenamefont {Meyer}, \citenamefont {Montgomery}, \citenamefont {Myers},
  \citenamefont {Natoli}, \citenamefont {Nguyen}, \citenamefont {Novosad},
  \citenamefont {Padin}, \citenamefont {Pan}, \citenamefont {Pearson},
  \citenamefont {Reichardt}, \citenamefont {Ruhl}, \citenamefont {Saliwanchik},
  \citenamefont {Simard}, \citenamefont {Smecher}, \citenamefont {Sayre},
  \citenamefont {Shirokoff}, \citenamefont {Stark}, \citenamefont {Story},
  \citenamefont {Suzuki}, \citenamefont {Thompson}, \citenamefont {Tucker},
  \citenamefont {Vanderlinde}, \citenamefont {Vieira}, \citenamefont
  {Vikhlinin}, \citenamefont {Wang}, \citenamefont {Yefremenko},\ and\
  \citenamefont {Yoon}}]{benson2014}%
  \BibitemOpen
  \bibfield  {author} {\bibinfo {author} {\bibfnamefont {B.~A.}\ \bibnamefont
  {Benson}}, \bibinfo {author} {\bibfnamefont {P.~A.~R.}\ \bibnamefont {Ade}},
  \bibinfo {author} {\bibfnamefont {Z.}~\bibnamefont {Ahmed}}, \bibinfo
  {author} {\bibfnamefont {S.~W.}\ \bibnamefont {Allen}}, \bibinfo {author}
  {\bibfnamefont {K.}~\bibnamefont {Arnold}}, \bibinfo {author} {\bibfnamefont
  {J.~E.}\ \bibnamefont {Austermann}}, \bibinfo {author} {\bibfnamefont
  {A.~N.}\ \bibnamefont {Bender}}, \bibinfo {author} {\bibfnamefont {L.~E.}\
  \bibnamefont {Bleem}}, \bibinfo {author} {\bibfnamefont {J.~E.}\ \bibnamefont
  {Carlstrom}}, \bibinfo {author} {\bibfnamefont {C.~L.}\ \bibnamefont
  {Chang}}, \bibinfo {author} {\bibfnamefont {H.~M.}\ \bibnamefont {Cho}},
  \bibinfo {author} {\bibfnamefont {S.~T.}\ \bibnamefont {Ciocys}}, \bibinfo
  {author} {\bibfnamefont {J.~F.}\ \bibnamefont {Cliche}}, \bibinfo {author}
  {\bibfnamefont {T.~M.}\ \bibnamefont {Crawford}}, \bibinfo {author}
  {\bibfnamefont {A.}~\bibnamefont {Cukierman}}, \bibinfo {author}
  {\bibfnamefont {T.}~\bibnamefont {{de Haan}}}, \bibinfo {author}
  {\bibfnamefont {M.~A.}\ \bibnamefont {Dobbs}}, \bibinfo {author}
  {\bibfnamefont {D.}~\bibnamefont {Dutcher}}, \bibinfo {author} {\bibfnamefont
  {W.}~\bibnamefont {Everett}}, \bibinfo {author} {\bibfnamefont
  {A.}~\bibnamefont {Gilbert}}, \bibinfo {author} {\bibfnamefont {N.~W.}\
  \bibnamefont {Halverson}}, \bibinfo {author} {\bibfnamefont {D.}~\bibnamefont
  {Hanson}}, \bibinfo {author} {\bibfnamefont {N.~L.}\ \bibnamefont
  {Harrington}}, \bibinfo {author} {\bibfnamefont {K.}~\bibnamefont {Hattori}},
  \bibinfo {author} {\bibfnamefont {J.~W.}\ \bibnamefont {Henning}}, \bibinfo
  {author} {\bibfnamefont {G.~C.}\ \bibnamefont {Hilton}}, \bibinfo {author}
  {\bibfnamefont {G.~P.}\ \bibnamefont {Holder}}, \bibinfo {author}
  {\bibfnamefont {W.~L.}\ \bibnamefont {Holzapfel}}, \bibinfo {author}
  {\bibfnamefont {K.~D.}\ \bibnamefont {Irwin}}, \bibinfo {author}
  {\bibfnamefont {R.}~\bibnamefont {Keisler}}, \bibinfo {author} {\bibfnamefont
  {L.}~\bibnamefont {Knox}}, \bibinfo {author} {\bibfnamefont {D.}~\bibnamefont
  {Kubik}}, \bibinfo {author} {\bibfnamefont {C.~L.}\ \bibnamefont {Kuo}},
  \bibinfo {author} {\bibfnamefont {A.~T.}\ \bibnamefont {Lee}}, \bibinfo
  {author} {\bibfnamefont {E.~M.}\ \bibnamefont {Leitch}}, \bibinfo {author}
  {\bibfnamefont {D.}~\bibnamefont {Li}}, \bibinfo {author} {\bibfnamefont
  {M.}~\bibnamefont {McDonald}}, \bibinfo {author} {\bibfnamefont {S.~S.}\
  \bibnamefont {Meyer}}, \bibinfo {author} {\bibfnamefont {J.}~\bibnamefont
  {Montgomery}}, \bibinfo {author} {\bibfnamefont {M.}~\bibnamefont {Myers}},
  \bibinfo {author} {\bibfnamefont {T.}~\bibnamefont {Natoli}}, \bibinfo
  {author} {\bibfnamefont {H.}~\bibnamefont {Nguyen}}, \bibinfo {author}
  {\bibfnamefont {V.}~\bibnamefont {Novosad}}, \bibinfo {author} {\bibfnamefont
  {S.}~\bibnamefont {Padin}}, \bibinfo {author} {\bibfnamefont
  {Z.}~\bibnamefont {Pan}}, \bibinfo {author} {\bibfnamefont {J.}~\bibnamefont
  {Pearson}}, \bibinfo {author} {\bibfnamefont {C.~L.}\ \bibnamefont
  {Reichardt}}, \bibinfo {author} {\bibfnamefont {J.~E.}\ \bibnamefont {Ruhl}},
  \bibinfo {author} {\bibfnamefont {B.~R.}\ \bibnamefont {Saliwanchik}},
  \bibinfo {author} {\bibfnamefont {G.}~\bibnamefont {Simard}}, \bibinfo
  {author} {\bibfnamefont {G.}~\bibnamefont {Smecher}}, \bibinfo {author}
  {\bibfnamefont {J.~T.}\ \bibnamefont {Sayre}}, \bibinfo {author}
  {\bibfnamefont {E.}~\bibnamefont {Shirokoff}}, \bibinfo {author}
  {\bibfnamefont {A.~A.}\ \bibnamefont {Stark}}, \bibinfo {author}
  {\bibfnamefont {K.}~\bibnamefont {Story}}, \bibinfo {author} {\bibfnamefont
  {A.}~\bibnamefont {Suzuki}}, \bibinfo {author} {\bibfnamefont {K.~L.}\
  \bibnamefont {Thompson}}, \bibinfo {author} {\bibfnamefont {C.}~\bibnamefont
  {Tucker}}, \bibinfo {author} {\bibfnamefont {K.}~\bibnamefont {Vanderlinde}},
  \bibinfo {author} {\bibfnamefont {J.~D.}\ \bibnamefont {Vieira}}, \bibinfo
  {author} {\bibfnamefont {A.}~\bibnamefont {Vikhlinin}}, \bibinfo {author}
  {\bibfnamefont {G.}~\bibnamefont {Wang}}, \bibinfo {author} {\bibfnamefont
  {V.}~\bibnamefont {Yefremenko}}, \ and\ \bibinfo {author} {\bibfnamefont
  {K.~W.}\ \bibnamefont {Yoon}},\ }\href {\doibase 10.1117/12.2057305}
  {\bibfield  {journal} {\bibinfo  {journal} {arXiv:1407.2973 [astro-ph]}\ ,\
  \bibinfo {pages} {91531P}} (\bibinfo {year} {2014})},\ \Eprint
  {http://arxiv.org/abs/1407.2973} {arXiv:1407.2973 [astro-ph]} \BibitemShut
  {NoStop}%
\bibitem [{\citenamefont {Anderson}\ \emph {et~al.}(2018)\citenamefont
  {Anderson}, \citenamefont {Ade}, \citenamefont {Ahmed}, \citenamefont
  {Austermann}, \citenamefont {Avva}, \citenamefont {Barry}, \citenamefont
  {Thakur}, \citenamefont {Bender}, \citenamefont {Benson}, \citenamefont
  {Bleem}, \citenamefont {Byrum}, \citenamefont {Carlstrom}, \citenamefont
  {Carter}, \citenamefont {Cecil}, \citenamefont {Chang}, \citenamefont {Cho},
  \citenamefont {Cliche}, \citenamefont {Crawford}, \citenamefont {Cukierman},
  \citenamefont {Denison}, \citenamefont {{de Haan}}, \citenamefont {Ding},
  \citenamefont {Dobbs}, \citenamefont {Dutcher}, \citenamefont {Everett},
  \citenamefont {Foster}, \citenamefont {Gannon}, \citenamefont {Gilbert},
  \citenamefont {Groh}, \citenamefont {Halverson}, \citenamefont
  {{Harke-Hosemann}}, \citenamefont {Harrington}, \citenamefont {Henning},
  \citenamefont {Hilton}, \citenamefont {Holder}, \citenamefont {Holzapfel},
  \citenamefont {Huang}, \citenamefont {Irwin}, \citenamefont {Jeong},
  \citenamefont {Jonas}, \citenamefont {Khaire}, \citenamefont {Knox},
  \citenamefont {Kofman}, \citenamefont {Korman}, \citenamefont {Kubik},
  \citenamefont {Kuhlmann}, \citenamefont {Kuklev}, \citenamefont {Kuo},
  \citenamefont {Lee}, \citenamefont {Leitch}, \citenamefont {Lowitz},
  \citenamefont {Meyer}, \citenamefont {Michalik}, \citenamefont {Montgomery},
  \citenamefont {Nadolski}, \citenamefont {Natoli}, \citenamefont {Nguyen},
  \citenamefont {Noble}, \citenamefont {Novosad}, \citenamefont {Padin},
  \citenamefont {Pan}, \citenamefont {Pearson}, \citenamefont {Posada},
  \citenamefont {Rahlin}, \citenamefont {Reichardt}, \citenamefont {Ruhl},
  \citenamefont {Saunders}, \citenamefont {Sayre}, \citenamefont {Shirley},
  \citenamefont {Shirokoff}, \citenamefont {Smecher}, \citenamefont {Sobrin},
  \citenamefont {Stark}, \citenamefont {Story}, \citenamefont {Suzuki},
  \citenamefont {Tang}, \citenamefont {Thompson}, \citenamefont {Tucker},
  \citenamefont {Vale}, \citenamefont {Vanderlinde}, \citenamefont {Vieira},
  \citenamefont {Wang}, \citenamefont {Whitehorn}, \citenamefont {Yefremenko},
  \citenamefont {Yoon},\ and\ \citenamefont {Young}}]{anderson2018}%
  \BibitemOpen
  \bibfield  {author} {\bibinfo {author} {\bibfnamefont {A.~J.}\ \bibnamefont
  {Anderson}}, \bibinfo {author} {\bibfnamefont {P.~A.~R.}\ \bibnamefont
  {Ade}}, \bibinfo {author} {\bibfnamefont {Z.}~\bibnamefont {Ahmed}}, \bibinfo
  {author} {\bibfnamefont {J.~E.}\ \bibnamefont {Austermann}}, \bibinfo
  {author} {\bibfnamefont {J.~S.}\ \bibnamefont {Avva}}, \bibinfo {author}
  {\bibfnamefont {P.~S.}\ \bibnamefont {Barry}}, \bibinfo {author}
  {\bibfnamefont {R.~B.}\ \bibnamefont {Thakur}}, \bibinfo {author}
  {\bibfnamefont {A.~N.}\ \bibnamefont {Bender}}, \bibinfo {author}
  {\bibfnamefont {B.~A.}\ \bibnamefont {Benson}}, \bibinfo {author}
  {\bibfnamefont {L.~E.}\ \bibnamefont {Bleem}}, \bibinfo {author}
  {\bibfnamefont {K.}~\bibnamefont {Byrum}}, \bibinfo {author} {\bibfnamefont
  {J.~E.}\ \bibnamefont {Carlstrom}}, \bibinfo {author} {\bibfnamefont {F.~W.}\
  \bibnamefont {Carter}}, \bibinfo {author} {\bibfnamefont {T.}~\bibnamefont
  {Cecil}}, \bibinfo {author} {\bibfnamefont {C.~L.}\ \bibnamefont {Chang}},
  \bibinfo {author} {\bibfnamefont {H.~M.}\ \bibnamefont {Cho}}, \bibinfo
  {author} {\bibfnamefont {J.~F.}\ \bibnamefont {Cliche}}, \bibinfo {author}
  {\bibfnamefont {T.~M.}\ \bibnamefont {Crawford}}, \bibinfo {author}
  {\bibfnamefont {A.}~\bibnamefont {Cukierman}}, \bibinfo {author}
  {\bibfnamefont {E.~V.}\ \bibnamefont {Denison}}, \bibinfo {author}
  {\bibfnamefont {T.}~\bibnamefont {{de Haan}}}, \bibinfo {author}
  {\bibfnamefont {J.}~\bibnamefont {Ding}}, \bibinfo {author} {\bibfnamefont
  {M.~A.}\ \bibnamefont {Dobbs}}, \bibinfo {author} {\bibfnamefont
  {D.}~\bibnamefont {Dutcher}}, \bibinfo {author} {\bibfnamefont
  {W.}~\bibnamefont {Everett}}, \bibinfo {author} {\bibfnamefont
  {A.}~\bibnamefont {Foster}}, \bibinfo {author} {\bibfnamefont {R.~N.}\
  \bibnamefont {Gannon}}, \bibinfo {author} {\bibfnamefont {A.}~\bibnamefont
  {Gilbert}}, \bibinfo {author} {\bibfnamefont {J.~C.}\ \bibnamefont {Groh}},
  \bibinfo {author} {\bibfnamefont {N.~W.}\ \bibnamefont {Halverson}}, \bibinfo
  {author} {\bibfnamefont {A.~H.}\ \bibnamefont {{Harke-Hosemann}}}, \bibinfo
  {author} {\bibfnamefont {N.~L.}\ \bibnamefont {Harrington}}, \bibinfo
  {author} {\bibfnamefont {J.~W.}\ \bibnamefont {Henning}}, \bibinfo {author}
  {\bibfnamefont {G.~C.}\ \bibnamefont {Hilton}}, \bibinfo {author}
  {\bibfnamefont {G.~P.}\ \bibnamefont {Holder}}, \bibinfo {author}
  {\bibfnamefont {W.~L.}\ \bibnamefont {Holzapfel}}, \bibinfo {author}
  {\bibfnamefont {N.}~\bibnamefont {Huang}}, \bibinfo {author} {\bibfnamefont
  {K.~D.}\ \bibnamefont {Irwin}}, \bibinfo {author} {\bibfnamefont {O.~B.}\
  \bibnamefont {Jeong}}, \bibinfo {author} {\bibfnamefont {M.}~\bibnamefont
  {Jonas}}, \bibinfo {author} {\bibfnamefont {T.}~\bibnamefont {Khaire}},
  \bibinfo {author} {\bibfnamefont {L.}~\bibnamefont {Knox}}, \bibinfo {author}
  {\bibfnamefont {A.~M.}\ \bibnamefont {Kofman}}, \bibinfo {author}
  {\bibfnamefont {M.}~\bibnamefont {Korman}}, \bibinfo {author} {\bibfnamefont
  {D.}~\bibnamefont {Kubik}}, \bibinfo {author} {\bibfnamefont
  {S.}~\bibnamefont {Kuhlmann}}, \bibinfo {author} {\bibfnamefont
  {N.}~\bibnamefont {Kuklev}}, \bibinfo {author} {\bibfnamefont {C.~L.}\
  \bibnamefont {Kuo}}, \bibinfo {author} {\bibfnamefont {A.~T.}\ \bibnamefont
  {Lee}}, \bibinfo {author} {\bibfnamefont {E.~M.}\ \bibnamefont {Leitch}},
  \bibinfo {author} {\bibfnamefont {A.~E.}\ \bibnamefont {Lowitz}}, \bibinfo
  {author} {\bibfnamefont {S.~S.}\ \bibnamefont {Meyer}}, \bibinfo {author}
  {\bibfnamefont {D.}~\bibnamefont {Michalik}}, \bibinfo {author}
  {\bibfnamefont {J.}~\bibnamefont {Montgomery}}, \bibinfo {author}
  {\bibfnamefont {A.}~\bibnamefont {Nadolski}}, \bibinfo {author}
  {\bibfnamefont {T.}~\bibnamefont {Natoli}}, \bibinfo {author} {\bibfnamefont
  {H.}~\bibnamefont {Nguyen}}, \bibinfo {author} {\bibfnamefont {G.~I.}\
  \bibnamefont {Noble}}, \bibinfo {author} {\bibfnamefont {V.}~\bibnamefont
  {Novosad}}, \bibinfo {author} {\bibfnamefont {S.}~\bibnamefont {Padin}},
  \bibinfo {author} {\bibfnamefont {Z.}~\bibnamefont {Pan}}, \bibinfo {author}
  {\bibfnamefont {J.}~\bibnamefont {Pearson}}, \bibinfo {author} {\bibfnamefont
  {C.~M.}\ \bibnamefont {Posada}}, \bibinfo {author} {\bibfnamefont
  {A.}~\bibnamefont {Rahlin}}, \bibinfo {author} {\bibfnamefont {C.~L.}\
  \bibnamefont {Reichardt}}, \bibinfo {author} {\bibfnamefont {J.~E.}\
  \bibnamefont {Ruhl}}, \bibinfo {author} {\bibfnamefont {L.~J.}\ \bibnamefont
  {Saunders}}, \bibinfo {author} {\bibfnamefont {J.~T.}\ \bibnamefont {Sayre}},
  \bibinfo {author} {\bibfnamefont {I.}~\bibnamefont {Shirley}}, \bibinfo
  {author} {\bibfnamefont {E.}~\bibnamefont {Shirokoff}}, \bibinfo {author}
  {\bibfnamefont {G.}~\bibnamefont {Smecher}}, \bibinfo {author} {\bibfnamefont
  {J.~A.}\ \bibnamefont {Sobrin}}, \bibinfo {author} {\bibfnamefont {A.~A.}\
  \bibnamefont {Stark}}, \bibinfo {author} {\bibfnamefont {K.~T.}\ \bibnamefont
  {Story}}, \bibinfo {author} {\bibfnamefont {A.}~\bibnamefont {Suzuki}},
  \bibinfo {author} {\bibfnamefont {Q.~Y.}\ \bibnamefont {Tang}}, \bibinfo
  {author} {\bibfnamefont {K.~L.}\ \bibnamefont {Thompson}}, \bibinfo {author}
  {\bibfnamefont {C.}~\bibnamefont {Tucker}}, \bibinfo {author} {\bibfnamefont
  {L.~R.}\ \bibnamefont {Vale}}, \bibinfo {author} {\bibfnamefont
  {K.}~\bibnamefont {Vanderlinde}}, \bibinfo {author} {\bibfnamefont {J.~D.}\
  \bibnamefont {Vieira}}, \bibinfo {author} {\bibfnamefont {G.}~\bibnamefont
  {Wang}}, \bibinfo {author} {\bibfnamefont {N.}~\bibnamefont {Whitehorn}},
  \bibinfo {author} {\bibfnamefont {V.}~\bibnamefont {Yefremenko}}, \bibinfo
  {author} {\bibfnamefont {K.~W.}\ \bibnamefont {Yoon}}, \ and\ \bibinfo
  {author} {\bibfnamefont {M.~R.}\ \bibnamefont {Young}},\ }\href {\doibase
  10.1007/s10909-018-2007-z} {\bibfield  {journal} {\bibinfo  {journal}
  {Journal of Low Temperature Physics}\ }\textbf {\bibinfo {volume} {193}},\
  \bibinfo {pages} {1057} (\bibinfo {year} {2018})}\BibitemShut {NoStop}%
\bibitem [{\citenamefont {Henderson}\ \emph {et~al.}(2016)\citenamefont
  {Henderson}, \citenamefont {Allison}, \citenamefont {Austermann},
  \citenamefont {Baildon}, \citenamefont {Battaglia}, \citenamefont {Beall},
  \citenamefont {Becker}, \citenamefont {De~Bernardis}, \citenamefont {Bond},
  \citenamefont {Calabrese}, \citenamefont {Choi}, \citenamefont {Coughlin},
  \citenamefont {Crowley}, \citenamefont {Datta}, \citenamefont {Devlin},
  \citenamefont {Duff}, \citenamefont {Dunkley}, \citenamefont {D{\"u}nner},
  \citenamefont {{van Engelen}}, \citenamefont {Gallardo}, \citenamefont
  {Grace}, \citenamefont {Hasselfield}, \citenamefont {Hills}, \citenamefont
  {Hilton}, \citenamefont {Hincks}, \citenamefont {Hlo{\^z}ek}, \citenamefont
  {Ho}, \citenamefont {Hubmayr}, \citenamefont {Huffenberger}, \citenamefont
  {Hughes}, \citenamefont {Irwin}, \citenamefont {Koopman}, \citenamefont
  {Kosowsky}, \citenamefont {Li}, \citenamefont {McMahon}, \citenamefont
  {Munson}, \citenamefont {Nati}, \citenamefont {Newburgh}, \citenamefont
  {Niemack}, \citenamefont {Niraula}, \citenamefont {Page}, \citenamefont
  {Pappas}, \citenamefont {Salatino}, \citenamefont {Schillaci}, \citenamefont
  {Schmitt}, \citenamefont {Sehgal}, \citenamefont {Sherwin}, \citenamefont
  {Sievers}, \citenamefont {Simon}, \citenamefont {Spergel}, \citenamefont
  {Staggs}, \citenamefont {Stevens}, \citenamefont {Thornton}, \citenamefont
  {Van~Lanen}, \citenamefont {Vavagiakis}, \citenamefont {Ward},\ and\
  \citenamefont {Wollack}}]{henderson2016}%
  \BibitemOpen
  \bibfield  {author} {\bibinfo {author} {\bibfnamefont {S.~W.}\ \bibnamefont
  {Henderson}}, \bibinfo {author} {\bibfnamefont {R.}~\bibnamefont {Allison}},
  \bibinfo {author} {\bibfnamefont {J.}~\bibnamefont {Austermann}}, \bibinfo
  {author} {\bibfnamefont {T.}~\bibnamefont {Baildon}}, \bibinfo {author}
  {\bibfnamefont {N.}~\bibnamefont {Battaglia}}, \bibinfo {author}
  {\bibfnamefont {J.~A.}\ \bibnamefont {Beall}}, \bibinfo {author}
  {\bibfnamefont {D.}~\bibnamefont {Becker}}, \bibinfo {author} {\bibfnamefont
  {F.}~\bibnamefont {De~Bernardis}}, \bibinfo {author} {\bibfnamefont {J.~R.}\
  \bibnamefont {Bond}}, \bibinfo {author} {\bibfnamefont {E.}~\bibnamefont
  {Calabrese}}, \bibinfo {author} {\bibfnamefont {S.~K.}\ \bibnamefont {Choi}},
  \bibinfo {author} {\bibfnamefont {K.~P.}\ \bibnamefont {Coughlin}}, \bibinfo
  {author} {\bibfnamefont {K.~T.}\ \bibnamefont {Crowley}}, \bibinfo {author}
  {\bibfnamefont {R.}~\bibnamefont {Datta}}, \bibinfo {author} {\bibfnamefont
  {M.~J.}\ \bibnamefont {Devlin}}, \bibinfo {author} {\bibfnamefont {S.~M.}\
  \bibnamefont {Duff}}, \bibinfo {author} {\bibfnamefont {J.}~\bibnamefont
  {Dunkley}}, \bibinfo {author} {\bibfnamefont {R.}~\bibnamefont {D{\"u}nner}},
  \bibinfo {author} {\bibfnamefont {A.}~\bibnamefont {{van Engelen}}}, \bibinfo
  {author} {\bibfnamefont {P.~A.}\ \bibnamefont {Gallardo}}, \bibinfo {author}
  {\bibfnamefont {E.}~\bibnamefont {Grace}}, \bibinfo {author} {\bibfnamefont
  {M.}~\bibnamefont {Hasselfield}}, \bibinfo {author} {\bibfnamefont
  {F.}~\bibnamefont {Hills}}, \bibinfo {author} {\bibfnamefont {G.~C.}\
  \bibnamefont {Hilton}}, \bibinfo {author} {\bibfnamefont {A.~D.}\
  \bibnamefont {Hincks}}, \bibinfo {author} {\bibfnamefont {R.}~\bibnamefont
  {Hlo{\^z}ek}}, \bibinfo {author} {\bibfnamefont {S.~P.}\ \bibnamefont {Ho}},
  \bibinfo {author} {\bibfnamefont {J.}~\bibnamefont {Hubmayr}}, \bibinfo
  {author} {\bibfnamefont {K.}~\bibnamefont {Huffenberger}}, \bibinfo {author}
  {\bibfnamefont {J.~P.}\ \bibnamefont {Hughes}}, \bibinfo {author}
  {\bibfnamefont {K.~D.}\ \bibnamefont {Irwin}}, \bibinfo {author}
  {\bibfnamefont {B.~J.}\ \bibnamefont {Koopman}}, \bibinfo {author}
  {\bibfnamefont {A.~B.}\ \bibnamefont {Kosowsky}}, \bibinfo {author}
  {\bibfnamefont {D.}~\bibnamefont {Li}}, \bibinfo {author} {\bibfnamefont
  {J.}~\bibnamefont {McMahon}}, \bibinfo {author} {\bibfnamefont
  {C.}~\bibnamefont {Munson}}, \bibinfo {author} {\bibfnamefont
  {F.}~\bibnamefont {Nati}}, \bibinfo {author} {\bibfnamefont {L.}~\bibnamefont
  {Newburgh}}, \bibinfo {author} {\bibfnamefont {M.~D.}\ \bibnamefont
  {Niemack}}, \bibinfo {author} {\bibfnamefont {P.}~\bibnamefont {Niraula}},
  \bibinfo {author} {\bibfnamefont {L.~A.}\ \bibnamefont {Page}}, \bibinfo
  {author} {\bibfnamefont {C.~G.}\ \bibnamefont {Pappas}}, \bibinfo {author}
  {\bibfnamefont {M.}~\bibnamefont {Salatino}}, \bibinfo {author}
  {\bibfnamefont {A.}~\bibnamefont {Schillaci}}, \bibinfo {author}
  {\bibfnamefont {B.~L.}\ \bibnamefont {Schmitt}}, \bibinfo {author}
  {\bibfnamefont {N.}~\bibnamefont {Sehgal}}, \bibinfo {author} {\bibfnamefont
  {B.~D.}\ \bibnamefont {Sherwin}}, \bibinfo {author} {\bibfnamefont {J.~L.}\
  \bibnamefont {Sievers}}, \bibinfo {author} {\bibfnamefont {S.~M.}\
  \bibnamefont {Simon}}, \bibinfo {author} {\bibfnamefont {D.~N.}\ \bibnamefont
  {Spergel}}, \bibinfo {author} {\bibfnamefont {S.~T.}\ \bibnamefont {Staggs}},
  \bibinfo {author} {\bibfnamefont {J.~R.}\ \bibnamefont {Stevens}}, \bibinfo
  {author} {\bibfnamefont {R.}~\bibnamefont {Thornton}}, \bibinfo {author}
  {\bibfnamefont {J.}~\bibnamefont {Van~Lanen}}, \bibinfo {author}
  {\bibfnamefont {E.~M.}\ \bibnamefont {Vavagiakis}}, \bibinfo {author}
  {\bibfnamefont {J.~T.}\ \bibnamefont {Ward}}, \ and\ \bibinfo {author}
  {\bibfnamefont {E.~J.}\ \bibnamefont {Wollack}},\ }\href {\doibase
  10.1007/s10909-016-1575-z} {\bibfield  {journal} {\bibinfo  {journal}
  {Journal of Low Temperature Physics}\ }\textbf {\bibinfo {volume} {184}},\
  \bibinfo {pages} {772} (\bibinfo {year} {2016})}\BibitemShut {NoStop}%
\bibitem [{\citenamefont {Suzuki}\ \emph {et~al.}(2016)\citenamefont {Suzuki},
  \citenamefont {Ade}, \citenamefont {Akiba}, \citenamefont {Aleman},
  \citenamefont {Arnold}, \citenamefont {Baccigalupi}, \citenamefont {Barch},
  \citenamefont {Barron}, \citenamefont {Bender}, \citenamefont {Boettger},
  \citenamefont {Borrill}, \citenamefont {Chapman}, \citenamefont {Chinone},
  \citenamefont {Cukierman}, \citenamefont {Dobbs}, \citenamefont {Ducout},
  \citenamefont {Dunner}, \citenamefont {Elleflot}, \citenamefont {Errard},
  \citenamefont {Fabbian}, \citenamefont {Feeney}, \citenamefont {Feng},
  \citenamefont {Fujino}, \citenamefont {Fuller}, \citenamefont {Gilbert},
  \citenamefont {{Goeckner-Wald}}, \citenamefont {Groh}, \citenamefont {Haan},
  \citenamefont {Hall}, \citenamefont {Halverson}, \citenamefont {Hamada},
  \citenamefont {Hasegawa}, \citenamefont {Hattori}, \citenamefont {Hazumi},
  \citenamefont {Hill}, \citenamefont {Holzapfel}, \citenamefont {Hori},
  \citenamefont {Howe}, \citenamefont {Inoue}, \citenamefont {Irie},
  \citenamefont {Jaehnig}, \citenamefont {Jaffe}, \citenamefont {Jeong},
  \citenamefont {Katayama}, \citenamefont {Kaufman}, \citenamefont
  {Kazemzadeh}, \citenamefont {Keating}, \citenamefont {Kermish}, \citenamefont
  {Keskitalo}, \citenamefont {Kisner}, \citenamefont {Kusaka}, \citenamefont
  {Jeune}, \citenamefont {Lee}, \citenamefont {Leon}, \citenamefont {Linder},
  \citenamefont {Lowry}, \citenamefont {Matsuda}, \citenamefont {Matsumura},
  \citenamefont {Miller}, \citenamefont {Mizukami}, \citenamefont {Montgomery},
  \citenamefont {Navaroli}, \citenamefont {Nishino}, \citenamefont {Peloton},
  \citenamefont {Poletti}, \citenamefont {Puglisi}, \citenamefont {Rebeiz},
  \citenamefont {Raum}, \citenamefont {Reichardt}, \citenamefont {Richards},
  \citenamefont {Ross}, \citenamefont {Rotermund}, \citenamefont {Segawa},
  \citenamefont {Sherwin}, \citenamefont {Shirley}, \citenamefont
  {Siritanasak}, \citenamefont {Stebor}, \citenamefont {Stompor}, \citenamefont
  {Suzuki}, \citenamefont {Tajima}, \citenamefont {Takada}, \citenamefont
  {Takakura}, \citenamefont {Takatori}, \citenamefont {Tikhomirov},
  \citenamefont {Tomaru}, \citenamefont {Westbrook}, \citenamefont {Whitehorn},
  \citenamefont {Yamashita}, \citenamefont {Zahn},\ and\ \citenamefont
  {Zahn}}]{suzuki2016}%
  \BibitemOpen
  \bibfield  {author} {\bibinfo {author} {\bibfnamefont {A.}~\bibnamefont
  {Suzuki}}, \bibinfo {author} {\bibfnamefont {P.}~\bibnamefont {Ade}},
  \bibinfo {author} {\bibfnamefont {Y.}~\bibnamefont {Akiba}}, \bibinfo
  {author} {\bibfnamefont {C.}~\bibnamefont {Aleman}}, \bibinfo {author}
  {\bibfnamefont {K.}~\bibnamefont {Arnold}}, \bibinfo {author} {\bibfnamefont
  {C.}~\bibnamefont {Baccigalupi}}, \bibinfo {author} {\bibfnamefont
  {B.}~\bibnamefont {Barch}}, \bibinfo {author} {\bibfnamefont
  {D.}~\bibnamefont {Barron}}, \bibinfo {author} {\bibfnamefont
  {A.}~\bibnamefont {Bender}}, \bibinfo {author} {\bibfnamefont
  {D.}~\bibnamefont {Boettger}}, \bibinfo {author} {\bibfnamefont
  {J.}~\bibnamefont {Borrill}}, \bibinfo {author} {\bibfnamefont
  {S.}~\bibnamefont {Chapman}}, \bibinfo {author} {\bibfnamefont
  {Y.}~\bibnamefont {Chinone}}, \bibinfo {author} {\bibfnamefont
  {A.}~\bibnamefont {Cukierman}}, \bibinfo {author} {\bibfnamefont
  {M.}~\bibnamefont {Dobbs}}, \bibinfo {author} {\bibfnamefont
  {A.}~\bibnamefont {Ducout}}, \bibinfo {author} {\bibfnamefont
  {R.}~\bibnamefont {Dunner}}, \bibinfo {author} {\bibfnamefont
  {T.}~\bibnamefont {Elleflot}}, \bibinfo {author} {\bibfnamefont
  {J.}~\bibnamefont {Errard}}, \bibinfo {author} {\bibfnamefont
  {G.}~\bibnamefont {Fabbian}}, \bibinfo {author} {\bibfnamefont
  {S.}~\bibnamefont {Feeney}}, \bibinfo {author} {\bibfnamefont
  {C.}~\bibnamefont {Feng}}, \bibinfo {author} {\bibfnamefont {T.}~\bibnamefont
  {Fujino}}, \bibinfo {author} {\bibfnamefont {G.}~\bibnamefont {Fuller}},
  \bibinfo {author} {\bibfnamefont {A.}~\bibnamefont {Gilbert}}, \bibinfo
  {author} {\bibfnamefont {N.}~\bibnamefont {{Goeckner-Wald}}}, \bibinfo
  {author} {\bibfnamefont {J.}~\bibnamefont {Groh}}, \bibinfo {author}
  {\bibfnamefont {T.~D.}\ \bibnamefont {Haan}}, \bibinfo {author}
  {\bibfnamefont {G.}~\bibnamefont {Hall}}, \bibinfo {author} {\bibfnamefont
  {N.}~\bibnamefont {Halverson}}, \bibinfo {author} {\bibfnamefont
  {T.}~\bibnamefont {Hamada}}, \bibinfo {author} {\bibfnamefont
  {M.}~\bibnamefont {Hasegawa}}, \bibinfo {author} {\bibfnamefont
  {K.}~\bibnamefont {Hattori}}, \bibinfo {author} {\bibfnamefont
  {M.}~\bibnamefont {Hazumi}}, \bibinfo {author} {\bibfnamefont
  {C.}~\bibnamefont {Hill}}, \bibinfo {author} {\bibfnamefont {W.}~\bibnamefont
  {Holzapfel}}, \bibinfo {author} {\bibfnamefont {Y.}~\bibnamefont {Hori}},
  \bibinfo {author} {\bibfnamefont {L.}~\bibnamefont {Howe}}, \bibinfo {author}
  {\bibfnamefont {Y.}~\bibnamefont {Inoue}}, \bibinfo {author} {\bibfnamefont
  {F.}~\bibnamefont {Irie}}, \bibinfo {author} {\bibfnamefont {G.}~\bibnamefont
  {Jaehnig}}, \bibinfo {author} {\bibfnamefont {A.}~\bibnamefont {Jaffe}},
  \bibinfo {author} {\bibfnamefont {O.}~\bibnamefont {Jeong}}, \bibinfo
  {author} {\bibfnamefont {N.}~\bibnamefont {Katayama}}, \bibinfo {author}
  {\bibfnamefont {J.}~\bibnamefont {Kaufman}}, \bibinfo {author} {\bibfnamefont
  {K.}~\bibnamefont {Kazemzadeh}}, \bibinfo {author} {\bibfnamefont
  {B.}~\bibnamefont {Keating}}, \bibinfo {author} {\bibfnamefont
  {Z.}~\bibnamefont {Kermish}}, \bibinfo {author} {\bibfnamefont
  {R.}~\bibnamefont {Keskitalo}}, \bibinfo {author} {\bibfnamefont
  {T.}~\bibnamefont {Kisner}}, \bibinfo {author} {\bibfnamefont
  {A.}~\bibnamefont {Kusaka}}, \bibinfo {author} {\bibfnamefont {M.~L.}\
  \bibnamefont {Jeune}}, \bibinfo {author} {\bibfnamefont {A.}~\bibnamefont
  {Lee}}, \bibinfo {author} {\bibfnamefont {D.}~\bibnamefont {Leon}}, \bibinfo
  {author} {\bibfnamefont {E.}~\bibnamefont {Linder}}, \bibinfo {author}
  {\bibfnamefont {L.}~\bibnamefont {Lowry}}, \bibinfo {author} {\bibfnamefont
  {F.}~\bibnamefont {Matsuda}}, \bibinfo {author} {\bibfnamefont
  {T.}~\bibnamefont {Matsumura}}, \bibinfo {author} {\bibfnamefont
  {N.}~\bibnamefont {Miller}}, \bibinfo {author} {\bibfnamefont
  {K.}~\bibnamefont {Mizukami}}, \bibinfo {author} {\bibfnamefont
  {J.}~\bibnamefont {Montgomery}}, \bibinfo {author} {\bibfnamefont
  {M.}~\bibnamefont {Navaroli}}, \bibinfo {author} {\bibfnamefont
  {H.}~\bibnamefont {Nishino}}, \bibinfo {author} {\bibfnamefont
  {J.}~\bibnamefont {Peloton}}, \bibinfo {author} {\bibfnamefont
  {D.}~\bibnamefont {Poletti}}, \bibinfo {author} {\bibfnamefont
  {G.}~\bibnamefont {Puglisi}}, \bibinfo {author} {\bibfnamefont
  {G.}~\bibnamefont {Rebeiz}}, \bibinfo {author} {\bibfnamefont
  {C.}~\bibnamefont {Raum}}, \bibinfo {author} {\bibfnamefont {C.}~\bibnamefont
  {Reichardt}}, \bibinfo {author} {\bibfnamefont {P.}~\bibnamefont {Richards}},
  \bibinfo {author} {\bibfnamefont {C.}~\bibnamefont {Ross}}, \bibinfo {author}
  {\bibfnamefont {K.}~\bibnamefont {Rotermund}}, \bibinfo {author}
  {\bibfnamefont {Y.}~\bibnamefont {Segawa}}, \bibinfo {author} {\bibfnamefont
  {B.}~\bibnamefont {Sherwin}}, \bibinfo {author} {\bibfnamefont
  {I.}~\bibnamefont {Shirley}}, \bibinfo {author} {\bibfnamefont
  {P.}~\bibnamefont {Siritanasak}}, \bibinfo {author} {\bibfnamefont
  {N.}~\bibnamefont {Stebor}}, \bibinfo {author} {\bibfnamefont
  {R.}~\bibnamefont {Stompor}}, \bibinfo {author} {\bibfnamefont
  {J.}~\bibnamefont {Suzuki}}, \bibinfo {author} {\bibfnamefont
  {O.}~\bibnamefont {Tajima}}, \bibinfo {author} {\bibfnamefont
  {S.}~\bibnamefont {Takada}}, \bibinfo {author} {\bibfnamefont
  {S.}~\bibnamefont {Takakura}}, \bibinfo {author} {\bibfnamefont
  {S.}~\bibnamefont {Takatori}}, \bibinfo {author} {\bibfnamefont
  {A.}~\bibnamefont {Tikhomirov}}, \bibinfo {author} {\bibfnamefont
  {T.}~\bibnamefont {Tomaru}}, \bibinfo {author} {\bibfnamefont
  {B.}~\bibnamefont {Westbrook}}, \bibinfo {author} {\bibfnamefont
  {N.}~\bibnamefont {Whitehorn}}, \bibinfo {author} {\bibfnamefont
  {T.}~\bibnamefont {Yamashita}}, \bibinfo {author} {\bibfnamefont
  {A.}~\bibnamefont {Zahn}}, \ and\ \bibinfo {author} {\bibfnamefont
  {O.}~\bibnamefont {Zahn}},\ }\href {\doibase 10.1007/s10909-015-1425-4}
  {\bibfield  {journal} {\bibinfo  {journal} {Journal of Low Temperature
  Physics}\ }\textbf {\bibinfo {volume} {184}},\ \bibinfo {pages} {805}
  (\bibinfo {year} {2016})}\BibitemShut {NoStop}%
\bibitem [{\citenamefont {Collaboration}\ \emph {et~al.}(2018)\citenamefont
  {Collaboration}, \citenamefont {Ade}, \citenamefont {Aguirre}, \citenamefont
  {Ahmed}, \citenamefont {Aiola}, \citenamefont {Ali}, \citenamefont {Alonso},
  \citenamefont {Alvarez}, \citenamefont {Arnold}, \citenamefont {Ashton},
  \citenamefont {Austermann}, \citenamefont {Awan}, \citenamefont
  {Baccigalupi}, \citenamefont {Baildon}, \citenamefont {Barron}, \citenamefont
  {Battaglia}, \citenamefont {Battye}, \citenamefont {Baxter}, \citenamefont
  {Bazarko}, \citenamefont {Beall}, \citenamefont {Bean}, \citenamefont {Beck},
  \citenamefont {Beckman}, \citenamefont {Beringue}, \citenamefont {Bianchini},
  \citenamefont {Boada}, \citenamefont {Boettger}, \citenamefont {Bond},
  \citenamefont {Borrill}, \citenamefont {Brown}, \citenamefont {Bruno},
  \citenamefont {Bryan}, \citenamefont {Calabrese}, \citenamefont {Calafut},
  \citenamefont {Calisse}, \citenamefont {Carron}, \citenamefont {Challinor},
  \citenamefont {Chesmore}, \citenamefont {Chinone}, \citenamefont {Chluba},
  \citenamefont {Cho}, \citenamefont {Choi}, \citenamefont {Coppi},
  \citenamefont {Cothard}, \citenamefont {Coughlin}, \citenamefont {Crichton},
  \citenamefont {Crowley}, \citenamefont {Crowley}, \citenamefont {Cukierman},
  \citenamefont {D`Ewart}, \citenamefont {D{\"u}nner}, \citenamefont {{de
  Haan}}, \citenamefont {Devlin}, \citenamefont {Dicker}, \citenamefont
  {Didier}, \citenamefont {Dobbs}, \citenamefont {Dober}, \citenamefont
  {Duell}, \citenamefont {Duff}, \citenamefont {Duivenvoorden}, \citenamefont
  {Dunkley}, \citenamefont {Dusatko}, \citenamefont {Errard}, \citenamefont
  {Fabbian}, \citenamefont {Feeney}, \citenamefont {Ferraro}, \citenamefont
  {Flux{\`a}}, \citenamefont {Freese}, \citenamefont {Frisch}, \citenamefont
  {Frolov}, \citenamefont {Fuller}, \citenamefont {Fuzia}, \citenamefont
  {Galitzki}, \citenamefont {Gallardo}, \citenamefont {Ghersi}, \citenamefont
  {Gao}, \citenamefont {Gawiser}, \citenamefont {Gerbino}, \citenamefont
  {Gluscevic}, \citenamefont {{Goeckner-Wald}}, \citenamefont {Golec},
  \citenamefont {Gordon}, \citenamefont {Gralla}, \citenamefont {Green},
  \citenamefont {Grigorian}, \citenamefont {Groh}, \citenamefont {Groppi},
  \citenamefont {Guan}, \citenamefont {Gudmundsson}, \citenamefont {Han},
  \citenamefont {Hargrave}, \citenamefont {Hasegawa}, \citenamefont
  {Hasselfield}, \citenamefont {Hattori}, \citenamefont {Haynes}, \citenamefont
  {Hazumi}, \citenamefont {He}, \citenamefont {Healy}, \citenamefont
  {Henderson}, \citenamefont {{Hervias-Caimapo}}, \citenamefont {Hill},
  \citenamefont {Hill}, \citenamefont {Hilton}, \citenamefont {Hilton},
  \citenamefont {Hincks}, \citenamefont {Hinshaw}, \citenamefont {Hlo{\v z}ek},
  \citenamefont {Ho}, \citenamefont {Ho}, \citenamefont {Howe}, \citenamefont
  {Huang}, \citenamefont {Hubmayr}, \citenamefont {Huffenberger}, \citenamefont
  {Hughes}, \citenamefont {Ijjas}, \citenamefont {Ikape}, \citenamefont
  {Irwin}, \citenamefont {Jaffe}, \citenamefont {Jain}, \citenamefont {Jeong},
  \citenamefont {Kaneko}, \citenamefont {Karpel}, \citenamefont {Katayama},
  \citenamefont {Keating}, \citenamefont {Kernasovski}, \citenamefont
  {Keskitalo}, \citenamefont {Kisner}, \citenamefont {Kiuchi}, \citenamefont
  {Klein}, \citenamefont {Knowles}, \citenamefont {Koopman}, \citenamefont
  {Kosowsky}, \citenamefont {Krachmalnicoff}, \citenamefont {Kuenstner},
  \citenamefont {Kuo}, \citenamefont {Kusaka}, \citenamefont {Lashner},
  \citenamefont {Lee}, \citenamefont {Lee}, \citenamefont {Leon}, \citenamefont
  {Leung}, \citenamefont {Lewis}, \citenamefont {Li}, \citenamefont {Li},
  \citenamefont {Limon}, \citenamefont {Linder}, \citenamefont
  {{Lopez-Caraballo}}, \citenamefont {Louis}, \citenamefont {Lowry},
  \citenamefont {Lungu}, \citenamefont {Madhavacheril}, \citenamefont {Mak},
  \citenamefont {Maldonado}, \citenamefont {Mani}, \citenamefont {Mates},
  \citenamefont {Matsuda}, \citenamefont {Maurin}, \citenamefont {Mauskopf},
  \citenamefont {May}, \citenamefont {McCallum}, \citenamefont {McKenney},
  \citenamefont {McMahon}, \citenamefont {Meerburg}, \citenamefont {Meyers},
  \citenamefont {Miller}, \citenamefont {Mirmelstein}, \citenamefont {Moodley},
  \citenamefont {Munchmeyer}, \citenamefont {Munson}, \citenamefont {Naess},
  \citenamefont {Nati}, \citenamefont {Navaroli}, \citenamefont {Newburgh},
  \citenamefont {Nguyen}, \citenamefont {Niemack}, \citenamefont {Nishino},
  \citenamefont {{Orlowski-Scherer}}, \citenamefont {Page}, \citenamefont
  {Partridge}, \citenamefont {Peloton}, \citenamefont {Perrotta}, \citenamefont
  {Piccirillo}, \citenamefont {Pisano}, \citenamefont {Poletti}, \citenamefont
  {Puddu}, \citenamefont {Puglisi}, \citenamefont {Raum}, \citenamefont
  {Reichardt}, \citenamefont {Remazeilles}, \citenamefont {Rephaeli},
  \citenamefont {Riechers}, \citenamefont {Rojas}, \citenamefont {Roy},
  \citenamefont {Sadeh}, \citenamefont {Sakurai}, \citenamefont {Salatino},
  \citenamefont {Rao}, \citenamefont {Schaan}, \citenamefont {Schmittfull},
  \citenamefont {Sehgal}, \citenamefont {Seibert}, \citenamefont {Seljak},
  \citenamefont {Sherwin}, \citenamefont {Shimon}, \citenamefont {Sierra},
  \citenamefont {Sievers}, \citenamefont {Sikhosana}, \citenamefont
  {{Silva-Feaver}}, \citenamefont {Simon}, \citenamefont {Sinclair},
  \citenamefont {Siritanasak}, \citenamefont {Smith}, \citenamefont {Smith},
  \citenamefont {Spergel}, \citenamefont {Staggs}, \citenamefont {Stein},
  \citenamefont {Stevens}, \citenamefont {Stompor}, \citenamefont {Sudiwala},
  \citenamefont {Suzuki}, \citenamefont {Tajima}, \citenamefont {Takakura},
  \citenamefont {Teply}, \citenamefont {Thomas}, \citenamefont {Thorne},
  \citenamefont {Thornton}, \citenamefont {Trac}, \citenamefont {Tsai},
  \citenamefont {Tucker}, \citenamefont {Ullom}, \citenamefont {Vagnozzi},
  \citenamefont {{van Engelen}}, \citenamefont {Van~Lanen}, \citenamefont {{van
  Winkle}}, \citenamefont {Vavagiakis}, \citenamefont {Verg{\`e}s},
  \citenamefont {Vissers}, \citenamefont {Wagoner}, \citenamefont {Ward},
  \citenamefont {Westbrook}, \citenamefont {Whitehorn}, \citenamefont
  {Williams}, \citenamefont {Williams}, \citenamefont {Wollack}, \citenamefont
  {Xu}, \citenamefont {Ye}, \citenamefont {Yu}, \citenamefont {Yu},
  \citenamefont {Zago}, \citenamefont {Zhang},\ and\ \citenamefont
  {Zhu}}]{thesimonsobservatorycollaboration2018}%
  \BibitemOpen
  \bibfield  {author} {\bibinfo {author} {\bibfnamefont {T.~S.~O.}\
  \bibnamefont {Collaboration}}, \bibinfo {author} {\bibfnamefont
  {P.}~\bibnamefont {Ade}}, \bibinfo {author} {\bibfnamefont {J.}~\bibnamefont
  {Aguirre}}, \bibinfo {author} {\bibfnamefont {Z.}~\bibnamefont {Ahmed}},
  \bibinfo {author} {\bibfnamefont {S.}~\bibnamefont {Aiola}}, \bibinfo
  {author} {\bibfnamefont {A.}~\bibnamefont {Ali}}, \bibinfo {author}
  {\bibfnamefont {D.}~\bibnamefont {Alonso}}, \bibinfo {author} {\bibfnamefont
  {M.~A.}\ \bibnamefont {Alvarez}}, \bibinfo {author} {\bibfnamefont
  {K.}~\bibnamefont {Arnold}}, \bibinfo {author} {\bibfnamefont
  {P.}~\bibnamefont {Ashton}}, \bibinfo {author} {\bibfnamefont
  {J.}~\bibnamefont {Austermann}}, \bibinfo {author} {\bibfnamefont
  {H.}~\bibnamefont {Awan}}, \bibinfo {author} {\bibfnamefont {C.}~\bibnamefont
  {Baccigalupi}}, \bibinfo {author} {\bibfnamefont {T.}~\bibnamefont
  {Baildon}}, \bibinfo {author} {\bibfnamefont {D.}~\bibnamefont {Barron}},
  \bibinfo {author} {\bibfnamefont {N.}~\bibnamefont {Battaglia}}, \bibinfo
  {author} {\bibfnamefont {R.}~\bibnamefont {Battye}}, \bibinfo {author}
  {\bibfnamefont {E.}~\bibnamefont {Baxter}}, \bibinfo {author} {\bibfnamefont
  {A.}~\bibnamefont {Bazarko}}, \bibinfo {author} {\bibfnamefont {J.~A.}\
  \bibnamefont {Beall}}, \bibinfo {author} {\bibfnamefont {R.}~\bibnamefont
  {Bean}}, \bibinfo {author} {\bibfnamefont {D.}~\bibnamefont {Beck}}, \bibinfo
  {author} {\bibfnamefont {S.}~\bibnamefont {Beckman}}, \bibinfo {author}
  {\bibfnamefont {B.}~\bibnamefont {Beringue}}, \bibinfo {author}
  {\bibfnamefont {F.}~\bibnamefont {Bianchini}}, \bibinfo {author}
  {\bibfnamefont {S.}~\bibnamefont {Boada}}, \bibinfo {author} {\bibfnamefont
  {D.}~\bibnamefont {Boettger}}, \bibinfo {author} {\bibfnamefont {J.~R.}\
  \bibnamefont {Bond}}, \bibinfo {author} {\bibfnamefont {J.}~\bibnamefont
  {Borrill}}, \bibinfo {author} {\bibfnamefont {M.~L.}\ \bibnamefont {Brown}},
  \bibinfo {author} {\bibfnamefont {S.~M.}\ \bibnamefont {Bruno}}, \bibinfo
  {author} {\bibfnamefont {S.}~\bibnamefont {Bryan}}, \bibinfo {author}
  {\bibfnamefont {E.}~\bibnamefont {Calabrese}}, \bibinfo {author}
  {\bibfnamefont {V.}~\bibnamefont {Calafut}}, \bibinfo {author} {\bibfnamefont
  {P.}~\bibnamefont {Calisse}}, \bibinfo {author} {\bibfnamefont
  {J.}~\bibnamefont {Carron}}, \bibinfo {author} {\bibfnamefont
  {A.}~\bibnamefont {Challinor}}, \bibinfo {author} {\bibfnamefont
  {G.}~\bibnamefont {Chesmore}}, \bibinfo {author} {\bibfnamefont
  {Y.}~\bibnamefont {Chinone}}, \bibinfo {author} {\bibfnamefont
  {J.}~\bibnamefont {Chluba}}, \bibinfo {author} {\bibfnamefont {H.-M.~S.}\
  \bibnamefont {Cho}}, \bibinfo {author} {\bibfnamefont {S.}~\bibnamefont
  {Choi}}, \bibinfo {author} {\bibfnamefont {G.}~\bibnamefont {Coppi}},
  \bibinfo {author} {\bibfnamefont {N.~F.}\ \bibnamefont {Cothard}}, \bibinfo
  {author} {\bibfnamefont {K.}~\bibnamefont {Coughlin}}, \bibinfo {author}
  {\bibfnamefont {D.}~\bibnamefont {Crichton}}, \bibinfo {author}
  {\bibfnamefont {K.~D.}\ \bibnamefont {Crowley}}, \bibinfo {author}
  {\bibfnamefont {K.~T.}\ \bibnamefont {Crowley}}, \bibinfo {author}
  {\bibfnamefont {A.}~\bibnamefont {Cukierman}}, \bibinfo {author}
  {\bibfnamefont {M.}~\bibnamefont {D`Ewart}}, \bibinfo {author} {\bibfnamefont
  {R.}~\bibnamefont {D{\"u}nner}}, \bibinfo {author} {\bibfnamefont
  {T.}~\bibnamefont {{de Haan}}}, \bibinfo {author} {\bibfnamefont
  {M.}~\bibnamefont {Devlin}}, \bibinfo {author} {\bibfnamefont
  {S.}~\bibnamefont {Dicker}}, \bibinfo {author} {\bibfnamefont
  {J.}~\bibnamefont {Didier}}, \bibinfo {author} {\bibfnamefont
  {M.}~\bibnamefont {Dobbs}}, \bibinfo {author} {\bibfnamefont
  {B.}~\bibnamefont {Dober}}, \bibinfo {author} {\bibfnamefont
  {C.}~\bibnamefont {Duell}}, \bibinfo {author} {\bibfnamefont
  {S.}~\bibnamefont {Duff}}, \bibinfo {author} {\bibfnamefont {A.}~\bibnamefont
  {Duivenvoorden}}, \bibinfo {author} {\bibfnamefont {J.}~\bibnamefont
  {Dunkley}}, \bibinfo {author} {\bibfnamefont {J.}~\bibnamefont {Dusatko}},
  \bibinfo {author} {\bibfnamefont {J.}~\bibnamefont {Errard}}, \bibinfo
  {author} {\bibfnamefont {G.}~\bibnamefont {Fabbian}}, \bibinfo {author}
  {\bibfnamefont {S.}~\bibnamefont {Feeney}}, \bibinfo {author} {\bibfnamefont
  {S.}~\bibnamefont {Ferraro}}, \bibinfo {author} {\bibfnamefont
  {P.}~\bibnamefont {Flux{\`a}}}, \bibinfo {author} {\bibfnamefont
  {K.}~\bibnamefont {Freese}}, \bibinfo {author} {\bibfnamefont
  {J.}~\bibnamefont {Frisch}}, \bibinfo {author} {\bibfnamefont
  {A.}~\bibnamefont {Frolov}}, \bibinfo {author} {\bibfnamefont
  {G.}~\bibnamefont {Fuller}}, \bibinfo {author} {\bibfnamefont
  {B.}~\bibnamefont {Fuzia}}, \bibinfo {author} {\bibfnamefont
  {N.}~\bibnamefont {Galitzki}}, \bibinfo {author} {\bibfnamefont {P.~A.}\
  \bibnamefont {Gallardo}}, \bibinfo {author} {\bibfnamefont {J.~T.~G.}\
  \bibnamefont {Ghersi}}, \bibinfo {author} {\bibfnamefont {J.}~\bibnamefont
  {Gao}}, \bibinfo {author} {\bibfnamefont {E.}~\bibnamefont {Gawiser}},
  \bibinfo {author} {\bibfnamefont {M.}~\bibnamefont {Gerbino}}, \bibinfo
  {author} {\bibfnamefont {V.}~\bibnamefont {Gluscevic}}, \bibinfo {author}
  {\bibfnamefont {N.}~\bibnamefont {{Goeckner-Wald}}}, \bibinfo {author}
  {\bibfnamefont {J.}~\bibnamefont {Golec}}, \bibinfo {author} {\bibfnamefont
  {S.}~\bibnamefont {Gordon}}, \bibinfo {author} {\bibfnamefont
  {M.}~\bibnamefont {Gralla}}, \bibinfo {author} {\bibfnamefont
  {D.}~\bibnamefont {Green}}, \bibinfo {author} {\bibfnamefont
  {A.}~\bibnamefont {Grigorian}}, \bibinfo {author} {\bibfnamefont
  {J.}~\bibnamefont {Groh}}, \bibinfo {author} {\bibfnamefont {C.}~\bibnamefont
  {Groppi}}, \bibinfo {author} {\bibfnamefont {Y.}~\bibnamefont {Guan}},
  \bibinfo {author} {\bibfnamefont {J.~E.}\ \bibnamefont {Gudmundsson}},
  \bibinfo {author} {\bibfnamefont {D.}~\bibnamefont {Han}}, \bibinfo {author}
  {\bibfnamefont {P.}~\bibnamefont {Hargrave}}, \bibinfo {author}
  {\bibfnamefont {M.}~\bibnamefont {Hasegawa}}, \bibinfo {author}
  {\bibfnamefont {M.}~\bibnamefont {Hasselfield}}, \bibinfo {author}
  {\bibfnamefont {M.}~\bibnamefont {Hattori}}, \bibinfo {author} {\bibfnamefont
  {V.}~\bibnamefont {Haynes}}, \bibinfo {author} {\bibfnamefont
  {M.}~\bibnamefont {Hazumi}}, \bibinfo {author} {\bibfnamefont
  {Y.}~\bibnamefont {He}}, \bibinfo {author} {\bibfnamefont {E.}~\bibnamefont
  {Healy}}, \bibinfo {author} {\bibfnamefont {S.}~\bibnamefont {Henderson}},
  \bibinfo {author} {\bibfnamefont {C.}~\bibnamefont {{Hervias-Caimapo}}},
  \bibinfo {author} {\bibfnamefont {C.~A.}\ \bibnamefont {Hill}}, \bibinfo
  {author} {\bibfnamefont {J.~C.}\ \bibnamefont {Hill}}, \bibinfo {author}
  {\bibfnamefont {G.}~\bibnamefont {Hilton}}, \bibinfo {author} {\bibfnamefont
  {M.}~\bibnamefont {Hilton}}, \bibinfo {author} {\bibfnamefont {A.~D.}\
  \bibnamefont {Hincks}}, \bibinfo {author} {\bibfnamefont {G.}~\bibnamefont
  {Hinshaw}}, \bibinfo {author} {\bibfnamefont {R.}~\bibnamefont {Hlo{\v
  z}ek}}, \bibinfo {author} {\bibfnamefont {S.}~\bibnamefont {Ho}}, \bibinfo
  {author} {\bibfnamefont {S.-P.~P.}\ \bibnamefont {Ho}}, \bibinfo {author}
  {\bibfnamefont {L.}~\bibnamefont {Howe}}, \bibinfo {author} {\bibfnamefont
  {Z.}~\bibnamefont {Huang}}, \bibinfo {author} {\bibfnamefont
  {J.}~\bibnamefont {Hubmayr}}, \bibinfo {author} {\bibfnamefont
  {K.}~\bibnamefont {Huffenberger}}, \bibinfo {author} {\bibfnamefont {J.~P.}\
  \bibnamefont {Hughes}}, \bibinfo {author} {\bibfnamefont {A.}~\bibnamefont
  {Ijjas}}, \bibinfo {author} {\bibfnamefont {M.}~\bibnamefont {Ikape}},
  \bibinfo {author} {\bibfnamefont {K.}~\bibnamefont {Irwin}}, \bibinfo
  {author} {\bibfnamefont {A.~H.}\ \bibnamefont {Jaffe}}, \bibinfo {author}
  {\bibfnamefont {B.}~\bibnamefont {Jain}}, \bibinfo {author} {\bibfnamefont
  {O.}~\bibnamefont {Jeong}}, \bibinfo {author} {\bibfnamefont
  {D.}~\bibnamefont {Kaneko}}, \bibinfo {author} {\bibfnamefont
  {E.}~\bibnamefont {Karpel}}, \bibinfo {author} {\bibfnamefont
  {N.}~\bibnamefont {Katayama}}, \bibinfo {author} {\bibfnamefont
  {B.}~\bibnamefont {Keating}}, \bibinfo {author} {\bibfnamefont
  {S.}~\bibnamefont {Kernasovski}}, \bibinfo {author} {\bibfnamefont
  {R.}~\bibnamefont {Keskitalo}}, \bibinfo {author} {\bibfnamefont
  {T.}~\bibnamefont {Kisner}}, \bibinfo {author} {\bibfnamefont
  {K.}~\bibnamefont {Kiuchi}}, \bibinfo {author} {\bibfnamefont
  {J.}~\bibnamefont {Klein}}, \bibinfo {author} {\bibfnamefont
  {K.}~\bibnamefont {Knowles}}, \bibinfo {author} {\bibfnamefont
  {B.}~\bibnamefont {Koopman}}, \bibinfo {author} {\bibfnamefont
  {A.}~\bibnamefont {Kosowsky}}, \bibinfo {author} {\bibfnamefont
  {N.}~\bibnamefont {Krachmalnicoff}}, \bibinfo {author} {\bibfnamefont
  {S.}~\bibnamefont {Kuenstner}}, \bibinfo {author} {\bibfnamefont {C.-L.}\
  \bibnamefont {Kuo}}, \bibinfo {author} {\bibfnamefont {A.}~\bibnamefont
  {Kusaka}}, \bibinfo {author} {\bibfnamefont {J.}~\bibnamefont {Lashner}},
  \bibinfo {author} {\bibfnamefont {A.}~\bibnamefont {Lee}}, \bibinfo {author}
  {\bibfnamefont {E.}~\bibnamefont {Lee}}, \bibinfo {author} {\bibfnamefont
  {D.}~\bibnamefont {Leon}}, \bibinfo {author} {\bibfnamefont {J.~S.-Y.}\
  \bibnamefont {Leung}}, \bibinfo {author} {\bibfnamefont {A.}~\bibnamefont
  {Lewis}}, \bibinfo {author} {\bibfnamefont {Y.}~\bibnamefont {Li}}, \bibinfo
  {author} {\bibfnamefont {Z.}~\bibnamefont {Li}}, \bibinfo {author}
  {\bibfnamefont {M.}~\bibnamefont {Limon}}, \bibinfo {author} {\bibfnamefont
  {E.}~\bibnamefont {Linder}}, \bibinfo {author} {\bibfnamefont
  {C.}~\bibnamefont {{Lopez-Caraballo}}}, \bibinfo {author} {\bibfnamefont
  {T.}~\bibnamefont {Louis}}, \bibinfo {author} {\bibfnamefont
  {L.}~\bibnamefont {Lowry}}, \bibinfo {author} {\bibfnamefont
  {M.}~\bibnamefont {Lungu}}, \bibinfo {author} {\bibfnamefont
  {M.}~\bibnamefont {Madhavacheril}}, \bibinfo {author} {\bibfnamefont
  {D.}~\bibnamefont {Mak}}, \bibinfo {author} {\bibfnamefont {F.}~\bibnamefont
  {Maldonado}}, \bibinfo {author} {\bibfnamefont {H.}~\bibnamefont {Mani}},
  \bibinfo {author} {\bibfnamefont {B.}~\bibnamefont {Mates}}, \bibinfo
  {author} {\bibfnamefont {F.}~\bibnamefont {Matsuda}}, \bibinfo {author}
  {\bibfnamefont {L.}~\bibnamefont {Maurin}}, \bibinfo {author} {\bibfnamefont
  {P.}~\bibnamefont {Mauskopf}}, \bibinfo {author} {\bibfnamefont
  {A.}~\bibnamefont {May}}, \bibinfo {author} {\bibfnamefont {N.}~\bibnamefont
  {McCallum}}, \bibinfo {author} {\bibfnamefont {C.}~\bibnamefont {McKenney}},
  \bibinfo {author} {\bibfnamefont {J.}~\bibnamefont {McMahon}}, \bibinfo
  {author} {\bibfnamefont {P.~D.}\ \bibnamefont {Meerburg}}, \bibinfo {author}
  {\bibfnamefont {J.}~\bibnamefont {Meyers}}, \bibinfo {author} {\bibfnamefont
  {A.}~\bibnamefont {Miller}}, \bibinfo {author} {\bibfnamefont
  {M.}~\bibnamefont {Mirmelstein}}, \bibinfo {author} {\bibfnamefont
  {K.}~\bibnamefont {Moodley}}, \bibinfo {author} {\bibfnamefont
  {M.}~\bibnamefont {Munchmeyer}}, \bibinfo {author} {\bibfnamefont
  {C.}~\bibnamefont {Munson}}, \bibinfo {author} {\bibfnamefont
  {S.}~\bibnamefont {Naess}}, \bibinfo {author} {\bibfnamefont
  {F.}~\bibnamefont {Nati}}, \bibinfo {author} {\bibfnamefont {M.}~\bibnamefont
  {Navaroli}}, \bibinfo {author} {\bibfnamefont {L.}~\bibnamefont {Newburgh}},
  \bibinfo {author} {\bibfnamefont {H.~N.}\ \bibnamefont {Nguyen}}, \bibinfo
  {author} {\bibfnamefont {M.}~\bibnamefont {Niemack}}, \bibinfo {author}
  {\bibfnamefont {H.}~\bibnamefont {Nishino}}, \bibinfo {author} {\bibfnamefont
  {J.}~\bibnamefont {{Orlowski-Scherer}}}, \bibinfo {author} {\bibfnamefont
  {L.}~\bibnamefont {Page}}, \bibinfo {author} {\bibfnamefont {B.}~\bibnamefont
  {Partridge}}, \bibinfo {author} {\bibfnamefont {J.}~\bibnamefont {Peloton}},
  \bibinfo {author} {\bibfnamefont {F.}~\bibnamefont {Perrotta}}, \bibinfo
  {author} {\bibfnamefont {L.}~\bibnamefont {Piccirillo}}, \bibinfo {author}
  {\bibfnamefont {G.}~\bibnamefont {Pisano}}, \bibinfo {author} {\bibfnamefont
  {D.}~\bibnamefont {Poletti}}, \bibinfo {author} {\bibfnamefont
  {R.}~\bibnamefont {Puddu}}, \bibinfo {author} {\bibfnamefont
  {G.}~\bibnamefont {Puglisi}}, \bibinfo {author} {\bibfnamefont
  {C.}~\bibnamefont {Raum}}, \bibinfo {author} {\bibfnamefont {C.~L.}\
  \bibnamefont {Reichardt}}, \bibinfo {author} {\bibfnamefont {M.}~\bibnamefont
  {Remazeilles}}, \bibinfo {author} {\bibfnamefont {Y.}~\bibnamefont
  {Rephaeli}}, \bibinfo {author} {\bibfnamefont {D.}~\bibnamefont {Riechers}},
  \bibinfo {author} {\bibfnamefont {F.}~\bibnamefont {Rojas}}, \bibinfo
  {author} {\bibfnamefont {A.}~\bibnamefont {Roy}}, \bibinfo {author}
  {\bibfnamefont {S.}~\bibnamefont {Sadeh}}, \bibinfo {author} {\bibfnamefont
  {Y.}~\bibnamefont {Sakurai}}, \bibinfo {author} {\bibfnamefont
  {M.}~\bibnamefont {Salatino}}, \bibinfo {author} {\bibfnamefont {M.~S.}\
  \bibnamefont {Rao}}, \bibinfo {author} {\bibfnamefont {E.}~\bibnamefont
  {Schaan}}, \bibinfo {author} {\bibfnamefont {M.}~\bibnamefont {Schmittfull}},
  \bibinfo {author} {\bibfnamefont {N.}~\bibnamefont {Sehgal}}, \bibinfo
  {author} {\bibfnamefont {J.}~\bibnamefont {Seibert}}, \bibinfo {author}
  {\bibfnamefont {U.}~\bibnamefont {Seljak}}, \bibinfo {author} {\bibfnamefont
  {B.}~\bibnamefont {Sherwin}}, \bibinfo {author} {\bibfnamefont
  {M.}~\bibnamefont {Shimon}}, \bibinfo {author} {\bibfnamefont
  {C.}~\bibnamefont {Sierra}}, \bibinfo {author} {\bibfnamefont
  {J.}~\bibnamefont {Sievers}}, \bibinfo {author} {\bibfnamefont
  {P.}~\bibnamefont {Sikhosana}}, \bibinfo {author} {\bibfnamefont
  {M.}~\bibnamefont {{Silva-Feaver}}}, \bibinfo {author} {\bibfnamefont
  {S.~M.}\ \bibnamefont {Simon}}, \bibinfo {author} {\bibfnamefont
  {A.}~\bibnamefont {Sinclair}}, \bibinfo {author} {\bibfnamefont
  {P.}~\bibnamefont {Siritanasak}}, \bibinfo {author} {\bibfnamefont
  {K.}~\bibnamefont {Smith}}, \bibinfo {author} {\bibfnamefont
  {S.}~\bibnamefont {Smith}}, \bibinfo {author} {\bibfnamefont
  {D.}~\bibnamefont {Spergel}}, \bibinfo {author} {\bibfnamefont
  {S.}~\bibnamefont {Staggs}}, \bibinfo {author} {\bibfnamefont
  {G.}~\bibnamefont {Stein}}, \bibinfo {author} {\bibfnamefont {J.~R.}\
  \bibnamefont {Stevens}}, \bibinfo {author} {\bibfnamefont {R.}~\bibnamefont
  {Stompor}}, \bibinfo {author} {\bibfnamefont {R.}~\bibnamefont {Sudiwala}},
  \bibinfo {author} {\bibfnamefont {A.}~\bibnamefont {Suzuki}}, \bibinfo
  {author} {\bibfnamefont {O.}~\bibnamefont {Tajima}}, \bibinfo {author}
  {\bibfnamefont {S.}~\bibnamefont {Takakura}}, \bibinfo {author}
  {\bibfnamefont {G.}~\bibnamefont {Teply}}, \bibinfo {author} {\bibfnamefont
  {D.~B.}\ \bibnamefont {Thomas}}, \bibinfo {author} {\bibfnamefont
  {B.}~\bibnamefont {Thorne}}, \bibinfo {author} {\bibfnamefont
  {R.}~\bibnamefont {Thornton}}, \bibinfo {author} {\bibfnamefont
  {H.}~\bibnamefont {Trac}}, \bibinfo {author} {\bibfnamefont {C.}~\bibnamefont
  {Tsai}}, \bibinfo {author} {\bibfnamefont {C.}~\bibnamefont {Tucker}},
  \bibinfo {author} {\bibfnamefont {J.}~\bibnamefont {Ullom}}, \bibinfo
  {author} {\bibfnamefont {S.}~\bibnamefont {Vagnozzi}}, \bibinfo {author}
  {\bibfnamefont {A.}~\bibnamefont {{van Engelen}}}, \bibinfo {author}
  {\bibfnamefont {J.}~\bibnamefont {Van~Lanen}}, \bibinfo {author}
  {\bibfnamefont {D.}~\bibnamefont {{van Winkle}}}, \bibinfo {author}
  {\bibfnamefont {E.~M.}\ \bibnamefont {Vavagiakis}}, \bibinfo {author}
  {\bibfnamefont {C.}~\bibnamefont {Verg{\`e}s}}, \bibinfo {author}
  {\bibfnamefont {M.}~\bibnamefont {Vissers}}, \bibinfo {author} {\bibfnamefont
  {K.}~\bibnamefont {Wagoner}}, \bibinfo {author} {\bibfnamefont
  {J.}~\bibnamefont {Ward}}, \bibinfo {author} {\bibfnamefont {B.}~\bibnamefont
  {Westbrook}}, \bibinfo {author} {\bibfnamefont {N.}~\bibnamefont
  {Whitehorn}}, \bibinfo {author} {\bibfnamefont {J.}~\bibnamefont {Williams}},
  \bibinfo {author} {\bibfnamefont {J.}~\bibnamefont {Williams}}, \bibinfo
  {author} {\bibfnamefont {E.~J.}\ \bibnamefont {Wollack}}, \bibinfo {author}
  {\bibfnamefont {Z.}~\bibnamefont {Xu}}, \bibinfo {author} {\bibfnamefont
  {J.}~\bibnamefont {Ye}}, \bibinfo {author} {\bibfnamefont {B.}~\bibnamefont
  {Yu}}, \bibinfo {author} {\bibfnamefont {C.}~\bibnamefont {Yu}}, \bibinfo
  {author} {\bibfnamefont {F.}~\bibnamefont {Zago}}, \bibinfo {author}
  {\bibfnamefont {H.}~\bibnamefont {Zhang}}, \ and\ \bibinfo {author}
  {\bibfnamefont {N.}~\bibnamefont {Zhu}},\ }\href
  {http://arxiv.org/abs/1808.07445} {\bibfield  {journal} {\bibinfo  {journal}
  {arXiv:1808.07445 [astro-ph]}\ } (\bibinfo {year} {2018})},\ \Eprint
  {http://arxiv.org/abs/1808.07445} {arXiv:1808.07445 [astro-ph]} \BibitemShut
  {NoStop}%
\bibitem [{\citenamefont {Abitbol}\ \emph {et~al.}(2017)\citenamefont
  {Abitbol}, \citenamefont {Ahmed}, \citenamefont {Barron}, \citenamefont
  {Basu~Thakur}, \citenamefont {Bender}, \citenamefont {Benson}, \citenamefont
  {Bischoff}, \citenamefont {Bryan}, \citenamefont {Carlstrom}, \citenamefont
  {Chang}, \citenamefont {Chuss}, \citenamefont {Cukierman}, \citenamefont {{de
  Haan}}, \citenamefont {Dobbs}, \citenamefont {{Essinger-Hileman}},
  \citenamefont {Filippini}, \citenamefont {Ganga}, \citenamefont
  {Gudmundsson}, \citenamefont {Halverson}, \citenamefont {Hanany},
  \citenamefont {Henderson}, \citenamefont {Hill}, \citenamefont {Ho},
  \citenamefont {Hubmayr}, \citenamefont {Irwin}, \citenamefont {Jeong},
  \citenamefont {Johnson}, \citenamefont {Kernasovskiy}, \citenamefont {Kovac},
  \citenamefont {Kusaka}, \citenamefont {Lee}, \citenamefont {Maria},
  \citenamefont {Mauskopf}, \citenamefont {McMahon}, \citenamefont {Moncelsi},
  \citenamefont {Nadolski}, \citenamefont {Nagy}, \citenamefont {Niemack},
  \citenamefont {O'Brient}, \citenamefont {Padin}, \citenamefont {Parshley},
  \citenamefont {Pryke}, \citenamefont {Roe}, \citenamefont {Rostem},
  \citenamefont {Ruhl}, \citenamefont {Simon}, \citenamefont {Staggs},
  \citenamefont {Suzuki}, \citenamefont {Switzer}, \citenamefont {Thompson},
  \citenamefont {Timbie}, \citenamefont {Tucker}, \citenamefont {Vieira},
  \citenamefont {Vieregg}, \citenamefont {Westbrook}, \citenamefont {Wollack},
  \citenamefont {Yoon}, \citenamefont {Young},\ and\ \citenamefont
  {Young}}]{abitbol2017}%
  \BibitemOpen
  \bibfield  {author} {\bibinfo {author} {\bibfnamefont {M.~H.}\ \bibnamefont
  {Abitbol}}, \bibinfo {author} {\bibfnamefont {Z.}~\bibnamefont {Ahmed}},
  \bibinfo {author} {\bibfnamefont {D.}~\bibnamefont {Barron}}, \bibinfo
  {author} {\bibfnamefont {R.}~\bibnamefont {Basu~Thakur}}, \bibinfo {author}
  {\bibfnamefont {A.~N.}\ \bibnamefont {Bender}}, \bibinfo {author}
  {\bibfnamefont {B.~A.}\ \bibnamefont {Benson}}, \bibinfo {author}
  {\bibfnamefont {C.~A.}\ \bibnamefont {Bischoff}}, \bibinfo {author}
  {\bibfnamefont {S.~A.}\ \bibnamefont {Bryan}}, \bibinfo {author}
  {\bibfnamefont {J.~E.}\ \bibnamefont {Carlstrom}}, \bibinfo {author}
  {\bibfnamefont {C.~L.}\ \bibnamefont {Chang}}, \bibinfo {author}
  {\bibfnamefont {D.~T.}\ \bibnamefont {Chuss}}, \bibinfo {author}
  {\bibfnamefont {A.}~\bibnamefont {Cukierman}}, \bibinfo {author}
  {\bibfnamefont {T.}~\bibnamefont {{de Haan}}}, \bibinfo {author}
  {\bibfnamefont {M.}~\bibnamefont {Dobbs}}, \bibinfo {author} {\bibfnamefont
  {T.}~\bibnamefont {{Essinger-Hileman}}}, \bibinfo {author} {\bibfnamefont
  {J.~P.}\ \bibnamefont {Filippini}}, \bibinfo {author} {\bibfnamefont
  {K.}~\bibnamefont {Ganga}}, \bibinfo {author} {\bibfnamefont {J.~E.}\
  \bibnamefont {Gudmundsson}}, \bibinfo {author} {\bibfnamefont {N.~W.}\
  \bibnamefont {Halverson}}, \bibinfo {author} {\bibfnamefont {S.}~\bibnamefont
  {Hanany}}, \bibinfo {author} {\bibfnamefont {S.~W.}\ \bibnamefont
  {Henderson}}, \bibinfo {author} {\bibfnamefont {C.~A.}\ \bibnamefont {Hill}},
  \bibinfo {author} {\bibfnamefont {S.-P.~P.}\ \bibnamefont {Ho}}, \bibinfo
  {author} {\bibfnamefont {J.}~\bibnamefont {Hubmayr}}, \bibinfo {author}
  {\bibfnamefont {K.}~\bibnamefont {Irwin}}, \bibinfo {author} {\bibfnamefont
  {O.}~\bibnamefont {Jeong}}, \bibinfo {author} {\bibfnamefont {B.~R.}\
  \bibnamefont {Johnson}}, \bibinfo {author} {\bibfnamefont {S.~A.}\
  \bibnamefont {Kernasovskiy}}, \bibinfo {author} {\bibfnamefont {J.~M.}\
  \bibnamefont {Kovac}}, \bibinfo {author} {\bibfnamefont {A.}~\bibnamefont
  {Kusaka}}, \bibinfo {author} {\bibfnamefont {A.~T.}\ \bibnamefont {Lee}},
  \bibinfo {author} {\bibfnamefont {S.}~\bibnamefont {Maria}}, \bibinfo
  {author} {\bibfnamefont {P.}~\bibnamefont {Mauskopf}}, \bibinfo {author}
  {\bibfnamefont {J.~J.}\ \bibnamefont {McMahon}}, \bibinfo {author}
  {\bibfnamefont {L.}~\bibnamefont {Moncelsi}}, \bibinfo {author}
  {\bibfnamefont {A.~W.}\ \bibnamefont {Nadolski}}, \bibinfo {author}
  {\bibfnamefont {J.~M.}\ \bibnamefont {Nagy}}, \bibinfo {author}
  {\bibfnamefont {M.~D.}\ \bibnamefont {Niemack}}, \bibinfo {author}
  {\bibfnamefont {R.~C.}\ \bibnamefont {O'Brient}}, \bibinfo {author}
  {\bibfnamefont {S.}~\bibnamefont {Padin}}, \bibinfo {author} {\bibfnamefont
  {S.~C.}\ \bibnamefont {Parshley}}, \bibinfo {author} {\bibfnamefont
  {C.}~\bibnamefont {Pryke}}, \bibinfo {author} {\bibfnamefont {N.~A.}\
  \bibnamefont {Roe}}, \bibinfo {author} {\bibfnamefont {K.}~\bibnamefont
  {Rostem}}, \bibinfo {author} {\bibfnamefont {J.}~\bibnamefont {Ruhl}},
  \bibinfo {author} {\bibfnamefont {S.~M.}\ \bibnamefont {Simon}}, \bibinfo
  {author} {\bibfnamefont {S.~T.}\ \bibnamefont {Staggs}}, \bibinfo {author}
  {\bibfnamefont {A.}~\bibnamefont {Suzuki}}, \bibinfo {author} {\bibfnamefont
  {E.~R.}\ \bibnamefont {Switzer}}, \bibinfo {author} {\bibfnamefont {K.~L.}\
  \bibnamefont {Thompson}}, \bibinfo {author} {\bibfnamefont {P.}~\bibnamefont
  {Timbie}}, \bibinfo {author} {\bibfnamefont {G.~S.}\ \bibnamefont {Tucker}},
  \bibinfo {author} {\bibfnamefont {J.~D.}\ \bibnamefont {Vieira}}, \bibinfo
  {author} {\bibfnamefont {A.~G.}\ \bibnamefont {Vieregg}}, \bibinfo {author}
  {\bibfnamefont {B.}~\bibnamefont {Westbrook}}, \bibinfo {author}
  {\bibfnamefont {E.~J.}\ \bibnamefont {Wollack}}, \bibinfo {author}
  {\bibfnamefont {K.~W.}\ \bibnamefont {Yoon}}, \bibinfo {author}
  {\bibfnamefont {K.~S.}\ \bibnamefont {Young}}, \ and\ \bibinfo {author}
  {\bibfnamefont {E.~Y.}\ \bibnamefont {Young}},\ }\href
  {http://adsabs.harvard.edu/abs/2017arXiv170602464A} {\bibfield  {journal}
  {\bibinfo  {journal} {ArXiv e-prints}\ }\textbf {\bibinfo {volume} {1706}},\
  \bibinfo {pages} {arXiv:1706.02464} (\bibinfo {year} {2017})}\BibitemShut
  {NoStop}%
\bibitem [{\citenamefont {Hanany}\ \emph {et~al.}(2019)\citenamefont {Hanany},
  \citenamefont {Alvarez}, \citenamefont {Artis}, \citenamefont {Ashton},
  \citenamefont {Aumont}, \citenamefont {Aurlien}, \citenamefont {Banerji},
  \citenamefont {Barreiro}, \citenamefont {Bartlett}, \citenamefont {Basak},
  \citenamefont {Battaglia}, \citenamefont {Bock}, \citenamefont {Boddy},
  \citenamefont {Bonato}, \citenamefont {Borrill}, \citenamefont {Bouchet},
  \citenamefont {Boulanger}, \citenamefont {Burkhart}, \citenamefont {Chluba},
  \citenamefont {Chuss}, \citenamefont {Clark}, \citenamefont {Cooperrider},
  \citenamefont {Crill}, \citenamefont {De~Zotti}, \citenamefont
  {Delabrouille}, \citenamefont {Di~Valentino}, \citenamefont {Didier},
  \citenamefont {Dor{\'e}}, \citenamefont {Eriksen}, \citenamefont {Errard},
  \citenamefont {{Essinger-Hileman}}, \citenamefont {Feeney}, \citenamefont
  {Filippini}, \citenamefont {Fissel}, \citenamefont {Flauger}, \citenamefont
  {Fuskeland}, \citenamefont {Gluscevic}, \citenamefont {Gorski}, \citenamefont
  {Green}, \citenamefont {Hensley}, \citenamefont {Herranz}, \citenamefont
  {Hill}, \citenamefont {Hivon}, \citenamefont {Hlo{\v z}ek}, \citenamefont
  {Hubmayr}, \citenamefont {Johnson}, \citenamefont {Jones}, \citenamefont
  {Jones}, \citenamefont {Knox}, \citenamefont {Kogut}, \citenamefont
  {{L{\'o}pez-Caniego}}, \citenamefont {Lawrence}, \citenamefont {Lazarian},
  \citenamefont {Li}, \citenamefont {Madhavacheril}, \citenamefont {Melin},
  \citenamefont {Meyers}, \citenamefont {Murray}, \citenamefont {Negrello},
  \citenamefont {Novak}, \citenamefont {O'Brient}, \citenamefont {Paine},
  \citenamefont {Pearson}, \citenamefont {Pogosian}, \citenamefont {Pryke},
  \citenamefont {Puglisi}, \citenamefont {Remazeilles}, \citenamefont {Rocha},
  \citenamefont {Schmittfull}, \citenamefont {Scott}, \citenamefont {Shirron},
  \citenamefont {Stephens}, \citenamefont {Sutin}, \citenamefont {Tomasi},
  \citenamefont {Trangsrud}, \citenamefont {{van Engelen}}, \citenamefont
  {Vansyngel}, \citenamefont {Wehus}, \citenamefont {Wen}, \citenamefont {Xu},
  \citenamefont {Young},\ and\ \citenamefont {Zonca}}]{hanany2019}%
  \BibitemOpen
  \bibfield  {author} {\bibinfo {author} {\bibfnamefont {S.}~\bibnamefont
  {Hanany}}, \bibinfo {author} {\bibfnamefont {M.}~\bibnamefont {Alvarez}},
  \bibinfo {author} {\bibfnamefont {E.}~\bibnamefont {Artis}}, \bibinfo
  {author} {\bibfnamefont {P.}~\bibnamefont {Ashton}}, \bibinfo {author}
  {\bibfnamefont {J.}~\bibnamefont {Aumont}}, \bibinfo {author} {\bibfnamefont
  {R.}~\bibnamefont {Aurlien}}, \bibinfo {author} {\bibfnamefont
  {R.}~\bibnamefont {Banerji}}, \bibinfo {author} {\bibfnamefont {R.~B.}\
  \bibnamefont {Barreiro}}, \bibinfo {author} {\bibfnamefont {J.~G.}\
  \bibnamefont {Bartlett}}, \bibinfo {author} {\bibfnamefont {S.}~\bibnamefont
  {Basak}}, \bibinfo {author} {\bibfnamefont {N.}~\bibnamefont {Battaglia}},
  \bibinfo {author} {\bibfnamefont {J.}~\bibnamefont {Bock}}, \bibinfo {author}
  {\bibfnamefont {K.~K.}\ \bibnamefont {Boddy}}, \bibinfo {author}
  {\bibfnamefont {M.}~\bibnamefont {Bonato}}, \bibinfo {author} {\bibfnamefont
  {J.}~\bibnamefont {Borrill}}, \bibinfo {author} {\bibfnamefont
  {F.}~\bibnamefont {Bouchet}}, \bibinfo {author} {\bibfnamefont
  {F.}~\bibnamefont {Boulanger}}, \bibinfo {author} {\bibfnamefont
  {B.}~\bibnamefont {Burkhart}}, \bibinfo {author} {\bibfnamefont
  {J.}~\bibnamefont {Chluba}}, \bibinfo {author} {\bibfnamefont
  {D.}~\bibnamefont {Chuss}}, \bibinfo {author} {\bibfnamefont {S.~E.}\
  \bibnamefont {Clark}}, \bibinfo {author} {\bibfnamefont {J.}~\bibnamefont
  {Cooperrider}}, \bibinfo {author} {\bibfnamefont {B.~P.}\ \bibnamefont
  {Crill}}, \bibinfo {author} {\bibfnamefont {G.}~\bibnamefont {De~Zotti}},
  \bibinfo {author} {\bibfnamefont {J.}~\bibnamefont {Delabrouille}}, \bibinfo
  {author} {\bibfnamefont {E.}~\bibnamefont {Di~Valentino}}, \bibinfo {author}
  {\bibfnamefont {J.}~\bibnamefont {Didier}}, \bibinfo {author} {\bibfnamefont
  {O.}~\bibnamefont {Dor{\'e}}}, \bibinfo {author} {\bibfnamefont {H.~K.}\
  \bibnamefont {Eriksen}}, \bibinfo {author} {\bibfnamefont {J.}~\bibnamefont
  {Errard}}, \bibinfo {author} {\bibfnamefont {T.}~\bibnamefont
  {{Essinger-Hileman}}}, \bibinfo {author} {\bibfnamefont {S.}~\bibnamefont
  {Feeney}}, \bibinfo {author} {\bibfnamefont {J.}~\bibnamefont {Filippini}},
  \bibinfo {author} {\bibfnamefont {L.}~\bibnamefont {Fissel}}, \bibinfo
  {author} {\bibfnamefont {R.}~\bibnamefont {Flauger}}, \bibinfo {author}
  {\bibfnamefont {U.}~\bibnamefont {Fuskeland}}, \bibinfo {author}
  {\bibfnamefont {V.}~\bibnamefont {Gluscevic}}, \bibinfo {author}
  {\bibfnamefont {K.~M.}\ \bibnamefont {Gorski}}, \bibinfo {author}
  {\bibfnamefont {D.}~\bibnamefont {Green}}, \bibinfo {author} {\bibfnamefont
  {B.}~\bibnamefont {Hensley}}, \bibinfo {author} {\bibfnamefont
  {D.}~\bibnamefont {Herranz}}, \bibinfo {author} {\bibfnamefont {J.~C.}\
  \bibnamefont {Hill}}, \bibinfo {author} {\bibfnamefont {E.}~\bibnamefont
  {Hivon}}, \bibinfo {author} {\bibfnamefont {R.}~\bibnamefont {Hlo{\v z}ek}},
  \bibinfo {author} {\bibfnamefont {J.}~\bibnamefont {Hubmayr}}, \bibinfo
  {author} {\bibfnamefont {B.~R.}\ \bibnamefont {Johnson}}, \bibinfo {author}
  {\bibfnamefont {W.}~\bibnamefont {Jones}}, \bibinfo {author} {\bibfnamefont
  {T.}~\bibnamefont {Jones}}, \bibinfo {author} {\bibfnamefont
  {L.}~\bibnamefont {Knox}}, \bibinfo {author} {\bibfnamefont {A.}~\bibnamefont
  {Kogut}}, \bibinfo {author} {\bibfnamefont {M.}~\bibnamefont
  {{L{\'o}pez-Caniego}}}, \bibinfo {author} {\bibfnamefont {C.}~\bibnamefont
  {Lawrence}}, \bibinfo {author} {\bibfnamefont {A.}~\bibnamefont {Lazarian}},
  \bibinfo {author} {\bibfnamefont {Z.}~\bibnamefont {Li}}, \bibinfo {author}
  {\bibfnamefont {M.}~\bibnamefont {Madhavacheril}}, \bibinfo {author}
  {\bibfnamefont {J.-B.}\ \bibnamefont {Melin}}, \bibinfo {author}
  {\bibfnamefont {J.}~\bibnamefont {Meyers}}, \bibinfo {author} {\bibfnamefont
  {C.}~\bibnamefont {Murray}}, \bibinfo {author} {\bibfnamefont
  {M.}~\bibnamefont {Negrello}}, \bibinfo {author} {\bibfnamefont
  {G.}~\bibnamefont {Novak}}, \bibinfo {author} {\bibfnamefont
  {R.}~\bibnamefont {O'Brient}}, \bibinfo {author} {\bibfnamefont
  {C.}~\bibnamefont {Paine}}, \bibinfo {author} {\bibfnamefont
  {T.}~\bibnamefont {Pearson}}, \bibinfo {author} {\bibfnamefont
  {L.}~\bibnamefont {Pogosian}}, \bibinfo {author} {\bibfnamefont
  {C.}~\bibnamefont {Pryke}}, \bibinfo {author} {\bibfnamefont
  {G.}~\bibnamefont {Puglisi}}, \bibinfo {author} {\bibfnamefont
  {M.}~\bibnamefont {Remazeilles}}, \bibinfo {author} {\bibfnamefont
  {G.}~\bibnamefont {Rocha}}, \bibinfo {author} {\bibfnamefont
  {M.}~\bibnamefont {Schmittfull}}, \bibinfo {author} {\bibfnamefont
  {D.}~\bibnamefont {Scott}}, \bibinfo {author} {\bibfnamefont
  {P.}~\bibnamefont {Shirron}}, \bibinfo {author} {\bibfnamefont
  {I.}~\bibnamefont {Stephens}}, \bibinfo {author} {\bibfnamefont
  {B.}~\bibnamefont {Sutin}}, \bibinfo {author} {\bibfnamefont
  {M.}~\bibnamefont {Tomasi}}, \bibinfo {author} {\bibfnamefont
  {A.}~\bibnamefont {Trangsrud}}, \bibinfo {author} {\bibfnamefont
  {A.}~\bibnamefont {{van Engelen}}}, \bibinfo {author} {\bibfnamefont
  {F.}~\bibnamefont {Vansyngel}}, \bibinfo {author} {\bibfnamefont {I.~K.}\
  \bibnamefont {Wehus}}, \bibinfo {author} {\bibfnamefont {Q.}~\bibnamefont
  {Wen}}, \bibinfo {author} {\bibfnamefont {S.}~\bibnamefont {Xu}}, \bibinfo
  {author} {\bibfnamefont {K.}~\bibnamefont {Young}}, \ and\ \bibinfo {author}
  {\bibfnamefont {A.}~\bibnamefont {Zonca}},\ }\href
  {http://arxiv.org/abs/1902.10541} {\bibfield  {journal} {\bibinfo  {journal}
  {arXiv:1902.10541 [astro-ph]}\ } (\bibinfo {year} {2019})},\ \Eprint
  {http://arxiv.org/abs/1902.10541} {arXiv:1902.10541 [astro-ph]} \BibitemShut
  {NoStop}%
\bibitem [{\citenamefont {Abazajian}\ \emph
  {et~al.}(2019{\natexlab{a}})\citenamefont {Abazajian}, \citenamefont
  {Addison}, \citenamefont {Adshead}, \citenamefont {Ahmed}, \citenamefont
  {Allen}, \citenamefont {Alonso}, \citenamefont {Alvarez}, \citenamefont
  {Amin}, \citenamefont {Anderson}, \citenamefont {Arnold}, \citenamefont
  {Baccigalupi}, \citenamefont {Bailey}, \citenamefont {Barkats}, \citenamefont
  {Barron}, \citenamefont {Barry}, \citenamefont {Bartlett}, \citenamefont
  {Thakur}, \citenamefont {Battaglia}, \citenamefont {Baxter}, \citenamefont
  {Bean}, \citenamefont {Bebek}, \citenamefont {Bender}, \citenamefont
  {Benson}, \citenamefont {Berger}, \citenamefont {Bhimani}, \citenamefont
  {Bischoff}, \citenamefont {Bleem}, \citenamefont {Bock}, \citenamefont
  {Bocquet}, \citenamefont {Boddy}, \citenamefont {Bonato}, \citenamefont
  {Bond}, \citenamefont {Borrill}, \citenamefont {Bouchet}, \citenamefont
  {Brown}, \citenamefont {Bryan}, \citenamefont {Burkhart}, \citenamefont
  {Buza}, \citenamefont {Byrum}, \citenamefont {Calabrese}, \citenamefont
  {Calafut}, \citenamefont {Caldwell}, \citenamefont {Carlstrom}, \citenamefont
  {Carron}, \citenamefont {Cecil}, \citenamefont {Challinor}, \citenamefont
  {Chang}, \citenamefont {Chinone}, \citenamefont {Cho}, \citenamefont
  {Cooray}, \citenamefont {Crawford}, \citenamefont {Crites}, \citenamefont
  {Cukierman}, \citenamefont {{Cyr-Racine}}, \citenamefont {{de Haan}},
  \citenamefont {{de Zotti}}, \citenamefont {Delabrouille}, \citenamefont
  {Demarteau}, \citenamefont {Devlin}, \citenamefont {Di~Valentino},
  \citenamefont {Dobbs}, \citenamefont {Duff}, \citenamefont {Duivenvoorden},
  \citenamefont {Dvorkin}, \citenamefont {Edwards}, \citenamefont {Eimer},
  \citenamefont {Errard}, \citenamefont {{Essinger-Hileman}}, \citenamefont
  {Fabbian}, \citenamefont {Feng}, \citenamefont {Ferraro}, \citenamefont
  {Filippini}, \citenamefont {Flauger}, \citenamefont {Flaugher}, \citenamefont
  {Fraisse}, \citenamefont {Frolov}, \citenamefont {Galitzki}, \citenamefont
  {Galli}, \citenamefont {Ganga}, \citenamefont {Gerbino}, \citenamefont
  {Gilchriese}, \citenamefont {Gluscevic}, \citenamefont {Green}, \citenamefont
  {Grin}, \citenamefont {Grohs}, \citenamefont {Gualtieri}, \citenamefont
  {Guarino}, \citenamefont {Gudmundsson}, \citenamefont {Habib}, \citenamefont
  {Haller}, \citenamefont {Halpern}, \citenamefont {Halverson}, \citenamefont
  {Hanany}, \citenamefont {Harrington}, \citenamefont {Hasegawa}, \citenamefont
  {Hasselfield}, \citenamefont {Hazumi}, \citenamefont {Heitmann},
  \citenamefont {Henderson}, \citenamefont {Henning}, \citenamefont {Hill},
  \citenamefont {Hlozek}, \citenamefont {Holder}, \citenamefont {Holzapfel},
  \citenamefont {Hubmayr}, \citenamefont {Huffenberger}, \citenamefont
  {Huffer}, \citenamefont {Hui}, \citenamefont {Irwin}, \citenamefont
  {Johnson}, \citenamefont {Johnstone}, \citenamefont {Jones}, \citenamefont
  {Karkare}, \citenamefont {Katayama}, \citenamefont {Kerby}, \citenamefont
  {Kernovsky}, \citenamefont {Keskitalo}, \citenamefont {Kisner}, \citenamefont
  {Knox}, \citenamefont {Kosowsky}, \citenamefont {Kovac}, \citenamefont
  {Kovetz}, \citenamefont {Kuhlmann}, \citenamefont {Kuo}, \citenamefont
  {Kurita}, \citenamefont {Kusaka}, \citenamefont {Lahteenmaki}, \citenamefont
  {Lawrence}, \citenamefont {Lee}, \citenamefont {Lewis}, \citenamefont {Li},
  \citenamefont {Linder}, \citenamefont {Loverde}, \citenamefont {Lowitz},
  \citenamefont {Madhavacheril}, \citenamefont {Mantz}, \citenamefont
  {Matsuda}, \citenamefont {Mauskopf}, \citenamefont {McMahon}, \citenamefont
  {Meerburg}, \citenamefont {Melin}, \citenamefont {Meyers}, \citenamefont
  {Millea}, \citenamefont {Mohr}, \citenamefont {Moncelsi}, \citenamefont
  {Mroczkowski}, \citenamefont {Mukherjee}, \citenamefont {M{\"u}nchmeyer},
  \citenamefont {Nagai}, \citenamefont {Nagy}, \citenamefont {Namikawa},
  \citenamefont {Nati}, \citenamefont {Natoli}, \citenamefont {Negrello},
  \citenamefont {Newburgh}, \citenamefont {Niemack}, \citenamefont {Nishino},
  \citenamefont {Nordby}, \citenamefont {Novosad}, \citenamefont {O'Connor},
  \citenamefont {Obied}, \citenamefont {Padin}, \citenamefont {Pandey},
  \citenamefont {Partridge}, \citenamefont {Pierpaoli}, \citenamefont
  {Pogosian}, \citenamefont {Pryke}, \citenamefont {Puglisi}, \citenamefont
  {Racine}, \citenamefont {Raghunathan}, \citenamefont {Rahlin}, \citenamefont
  {Rajagopalan}, \citenamefont {Raveri}, \citenamefont {Reichanadter},
  \citenamefont {Reichardt}, \citenamefont {Remazeilles}, \citenamefont
  {Rocha}, \citenamefont {Roe}, \citenamefont {Roy}, \citenamefont {Ruhl},
  \citenamefont {Salatino}, \citenamefont {Saliwanchik}, \citenamefont
  {Schaan}, \citenamefont {Schillaci}, \citenamefont {Schmittfull},
  \citenamefont {Scott}, \citenamefont {Sehgal}, \citenamefont {Shandera},
  \citenamefont {Sheehy}, \citenamefont {Sherwin}, \citenamefont {Shirokoff},
  \citenamefont {Simon}, \citenamefont {Slosar}, \citenamefont {Somerville},
  \citenamefont {Staggs}, \citenamefont {Stark}, \citenamefont {Stompor},
  \citenamefont {Story}, \citenamefont {Stoughton}, \citenamefont {Suzuki},
  \citenamefont {Tajima}, \citenamefont {Teply}, \citenamefont {Thompson},
  \citenamefont {Timbie}, \citenamefont {Tomasi}, \citenamefont {Treu},
  \citenamefont {Tristram}, \citenamefont {Tucker}, \citenamefont {Umilt{\`a}},
  \citenamefont {{van Engelen}}, \citenamefont {Vieira}, \citenamefont
  {Vieregg}, \citenamefont {Vogelsberger}, \citenamefont {Wang}, \citenamefont
  {Watson}, \citenamefont {White}, \citenamefont {Whitehorn}, \citenamefont
  {Wollack}, \citenamefont {Wu}, \citenamefont {Xu}, \citenamefont {Yasini},
  \citenamefont {Yeck}, \citenamefont {Yoon}, \citenamefont {Young},\ and\
  \citenamefont {Zonca}}]{abazajian2019}%
  \BibitemOpen
  \bibfield  {author} {\bibinfo {author} {\bibfnamefont {K.}~\bibnamefont
  {Abazajian}}, \bibinfo {author} {\bibfnamefont {G.}~\bibnamefont {Addison}},
  \bibinfo {author} {\bibfnamefont {P.}~\bibnamefont {Adshead}}, \bibinfo
  {author} {\bibfnamefont {Z.}~\bibnamefont {Ahmed}}, \bibinfo {author}
  {\bibfnamefont {S.~W.}\ \bibnamefont {Allen}}, \bibinfo {author}
  {\bibfnamefont {D.}~\bibnamefont {Alonso}}, \bibinfo {author} {\bibfnamefont
  {M.}~\bibnamefont {Alvarez}}, \bibinfo {author} {\bibfnamefont {M.~A.}\
  \bibnamefont {Amin}}, \bibinfo {author} {\bibfnamefont {A.}~\bibnamefont
  {Anderson}}, \bibinfo {author} {\bibfnamefont {K.~S.}\ \bibnamefont
  {Arnold}}, \bibinfo {author} {\bibfnamefont {C.}~\bibnamefont {Baccigalupi}},
  \bibinfo {author} {\bibfnamefont {K.}~\bibnamefont {Bailey}}, \bibinfo
  {author} {\bibfnamefont {D.}~\bibnamefont {Barkats}}, \bibinfo {author}
  {\bibfnamefont {D.}~\bibnamefont {Barron}}, \bibinfo {author} {\bibfnamefont
  {P.~S.}\ \bibnamefont {Barry}}, \bibinfo {author} {\bibfnamefont {J.~G.}\
  \bibnamefont {Bartlett}}, \bibinfo {author} {\bibfnamefont {R.~B.}\
  \bibnamefont {Thakur}}, \bibinfo {author} {\bibfnamefont {N.}~\bibnamefont
  {Battaglia}}, \bibinfo {author} {\bibfnamefont {E.}~\bibnamefont {Baxter}},
  \bibinfo {author} {\bibfnamefont {R.}~\bibnamefont {Bean}}, \bibinfo {author}
  {\bibfnamefont {C.}~\bibnamefont {Bebek}}, \bibinfo {author} {\bibfnamefont
  {A.~N.}\ \bibnamefont {Bender}}, \bibinfo {author} {\bibfnamefont {B.~A.}\
  \bibnamefont {Benson}}, \bibinfo {author} {\bibfnamefont {E.}~\bibnamefont
  {Berger}}, \bibinfo {author} {\bibfnamefont {S.}~\bibnamefont {Bhimani}},
  \bibinfo {author} {\bibfnamefont {C.~A.}\ \bibnamefont {Bischoff}}, \bibinfo
  {author} {\bibfnamefont {L.}~\bibnamefont {Bleem}}, \bibinfo {author}
  {\bibfnamefont {J.~J.}\ \bibnamefont {Bock}}, \bibinfo {author}
  {\bibfnamefont {S.}~\bibnamefont {Bocquet}}, \bibinfo {author} {\bibfnamefont
  {K.}~\bibnamefont {Boddy}}, \bibinfo {author} {\bibfnamefont
  {M.}~\bibnamefont {Bonato}}, \bibinfo {author} {\bibfnamefont {J.~R.}\
  \bibnamefont {Bond}}, \bibinfo {author} {\bibfnamefont {J.}~\bibnamefont
  {Borrill}}, \bibinfo {author} {\bibfnamefont {F.~R.}\ \bibnamefont
  {Bouchet}}, \bibinfo {author} {\bibfnamefont {M.~L.}\ \bibnamefont {Brown}},
  \bibinfo {author} {\bibfnamefont {S.}~\bibnamefont {Bryan}}, \bibinfo
  {author} {\bibfnamefont {B.}~\bibnamefont {Burkhart}}, \bibinfo {author}
  {\bibfnamefont {V.}~\bibnamefont {Buza}}, \bibinfo {author} {\bibfnamefont
  {K.}~\bibnamefont {Byrum}}, \bibinfo {author} {\bibfnamefont
  {E.}~\bibnamefont {Calabrese}}, \bibinfo {author} {\bibfnamefont
  {V.}~\bibnamefont {Calafut}}, \bibinfo {author} {\bibfnamefont
  {R.}~\bibnamefont {Caldwell}}, \bibinfo {author} {\bibfnamefont {J.~E.}\
  \bibnamefont {Carlstrom}}, \bibinfo {author} {\bibfnamefont {J.}~\bibnamefont
  {Carron}}, \bibinfo {author} {\bibfnamefont {T.}~\bibnamefont {Cecil}},
  \bibinfo {author} {\bibfnamefont {A.}~\bibnamefont {Challinor}}, \bibinfo
  {author} {\bibfnamefont {C.~L.}\ \bibnamefont {Chang}}, \bibinfo {author}
  {\bibfnamefont {Y.}~\bibnamefont {Chinone}}, \bibinfo {author} {\bibfnamefont
  {H.-M.~S.}\ \bibnamefont {Cho}}, \bibinfo {author} {\bibfnamefont
  {A.}~\bibnamefont {Cooray}}, \bibinfo {author} {\bibfnamefont {T.~M.}\
  \bibnamefont {Crawford}}, \bibinfo {author} {\bibfnamefont {A.}~\bibnamefont
  {Crites}}, \bibinfo {author} {\bibfnamefont {A.}~\bibnamefont {Cukierman}},
  \bibinfo {author} {\bibfnamefont {F.-Y.}\ \bibnamefont {{Cyr-Racine}}},
  \bibinfo {author} {\bibfnamefont {T.}~\bibnamefont {{de Haan}}}, \bibinfo
  {author} {\bibfnamefont {G.}~\bibnamefont {{de Zotti}}}, \bibinfo {author}
  {\bibfnamefont {J.}~\bibnamefont {Delabrouille}}, \bibinfo {author}
  {\bibfnamefont {M.}~\bibnamefont {Demarteau}}, \bibinfo {author}
  {\bibfnamefont {M.}~\bibnamefont {Devlin}}, \bibinfo {author} {\bibfnamefont
  {E.}~\bibnamefont {Di~Valentino}}, \bibinfo {author} {\bibfnamefont
  {M.}~\bibnamefont {Dobbs}}, \bibinfo {author} {\bibfnamefont
  {S.}~\bibnamefont {Duff}}, \bibinfo {author} {\bibfnamefont {A.}~\bibnamefont
  {Duivenvoorden}}, \bibinfo {author} {\bibfnamefont {C.}~\bibnamefont
  {Dvorkin}}, \bibinfo {author} {\bibfnamefont {W.}~\bibnamefont {Edwards}},
  \bibinfo {author} {\bibfnamefont {J.}~\bibnamefont {Eimer}}, \bibinfo
  {author} {\bibfnamefont {J.}~\bibnamefont {Errard}}, \bibinfo {author}
  {\bibfnamefont {T.}~\bibnamefont {{Essinger-Hileman}}}, \bibinfo {author}
  {\bibfnamefont {G.}~\bibnamefont {Fabbian}}, \bibinfo {author} {\bibfnamefont
  {C.}~\bibnamefont {Feng}}, \bibinfo {author} {\bibfnamefont {S.}~\bibnamefont
  {Ferraro}}, \bibinfo {author} {\bibfnamefont {J.~P.}\ \bibnamefont
  {Filippini}}, \bibinfo {author} {\bibfnamefont {R.}~\bibnamefont {Flauger}},
  \bibinfo {author} {\bibfnamefont {B.}~\bibnamefont {Flaugher}}, \bibinfo
  {author} {\bibfnamefont {A.~A.}\ \bibnamefont {Fraisse}}, \bibinfo {author}
  {\bibfnamefont {A.}~\bibnamefont {Frolov}}, \bibinfo {author} {\bibfnamefont
  {N.}~\bibnamefont {Galitzki}}, \bibinfo {author} {\bibfnamefont
  {S.}~\bibnamefont {Galli}}, \bibinfo {author} {\bibfnamefont
  {K.}~\bibnamefont {Ganga}}, \bibinfo {author} {\bibfnamefont
  {M.}~\bibnamefont {Gerbino}}, \bibinfo {author} {\bibfnamefont
  {M.}~\bibnamefont {Gilchriese}}, \bibinfo {author} {\bibfnamefont
  {V.}~\bibnamefont {Gluscevic}}, \bibinfo {author} {\bibfnamefont
  {D.}~\bibnamefont {Green}}, \bibinfo {author} {\bibfnamefont
  {D.}~\bibnamefont {Grin}}, \bibinfo {author} {\bibfnamefont {E.}~\bibnamefont
  {Grohs}}, \bibinfo {author} {\bibfnamefont {R.}~\bibnamefont {Gualtieri}},
  \bibinfo {author} {\bibfnamefont {V.}~\bibnamefont {Guarino}}, \bibinfo
  {author} {\bibfnamefont {J.~E.}\ \bibnamefont {Gudmundsson}}, \bibinfo
  {author} {\bibfnamefont {S.}~\bibnamefont {Habib}}, \bibinfo {author}
  {\bibfnamefont {G.}~\bibnamefont {Haller}}, \bibinfo {author} {\bibfnamefont
  {M.}~\bibnamefont {Halpern}}, \bibinfo {author} {\bibfnamefont {N.~W.}\
  \bibnamefont {Halverson}}, \bibinfo {author} {\bibfnamefont {S.}~\bibnamefont
  {Hanany}}, \bibinfo {author} {\bibfnamefont {K.}~\bibnamefont {Harrington}},
  \bibinfo {author} {\bibfnamefont {M.}~\bibnamefont {Hasegawa}}, \bibinfo
  {author} {\bibfnamefont {M.}~\bibnamefont {Hasselfield}}, \bibinfo {author}
  {\bibfnamefont {M.}~\bibnamefont {Hazumi}}, \bibinfo {author} {\bibfnamefont
  {K.}~\bibnamefont {Heitmann}}, \bibinfo {author} {\bibfnamefont
  {S.}~\bibnamefont {Henderson}}, \bibinfo {author} {\bibfnamefont {J.~W.}\
  \bibnamefont {Henning}}, \bibinfo {author} {\bibfnamefont {J.~C.}\
  \bibnamefont {Hill}}, \bibinfo {author} {\bibfnamefont {R.}~\bibnamefont
  {Hlozek}}, \bibinfo {author} {\bibfnamefont {G.}~\bibnamefont {Holder}},
  \bibinfo {author} {\bibfnamefont {W.}~\bibnamefont {Holzapfel}}, \bibinfo
  {author} {\bibfnamefont {J.}~\bibnamefont {Hubmayr}}, \bibinfo {author}
  {\bibfnamefont {K.~M.}\ \bibnamefont {Huffenberger}}, \bibinfo {author}
  {\bibfnamefont {M.}~\bibnamefont {Huffer}}, \bibinfo {author} {\bibfnamefont
  {H.}~\bibnamefont {Hui}}, \bibinfo {author} {\bibfnamefont {K.}~\bibnamefont
  {Irwin}}, \bibinfo {author} {\bibfnamefont {B.~R.}\ \bibnamefont {Johnson}},
  \bibinfo {author} {\bibfnamefont {D.}~\bibnamefont {Johnstone}}, \bibinfo
  {author} {\bibfnamefont {W.~C.}\ \bibnamefont {Jones}}, \bibinfo {author}
  {\bibfnamefont {K.}~\bibnamefont {Karkare}}, \bibinfo {author} {\bibfnamefont
  {N.}~\bibnamefont {Katayama}}, \bibinfo {author} {\bibfnamefont
  {J.}~\bibnamefont {Kerby}}, \bibinfo {author} {\bibfnamefont
  {S.}~\bibnamefont {Kernovsky}}, \bibinfo {author} {\bibfnamefont
  {R.}~\bibnamefont {Keskitalo}}, \bibinfo {author} {\bibfnamefont
  {T.}~\bibnamefont {Kisner}}, \bibinfo {author} {\bibfnamefont
  {L.}~\bibnamefont {Knox}}, \bibinfo {author} {\bibfnamefont {A.}~\bibnamefont
  {Kosowsky}}, \bibinfo {author} {\bibfnamefont {J.}~\bibnamefont {Kovac}},
  \bibinfo {author} {\bibfnamefont {E.~D.}\ \bibnamefont {Kovetz}}, \bibinfo
  {author} {\bibfnamefont {S.}~\bibnamefont {Kuhlmann}}, \bibinfo {author}
  {\bibfnamefont {C.-l.}\ \bibnamefont {Kuo}}, \bibinfo {author} {\bibfnamefont
  {N.}~\bibnamefont {Kurita}}, \bibinfo {author} {\bibfnamefont
  {A.}~\bibnamefont {Kusaka}}, \bibinfo {author} {\bibfnamefont
  {A.}~\bibnamefont {Lahteenmaki}}, \bibinfo {author} {\bibfnamefont {C.~R.}\
  \bibnamefont {Lawrence}}, \bibinfo {author} {\bibfnamefont {A.~T.}\
  \bibnamefont {Lee}}, \bibinfo {author} {\bibfnamefont {A.}~\bibnamefont
  {Lewis}}, \bibinfo {author} {\bibfnamefont {D.}~\bibnamefont {Li}}, \bibinfo
  {author} {\bibfnamefont {E.}~\bibnamefont {Linder}}, \bibinfo {author}
  {\bibfnamefont {M.}~\bibnamefont {Loverde}}, \bibinfo {author} {\bibfnamefont
  {A.}~\bibnamefont {Lowitz}}, \bibinfo {author} {\bibfnamefont {M.~S.}\
  \bibnamefont {Madhavacheril}}, \bibinfo {author} {\bibfnamefont
  {A.}~\bibnamefont {Mantz}}, \bibinfo {author} {\bibfnamefont
  {F.}~\bibnamefont {Matsuda}}, \bibinfo {author} {\bibfnamefont
  {P.}~\bibnamefont {Mauskopf}}, \bibinfo {author} {\bibfnamefont
  {J.}~\bibnamefont {McMahon}}, \bibinfo {author} {\bibfnamefont {P.~D.}\
  \bibnamefont {Meerburg}}, \bibinfo {author} {\bibfnamefont {J.-B.}\
  \bibnamefont {Melin}}, \bibinfo {author} {\bibfnamefont {J.}~\bibnamefont
  {Meyers}}, \bibinfo {author} {\bibfnamefont {M.}~\bibnamefont {Millea}},
  \bibinfo {author} {\bibfnamefont {J.}~\bibnamefont {Mohr}}, \bibinfo {author}
  {\bibfnamefont {L.}~\bibnamefont {Moncelsi}}, \bibinfo {author}
  {\bibfnamefont {T.}~\bibnamefont {Mroczkowski}}, \bibinfo {author}
  {\bibfnamefont {S.}~\bibnamefont {Mukherjee}}, \bibinfo {author}
  {\bibfnamefont {M.}~\bibnamefont {M{\"u}nchmeyer}}, \bibinfo {author}
  {\bibfnamefont {D.}~\bibnamefont {Nagai}}, \bibinfo {author} {\bibfnamefont
  {J.}~\bibnamefont {Nagy}}, \bibinfo {author} {\bibfnamefont {T.}~\bibnamefont
  {Namikawa}}, \bibinfo {author} {\bibfnamefont {F.}~\bibnamefont {Nati}},
  \bibinfo {author} {\bibfnamefont {T.}~\bibnamefont {Natoli}}, \bibinfo
  {author} {\bibfnamefont {M.}~\bibnamefont {Negrello}}, \bibinfo {author}
  {\bibfnamefont {L.}~\bibnamefont {Newburgh}}, \bibinfo {author}
  {\bibfnamefont {M.~D.}\ \bibnamefont {Niemack}}, \bibinfo {author}
  {\bibfnamefont {H.}~\bibnamefont {Nishino}}, \bibinfo {author} {\bibfnamefont
  {M.}~\bibnamefont {Nordby}}, \bibinfo {author} {\bibfnamefont
  {V.}~\bibnamefont {Novosad}}, \bibinfo {author} {\bibfnamefont
  {P.}~\bibnamefont {O'Connor}}, \bibinfo {author} {\bibfnamefont
  {G.}~\bibnamefont {Obied}}, \bibinfo {author} {\bibfnamefont
  {S.}~\bibnamefont {Padin}}, \bibinfo {author} {\bibfnamefont
  {S.}~\bibnamefont {Pandey}}, \bibinfo {author} {\bibfnamefont
  {B.}~\bibnamefont {Partridge}}, \bibinfo {author} {\bibfnamefont
  {E.}~\bibnamefont {Pierpaoli}}, \bibinfo {author} {\bibfnamefont
  {L.}~\bibnamefont {Pogosian}}, \bibinfo {author} {\bibfnamefont
  {C.}~\bibnamefont {Pryke}}, \bibinfo {author} {\bibfnamefont
  {G.}~\bibnamefont {Puglisi}}, \bibinfo {author} {\bibfnamefont
  {B.}~\bibnamefont {Racine}}, \bibinfo {author} {\bibfnamefont
  {S.}~\bibnamefont {Raghunathan}}, \bibinfo {author} {\bibfnamefont
  {A.}~\bibnamefont {Rahlin}}, \bibinfo {author} {\bibfnamefont
  {S.}~\bibnamefont {Rajagopalan}}, \bibinfo {author} {\bibfnamefont
  {M.}~\bibnamefont {Raveri}}, \bibinfo {author} {\bibfnamefont
  {M.}~\bibnamefont {Reichanadter}}, \bibinfo {author} {\bibfnamefont {C.~L.}\
  \bibnamefont {Reichardt}}, \bibinfo {author} {\bibfnamefont {M.}~\bibnamefont
  {Remazeilles}}, \bibinfo {author} {\bibfnamefont {G.}~\bibnamefont {Rocha}},
  \bibinfo {author} {\bibfnamefont {N.~A.}\ \bibnamefont {Roe}}, \bibinfo
  {author} {\bibfnamefont {A.}~\bibnamefont {Roy}}, \bibinfo {author}
  {\bibfnamefont {J.}~\bibnamefont {Ruhl}}, \bibinfo {author} {\bibfnamefont
  {M.}~\bibnamefont {Salatino}}, \bibinfo {author} {\bibfnamefont
  {B.}~\bibnamefont {Saliwanchik}}, \bibinfo {author} {\bibfnamefont
  {E.}~\bibnamefont {Schaan}}, \bibinfo {author} {\bibfnamefont
  {A.}~\bibnamefont {Schillaci}}, \bibinfo {author} {\bibfnamefont {M.~M.}\
  \bibnamefont {Schmittfull}}, \bibinfo {author} {\bibfnamefont
  {D.}~\bibnamefont {Scott}}, \bibinfo {author} {\bibfnamefont
  {N.}~\bibnamefont {Sehgal}}, \bibinfo {author} {\bibfnamefont
  {S.}~\bibnamefont {Shandera}}, \bibinfo {author} {\bibfnamefont
  {C.}~\bibnamefont {Sheehy}}, \bibinfo {author} {\bibfnamefont {B.~D.}\
  \bibnamefont {Sherwin}}, \bibinfo {author} {\bibfnamefont {E.}~\bibnamefont
  {Shirokoff}}, \bibinfo {author} {\bibfnamefont {S.~M.}\ \bibnamefont
  {Simon}}, \bibinfo {author} {\bibfnamefont {A.}~\bibnamefont {Slosar}},
  \bibinfo {author} {\bibfnamefont {R.}~\bibnamefont {Somerville}}, \bibinfo
  {author} {\bibfnamefont {S.~T.}\ \bibnamefont {Staggs}}, \bibinfo {author}
  {\bibfnamefont {A.}~\bibnamefont {Stark}}, \bibinfo {author} {\bibfnamefont
  {R.}~\bibnamefont {Stompor}}, \bibinfo {author} {\bibfnamefont {K.~T.}\
  \bibnamefont {Story}}, \bibinfo {author} {\bibfnamefont {C.}~\bibnamefont
  {Stoughton}}, \bibinfo {author} {\bibfnamefont {A.}~\bibnamefont {Suzuki}},
  \bibinfo {author} {\bibfnamefont {O.}~\bibnamefont {Tajima}}, \bibinfo
  {author} {\bibfnamefont {G.~P.}\ \bibnamefont {Teply}}, \bibinfo {author}
  {\bibfnamefont {K.}~\bibnamefont {Thompson}}, \bibinfo {author}
  {\bibfnamefont {P.}~\bibnamefont {Timbie}}, \bibinfo {author} {\bibfnamefont
  {M.}~\bibnamefont {Tomasi}}, \bibinfo {author} {\bibfnamefont {J.~I.}\
  \bibnamefont {Treu}}, \bibinfo {author} {\bibfnamefont {M.}~\bibnamefont
  {Tristram}}, \bibinfo {author} {\bibfnamefont {G.}~\bibnamefont {Tucker}},
  \bibinfo {author} {\bibfnamefont {C.}~\bibnamefont {Umilt{\`a}}}, \bibinfo
  {author} {\bibfnamefont {A.}~\bibnamefont {{van Engelen}}}, \bibinfo {author}
  {\bibfnamefont {J.~D.}\ \bibnamefont {Vieira}}, \bibinfo {author}
  {\bibfnamefont {A.~G.}\ \bibnamefont {Vieregg}}, \bibinfo {author}
  {\bibfnamefont {M.}~\bibnamefont {Vogelsberger}}, \bibinfo {author}
  {\bibfnamefont {G.}~\bibnamefont {Wang}}, \bibinfo {author} {\bibfnamefont
  {S.}~\bibnamefont {Watson}}, \bibinfo {author} {\bibfnamefont
  {M.}~\bibnamefont {White}}, \bibinfo {author} {\bibfnamefont
  {N.}~\bibnamefont {Whitehorn}}, \bibinfo {author} {\bibfnamefont {E.~J.}\
  \bibnamefont {Wollack}}, \bibinfo {author} {\bibfnamefont {W.~L.~K.}\
  \bibnamefont {Wu}}, \bibinfo {author} {\bibfnamefont {Z.}~\bibnamefont {Xu}},
  \bibinfo {author} {\bibfnamefont {S.}~\bibnamefont {Yasini}}, \bibinfo
  {author} {\bibfnamefont {J.}~\bibnamefont {Yeck}}, \bibinfo {author}
  {\bibfnamefont {K.~W.}\ \bibnamefont {Yoon}}, \bibinfo {author}
  {\bibfnamefont {E.}~\bibnamefont {Young}}, \ and\ \bibinfo {author}
  {\bibfnamefont {A.}~\bibnamefont {Zonca}},\ }\href
  {http://arxiv.org/abs/1908.01062} {\bibfield  {journal} {\bibinfo  {journal}
  {arXiv:1908.01062 [astro-ph]}\ } (\bibinfo {year} {2019}{\natexlab{a}})},\
  \Eprint {http://arxiv.org/abs/1908.01062} {arXiv:1908.01062 [astro-ph]}
  \BibitemShut {NoStop}%
\bibitem [{\citenamefont {Hu}\ and\ \citenamefont {Okamoto}(2002)}]{hu2002}%
  \BibitemOpen
  \bibfield  {author} {\bibinfo {author} {\bibfnamefont {W.}~\bibnamefont
  {Hu}}\ and\ \bibinfo {author} {\bibfnamefont {T.}~\bibnamefont {Okamoto}},\
  }\href {\doibase 10.1086/341110} {\bibfield  {journal} {\bibinfo  {journal}
  {The Astrophysical Journal}\ }\textbf {\bibinfo {volume} {574}},\ \bibinfo
  {pages} {566} (\bibinfo {year} {2002})}\BibitemShut {NoStop}%
\bibitem [{\citenamefont {Smith}\ \emph {et~al.}(2007)\citenamefont {Smith},
  \citenamefont {Zahn},\ and\ \citenamefont {Dore}}]{smith2007}%
  \BibitemOpen
  \bibfield  {author} {\bibinfo {author} {\bibfnamefont {K.~M.}\ \bibnamefont
  {Smith}}, \bibinfo {author} {\bibfnamefont {O.}~\bibnamefont {Zahn}}, \ and\
  \bibinfo {author} {\bibfnamefont {O.}~\bibnamefont {Dore}},\ }\href {\doibase
  10.1103/PhysRevD.76.043510} {\bibfield  {journal} {\bibinfo  {journal}
  {Physical Review D}\ }\textbf {\bibinfo {volume} {76}} (\bibinfo {year}
  {2007}),\ 10.1103/PhysRevD.76.043510},\ \Eprint
  {http://arxiv.org/abs/0705.3980} {arXiv:0705.3980} \BibitemShut {NoStop}%
\bibitem [{\citenamefont {Das}\ \emph {et~al.}(2011)\citenamefont {Das},
  \citenamefont {Sherwin}, \citenamefont {Aguirre}, \citenamefont {Appel},
  \citenamefont {Bond}, \citenamefont {Carvalho}, \citenamefont {Devlin},
  \citenamefont {Dunkley}, \citenamefont {D{\"u}nner}, \citenamefont
  {{Essinger-Hileman}}, \citenamefont {Fowler}, \citenamefont {Hajian},
  \citenamefont {Halpern}, \citenamefont {Hasselfield}, \citenamefont {Hincks},
  \citenamefont {Hlozek}, \citenamefont {Huffenberger}, \citenamefont {Hughes},
  \citenamefont {Irwin}, \citenamefont {Klein}, \citenamefont {Kosowsky},
  \citenamefont {Lupton}, \citenamefont {Marriage}, \citenamefont {Marsden},
  \citenamefont {Menanteau}, \citenamefont {Moodley}, \citenamefont {Niemack},
  \citenamefont {Nolta}, \citenamefont {Page}, \citenamefont {Parker},
  \citenamefont {Reese}, \citenamefont {Schmitt}, \citenamefont {Sehgal},
  \citenamefont {Sievers}, \citenamefont {Spergel}, \citenamefont {Staggs},
  \citenamefont {Swetz}, \citenamefont {Switzer}, \citenamefont {Thornton},
  \citenamefont {Visnjic},\ and\ \citenamefont {Wollack}}]{das2011a}%
  \BibitemOpen
  \bibfield  {author} {\bibinfo {author} {\bibfnamefont {S.}~\bibnamefont
  {Das}}, \bibinfo {author} {\bibfnamefont {B.~D.}\ \bibnamefont {Sherwin}},
  \bibinfo {author} {\bibfnamefont {P.}~\bibnamefont {Aguirre}}, \bibinfo
  {author} {\bibfnamefont {J.~W.}\ \bibnamefont {Appel}}, \bibinfo {author}
  {\bibfnamefont {J.~R.}\ \bibnamefont {Bond}}, \bibinfo {author}
  {\bibfnamefont {C.~S.}\ \bibnamefont {Carvalho}}, \bibinfo {author}
  {\bibfnamefont {M.~J.}\ \bibnamefont {Devlin}}, \bibinfo {author}
  {\bibfnamefont {J.}~\bibnamefont {Dunkley}}, \bibinfo {author} {\bibfnamefont
  {R.}~\bibnamefont {D{\"u}nner}}, \bibinfo {author} {\bibfnamefont
  {T.}~\bibnamefont {{Essinger-Hileman}}}, \bibinfo {author} {\bibfnamefont
  {J.~W.}\ \bibnamefont {Fowler}}, \bibinfo {author} {\bibfnamefont
  {A.}~\bibnamefont {Hajian}}, \bibinfo {author} {\bibfnamefont
  {M.}~\bibnamefont {Halpern}}, \bibinfo {author} {\bibfnamefont
  {M.}~\bibnamefont {Hasselfield}}, \bibinfo {author} {\bibfnamefont {A.~D.}\
  \bibnamefont {Hincks}}, \bibinfo {author} {\bibfnamefont {R.}~\bibnamefont
  {Hlozek}}, \bibinfo {author} {\bibfnamefont {K.~M.}\ \bibnamefont
  {Huffenberger}}, \bibinfo {author} {\bibfnamefont {J.~P.}\ \bibnamefont
  {Hughes}}, \bibinfo {author} {\bibfnamefont {K.~D.}\ \bibnamefont {Irwin}},
  \bibinfo {author} {\bibfnamefont {J.}~\bibnamefont {Klein}}, \bibinfo
  {author} {\bibfnamefont {A.}~\bibnamefont {Kosowsky}}, \bibinfo {author}
  {\bibfnamefont {R.~H.}\ \bibnamefont {Lupton}}, \bibinfo {author}
  {\bibfnamefont {T.~A.}\ \bibnamefont {Marriage}}, \bibinfo {author}
  {\bibfnamefont {D.}~\bibnamefont {Marsden}}, \bibinfo {author} {\bibfnamefont
  {F.}~\bibnamefont {Menanteau}}, \bibinfo {author} {\bibfnamefont
  {K.}~\bibnamefont {Moodley}}, \bibinfo {author} {\bibfnamefont {M.~D.}\
  \bibnamefont {Niemack}}, \bibinfo {author} {\bibfnamefont {M.~R.}\
  \bibnamefont {Nolta}}, \bibinfo {author} {\bibfnamefont {L.~A.}\ \bibnamefont
  {Page}}, \bibinfo {author} {\bibfnamefont {L.}~\bibnamefont {Parker}},
  \bibinfo {author} {\bibfnamefont {E.~D.}\ \bibnamefont {Reese}}, \bibinfo
  {author} {\bibfnamefont {B.~L.}\ \bibnamefont {Schmitt}}, \bibinfo {author}
  {\bibfnamefont {N.}~\bibnamefont {Sehgal}}, \bibinfo {author} {\bibfnamefont
  {J.}~\bibnamefont {Sievers}}, \bibinfo {author} {\bibfnamefont {D.~N.}\
  \bibnamefont {Spergel}}, \bibinfo {author} {\bibfnamefont {S.~T.}\
  \bibnamefont {Staggs}}, \bibinfo {author} {\bibfnamefont {D.~S.}\
  \bibnamefont {Swetz}}, \bibinfo {author} {\bibfnamefont {E.~R.}\ \bibnamefont
  {Switzer}}, \bibinfo {author} {\bibfnamefont {R.}~\bibnamefont {Thornton}},
  \bibinfo {author} {\bibfnamefont {K.}~\bibnamefont {Visnjic}}, \ and\
  \bibinfo {author} {\bibfnamefont {E.}~\bibnamefont {Wollack}},\ }\href
  {\doibase 10.1103/PhysRevLett.107.021301} {\bibfield  {journal} {\bibinfo
  {journal} {Physical Review Letters}\ }\textbf {\bibinfo {volume} {107}},\
  \bibinfo {pages} {021301} (\bibinfo {year} {2011})}\BibitemShut {NoStop}%
\bibitem [{\citenamefont {Hanson}\ \emph {et~al.}(2013)\citenamefont {Hanson},
  \citenamefont {Hoover}, \citenamefont {Crites}, \citenamefont {Ade},
  \citenamefont {Aird}, \citenamefont {Austermann}, \citenamefont {Beall},
  \citenamefont {Bender}, \citenamefont {Benson}, \citenamefont {Bleem},
  \citenamefont {Bock}, \citenamefont {Carlstrom}, \citenamefont {Chang},
  \citenamefont {Chiang}, \citenamefont {Cho}, \citenamefont {Conley},
  \citenamefont {Crawford}, \citenamefont {{de Haan}}, \citenamefont {Dobbs},
  \citenamefont {Everett}, \citenamefont {Gallicchio}, \citenamefont {Gao},
  \citenamefont {George}, \citenamefont {Halverson}, \citenamefont
  {Harrington}, \citenamefont {Henning}, \citenamefont {Hilton}, \citenamefont
  {Holder}, \citenamefont {Holzapfel}, \citenamefont {Hrubes}, \citenamefont
  {Huang}, \citenamefont {Hubmayr}, \citenamefont {Irwin}, \citenamefont
  {Keisler}, \citenamefont {Knox}, \citenamefont {Lee}, \citenamefont {Leitch},
  \citenamefont {Li}, \citenamefont {Liang}, \citenamefont {{Luong-Van}},
  \citenamefont {Marsden}, \citenamefont {McMahon}, \citenamefont {Mehl},
  \citenamefont {Meyer}, \citenamefont {Mocanu}, \citenamefont {Montroy},
  \citenamefont {Natoli}, \citenamefont {Nibarger}, \citenamefont {Novosad},
  \citenamefont {Padin}, \citenamefont {Pryke}, \citenamefont {Reichardt},
  \citenamefont {Ruhl}, \citenamefont {Saliwanchik}, \citenamefont {Sayre},
  \citenamefont {Schaffer}, \citenamefont {Schulz}, \citenamefont {Smecher},
  \citenamefont {Stark}, \citenamefont {Story}, \citenamefont {Tucker},
  \citenamefont {Vanderlinde}, \citenamefont {Vieira}, \citenamefont {Viero},
  \citenamefont {Wang}, \citenamefont {Yefremenko}, \citenamefont {Zahn},\ and\
  \citenamefont {Zemcov}}]{hanson2013}%
  \BibitemOpen
  \bibfield  {author} {\bibinfo {author} {\bibfnamefont {D.}~\bibnamefont
  {Hanson}}, \bibinfo {author} {\bibfnamefont {S.}~\bibnamefont {Hoover}},
  \bibinfo {author} {\bibfnamefont {A.}~\bibnamefont {Crites}}, \bibinfo
  {author} {\bibfnamefont {P.~a.~R.}\ \bibnamefont {Ade}}, \bibinfo {author}
  {\bibfnamefont {K.~A.}\ \bibnamefont {Aird}}, \bibinfo {author}
  {\bibfnamefont {J.~E.}\ \bibnamefont {Austermann}}, \bibinfo {author}
  {\bibfnamefont {J.~A.}\ \bibnamefont {Beall}}, \bibinfo {author}
  {\bibfnamefont {A.~N.}\ \bibnamefont {Bender}}, \bibinfo {author}
  {\bibfnamefont {B.~A.}\ \bibnamefont {Benson}}, \bibinfo {author}
  {\bibfnamefont {L.~E.}\ \bibnamefont {Bleem}}, \bibinfo {author}
  {\bibfnamefont {J.~J.}\ \bibnamefont {Bock}}, \bibinfo {author}
  {\bibfnamefont {J.~E.}\ \bibnamefont {Carlstrom}}, \bibinfo {author}
  {\bibfnamefont {C.~L.}\ \bibnamefont {Chang}}, \bibinfo {author}
  {\bibfnamefont {H.~C.}\ \bibnamefont {Chiang}}, \bibinfo {author}
  {\bibfnamefont {H.-M.}\ \bibnamefont {Cho}}, \bibinfo {author} {\bibfnamefont
  {A.}~\bibnamefont {Conley}}, \bibinfo {author} {\bibfnamefont {T.~M.}\
  \bibnamefont {Crawford}}, \bibinfo {author} {\bibfnamefont {T.}~\bibnamefont
  {{de Haan}}}, \bibinfo {author} {\bibfnamefont {M.~A.}\ \bibnamefont
  {Dobbs}}, \bibinfo {author} {\bibfnamefont {W.}~\bibnamefont {Everett}},
  \bibinfo {author} {\bibfnamefont {J.}~\bibnamefont {Gallicchio}}, \bibinfo
  {author} {\bibfnamefont {J.}~\bibnamefont {Gao}}, \bibinfo {author}
  {\bibfnamefont {E.~M.}\ \bibnamefont {George}}, \bibinfo {author}
  {\bibfnamefont {N.~W.}\ \bibnamefont {Halverson}}, \bibinfo {author}
  {\bibfnamefont {N.}~\bibnamefont {Harrington}}, \bibinfo {author}
  {\bibfnamefont {J.~W.}\ \bibnamefont {Henning}}, \bibinfo {author}
  {\bibfnamefont {G.~C.}\ \bibnamefont {Hilton}}, \bibinfo {author}
  {\bibfnamefont {G.~P.}\ \bibnamefont {Holder}}, \bibinfo {author}
  {\bibfnamefont {W.~L.}\ \bibnamefont {Holzapfel}}, \bibinfo {author}
  {\bibfnamefont {J.~D.}\ \bibnamefont {Hrubes}}, \bibinfo {author}
  {\bibfnamefont {N.}~\bibnamefont {Huang}}, \bibinfo {author} {\bibfnamefont
  {J.}~\bibnamefont {Hubmayr}}, \bibinfo {author} {\bibfnamefont {K.~D.}\
  \bibnamefont {Irwin}}, \bibinfo {author} {\bibfnamefont {R.}~\bibnamefont
  {Keisler}}, \bibinfo {author} {\bibfnamefont {L.}~\bibnamefont {Knox}},
  \bibinfo {author} {\bibfnamefont {A.~T.}\ \bibnamefont {Lee}}, \bibinfo
  {author} {\bibfnamefont {E.}~\bibnamefont {Leitch}}, \bibinfo {author}
  {\bibfnamefont {D.}~\bibnamefont {Li}}, \bibinfo {author} {\bibfnamefont
  {C.}~\bibnamefont {Liang}}, \bibinfo {author} {\bibfnamefont
  {D.}~\bibnamefont {{Luong-Van}}}, \bibinfo {author} {\bibfnamefont
  {G.}~\bibnamefont {Marsden}}, \bibinfo {author} {\bibfnamefont {J.~J.}\
  \bibnamefont {McMahon}}, \bibinfo {author} {\bibfnamefont {J.}~\bibnamefont
  {Mehl}}, \bibinfo {author} {\bibfnamefont {S.~S.}\ \bibnamefont {Meyer}},
  \bibinfo {author} {\bibfnamefont {L.}~\bibnamefont {Mocanu}}, \bibinfo
  {author} {\bibfnamefont {T.~E.}\ \bibnamefont {Montroy}}, \bibinfo {author}
  {\bibfnamefont {T.}~\bibnamefont {Natoli}}, \bibinfo {author} {\bibfnamefont
  {J.~P.}\ \bibnamefont {Nibarger}}, \bibinfo {author} {\bibfnamefont
  {V.}~\bibnamefont {Novosad}}, \bibinfo {author} {\bibfnamefont
  {S.}~\bibnamefont {Padin}}, \bibinfo {author} {\bibfnamefont
  {C.}~\bibnamefont {Pryke}}, \bibinfo {author} {\bibfnamefont {C.~L.}\
  \bibnamefont {Reichardt}}, \bibinfo {author} {\bibfnamefont {J.~E.}\
  \bibnamefont {Ruhl}}, \bibinfo {author} {\bibfnamefont {B.~R.}\ \bibnamefont
  {Saliwanchik}}, \bibinfo {author} {\bibfnamefont {J.~T.}\ \bibnamefont
  {Sayre}}, \bibinfo {author} {\bibfnamefont {K.~K.}\ \bibnamefont {Schaffer}},
  \bibinfo {author} {\bibfnamefont {B.}~\bibnamefont {Schulz}}, \bibinfo
  {author} {\bibfnamefont {G.}~\bibnamefont {Smecher}}, \bibinfo {author}
  {\bibfnamefont {A.~A.}\ \bibnamefont {Stark}}, \bibinfo {author}
  {\bibfnamefont {K.~T.}\ \bibnamefont {Story}}, \bibinfo {author}
  {\bibfnamefont {C.}~\bibnamefont {Tucker}}, \bibinfo {author} {\bibfnamefont
  {K.}~\bibnamefont {Vanderlinde}}, \bibinfo {author} {\bibfnamefont {J.~D.}\
  \bibnamefont {Vieira}}, \bibinfo {author} {\bibfnamefont {M.~P.}\
  \bibnamefont {Viero}}, \bibinfo {author} {\bibfnamefont {G.}~\bibnamefont
  {Wang}}, \bibinfo {author} {\bibfnamefont {V.}~\bibnamefont {Yefremenko}},
  \bibinfo {author} {\bibfnamefont {O.}~\bibnamefont {Zahn}}, \ and\ \bibinfo
  {author} {\bibfnamefont {M.}~\bibnamefont {Zemcov}},\ }\href {\doibase
  10.1103/PhysRevLett.111.141301} {\bibfield  {journal} {\bibinfo  {journal}
  {Physical Review Letters}\ }\textbf {\bibinfo {volume} {111}},\ \bibinfo
  {pages} {141301} (\bibinfo {year} {2013})}\BibitemShut {NoStop}%
\bibitem [{\citenamefont {{\sorthelp{Planck Collaboration 2014Q}}{Planck
  Collaboration XVII}}(2014)}]{planck2013-p12}%
  \BibitemOpen
  \bibfield  {author} {\bibinfo {author} {\bibnamefont {{\sorthelp{Planck
  Collaboration 2014Q}}{Planck Collaboration XVII}}},\ }\href {\doibase
  10.1051/0004-6361/201321543} {\bibfield  {journal} {\bibinfo  {journal}
  {\aap}\ }\textbf {\bibinfo {volume} {571}},\ \bibinfo {pages} {A17} (\bibinfo
  {year} {2014})},\ \Eprint {http://arxiv.org/abs/1303.5077} {arXiv:1303.5077}
  \BibitemShut {NoStop}%
\bibitem [{\citenamefont {{\sorthelp{Planck Collaboration 2014R}}{Planck
  Collaboration XVIII}}(2014)}]{planck2013-p13}%
  \BibitemOpen
  \bibfield  {author} {\bibinfo {author} {\bibnamefont {{\sorthelp{Planck
  Collaboration 2014R}}{Planck Collaboration XVIII}}},\ }\href {\doibase
  10.1051/0004-6361/201321540} {\bibfield  {journal} {\bibinfo  {journal}
  {\aap}\ }\textbf {\bibinfo {volume} {571}},\ \bibinfo {pages} {A18} (\bibinfo
  {year} {2014})},\ \Eprint {http://arxiv.org/abs/1303.5078} {arXiv:1303.5078}
  \BibitemShut {NoStop}%
\bibitem [{\citenamefont {{\sorthelp{Planck Collaboration 2015O}}{Planck
  Collaboration XV}}(2016)}]{planck2014-a17}%
  \BibitemOpen
  \bibfield  {author} {\bibinfo {author} {\bibnamefont {{\sorthelp{Planck
  Collaboration 2015O}}{Planck Collaboration XV}}},\ }\href {\doibase
  10.1051/0004-6361/201525941} {\bibfield  {journal} {\bibinfo  {journal}
  {\aap}\ }\textbf {\bibinfo {volume} {594}},\ \bibinfo {pages} {A15} (\bibinfo
  {year} {2016})},\ \Eprint {http://arxiv.org/abs/1502.01591}
  {arXiv:1502.01591} \BibitemShut {NoStop}%
\bibitem [{\citenamefont {{\sorthelp{Planck Collaboration IntZP}}{Planck
  Collaboration Int. XLI}}(2016)}]{planck2015-XLI}%
  \BibitemOpen
  \bibfield  {author} {\bibinfo {author} {\bibnamefont {{\sorthelp{Planck
  Collaboration IntZP}}{Planck Collaboration Int. XLI}}},\ }\href {\doibase
  10.1051/0004-6361/201527932} {\bibfield  {journal} {\bibinfo  {journal}
  {\aap}\ }\textbf {\bibinfo {volume} {596}},\ \bibinfo {pages} {A102}
  (\bibinfo {year} {2016})},\ \Eprint {http://arxiv.org/abs/1512.02882}
  {arXiv:1512.02882} \BibitemShut {NoStop}%
\bibitem [{\citenamefont {{\sorthelp{Planck Collaboration 2018H}}{Planck
  Collaboration VIII}}(2018)}]{planck2016-l08}%
  \BibitemOpen
  \bibfield  {author} {\bibinfo {author} {\bibnamefont {{\sorthelp{Planck
  Collaboration 2018H}}{Planck Collaboration VIII}}},\ }\href@noop {}
  {\bibfield  {journal} {\bibinfo  {journal} {\aap, submitted}\ } (\bibinfo
  {year} {2018})},\ \Eprint {http://arxiv.org/abs/1807.06210}
  {arXiv:1807.06210} \BibitemShut {NoStop}%
\bibitem [{\citenamefont {{van Engelen}}\ \emph {et~al.}(2012)\citenamefont
  {{van Engelen}}, \citenamefont {Keisler}, \citenamefont {Zahn}, \citenamefont
  {Aird}, \citenamefont {Benson}, \citenamefont {Bleem}, \citenamefont
  {Carlstrom}, \citenamefont {Chang}, \citenamefont {Cho}, \citenamefont
  {Crawford}, \citenamefont {Crites}, \citenamefont {{de Haan}}, \citenamefont
  {Dobbs}, \citenamefont {Dudley}, \citenamefont {George}, \citenamefont
  {Halverson}, \citenamefont {Holder}, \citenamefont {Holzapfel}, \citenamefont
  {Hoover}, \citenamefont {Hou}, \citenamefont {Hrubes}, \citenamefont {Joy},
  \citenamefont {Knox}, \citenamefont {Lee}, \citenamefont {Leitch},
  \citenamefont {Lueker}, \citenamefont {{Luong-Van}}, \citenamefont {McMahon},
  \citenamefont {Mehl}, \citenamefont {Meyer}, \citenamefont {Millea},
  \citenamefont {Mohr}, \citenamefont {Montroy}, \citenamefont {Natoli},
  \citenamefont {Padin}, \citenamefont {Plagge}, \citenamefont {Pryke},
  \citenamefont {Reichardt}, \citenamefont {Ruhl}, \citenamefont {Sayre},
  \citenamefont {Schaffer}, \citenamefont {Shaw}, \citenamefont {Shirokoff},
  \citenamefont {Spieler}, \citenamefont {Staniszewski}, \citenamefont {Stark},
  \citenamefont {Story}, \citenamefont {Vanderlinde}, \citenamefont {Vieira},\
  and\ \citenamefont {Williamson}}]{vanengelen2012}%
  \BibitemOpen
  \bibfield  {author} {\bibinfo {author} {\bibfnamefont {A.}~\bibnamefont {{van
  Engelen}}}, \bibinfo {author} {\bibfnamefont {R.}~\bibnamefont {Keisler}},
  \bibinfo {author} {\bibfnamefont {O.}~\bibnamefont {Zahn}}, \bibinfo {author}
  {\bibfnamefont {K.~A.}\ \bibnamefont {Aird}}, \bibinfo {author}
  {\bibfnamefont {B.~A.}\ \bibnamefont {Benson}}, \bibinfo {author}
  {\bibfnamefont {L.~E.}\ \bibnamefont {Bleem}}, \bibinfo {author}
  {\bibfnamefont {J.~E.}\ \bibnamefont {Carlstrom}}, \bibinfo {author}
  {\bibfnamefont {C.~L.}\ \bibnamefont {Chang}}, \bibinfo {author}
  {\bibfnamefont {H.~M.}\ \bibnamefont {Cho}}, \bibinfo {author} {\bibfnamefont
  {T.~M.}\ \bibnamefont {Crawford}}, \bibinfo {author} {\bibfnamefont {A.~T.}\
  \bibnamefont {Crites}}, \bibinfo {author} {\bibfnamefont {T.}~\bibnamefont
  {{de Haan}}}, \bibinfo {author} {\bibfnamefont {M.~A.}\ \bibnamefont
  {Dobbs}}, \bibinfo {author} {\bibfnamefont {J.}~\bibnamefont {Dudley}},
  \bibinfo {author} {\bibfnamefont {E.~M.}\ \bibnamefont {George}}, \bibinfo
  {author} {\bibfnamefont {N.~W.}\ \bibnamefont {Halverson}}, \bibinfo {author}
  {\bibfnamefont {G.~P.}\ \bibnamefont {Holder}}, \bibinfo {author}
  {\bibfnamefont {W.~L.}\ \bibnamefont {Holzapfel}}, \bibinfo {author}
  {\bibfnamefont {S.}~\bibnamefont {Hoover}}, \bibinfo {author} {\bibfnamefont
  {Z.}~\bibnamefont {Hou}}, \bibinfo {author} {\bibfnamefont {J.~D.}\
  \bibnamefont {Hrubes}}, \bibinfo {author} {\bibfnamefont {M.}~\bibnamefont
  {Joy}}, \bibinfo {author} {\bibfnamefont {L.}~\bibnamefont {Knox}}, \bibinfo
  {author} {\bibfnamefont {A.~T.}\ \bibnamefont {Lee}}, \bibinfo {author}
  {\bibfnamefont {E.~M.}\ \bibnamefont {Leitch}}, \bibinfo {author}
  {\bibfnamefont {M.}~\bibnamefont {Lueker}}, \bibinfo {author} {\bibfnamefont
  {D.}~\bibnamefont {{Luong-Van}}}, \bibinfo {author} {\bibfnamefont {J.~J.}\
  \bibnamefont {McMahon}}, \bibinfo {author} {\bibfnamefont {J.}~\bibnamefont
  {Mehl}}, \bibinfo {author} {\bibfnamefont {S.~S.}\ \bibnamefont {Meyer}},
  \bibinfo {author} {\bibfnamefont {M.}~\bibnamefont {Millea}}, \bibinfo
  {author} {\bibfnamefont {J.~J.}\ \bibnamefont {Mohr}}, \bibinfo {author}
  {\bibfnamefont {T.~E.}\ \bibnamefont {Montroy}}, \bibinfo {author}
  {\bibfnamefont {T.}~\bibnamefont {Natoli}}, \bibinfo {author} {\bibfnamefont
  {S.}~\bibnamefont {Padin}}, \bibinfo {author} {\bibfnamefont
  {T.}~\bibnamefont {Plagge}}, \bibinfo {author} {\bibfnamefont
  {C.}~\bibnamefont {Pryke}}, \bibinfo {author} {\bibfnamefont {C.~L.}\
  \bibnamefont {Reichardt}}, \bibinfo {author} {\bibfnamefont {J.~E.}\
  \bibnamefont {Ruhl}}, \bibinfo {author} {\bibfnamefont {J.~T.}\ \bibnamefont
  {Sayre}}, \bibinfo {author} {\bibfnamefont {K.~K.}\ \bibnamefont {Schaffer}},
  \bibinfo {author} {\bibfnamefont {L.}~\bibnamefont {Shaw}}, \bibinfo {author}
  {\bibfnamefont {E.}~\bibnamefont {Shirokoff}}, \bibinfo {author}
  {\bibfnamefont {H.~G.}\ \bibnamefont {Spieler}}, \bibinfo {author}
  {\bibfnamefont {Z.}~\bibnamefont {Staniszewski}}, \bibinfo {author}
  {\bibfnamefont {A.~A.}\ \bibnamefont {Stark}}, \bibinfo {author}
  {\bibfnamefont {K.}~\bibnamefont {Story}}, \bibinfo {author} {\bibfnamefont
  {K.}~\bibnamefont {Vanderlinde}}, \bibinfo {author} {\bibfnamefont {J.~D.}\
  \bibnamefont {Vieira}}, \ and\ \bibinfo {author} {\bibfnamefont
  {R.}~\bibnamefont {Williamson}},\ }\href {\doibase
  10.1088/0004-637X/756/2/142} {\bibfield  {journal} {\bibinfo  {journal} {The
  Astrophysical Journal}\ }\textbf {\bibinfo {volume} {756}},\ \bibinfo {pages}
  {142} (\bibinfo {year} {2012})}\BibitemShut {NoStop}%
\bibitem [{\citenamefont {Ade}\ \emph {et~al.}(2014)\citenamefont {Ade},
  \citenamefont {Akiba}, \citenamefont {Anthony}, \citenamefont {Arnold},
  \citenamefont {Atlas}, \citenamefont {Barron}, \citenamefont {Boettger},
  \citenamefont {Borrill}, \citenamefont {Chapman}, \citenamefont {Chinone},
  \citenamefont {Dobbs}, \citenamefont {Elleflot}, \citenamefont {Errard},
  \citenamefont {Fabbian}, \citenamefont {Feng}, \citenamefont {Flanigan},
  \citenamefont {Gilbert}, \citenamefont {Grainger}, \citenamefont {Halverson},
  \citenamefont {Hasegawa}, \citenamefont {Hattori}, \citenamefont {Hazumi},
  \citenamefont {Holzapfel}, \citenamefont {Hori}, \citenamefont {Howard},
  \citenamefont {Hyland}, \citenamefont {Inoue}, \citenamefont {Jaehnig},
  \citenamefont {Jaffe}, \citenamefont {Keating}, \citenamefont {Kermish},
  \citenamefont {Keskitalo}, \citenamefont {Kisner}, \citenamefont {Le~Jeune},
  \citenamefont {Lee}, \citenamefont {Linder}, \citenamefont {Leitch},
  \citenamefont {Lungu}, \citenamefont {Matsuda}, \citenamefont {Matsumura},
  \citenamefont {Meng}, \citenamefont {Miller}, \citenamefont {Morii},
  \citenamefont {Moyerman}, \citenamefont {Myers}, \citenamefont {Navaroli},
  \citenamefont {Nishino}, \citenamefont {Paar}, \citenamefont {Peloton},
  \citenamefont {Quealy}, \citenamefont {Rebeiz}, \citenamefont {Reichardt},
  \citenamefont {Richards}, \citenamefont {Ross}, \citenamefont {Schanning},
  \citenamefont {Schenck}, \citenamefont {Sherwin}, \citenamefont {Shimizu},
  \citenamefont {Shimmin}, \citenamefont {Shimon}, \citenamefont {Siritanasak},
  \citenamefont {Smecher}, \citenamefont {Spieler}, \citenamefont {Stebor},
  \citenamefont {Steinbach}, \citenamefont {Stompor}, \citenamefont {Suzuki},
  \citenamefont {Takakura}, \citenamefont {Tomaru}, \citenamefont {Wilson},
  \citenamefont {Yadav}, \citenamefont {Zahn},\ and\ \citenamefont {{Polarbear
  Collaboration}}}]{ade2014}%
  \BibitemOpen
  \bibfield  {author} {\bibinfo {author} {\bibfnamefont {P.~A.~R.}\
  \bibnamefont {Ade}}, \bibinfo {author} {\bibfnamefont {Y.}~\bibnamefont
  {Akiba}}, \bibinfo {author} {\bibfnamefont {A.~E.}\ \bibnamefont {Anthony}},
  \bibinfo {author} {\bibfnamefont {K.}~\bibnamefont {Arnold}}, \bibinfo
  {author} {\bibfnamefont {M.}~\bibnamefont {Atlas}}, \bibinfo {author}
  {\bibfnamefont {D.}~\bibnamefont {Barron}}, \bibinfo {author} {\bibfnamefont
  {D.}~\bibnamefont {Boettger}}, \bibinfo {author} {\bibfnamefont
  {J.}~\bibnamefont {Borrill}}, \bibinfo {author} {\bibfnamefont
  {S.}~\bibnamefont {Chapman}}, \bibinfo {author} {\bibfnamefont
  {Y.}~\bibnamefont {Chinone}}, \bibinfo {author} {\bibfnamefont
  {M.}~\bibnamefont {Dobbs}}, \bibinfo {author} {\bibfnamefont
  {T.}~\bibnamefont {Elleflot}}, \bibinfo {author} {\bibfnamefont
  {J.}~\bibnamefont {Errard}}, \bibinfo {author} {\bibfnamefont
  {G.}~\bibnamefont {Fabbian}}, \bibinfo {author} {\bibfnamefont
  {C.}~\bibnamefont {Feng}}, \bibinfo {author} {\bibfnamefont {D.}~\bibnamefont
  {Flanigan}}, \bibinfo {author} {\bibfnamefont {A.}~\bibnamefont {Gilbert}},
  \bibinfo {author} {\bibfnamefont {W.}~\bibnamefont {Grainger}}, \bibinfo
  {author} {\bibfnamefont {N.~W.}\ \bibnamefont {Halverson}}, \bibinfo {author}
  {\bibfnamefont {M.}~\bibnamefont {Hasegawa}}, \bibinfo {author}
  {\bibfnamefont {K.}~\bibnamefont {Hattori}}, \bibinfo {author} {\bibfnamefont
  {M.}~\bibnamefont {Hazumi}}, \bibinfo {author} {\bibfnamefont {W.~L.}\
  \bibnamefont {Holzapfel}}, \bibinfo {author} {\bibfnamefont {Y.}~\bibnamefont
  {Hori}}, \bibinfo {author} {\bibfnamefont {J.}~\bibnamefont {Howard}},
  \bibinfo {author} {\bibfnamefont {P.}~\bibnamefont {Hyland}}, \bibinfo
  {author} {\bibfnamefont {Y.}~\bibnamefont {Inoue}}, \bibinfo {author}
  {\bibfnamefont {G.~C.}\ \bibnamefont {Jaehnig}}, \bibinfo {author}
  {\bibfnamefont {A.}~\bibnamefont {Jaffe}}, \bibinfo {author} {\bibfnamefont
  {B.}~\bibnamefont {Keating}}, \bibinfo {author} {\bibfnamefont
  {Z.}~\bibnamefont {Kermish}}, \bibinfo {author} {\bibfnamefont
  {R.}~\bibnamefont {Keskitalo}}, \bibinfo {author} {\bibfnamefont
  {T.}~\bibnamefont {Kisner}}, \bibinfo {author} {\bibfnamefont
  {M.}~\bibnamefont {Le~Jeune}}, \bibinfo {author} {\bibfnamefont {A.~T.}\
  \bibnamefont {Lee}}, \bibinfo {author} {\bibfnamefont {E.}~\bibnamefont
  {Linder}}, \bibinfo {author} {\bibfnamefont {E.~M.}\ \bibnamefont {Leitch}},
  \bibinfo {author} {\bibfnamefont {M.}~\bibnamefont {Lungu}}, \bibinfo
  {author} {\bibfnamefont {F.}~\bibnamefont {Matsuda}}, \bibinfo {author}
  {\bibfnamefont {T.}~\bibnamefont {Matsumura}}, \bibinfo {author}
  {\bibfnamefont {X.}~\bibnamefont {Meng}}, \bibinfo {author} {\bibfnamefont
  {N.~J.}\ \bibnamefont {Miller}}, \bibinfo {author} {\bibfnamefont
  {H.}~\bibnamefont {Morii}}, \bibinfo {author} {\bibfnamefont
  {S.}~\bibnamefont {Moyerman}}, \bibinfo {author} {\bibfnamefont {M.~J.}\
  \bibnamefont {Myers}}, \bibinfo {author} {\bibfnamefont {M.}~\bibnamefont
  {Navaroli}}, \bibinfo {author} {\bibfnamefont {H.}~\bibnamefont {Nishino}},
  \bibinfo {author} {\bibfnamefont {H.}~\bibnamefont {Paar}}, \bibinfo {author}
  {\bibfnamefont {J.}~\bibnamefont {Peloton}}, \bibinfo {author} {\bibfnamefont
  {E.}~\bibnamefont {Quealy}}, \bibinfo {author} {\bibfnamefont
  {G.}~\bibnamefont {Rebeiz}}, \bibinfo {author} {\bibfnamefont {C.~L.}\
  \bibnamefont {Reichardt}}, \bibinfo {author} {\bibfnamefont {P.~L.}\
  \bibnamefont {Richards}}, \bibinfo {author} {\bibfnamefont {C.}~\bibnamefont
  {Ross}}, \bibinfo {author} {\bibfnamefont {I.}~\bibnamefont {Schanning}},
  \bibinfo {author} {\bibfnamefont {D.~E.}\ \bibnamefont {Schenck}}, \bibinfo
  {author} {\bibfnamefont {B.}~\bibnamefont {Sherwin}}, \bibinfo {author}
  {\bibfnamefont {A.}~\bibnamefont {Shimizu}}, \bibinfo {author} {\bibfnamefont
  {C.}~\bibnamefont {Shimmin}}, \bibinfo {author} {\bibfnamefont
  {M.}~\bibnamefont {Shimon}}, \bibinfo {author} {\bibfnamefont
  {P.}~\bibnamefont {Siritanasak}}, \bibinfo {author} {\bibfnamefont
  {G.}~\bibnamefont {Smecher}}, \bibinfo {author} {\bibfnamefont
  {H.}~\bibnamefont {Spieler}}, \bibinfo {author} {\bibfnamefont
  {N.}~\bibnamefont {Stebor}}, \bibinfo {author} {\bibfnamefont
  {B.}~\bibnamefont {Steinbach}}, \bibinfo {author} {\bibfnamefont
  {R.}~\bibnamefont {Stompor}}, \bibinfo {author} {\bibfnamefont
  {A.}~\bibnamefont {Suzuki}}, \bibinfo {author} {\bibfnamefont
  {S.}~\bibnamefont {Takakura}}, \bibinfo {author} {\bibfnamefont
  {T.}~\bibnamefont {Tomaru}}, \bibinfo {author} {\bibfnamefont
  {B.}~\bibnamefont {Wilson}}, \bibinfo {author} {\bibfnamefont
  {A.}~\bibnamefont {Yadav}}, \bibinfo {author} {\bibfnamefont
  {O.}~\bibnamefont {Zahn}}, \ and\ \bibinfo {author} {\bibnamefont {{Polarbear
  Collaboration}}},\ }\href {\doibase 10.1103/PhysRevLett.113.021301}
  {\bibfield  {journal} {\bibinfo  {journal} {Physical Review Letters}\
  }\textbf {\bibinfo {volume} {113}},\ \bibinfo {pages} {021301} (\bibinfo
  {year} {2014})}\BibitemShut {NoStop}%
\bibitem [{\citenamefont {Story}\ \emph {et~al.}(2015)\citenamefont {Story},
  \citenamefont {Hanson}, \citenamefont {Ade}, \citenamefont {Aird},
  \citenamefont {Austermann}, \citenamefont {Beall}, \citenamefont {Bender},
  \citenamefont {Benson}, \citenamefont {Bleem}, \citenamefont {Carlstrom},
  \citenamefont {Chang}, \citenamefont {Chiang}, \citenamefont {Cho},
  \citenamefont {Citron}, \citenamefont {Crawford}, \citenamefont {Crites},
  \citenamefont {{de Haan}}, \citenamefont {Dobbs}, \citenamefont {Everett},
  \citenamefont {Gallicchio}, \citenamefont {Gao}, \citenamefont {George},
  \citenamefont {Gilbert}, \citenamefont {Halverson}, \citenamefont
  {Harrington}, \citenamefont {Henning}, \citenamefont {Hilton}, \citenamefont
  {Holder}, \citenamefont {Holzapfel}, \citenamefont {Hoover}, \citenamefont
  {Hou}, \citenamefont {Hrubes}, \citenamefont {Huang}, \citenamefont
  {Hubmayr}, \citenamefont {Irwin}, \citenamefont {Keisler}, \citenamefont
  {Knox}, \citenamefont {Lee}, \citenamefont {Leitch}, \citenamefont {Li},
  \citenamefont {Liang}, \citenamefont {{Luong-Van}}, \citenamefont {McMahon},
  \citenamefont {Mehl}, \citenamefont {Meyer}, \citenamefont {Mocanu},
  \citenamefont {Montroy}, \citenamefont {Natoli}, \citenamefont {Nibarger},
  \citenamefont {Novosad}, \citenamefont {Padin}, \citenamefont {Pryke},
  \citenamefont {Reichardt}, \citenamefont {Ruhl}, \citenamefont {Saliwanchik},
  \citenamefont {Sayre}, \citenamefont {Schaffer}, \citenamefont {Smecher},
  \citenamefont {Stark}, \citenamefont {Tucker}, \citenamefont {Vanderlinde},
  \citenamefont {Vieira}, \citenamefont {Wang}, \citenamefont {Whitehorn},
  \citenamefont {Yefremenko},\ and\ \citenamefont {Zahn}}]{story2015}%
  \BibitemOpen
  \bibfield  {author} {\bibinfo {author} {\bibfnamefont {K.~T.}\ \bibnamefont
  {Story}}, \bibinfo {author} {\bibfnamefont {D.}~\bibnamefont {Hanson}},
  \bibinfo {author} {\bibfnamefont {P.~A.~R.}\ \bibnamefont {Ade}}, \bibinfo
  {author} {\bibfnamefont {K.~A.}\ \bibnamefont {Aird}}, \bibinfo {author}
  {\bibfnamefont {J.~E.}\ \bibnamefont {Austermann}}, \bibinfo {author}
  {\bibfnamefont {J.~A.}\ \bibnamefont {Beall}}, \bibinfo {author}
  {\bibfnamefont {A.~N.}\ \bibnamefont {Bender}}, \bibinfo {author}
  {\bibfnamefont {B.~A.}\ \bibnamefont {Benson}}, \bibinfo {author}
  {\bibfnamefont {L.~E.}\ \bibnamefont {Bleem}}, \bibinfo {author}
  {\bibfnamefont {J.~E.}\ \bibnamefont {Carlstrom}}, \bibinfo {author}
  {\bibfnamefont {C.~L.}\ \bibnamefont {Chang}}, \bibinfo {author}
  {\bibfnamefont {H.~C.}\ \bibnamefont {Chiang}}, \bibinfo {author}
  {\bibfnamefont {H.-M.}\ \bibnamefont {Cho}}, \bibinfo {author} {\bibfnamefont
  {R.}~\bibnamefont {Citron}}, \bibinfo {author} {\bibfnamefont {T.~M.}\
  \bibnamefont {Crawford}}, \bibinfo {author} {\bibfnamefont {A.~T.}\
  \bibnamefont {Crites}}, \bibinfo {author} {\bibfnamefont {T.}~\bibnamefont
  {{de Haan}}}, \bibinfo {author} {\bibfnamefont {M.~A.}\ \bibnamefont
  {Dobbs}}, \bibinfo {author} {\bibfnamefont {W.}~\bibnamefont {Everett}},
  \bibinfo {author} {\bibfnamefont {J.}~\bibnamefont {Gallicchio}}, \bibinfo
  {author} {\bibfnamefont {J.}~\bibnamefont {Gao}}, \bibinfo {author}
  {\bibfnamefont {E.~M.}\ \bibnamefont {George}}, \bibinfo {author}
  {\bibfnamefont {A.}~\bibnamefont {Gilbert}}, \bibinfo {author} {\bibfnamefont
  {N.~W.}\ \bibnamefont {Halverson}}, \bibinfo {author} {\bibfnamefont
  {N.}~\bibnamefont {Harrington}}, \bibinfo {author} {\bibfnamefont {J.~W.}\
  \bibnamefont {Henning}}, \bibinfo {author} {\bibfnamefont {G.~C.}\
  \bibnamefont {Hilton}}, \bibinfo {author} {\bibfnamefont {G.~P.}\
  \bibnamefont {Holder}}, \bibinfo {author} {\bibfnamefont {W.~L.}\
  \bibnamefont {Holzapfel}}, \bibinfo {author} {\bibfnamefont {S.}~\bibnamefont
  {Hoover}}, \bibinfo {author} {\bibfnamefont {Z.}~\bibnamefont {Hou}},
  \bibinfo {author} {\bibfnamefont {J.~D.}\ \bibnamefont {Hrubes}}, \bibinfo
  {author} {\bibfnamefont {N.}~\bibnamefont {Huang}}, \bibinfo {author}
  {\bibfnamefont {J.}~\bibnamefont {Hubmayr}}, \bibinfo {author} {\bibfnamefont
  {K.~D.}\ \bibnamefont {Irwin}}, \bibinfo {author} {\bibfnamefont
  {R.}~\bibnamefont {Keisler}}, \bibinfo {author} {\bibfnamefont
  {L.}~\bibnamefont {Knox}}, \bibinfo {author} {\bibfnamefont {A.~T.}\
  \bibnamefont {Lee}}, \bibinfo {author} {\bibfnamefont {E.~M.}\ \bibnamefont
  {Leitch}}, \bibinfo {author} {\bibfnamefont {D.}~\bibnamefont {Li}}, \bibinfo
  {author} {\bibfnamefont {C.}~\bibnamefont {Liang}}, \bibinfo {author}
  {\bibfnamefont {D.}~\bibnamefont {{Luong-Van}}}, \bibinfo {author}
  {\bibfnamefont {J.~J.}\ \bibnamefont {McMahon}}, \bibinfo {author}
  {\bibfnamefont {J.}~\bibnamefont {Mehl}}, \bibinfo {author} {\bibfnamefont
  {S.~S.}\ \bibnamefont {Meyer}}, \bibinfo {author} {\bibfnamefont
  {L.}~\bibnamefont {Mocanu}}, \bibinfo {author} {\bibfnamefont {T.~E.}\
  \bibnamefont {Montroy}}, \bibinfo {author} {\bibfnamefont {T.}~\bibnamefont
  {Natoli}}, \bibinfo {author} {\bibfnamefont {J.~P.}\ \bibnamefont
  {Nibarger}}, \bibinfo {author} {\bibfnamefont {V.}~\bibnamefont {Novosad}},
  \bibinfo {author} {\bibfnamefont {S.}~\bibnamefont {Padin}}, \bibinfo
  {author} {\bibfnamefont {C.}~\bibnamefont {Pryke}}, \bibinfo {author}
  {\bibfnamefont {C.~L.}\ \bibnamefont {Reichardt}}, \bibinfo {author}
  {\bibfnamefont {J.~E.}\ \bibnamefont {Ruhl}}, \bibinfo {author}
  {\bibfnamefont {B.~R.}\ \bibnamefont {Saliwanchik}}, \bibinfo {author}
  {\bibfnamefont {J.~T.}\ \bibnamefont {Sayre}}, \bibinfo {author}
  {\bibfnamefont {K.~K.}\ \bibnamefont {Schaffer}}, \bibinfo {author}
  {\bibfnamefont {G.}~\bibnamefont {Smecher}}, \bibinfo {author} {\bibfnamefont
  {A.~A.}\ \bibnamefont {Stark}}, \bibinfo {author} {\bibfnamefont
  {C.}~\bibnamefont {Tucker}}, \bibinfo {author} {\bibfnamefont
  {K.}~\bibnamefont {Vanderlinde}}, \bibinfo {author} {\bibfnamefont {J.~D.}\
  \bibnamefont {Vieira}}, \bibinfo {author} {\bibfnamefont {G.}~\bibnamefont
  {Wang}}, \bibinfo {author} {\bibfnamefont {N.}~\bibnamefont {Whitehorn}},
  \bibinfo {author} {\bibfnamefont {V.}~\bibnamefont {Yefremenko}}, \ and\
  \bibinfo {author} {\bibfnamefont {O.}~\bibnamefont {Zahn}},\ }\href {\doibase
  10.1088/0004-637X/810/1/50} {\bibfield  {journal} {\bibinfo  {journal} {The
  Astrophysical Journal}\ }\textbf {\bibinfo {volume} {810}},\ \bibinfo {pages}
  {50} (\bibinfo {year} {2015})},\ \Eprint {http://arxiv.org/abs/1412.4760}
  {arXiv:1412.4760} \BibitemShut {NoStop}%
\bibitem [{\citenamefont {{BICEP2 Collaboration}}\ \emph
  {et~al.}(2016)\citenamefont {{BICEP2 Collaboration}}, \citenamefont {{Keck
  Array Collaboration}}, \citenamefont {Ade}, \citenamefont {Ahmed},
  \citenamefont {Aikin}, \citenamefont {Alexander}, \citenamefont {Barkats},
  \citenamefont {Benton}, \citenamefont {Bischoff}, \citenamefont {Bock},
  \citenamefont {{Bowens-Rubin}}, \citenamefont {Brevik}, \citenamefont
  {Buder}, \citenamefont {Bullock}, \citenamefont {Buza}, \citenamefont
  {Connors}, \citenamefont {Crill}, \citenamefont {Duband}, \citenamefont
  {Dvorkin}, \citenamefont {Filippini}, \citenamefont {Fliescher},
  \citenamefont {Grayson}, \citenamefont {Halpern}, \citenamefont {Harrison},
  \citenamefont {Hildebrandt}, \citenamefont {Hilton}, \citenamefont {Hui},
  \citenamefont {Irwin}, \citenamefont {Kang}, \citenamefont {Karkare},
  \citenamefont {Karpel}, \citenamefont {Kaufman}, \citenamefont {Keating},
  \citenamefont {Kefeli}, \citenamefont {Kernasovskiy}, \citenamefont {Kovac},
  \citenamefont {Kuo}, \citenamefont {Leitch}, \citenamefont {Lueker},
  \citenamefont {Megerian}, \citenamefont {Namikawa}, \citenamefont
  {Netterfield}, \citenamefont {Nguyen}, \citenamefont {O'Brient},
  \citenamefont {Ogburn}, \citenamefont {Orlando}, \citenamefont {Pryke},
  \citenamefont {Richter}, \citenamefont {Schwarz}, \citenamefont {Sheehy},
  \citenamefont {Staniszewski}, \citenamefont {Steinbach}, \citenamefont
  {Sudiwala}, \citenamefont {Teply}, \citenamefont {Thompson}, \citenamefont
  {Tolan}, \citenamefont {Tucker}, \citenamefont {Turner}, \citenamefont
  {Vieregg}, \citenamefont {Weber}, \citenamefont {Wiebe}, \citenamefont
  {Willmert}, \citenamefont {Wong}, \citenamefont {Wu},\ and\ \citenamefont
  {Yoon}}]{bicep2collaboration2016a}%
  \BibitemOpen
  \bibfield  {author} {\bibinfo {author} {\bibnamefont {{BICEP2
  Collaboration}}}, \bibinfo {author} {\bibnamefont {{Keck Array
  Collaboration}}}, \bibinfo {author} {\bibfnamefont {P.~A.~R.}\ \bibnamefont
  {Ade}}, \bibinfo {author} {\bibfnamefont {Z.}~\bibnamefont {Ahmed}}, \bibinfo
  {author} {\bibfnamefont {R.~W.}\ \bibnamefont {Aikin}}, \bibinfo {author}
  {\bibfnamefont {K.~D.}\ \bibnamefont {Alexander}}, \bibinfo {author}
  {\bibfnamefont {D.}~\bibnamefont {Barkats}}, \bibinfo {author} {\bibfnamefont
  {S.~J.}\ \bibnamefont {Benton}}, \bibinfo {author} {\bibfnamefont {C.~A.}\
  \bibnamefont {Bischoff}}, \bibinfo {author} {\bibfnamefont {J.~J.}\
  \bibnamefont {Bock}}, \bibinfo {author} {\bibfnamefont {R.}~\bibnamefont
  {{Bowens-Rubin}}}, \bibinfo {author} {\bibfnamefont {J.~A.}\ \bibnamefont
  {Brevik}}, \bibinfo {author} {\bibfnamefont {I.}~\bibnamefont {Buder}},
  \bibinfo {author} {\bibfnamefont {E.}~\bibnamefont {Bullock}}, \bibinfo
  {author} {\bibfnamefont {V.}~\bibnamefont {Buza}}, \bibinfo {author}
  {\bibfnamefont {J.}~\bibnamefont {Connors}}, \bibinfo {author} {\bibfnamefont
  {B.~P.}\ \bibnamefont {Crill}}, \bibinfo {author} {\bibfnamefont
  {L.}~\bibnamefont {Duband}}, \bibinfo {author} {\bibfnamefont
  {C.}~\bibnamefont {Dvorkin}}, \bibinfo {author} {\bibfnamefont {J.~P.}\
  \bibnamefont {Filippini}}, \bibinfo {author} {\bibfnamefont {S.}~\bibnamefont
  {Fliescher}}, \bibinfo {author} {\bibfnamefont {J.}~\bibnamefont {Grayson}},
  \bibinfo {author} {\bibfnamefont {M.}~\bibnamefont {Halpern}}, \bibinfo
  {author} {\bibfnamefont {S.}~\bibnamefont {Harrison}}, \bibinfo {author}
  {\bibfnamefont {S.~R.}\ \bibnamefont {Hildebrandt}}, \bibinfo {author}
  {\bibfnamefont {G.~C.}\ \bibnamefont {Hilton}}, \bibinfo {author}
  {\bibfnamefont {H.}~\bibnamefont {Hui}}, \bibinfo {author} {\bibfnamefont
  {K.~D.}\ \bibnamefont {Irwin}}, \bibinfo {author} {\bibfnamefont
  {J.}~\bibnamefont {Kang}}, \bibinfo {author} {\bibfnamefont {K.~S.}\
  \bibnamefont {Karkare}}, \bibinfo {author} {\bibfnamefont {E.}~\bibnamefont
  {Karpel}}, \bibinfo {author} {\bibfnamefont {J.~P.}\ \bibnamefont {Kaufman}},
  \bibinfo {author} {\bibfnamefont {B.~G.}\ \bibnamefont {Keating}}, \bibinfo
  {author} {\bibfnamefont {S.}~\bibnamefont {Kefeli}}, \bibinfo {author}
  {\bibfnamefont {S.~A.}\ \bibnamefont {Kernasovskiy}}, \bibinfo {author}
  {\bibfnamefont {J.~M.}\ \bibnamefont {Kovac}}, \bibinfo {author}
  {\bibfnamefont {C.~L.}\ \bibnamefont {Kuo}}, \bibinfo {author} {\bibfnamefont
  {E.~M.}\ \bibnamefont {Leitch}}, \bibinfo {author} {\bibfnamefont
  {M.}~\bibnamefont {Lueker}}, \bibinfo {author} {\bibfnamefont {K.~G.}\
  \bibnamefont {Megerian}}, \bibinfo {author} {\bibfnamefont {T.}~\bibnamefont
  {Namikawa}}, \bibinfo {author} {\bibfnamefont {C.~B.}\ \bibnamefont
  {Netterfield}}, \bibinfo {author} {\bibfnamefont {H.~T.}\ \bibnamefont
  {Nguyen}}, \bibinfo {author} {\bibfnamefont {R.}~\bibnamefont {O'Brient}},
  \bibinfo {author} {\bibfnamefont {R.~W.}\ \bibnamefont {Ogburn},
  \bibfnamefont {IV}}, \bibinfo {author} {\bibfnamefont {A.}~\bibnamefont
  {Orlando}}, \bibinfo {author} {\bibfnamefont {C.}~\bibnamefont {Pryke}},
  \bibinfo {author} {\bibfnamefont {S.}~\bibnamefont {Richter}}, \bibinfo
  {author} {\bibfnamefont {R.}~\bibnamefont {Schwarz}}, \bibinfo {author}
  {\bibfnamefont {C.~D.}\ \bibnamefont {Sheehy}}, \bibinfo {author}
  {\bibfnamefont {Z.~K.}\ \bibnamefont {Staniszewski}}, \bibinfo {author}
  {\bibfnamefont {B.}~\bibnamefont {Steinbach}}, \bibinfo {author}
  {\bibfnamefont {R.~V.}\ \bibnamefont {Sudiwala}}, \bibinfo {author}
  {\bibfnamefont {G.~P.}\ \bibnamefont {Teply}}, \bibinfo {author}
  {\bibfnamefont {K.~L.}\ \bibnamefont {Thompson}}, \bibinfo {author}
  {\bibfnamefont {J.~E.}\ \bibnamefont {Tolan}}, \bibinfo {author}
  {\bibfnamefont {C.}~\bibnamefont {Tucker}}, \bibinfo {author} {\bibfnamefont
  {A.~D.}\ \bibnamefont {Turner}}, \bibinfo {author} {\bibfnamefont {A.~G.}\
  \bibnamefont {Vieregg}}, \bibinfo {author} {\bibfnamefont {A.~C.}\
  \bibnamefont {Weber}}, \bibinfo {author} {\bibfnamefont {D.~V.}\ \bibnamefont
  {Wiebe}}, \bibinfo {author} {\bibfnamefont {J.}~\bibnamefont {Willmert}},
  \bibinfo {author} {\bibfnamefont {C.~L.}\ \bibnamefont {Wong}}, \bibinfo
  {author} {\bibfnamefont {W.~L.~K.}\ \bibnamefont {Wu}}, \ and\ \bibinfo
  {author} {\bibfnamefont {K.~W.}\ \bibnamefont {Yoon}},\ }\href {\doibase
  10.3847/1538-4357/833/2/228} {\bibfield  {journal} {\bibinfo  {journal} {The
  Astrophysical Journal}\ }\textbf {\bibinfo {volume} {833}},\ \bibinfo {pages}
  {228} (\bibinfo {year} {2016})}\BibitemShut {NoStop}%
\bibitem [{\citenamefont {Omori}\ \emph {et~al.}(2017)\citenamefont {Omori},
  \citenamefont {Chown}, \citenamefont {Simard}, \citenamefont {Story},
  \citenamefont {Aylor}, \citenamefont {Baxter}, \citenamefont {Benson},
  \citenamefont {Bleem}, \citenamefont {Carlstrom}, \citenamefont {Chang},
  \citenamefont {Cho}, \citenamefont {Crawford}, \citenamefont {Crites},
  \citenamefont {{de Haan}}, \citenamefont {Dobbs}, \citenamefont {Everett},
  \citenamefont {George}, \citenamefont {Halverson}, \citenamefont
  {Harrington}, \citenamefont {Holder}, \citenamefont {Hou}, \citenamefont
  {Holzapfel}, \citenamefont {Hrubes}, \citenamefont {Knox}, \citenamefont
  {Lee}, \citenamefont {Leitch}, \citenamefont {{Luong-Van}}, \citenamefont
  {Manzotti}, \citenamefont {Marrone}, \citenamefont {McMahon}, \citenamefont
  {Meyer}, \citenamefont {Mocanu}, \citenamefont {Mohr}, \citenamefont
  {Natoli}, \citenamefont {Padin}, \citenamefont {Pryke}, \citenamefont
  {Reichardt}, \citenamefont {Ruhl}, \citenamefont {Sayre}, \citenamefont
  {Schaffer}, \citenamefont {Shirokoff}, \citenamefont {Staniszewski},
  \citenamefont {Stark}, \citenamefont {Vanderlinde}, \citenamefont {Vieira},
  \citenamefont {Williamson},\ and\ \citenamefont {Zahn}}]{omori2017}%
  \BibitemOpen
  \bibfield  {author} {\bibinfo {author} {\bibfnamefont {Y.}~\bibnamefont
  {Omori}}, \bibinfo {author} {\bibfnamefont {R.}~\bibnamefont {Chown}},
  \bibinfo {author} {\bibfnamefont {G.}~\bibnamefont {Simard}}, \bibinfo
  {author} {\bibfnamefont {K.~T.}\ \bibnamefont {Story}}, \bibinfo {author}
  {\bibfnamefont {K.}~\bibnamefont {Aylor}}, \bibinfo {author} {\bibfnamefont
  {E.~J.}\ \bibnamefont {Baxter}}, \bibinfo {author} {\bibfnamefont {B.~A.}\
  \bibnamefont {Benson}}, \bibinfo {author} {\bibfnamefont {L.~E.}\
  \bibnamefont {Bleem}}, \bibinfo {author} {\bibfnamefont {J.~E.}\ \bibnamefont
  {Carlstrom}}, \bibinfo {author} {\bibfnamefont {C.~L.}\ \bibnamefont
  {Chang}}, \bibinfo {author} {\bibfnamefont {H.-M.}\ \bibnamefont {Cho}},
  \bibinfo {author} {\bibfnamefont {T.~M.}\ \bibnamefont {Crawford}}, \bibinfo
  {author} {\bibfnamefont {A.~T.}\ \bibnamefont {Crites}}, \bibinfo {author}
  {\bibfnamefont {T.}~\bibnamefont {{de Haan}}}, \bibinfo {author}
  {\bibfnamefont {M.~A.}\ \bibnamefont {Dobbs}}, \bibinfo {author}
  {\bibfnamefont {W.~B.}\ \bibnamefont {Everett}}, \bibinfo {author}
  {\bibfnamefont {E.~M.}\ \bibnamefont {George}}, \bibinfo {author}
  {\bibfnamefont {N.~W.}\ \bibnamefont {Halverson}}, \bibinfo {author}
  {\bibfnamefont {N.~L.}\ \bibnamefont {Harrington}}, \bibinfo {author}
  {\bibfnamefont {G.~P.}\ \bibnamefont {Holder}}, \bibinfo {author}
  {\bibfnamefont {Z.}~\bibnamefont {Hou}}, \bibinfo {author} {\bibfnamefont
  {W.~L.}\ \bibnamefont {Holzapfel}}, \bibinfo {author} {\bibfnamefont {J.~D.}\
  \bibnamefont {Hrubes}}, \bibinfo {author} {\bibfnamefont {L.}~\bibnamefont
  {Knox}}, \bibinfo {author} {\bibfnamefont {A.~T.}\ \bibnamefont {Lee}},
  \bibinfo {author} {\bibfnamefont {E.~M.}\ \bibnamefont {Leitch}}, \bibinfo
  {author} {\bibfnamefont {D.}~\bibnamefont {{Luong-Van}}}, \bibinfo {author}
  {\bibfnamefont {A.}~\bibnamefont {Manzotti}}, \bibinfo {author}
  {\bibfnamefont {D.~P.}\ \bibnamefont {Marrone}}, \bibinfo {author}
  {\bibfnamefont {J.~J.}\ \bibnamefont {McMahon}}, \bibinfo {author}
  {\bibfnamefont {S.~S.}\ \bibnamefont {Meyer}}, \bibinfo {author}
  {\bibfnamefont {L.~M.}\ \bibnamefont {Mocanu}}, \bibinfo {author}
  {\bibfnamefont {J.~J.}\ \bibnamefont {Mohr}}, \bibinfo {author}
  {\bibfnamefont {T.}~\bibnamefont {Natoli}}, \bibinfo {author} {\bibfnamefont
  {S.}~\bibnamefont {Padin}}, \bibinfo {author} {\bibfnamefont
  {C.}~\bibnamefont {Pryke}}, \bibinfo {author} {\bibfnamefont {C.~L.}\
  \bibnamefont {Reichardt}}, \bibinfo {author} {\bibfnamefont {J.~E.}\
  \bibnamefont {Ruhl}}, \bibinfo {author} {\bibfnamefont {J.~T.}\ \bibnamefont
  {Sayre}}, \bibinfo {author} {\bibfnamefont {K.~K.}\ \bibnamefont {Schaffer}},
  \bibinfo {author} {\bibfnamefont {E.}~\bibnamefont {Shirokoff}}, \bibinfo
  {author} {\bibfnamefont {Z.}~\bibnamefont {Staniszewski}}, \bibinfo {author}
  {\bibfnamefont {A.~A.}\ \bibnamefont {Stark}}, \bibinfo {author}
  {\bibfnamefont {K.}~\bibnamefont {Vanderlinde}}, \bibinfo {author}
  {\bibfnamefont {J.~D.}\ \bibnamefont {Vieira}}, \bibinfo {author}
  {\bibfnamefont {R.}~\bibnamefont {Williamson}}, \ and\ \bibinfo {author}
  {\bibfnamefont {O.}~\bibnamefont {Zahn}},\ }\href {\doibase
  10.3847/1538-4357/aa8d1d} {\bibfield  {journal} {\bibinfo  {journal} {The
  Astrophysical Journal}\ }\textbf {\bibinfo {volume} {849}},\ \bibinfo {pages}
  {124} (\bibinfo {year} {2017})},\ \Eprint {http://arxiv.org/abs/1705.00743}
  {arXiv:1705.00743} \BibitemShut {NoStop}%
\bibitem [{\citenamefont {Wu}\ \emph {et~al.}(2019)\citenamefont {Wu},
  \citenamefont {Mocanu}, \citenamefont {Ade}, \citenamefont {Anderson},
  \citenamefont {Austermann}, \citenamefont {Avva}, \citenamefont {Beall},
  \citenamefont {Bender}, \citenamefont {Benson}, \citenamefont {Bianchini},
  \citenamefont {Bleem}, \citenamefont {Carlstrom}, \citenamefont {Chang},
  \citenamefont {Chiang}, \citenamefont {Citron}, \citenamefont {Moran},
  \citenamefont {Crawford}, \citenamefont {Crites}, \citenamefont {{de Haan}},
  \citenamefont {Dobbs}, \citenamefont {Everett}, \citenamefont {Gallicchio},
  \citenamefont {George}, \citenamefont {Gilbert}, \citenamefont {Gupta},
  \citenamefont {Halverson}, \citenamefont {Harrington}, \citenamefont
  {Henning}, \citenamefont {Hilton}, \citenamefont {Holder}, \citenamefont
  {Holzapfel}, \citenamefont {Hou}, \citenamefont {Hrubes}, \citenamefont
  {Huang}, \citenamefont {Hubmayr}, \citenamefont {Irwin}, \citenamefont
  {Knox}, \citenamefont {Lee}, \citenamefont {Li}, \citenamefont {Lowitz},
  \citenamefont {Manzotti}, \citenamefont {McMahon}, \citenamefont {Meyer},
  \citenamefont {Millea}, \citenamefont {Montgomery}, \citenamefont {Nadolski},
  \citenamefont {Natoli}, \citenamefont {Nibarger}, \citenamefont {Noble},
  \citenamefont {Novosad}, \citenamefont {Omori}, \citenamefont {Padin},
  \citenamefont {Patil}, \citenamefont {Pryke}, \citenamefont {Reichardt},
  \citenamefont {Ruhl}, \citenamefont {Saliwanchik}, \citenamefont {Sayre},
  \citenamefont {Schaffer}, \citenamefont {Sievers}, \citenamefont {Simard},
  \citenamefont {Smecher}, \citenamefont {Stark}, \citenamefont {Story},
  \citenamefont {Tucker}, \citenamefont {Vanderlinde}, \citenamefont {Veach},
  \citenamefont {Vieira}, \citenamefont {Wang}, \citenamefont {Whitehorn},\
  and\ \citenamefont {Yefremenko}}]{wu2019}%
  \BibitemOpen
  \bibfield  {author} {\bibinfo {author} {\bibfnamefont {W.~L.~K.}\
  \bibnamefont {Wu}}, \bibinfo {author} {\bibfnamefont {L.~M.}\ \bibnamefont
  {Mocanu}}, \bibinfo {author} {\bibfnamefont {P.~A.~R.}\ \bibnamefont {Ade}},
  \bibinfo {author} {\bibfnamefont {A.~J.}\ \bibnamefont {Anderson}}, \bibinfo
  {author} {\bibfnamefont {J.~E.}\ \bibnamefont {Austermann}}, \bibinfo
  {author} {\bibfnamefont {J.~S.}\ \bibnamefont {Avva}}, \bibinfo {author}
  {\bibfnamefont {J.~A.}\ \bibnamefont {Beall}}, \bibinfo {author}
  {\bibfnamefont {A.~N.}\ \bibnamefont {Bender}}, \bibinfo {author}
  {\bibfnamefont {B.~A.}\ \bibnamefont {Benson}}, \bibinfo {author}
  {\bibfnamefont {F.}~\bibnamefont {Bianchini}}, \bibinfo {author}
  {\bibfnamefont {L.~E.}\ \bibnamefont {Bleem}}, \bibinfo {author}
  {\bibfnamefont {J.~E.}\ \bibnamefont {Carlstrom}}, \bibinfo {author}
  {\bibfnamefont {C.~L.}\ \bibnamefont {Chang}}, \bibinfo {author}
  {\bibfnamefont {H.~C.}\ \bibnamefont {Chiang}}, \bibinfo {author}
  {\bibfnamefont {R.}~\bibnamefont {Citron}}, \bibinfo {author} {\bibfnamefont
  {C.~C.}\ \bibnamefont {Moran}}, \bibinfo {author} {\bibfnamefont {T.~M.}\
  \bibnamefont {Crawford}}, \bibinfo {author} {\bibfnamefont {A.~T.}\
  \bibnamefont {Crites}}, \bibinfo {author} {\bibfnamefont {T.}~\bibnamefont
  {{de Haan}}}, \bibinfo {author} {\bibfnamefont {M.~A.}\ \bibnamefont
  {Dobbs}}, \bibinfo {author} {\bibfnamefont {W.}~\bibnamefont {Everett}},
  \bibinfo {author} {\bibfnamefont {J.}~\bibnamefont {Gallicchio}}, \bibinfo
  {author} {\bibfnamefont {E.~M.}\ \bibnamefont {George}}, \bibinfo {author}
  {\bibfnamefont {A.}~\bibnamefont {Gilbert}}, \bibinfo {author} {\bibfnamefont
  {N.}~\bibnamefont {Gupta}}, \bibinfo {author} {\bibfnamefont {N.~W.}\
  \bibnamefont {Halverson}}, \bibinfo {author} {\bibfnamefont {N.}~\bibnamefont
  {Harrington}}, \bibinfo {author} {\bibfnamefont {J.~W.}\ \bibnamefont
  {Henning}}, \bibinfo {author} {\bibfnamefont {G.~C.}\ \bibnamefont {Hilton}},
  \bibinfo {author} {\bibfnamefont {G.~P.}\ \bibnamefont {Holder}}, \bibinfo
  {author} {\bibfnamefont {W.~L.}\ \bibnamefont {Holzapfel}}, \bibinfo {author}
  {\bibfnamefont {Z.}~\bibnamefont {Hou}}, \bibinfo {author} {\bibfnamefont
  {J.~D.}\ \bibnamefont {Hrubes}}, \bibinfo {author} {\bibfnamefont
  {N.}~\bibnamefont {Huang}}, \bibinfo {author} {\bibfnamefont
  {J.}~\bibnamefont {Hubmayr}}, \bibinfo {author} {\bibfnamefont {K.~D.}\
  \bibnamefont {Irwin}}, \bibinfo {author} {\bibfnamefont {L.}~\bibnamefont
  {Knox}}, \bibinfo {author} {\bibfnamefont {A.~T.}\ \bibnamefont {Lee}},
  \bibinfo {author} {\bibfnamefont {D.}~\bibnamefont {Li}}, \bibinfo {author}
  {\bibfnamefont {A.}~\bibnamefont {Lowitz}}, \bibinfo {author} {\bibfnamefont
  {A.}~\bibnamefont {Manzotti}}, \bibinfo {author} {\bibfnamefont {J.~J.}\
  \bibnamefont {McMahon}}, \bibinfo {author} {\bibfnamefont {S.~S.}\
  \bibnamefont {Meyer}}, \bibinfo {author} {\bibfnamefont {M.}~\bibnamefont
  {Millea}}, \bibinfo {author} {\bibfnamefont {J.}~\bibnamefont {Montgomery}},
  \bibinfo {author} {\bibfnamefont {A.}~\bibnamefont {Nadolski}}, \bibinfo
  {author} {\bibfnamefont {T.}~\bibnamefont {Natoli}}, \bibinfo {author}
  {\bibfnamefont {J.~P.}\ \bibnamefont {Nibarger}}, \bibinfo {author}
  {\bibfnamefont {G.~I.}\ \bibnamefont {Noble}}, \bibinfo {author}
  {\bibfnamefont {V.}~\bibnamefont {Novosad}}, \bibinfo {author} {\bibfnamefont
  {Y.}~\bibnamefont {Omori}}, \bibinfo {author} {\bibfnamefont
  {S.}~\bibnamefont {Padin}}, \bibinfo {author} {\bibfnamefont
  {S.}~\bibnamefont {Patil}}, \bibinfo {author} {\bibfnamefont
  {C.}~\bibnamefont {Pryke}}, \bibinfo {author} {\bibfnamefont {C.~L.}\
  \bibnamefont {Reichardt}}, \bibinfo {author} {\bibfnamefont {J.~E.}\
  \bibnamefont {Ruhl}}, \bibinfo {author} {\bibfnamefont {B.~R.}\ \bibnamefont
  {Saliwanchik}}, \bibinfo {author} {\bibfnamefont {J.~T.}\ \bibnamefont
  {Sayre}}, \bibinfo {author} {\bibfnamefont {K.~K.}\ \bibnamefont {Schaffer}},
  \bibinfo {author} {\bibfnamefont {C.}~\bibnamefont {Sievers}}, \bibinfo
  {author} {\bibfnamefont {G.}~\bibnamefont {Simard}}, \bibinfo {author}
  {\bibfnamefont {G.}~\bibnamefont {Smecher}}, \bibinfo {author} {\bibfnamefont
  {A.~A.}\ \bibnamefont {Stark}}, \bibinfo {author} {\bibfnamefont {K.~T.}\
  \bibnamefont {Story}}, \bibinfo {author} {\bibfnamefont {C.}~\bibnamefont
  {Tucker}}, \bibinfo {author} {\bibfnamefont {K.}~\bibnamefont {Vanderlinde}},
  \bibinfo {author} {\bibfnamefont {T.}~\bibnamefont {Veach}}, \bibinfo
  {author} {\bibfnamefont {J.~D.}\ \bibnamefont {Vieira}}, \bibinfo {author}
  {\bibfnamefont {G.}~\bibnamefont {Wang}}, \bibinfo {author} {\bibfnamefont
  {N.}~\bibnamefont {Whitehorn}}, \ and\ \bibinfo {author} {\bibfnamefont
  {V.}~\bibnamefont {Yefremenko}},\ }\href {http://arxiv.org/abs/1905.05777}
  {\bibfield  {journal} {\bibinfo  {journal} {arXiv:1905.05777 [astro-ph]}\ }
  (\bibinfo {year} {2019})},\ \Eprint {http://arxiv.org/abs/1905.05777}
  {arXiv:1905.05777 [astro-ph]} \BibitemShut {NoStop}%
\bibitem [{\citenamefont {Carron}\ \emph {et~al.}(2017)\citenamefont {Carron},
  \citenamefont {Lewis},\ and\ \citenamefont {Challinor}}]{carron2017a}%
  \BibitemOpen
  \bibfield  {author} {\bibinfo {author} {\bibfnamefont {J.}~\bibnamefont
  {Carron}}, \bibinfo {author} {\bibfnamefont {A.}~\bibnamefont {Lewis}}, \
  and\ \bibinfo {author} {\bibfnamefont {A.}~\bibnamefont {Challinor}},\ }\href
  {http://arxiv.org/abs/1701.01712} {\bibfield  {journal} {\bibinfo  {journal}
  {arXiv:1701.01712 [astro-ph]}\ } (\bibinfo {year} {2017})},\ \Eprint
  {http://arxiv.org/abs/1701.01712} {arXiv:1701.01712 [astro-ph]} \BibitemShut
  {NoStop}%
\bibitem [{\citenamefont {Seljak}\ and\ \citenamefont
  {Hirata}(2004)}]{seljak2004}%
  \BibitemOpen
  \bibfield  {author} {\bibinfo {author} {\bibfnamefont {U.}~\bibnamefont
  {Seljak}}\ and\ \bibinfo {author} {\bibfnamefont {C.~M.}\ \bibnamefont
  {Hirata}},\ }\href {\doibase 10.1103/PhysRevD.69.043005} {\bibfield
  {journal} {\bibinfo  {journal} {Physical Review D}\ }\textbf {\bibinfo
  {volume} {69}},\ \bibinfo {pages} {043005} (\bibinfo {year}
  {2004})}\BibitemShut {NoStop}%
\bibitem [{\citenamefont {Smith}\ \emph {et~al.}(2012)\citenamefont {Smith},
  \citenamefont {Hanson}, \citenamefont {LoVerde}, \citenamefont {Hirata},\
  and\ \citenamefont {Zahn}}]{smith2012}%
  \BibitemOpen
  \bibfield  {author} {\bibinfo {author} {\bibfnamefont {K.~M.}\ \bibnamefont
  {Smith}}, \bibinfo {author} {\bibfnamefont {D.}~\bibnamefont {Hanson}},
  \bibinfo {author} {\bibfnamefont {M.}~\bibnamefont {LoVerde}}, \bibinfo
  {author} {\bibfnamefont {C.~M.}\ \bibnamefont {Hirata}}, \ and\ \bibinfo
  {author} {\bibfnamefont {O.}~\bibnamefont {Zahn}},\ }\href {\doibase
  10.1088/1475-7516/2012/06/014} {\bibfield  {journal} {\bibinfo  {journal}
  {Journal of Cosmology and Astroparticle Physics}\ }\textbf {\bibinfo {volume}
  {2012}},\ \bibinfo {pages} {014} (\bibinfo {year} {2012})},\ \Eprint
  {http://arxiv.org/abs/1010.0048} {arXiv:1010.0048} \BibitemShut {NoStop}%
\bibitem [{\citenamefont {Horowitz}\ \emph {et~al.}(2017)\citenamefont
  {Horowitz}, \citenamefont {Ferraro},\ and\ \citenamefont
  {Sherwin}}]{horowitz2017}%
  \BibitemOpen
  \bibfield  {author} {\bibinfo {author} {\bibfnamefont {B.}~\bibnamefont
  {Horowitz}}, \bibinfo {author} {\bibfnamefont {S.}~\bibnamefont {Ferraro}}, \
  and\ \bibinfo {author} {\bibfnamefont {B.~D.}\ \bibnamefont {Sherwin}},\
  }\href {http://arxiv.org/abs/1710.10236} {\bibfield  {journal} {\bibinfo
  {journal} {arXiv:1710.10236 [astro-ph]}\ } (\bibinfo {year} {2017})},\
  \Eprint {http://arxiv.org/abs/1710.10236} {arXiv:1710.10236 [astro-ph]}
  \BibitemShut {NoStop}%
\bibitem [{\citenamefont {Mirmelstein}\ \emph {et~al.}(2019)\citenamefont
  {Mirmelstein}, \citenamefont {Carron},\ and\ \citenamefont
  {Lewis}}]{mirmelstein2019}%
  \BibitemOpen
  \bibfield  {author} {\bibinfo {author} {\bibfnamefont {M.}~\bibnamefont
  {Mirmelstein}}, \bibinfo {author} {\bibfnamefont {J.}~\bibnamefont {Carron}},
  \ and\ \bibinfo {author} {\bibfnamefont {A.}~\bibnamefont {Lewis}},\ }\href
  {http://arxiv.org/abs/1909.02653} {\bibfield  {journal} {\bibinfo  {journal}
  {arXiv:1909.02653 [astro-ph]}\ } (\bibinfo {year} {2019})},\ \Eprint
  {http://arxiv.org/abs/1909.02653} {arXiv:1909.02653 [astro-ph]} \BibitemShut
  {NoStop}%
\bibitem [{\citenamefont {Hadzhiyska}\ \emph {et~al.}(2019)\citenamefont
  {Hadzhiyska}, \citenamefont {Sherwin}, \citenamefont {Madhavacheril},\ and\
  \citenamefont {Ferraro}}]{hadzhiyska2019}%
  \BibitemOpen
  \bibfield  {author} {\bibinfo {author} {\bibfnamefont {B.}~\bibnamefont
  {Hadzhiyska}}, \bibinfo {author} {\bibfnamefont {B.~D.}\ \bibnamefont
  {Sherwin}}, \bibinfo {author} {\bibfnamefont {M.}~\bibnamefont
  {Madhavacheril}}, \ and\ \bibinfo {author} {\bibfnamefont {S.}~\bibnamefont
  {Ferraro}},\ }\href {\doibase 10.1103/PhysRevD.100.023547} {\bibfield
  {journal} {\bibinfo  {journal} {Physical Review D}\ }\textbf {\bibinfo
  {volume} {100}},\ \bibinfo {pages} {023547} (\bibinfo {year}
  {2019})}\BibitemShut {NoStop}%
\bibitem [{\citenamefont {Caldeira}\ \emph {et~al.}(2018)\citenamefont
  {Caldeira}, \citenamefont {Wu}, \citenamefont {Nord}, \citenamefont
  {Avestruz}, \citenamefont {Trivedi},\ and\ \citenamefont
  {Story}}]{caldeira2018}%
  \BibitemOpen
  \bibfield  {author} {\bibinfo {author} {\bibfnamefont {J.}~\bibnamefont
  {Caldeira}}, \bibinfo {author} {\bibfnamefont {W.~L.~K.}\ \bibnamefont {Wu}},
  \bibinfo {author} {\bibfnamefont {B.}~\bibnamefont {Nord}}, \bibinfo {author}
  {\bibfnamefont {C.}~\bibnamefont {Avestruz}}, \bibinfo {author}
  {\bibfnamefont {S.}~\bibnamefont {Trivedi}}, \ and\ \bibinfo {author}
  {\bibfnamefont {K.~T.}\ \bibnamefont {Story}},\ }\href
  {http://arxiv.org/abs/1810.01483} {\bibfield  {journal} {\bibinfo  {journal}
  {arXiv:1810.01483 [astro-ph]}\ } (\bibinfo {year} {2018})},\ \Eprint
  {http://arxiv.org/abs/1810.01483} {arXiv:1810.01483 [astro-ph]} \BibitemShut
  {NoStop}%
\bibitem [{\citenamefont {Hirata}\ and\ \citenamefont
  {Seljak}(2003{\natexlab{a}})}]{hirata2003}%
  \BibitemOpen
  \bibfield  {author} {\bibinfo {author} {\bibfnamefont {C.~M.}\ \bibnamefont
  {Hirata}}\ and\ \bibinfo {author} {\bibfnamefont {U.}~\bibnamefont
  {Seljak}},\ }\href {\doibase 10.1103/PhysRevD.68.083002} {\bibfield
  {journal} {\bibinfo  {journal} {Physical Review D}\ }\textbf {\bibinfo
  {volume} {68}} (\bibinfo {year} {2003}{\natexlab{a}}),\
  10.1103/PhysRevD.68.083002},\ \Eprint {http://arxiv.org/abs/astro-ph/0306354}
  {arXiv:astro-ph/0306354} \BibitemShut {NoStop}%
\bibitem [{\citenamefont {Hirata}\ and\ \citenamefont
  {Seljak}(2003{\natexlab{b}})}]{hirata2003b}%
  \BibitemOpen
  \bibfield  {author} {\bibinfo {author} {\bibfnamefont {C.~M.}\ \bibnamefont
  {Hirata}}\ and\ \bibinfo {author} {\bibfnamefont {U.}~\bibnamefont
  {Seljak}},\ }\href {\doibase 10.1103/PhysRevD.67.043001} {\bibfield
  {journal} {\bibinfo  {journal} {Physical Review D}\ }\textbf {\bibinfo
  {volume} {67}} (\bibinfo {year} {2003}{\natexlab{b}}),\
  10.1103/PhysRevD.67.043001},\ \Eprint {http://arxiv.org/abs/astro-ph/0209489}
  {arXiv:astro-ph/0209489} \BibitemShut {NoStop}%
\bibitem [{\citenamefont {Carron}\ and\ \citenamefont
  {Lewis}(2017)}]{carron2017}%
  \BibitemOpen
  \bibfield  {author} {\bibinfo {author} {\bibfnamefont {J.}~\bibnamefont
  {Carron}}\ and\ \bibinfo {author} {\bibfnamefont {A.}~\bibnamefont {Lewis}},\
  }\href {\doibase 10.1103/PhysRevD.96.063510} {\bibfield  {journal} {\bibinfo
  {journal} {Physical Review D}\ }\textbf {\bibinfo {volume} {96}},\ \bibinfo
  {pages} {063510} (\bibinfo {year} {2017})}\BibitemShut {NoStop}%
\bibitem [{\citenamefont {Adachi}\ \emph {et~al.}(2019)\citenamefont {Adachi},
  \citenamefont {Aguilar~Fa{\'u}ndez}, \citenamefont {Akiba}, \citenamefont
  {Ali}, \citenamefont {Arnold}, \citenamefont {Baccigalupi}, \citenamefont
  {Barron}, \citenamefont {Beck}, \citenamefont {Bianchini}, \citenamefont
  {Borrill}, \citenamefont {Carron}, \citenamefont {Cheung}, \citenamefont
  {Chinone}, \citenamefont {Crowley}, \citenamefont {El~Bouhargani},
  \citenamefont {Elleflot}, \citenamefont {Errard}, \citenamefont {Fabbian},
  \citenamefont {Feng}, \citenamefont {Fujino}, \citenamefont
  {{Goeckner-Wald}}, \citenamefont {Hasegawa}, \citenamefont {Hazumi},
  \citenamefont {Hill}, \citenamefont {Howe}, \citenamefont {Katayama},
  \citenamefont {Keating}, \citenamefont {Kikuchi}, \citenamefont {Kusaka},
  \citenamefont {Lee}, \citenamefont {Leon}, \citenamefont {Linder},
  \citenamefont {Lowry}, \citenamefont {Matsuda}, \citenamefont {Matsumura},
  \citenamefont {Minami}, \citenamefont {Namikawa}, \citenamefont {Nishino},
  \citenamefont {Peloton}, \citenamefont {Pham}, \citenamefont {Poletti},
  \citenamefont {Puglisi}, \citenamefont {Reichardt}, \citenamefont {Segawa},
  \citenamefont {Sherwin}, \citenamefont {{Silva-Feaver}}, \citenamefont
  {Siritanasak}, \citenamefont {Stompor}, \citenamefont {Tajima}, \citenamefont
  {Takatori}, \citenamefont {Tanabe}, \citenamefont {Teply},\ and\
  \citenamefont {Verges}}]{adachi2019}%
  \BibitemOpen
  \bibfield  {author} {\bibinfo {author} {\bibfnamefont {S.}~\bibnamefont
  {Adachi}}, \bibinfo {author} {\bibfnamefont {M.~a.~O.}\ \bibnamefont
  {Aguilar~Fa{\'u}ndez}}, \bibinfo {author} {\bibfnamefont {Y.}~\bibnamefont
  {Akiba}}, \bibinfo {author} {\bibfnamefont {A.}~\bibnamefont {Ali}}, \bibinfo
  {author} {\bibfnamefont {K.}~\bibnamefont {Arnold}}, \bibinfo {author}
  {\bibfnamefont {C.}~\bibnamefont {Baccigalupi}}, \bibinfo {author}
  {\bibfnamefont {D.}~\bibnamefont {Barron}}, \bibinfo {author} {\bibfnamefont
  {D.}~\bibnamefont {Beck}}, \bibinfo {author} {\bibfnamefont {F.}~\bibnamefont
  {Bianchini}}, \bibinfo {author} {\bibfnamefont {J.}~\bibnamefont {Borrill}},
  \bibinfo {author} {\bibfnamefont {J.}~\bibnamefont {Carron}}, \bibinfo
  {author} {\bibfnamefont {K.}~\bibnamefont {Cheung}}, \bibinfo {author}
  {\bibfnamefont {Y.}~\bibnamefont {Chinone}}, \bibinfo {author} {\bibfnamefont
  {K.}~\bibnamefont {Crowley}}, \bibinfo {author} {\bibfnamefont
  {H.}~\bibnamefont {El~Bouhargani}}, \bibinfo {author} {\bibfnamefont
  {T.}~\bibnamefont {Elleflot}}, \bibinfo {author} {\bibfnamefont
  {J.}~\bibnamefont {Errard}}, \bibinfo {author} {\bibfnamefont
  {G.}~\bibnamefont {Fabbian}}, \bibinfo {author} {\bibfnamefont
  {C.}~\bibnamefont {Feng}}, \bibinfo {author} {\bibfnamefont {T.}~\bibnamefont
  {Fujino}}, \bibinfo {author} {\bibfnamefont {N.}~\bibnamefont
  {{Goeckner-Wald}}}, \bibinfo {author} {\bibfnamefont {M.}~\bibnamefont
  {Hasegawa}}, \bibinfo {author} {\bibfnamefont {M.}~\bibnamefont {Hazumi}},
  \bibinfo {author} {\bibfnamefont {C.~A.}\ \bibnamefont {Hill}}, \bibinfo
  {author} {\bibfnamefont {L.}~\bibnamefont {Howe}}, \bibinfo {author}
  {\bibfnamefont {N.}~\bibnamefont {Katayama}}, \bibinfo {author}
  {\bibfnamefont {B.}~\bibnamefont {Keating}}, \bibinfo {author} {\bibfnamefont
  {S.}~\bibnamefont {Kikuchi}}, \bibinfo {author} {\bibfnamefont
  {A.}~\bibnamefont {Kusaka}}, \bibinfo {author} {\bibfnamefont {A.~T.}\
  \bibnamefont {Lee}}, \bibinfo {author} {\bibfnamefont {D.}~\bibnamefont
  {Leon}}, \bibinfo {author} {\bibfnamefont {E.}~\bibnamefont {Linder}},
  \bibinfo {author} {\bibfnamefont {L.~N.}\ \bibnamefont {Lowry}}, \bibinfo
  {author} {\bibfnamefont {F.}~\bibnamefont {Matsuda}}, \bibinfo {author}
  {\bibfnamefont {T.}~\bibnamefont {Matsumura}}, \bibinfo {author}
  {\bibfnamefont {Y.}~\bibnamefont {Minami}}, \bibinfo {author} {\bibfnamefont
  {T.}~\bibnamefont {Namikawa}}, \bibinfo {author} {\bibfnamefont
  {H.}~\bibnamefont {Nishino}}, \bibinfo {author} {\bibfnamefont
  {J.}~\bibnamefont {Peloton}}, \bibinfo {author} {\bibfnamefont {A.~T.~P.}\
  \bibnamefont {Pham}}, \bibinfo {author} {\bibfnamefont {D.}~\bibnamefont
  {Poletti}}, \bibinfo {author} {\bibfnamefont {G.}~\bibnamefont {Puglisi}},
  \bibinfo {author} {\bibfnamefont {C.~L.}\ \bibnamefont {Reichardt}}, \bibinfo
  {author} {\bibfnamefont {Y.}~\bibnamefont {Segawa}}, \bibinfo {author}
  {\bibfnamefont {B.~D.}\ \bibnamefont {Sherwin}}, \bibinfo {author}
  {\bibfnamefont {M.}~\bibnamefont {{Silva-Feaver}}}, \bibinfo {author}
  {\bibfnamefont {P.}~\bibnamefont {Siritanasak}}, \bibinfo {author}
  {\bibfnamefont {R.}~\bibnamefont {Stompor}}, \bibinfo {author} {\bibfnamefont
  {O.}~\bibnamefont {Tajima}}, \bibinfo {author} {\bibfnamefont
  {S.}~\bibnamefont {Takatori}}, \bibinfo {author} {\bibfnamefont
  {D.}~\bibnamefont {Tanabe}}, \bibinfo {author} {\bibfnamefont {G.~P.}\
  \bibnamefont {Teply}}, \ and\ \bibinfo {author} {\bibfnamefont
  {C.}~\bibnamefont {Verges}},\ }\href
  {https://ui.adsabs.harvard.edu/abs/2019arXiv190913832A/abstract} {\bibfield
  {journal} {\bibinfo  {journal} {arXiv e-prints}\ ,\ \bibinfo {pages}
  {arXiv:1909.13832}} (\bibinfo {year} {2019})}\BibitemShut {NoStop}%
\bibitem [{\citenamefont {Carron}(2018)}]{carron2018}%
  \BibitemOpen
  \bibfield  {author} {\bibinfo {author} {\bibfnamefont {J.}~\bibnamefont
  {Carron}},\ }\href {http://arxiv.org/abs/1808.10349} {\bibfield  {journal}
  {\bibinfo  {journal} {arXiv:1808.10349 [astro-ph]}\ } (\bibinfo {year}
  {2018})},\ \Eprint {http://arxiv.org/abs/1808.10349} {arXiv:1808.10349
  [astro-ph]} \BibitemShut {NoStop}%
\bibitem [{\citenamefont {Betancourt}(2017)}]{betancourt2017}%
  \BibitemOpen
  \bibfield  {author} {\bibinfo {author} {\bibfnamefont {M.}~\bibnamefont
  {Betancourt}},\ }\href {http://arxiv.org/abs/1701.02434} {\bibfield
  {journal} {\bibinfo  {journal} {arXiv:1701.02434 [stat]}\ } (\bibinfo {year}
  {2017})},\ \Eprint {http://arxiv.org/abs/1701.02434} {arXiv:1701.02434
  [stat]} \BibitemShut {NoStop}%
\bibitem [{\citenamefont {Millea}\ \emph {et~al.}(2019)\citenamefont {Millea},
  \citenamefont {Anderes},\ and\ \citenamefont {Wandelt}}]{millea2019}%
  \BibitemOpen
  \bibfield  {author} {\bibinfo {author} {\bibfnamefont {M.}~\bibnamefont
  {Millea}}, \bibinfo {author} {\bibfnamefont {E.}~\bibnamefont {Anderes}}, \
  and\ \bibinfo {author} {\bibfnamefont {B.~D.}\ \bibnamefont {Wandelt}},\
  }\href {\doibase 10.1103/PhysRevD.100.023509} {\bibfield  {journal} {\bibinfo
   {journal} {Physical Review D}\ }\textbf {\bibinfo {volume} {100}},\ \bibinfo
  {pages} {023509} (\bibinfo {year} {2019})}\BibitemShut {NoStop}%
\bibitem [{\citenamefont {Abazajian}\ \emph {et~al.}(2016)\citenamefont
  {Abazajian}, \citenamefont {Adshead}, \citenamefont {Ahmed}, \citenamefont
  {Allen}, \citenamefont {Alonso}, \citenamefont {Arnold}, \citenamefont
  {Baccigalupi}, \citenamefont {Bartlett}, \citenamefont {Battaglia},
  \citenamefont {Benson}, \citenamefont {Bischoff}, \citenamefont {Borrill},
  \citenamefont {Buza}, \citenamefont {Calabrese}, \citenamefont {Caldwell},
  \citenamefont {Carlstrom}, \citenamefont {Chang}, \citenamefont {Crawford},
  \citenamefont {{Cyr-Racine}}, \citenamefont {De~Bernardis}, \citenamefont
  {{de Haan}}, \citenamefont {{di Serego Alighieri}}, \citenamefont {Dunkley},
  \citenamefont {Dvorkin}, \citenamefont {Errard}, \citenamefont {Fabbian},
  \citenamefont {Feeney}, \citenamefont {Ferraro}, \citenamefont {Filippini},
  \citenamefont {Flauger}, \citenamefont {Fuller}, \citenamefont {Gluscevic},
  \citenamefont {Green}, \citenamefont {Grin}, \citenamefont {Grohs},
  \citenamefont {Henning}, \citenamefont {Hill}, \citenamefont {Hlozek},
  \citenamefont {Holder}, \citenamefont {Holzapfel}, \citenamefont {Hu},
  \citenamefont {Huffenberger}, \citenamefont {Keskitalo}, \citenamefont
  {Knox}, \citenamefont {Kosowsky}, \citenamefont {Kovac}, \citenamefont
  {Kovetz}, \citenamefont {Kuo}, \citenamefont {Kusaka}, \citenamefont
  {Le~Jeune}, \citenamefont {Lee}, \citenamefont {Lilley}, \citenamefont
  {Loverde}, \citenamefont {Madhavacheril}, \citenamefont {Mantz},
  \citenamefont {Marsh}, \citenamefont {McMahon}, \citenamefont {Meerburg},
  \citenamefont {Meyers}, \citenamefont {Miller}, \citenamefont {Munoz},
  \citenamefont {Nguyen}, \citenamefont {Niemack}, \citenamefont {Peloso},
  \citenamefont {Peloton}, \citenamefont {Pogosian}, \citenamefont {Pryke},
  \citenamefont {Raveri}, \citenamefont {Reichardt}, \citenamefont {Rocha},
  \citenamefont {Rotti}, \citenamefont {Schaan}, \citenamefont {Schmittfull},
  \citenamefont {Scott}, \citenamefont {Sehgal}, \citenamefont {Shandera},
  \citenamefont {Sherwin}, \citenamefont {Smith}, \citenamefont {Sorbo},
  \citenamefont {Starkman}, \citenamefont {Story}, \citenamefont {{van
  Engelen}}, \citenamefont {Vieira}, \citenamefont {Watson}, \citenamefont
  {Whitehorn},\ and\ \citenamefont {Kimmy~Wu}}]{abazajian2016}%
  \BibitemOpen
  \bibfield  {author} {\bibinfo {author} {\bibfnamefont {K.~N.}\ \bibnamefont
  {Abazajian}}, \bibinfo {author} {\bibfnamefont {P.}~\bibnamefont {Adshead}},
  \bibinfo {author} {\bibfnamefont {Z.}~\bibnamefont {Ahmed}}, \bibinfo
  {author} {\bibfnamefont {S.~W.}\ \bibnamefont {Allen}}, \bibinfo {author}
  {\bibfnamefont {D.}~\bibnamefont {Alonso}}, \bibinfo {author} {\bibfnamefont
  {K.~S.}\ \bibnamefont {Arnold}}, \bibinfo {author} {\bibfnamefont
  {C.}~\bibnamefont {Baccigalupi}}, \bibinfo {author} {\bibfnamefont {J.~G.}\
  \bibnamefont {Bartlett}}, \bibinfo {author} {\bibfnamefont {N.}~\bibnamefont
  {Battaglia}}, \bibinfo {author} {\bibfnamefont {B.~A.}\ \bibnamefont
  {Benson}}, \bibinfo {author} {\bibfnamefont {C.~A.}\ \bibnamefont
  {Bischoff}}, \bibinfo {author} {\bibfnamefont {J.}~\bibnamefont {Borrill}},
  \bibinfo {author} {\bibfnamefont {V.}~\bibnamefont {Buza}}, \bibinfo {author}
  {\bibfnamefont {E.}~\bibnamefont {Calabrese}}, \bibinfo {author}
  {\bibfnamefont {R.}~\bibnamefont {Caldwell}}, \bibinfo {author}
  {\bibfnamefont {J.~E.}\ \bibnamefont {Carlstrom}}, \bibinfo {author}
  {\bibfnamefont {C.~L.}\ \bibnamefont {Chang}}, \bibinfo {author}
  {\bibfnamefont {T.~M.}\ \bibnamefont {Crawford}}, \bibinfo {author}
  {\bibfnamefont {F.-Y.}\ \bibnamefont {{Cyr-Racine}}}, \bibinfo {author}
  {\bibfnamefont {F.}~\bibnamefont {De~Bernardis}}, \bibinfo {author}
  {\bibfnamefont {T.}~\bibnamefont {{de Haan}}}, \bibinfo {author}
  {\bibfnamefont {S.}~\bibnamefont {{di Serego Alighieri}}}, \bibinfo {author}
  {\bibfnamefont {J.}~\bibnamefont {Dunkley}}, \bibinfo {author} {\bibfnamefont
  {C.}~\bibnamefont {Dvorkin}}, \bibinfo {author} {\bibfnamefont
  {J.}~\bibnamefont {Errard}}, \bibinfo {author} {\bibfnamefont
  {G.}~\bibnamefont {Fabbian}}, \bibinfo {author} {\bibfnamefont
  {S.}~\bibnamefont {Feeney}}, \bibinfo {author} {\bibfnamefont
  {S.}~\bibnamefont {Ferraro}}, \bibinfo {author} {\bibfnamefont {J.~P.}\
  \bibnamefont {Filippini}}, \bibinfo {author} {\bibfnamefont {R.}~\bibnamefont
  {Flauger}}, \bibinfo {author} {\bibfnamefont {G.~M.}\ \bibnamefont {Fuller}},
  \bibinfo {author} {\bibfnamefont {V.}~\bibnamefont {Gluscevic}}, \bibinfo
  {author} {\bibfnamefont {D.}~\bibnamefont {Green}}, \bibinfo {author}
  {\bibfnamefont {D.}~\bibnamefont {Grin}}, \bibinfo {author} {\bibfnamefont
  {E.}~\bibnamefont {Grohs}}, \bibinfo {author} {\bibfnamefont {J.~W.}\
  \bibnamefont {Henning}}, \bibinfo {author} {\bibfnamefont {J.~C.}\
  \bibnamefont {Hill}}, \bibinfo {author} {\bibfnamefont {R.}~\bibnamefont
  {Hlozek}}, \bibinfo {author} {\bibfnamefont {G.}~\bibnamefont {Holder}},
  \bibinfo {author} {\bibfnamefont {W.}~\bibnamefont {Holzapfel}}, \bibinfo
  {author} {\bibfnamefont {W.}~\bibnamefont {Hu}}, \bibinfo {author}
  {\bibfnamefont {K.~M.}\ \bibnamefont {Huffenberger}}, \bibinfo {author}
  {\bibfnamefont {R.}~\bibnamefont {Keskitalo}}, \bibinfo {author}
  {\bibfnamefont {L.}~\bibnamefont {Knox}}, \bibinfo {author} {\bibfnamefont
  {A.}~\bibnamefont {Kosowsky}}, \bibinfo {author} {\bibfnamefont
  {J.}~\bibnamefont {Kovac}}, \bibinfo {author} {\bibfnamefont {E.~D.}\
  \bibnamefont {Kovetz}}, \bibinfo {author} {\bibfnamefont {C.-L.}\
  \bibnamefont {Kuo}}, \bibinfo {author} {\bibfnamefont {A.}~\bibnamefont
  {Kusaka}}, \bibinfo {author} {\bibfnamefont {M.}~\bibnamefont {Le~Jeune}},
  \bibinfo {author} {\bibfnamefont {A.~T.}\ \bibnamefont {Lee}}, \bibinfo
  {author} {\bibfnamefont {M.}~\bibnamefont {Lilley}}, \bibinfo {author}
  {\bibfnamefont {M.}~\bibnamefont {Loverde}}, \bibinfo {author} {\bibfnamefont
  {M.~S.}\ \bibnamefont {Madhavacheril}}, \bibinfo {author} {\bibfnamefont
  {A.}~\bibnamefont {Mantz}}, \bibinfo {author} {\bibfnamefont {D.~J.~E.}\
  \bibnamefont {Marsh}}, \bibinfo {author} {\bibfnamefont {J.}~\bibnamefont
  {McMahon}}, \bibinfo {author} {\bibfnamefont {P.~D.}\ \bibnamefont
  {Meerburg}}, \bibinfo {author} {\bibfnamefont {J.}~\bibnamefont {Meyers}},
  \bibinfo {author} {\bibfnamefont {A.~D.}\ \bibnamefont {Miller}}, \bibinfo
  {author} {\bibfnamefont {J.~B.}\ \bibnamefont {Munoz}}, \bibinfo {author}
  {\bibfnamefont {H.~N.}\ \bibnamefont {Nguyen}}, \bibinfo {author}
  {\bibfnamefont {M.~D.}\ \bibnamefont {Niemack}}, \bibinfo {author}
  {\bibfnamefont {M.}~\bibnamefont {Peloso}}, \bibinfo {author} {\bibfnamefont
  {J.}~\bibnamefont {Peloton}}, \bibinfo {author} {\bibfnamefont
  {L.}~\bibnamefont {Pogosian}}, \bibinfo {author} {\bibfnamefont
  {C.}~\bibnamefont {Pryke}}, \bibinfo {author} {\bibfnamefont
  {M.}~\bibnamefont {Raveri}}, \bibinfo {author} {\bibfnamefont {C.~L.}\
  \bibnamefont {Reichardt}}, \bibinfo {author} {\bibfnamefont {G.}~\bibnamefont
  {Rocha}}, \bibinfo {author} {\bibfnamefont {A.}~\bibnamefont {Rotti}},
  \bibinfo {author} {\bibfnamefont {E.}~\bibnamefont {Schaan}}, \bibinfo
  {author} {\bibfnamefont {M.~M.}\ \bibnamefont {Schmittfull}}, \bibinfo
  {author} {\bibfnamefont {D.}~\bibnamefont {Scott}}, \bibinfo {author}
  {\bibfnamefont {N.}~\bibnamefont {Sehgal}}, \bibinfo {author} {\bibfnamefont
  {S.}~\bibnamefont {Shandera}}, \bibinfo {author} {\bibfnamefont {B.~D.}\
  \bibnamefont {Sherwin}}, \bibinfo {author} {\bibfnamefont {T.~L.}\
  \bibnamefont {Smith}}, \bibinfo {author} {\bibfnamefont {L.}~\bibnamefont
  {Sorbo}}, \bibinfo {author} {\bibfnamefont {G.~D.}\ \bibnamefont {Starkman}},
  \bibinfo {author} {\bibfnamefont {K.~T.}\ \bibnamefont {Story}}, \bibinfo
  {author} {\bibfnamefont {A.}~\bibnamefont {{van Engelen}}}, \bibinfo {author}
  {\bibfnamefont {J.~D.}\ \bibnamefont {Vieira}}, \bibinfo {author}
  {\bibfnamefont {S.}~\bibnamefont {Watson}}, \bibinfo {author} {\bibfnamefont
  {N.}~\bibnamefont {Whitehorn}}, \ and\ \bibinfo {author} {\bibfnamefont
  {W.~L.}\ \bibnamefont {Kimmy~Wu}},\ }\href
  {http://adsabs.harvard.edu/abs/2016arXiv161002743A} {\bibfield  {journal}
  {\bibinfo  {journal} {ArXiv e-prints}\ }\textbf {\bibinfo {volume} {1610}},\
  \bibinfo {pages} {arXiv:1610.02743} (\bibinfo {year} {2016})}\BibitemShut
  {NoStop}%
\bibitem [{\citenamefont {Abazajian}\ \emph
  {et~al.}(2019{\natexlab{b}})\citenamefont {Abazajian}, \citenamefont
  {Addison}, \citenamefont {Adshead}, \citenamefont {Ahmed}, \citenamefont
  {Allen}, \citenamefont {Alonso}, \citenamefont {Alvarez}, \citenamefont
  {Anderson}, \citenamefont {Arnold}, \citenamefont {Baccigalupi},
  \citenamefont {Bailey}, \citenamefont {Barkats}, \citenamefont {Barron},
  \citenamefont {Barry}, \citenamefont {Bartlett}, \citenamefont {Thakur},
  \citenamefont {Battaglia}, \citenamefont {Baxter}, \citenamefont {Bean},
  \citenamefont {Bebek}, \citenamefont {Bender}, \citenamefont {Benson},
  \citenamefont {Berger}, \citenamefont {Bhimani}, \citenamefont {Bischoff},
  \citenamefont {Bleem}, \citenamefont {Bocquet}, \citenamefont {Boddy},
  \citenamefont {Bonato}, \citenamefont {Bond}, \citenamefont {Borrill},
  \citenamefont {Bouchet}, \citenamefont {Brown}, \citenamefont {Bryan},
  \citenamefont {Burkhart}, \citenamefont {Buza}, \citenamefont {Byrum},
  \citenamefont {Calabrese}, \citenamefont {Calafut}, \citenamefont {Caldwell},
  \citenamefont {Carlstrom}, \citenamefont {Carron}, \citenamefont {Cecil},
  \citenamefont {Challinor}, \citenamefont {Chang}, \citenamefont {Chinone},
  \citenamefont {Cho}, \citenamefont {Cooray}, \citenamefont {Crawford},
  \citenamefont {Crites}, \citenamefont {Cukierman}, \citenamefont
  {{Cyr-Racine}}, \citenamefont {{de Haan}}, \citenamefont {{de Zotti}},
  \citenamefont {Delabrouille}, \citenamefont {Demarteau}, \citenamefont
  {Devlin}, \citenamefont {Di~Valentino}, \citenamefont {Dobbs}, \citenamefont
  {Duff}, \citenamefont {Duivenvoorden}, \citenamefont {Dvorkin}, \citenamefont
  {Edwards}, \citenamefont {Eimer}, \citenamefont {Errard}, \citenamefont
  {{Essinger-Hileman}}, \citenamefont {Fabbian}, \citenamefont {Feng},
  \citenamefont {Ferraro}, \citenamefont {Filippini}, \citenamefont {Flauger},
  \citenamefont {Flaugher}, \citenamefont {Fraisse}, \citenamefont {Frolov},
  \citenamefont {Galitzki}, \citenamefont {Galli}, \citenamefont {Ganga},
  \citenamefont {Gerbino}, \citenamefont {Gilchriese}, \citenamefont
  {Gluscevic}, \citenamefont {Green}, \citenamefont {Grin}, \citenamefont
  {Grohs}, \citenamefont {Gualtieri}, \citenamefont {Guarino}, \citenamefont
  {Gudmundsson}, \citenamefont {Habib}, \citenamefont {Haller}, \citenamefont
  {Halpern}, \citenamefont {Halverson}, \citenamefont {Hanany}, \citenamefont
  {Harrington}, \citenamefont {Hasegawa}, \citenamefont {Hasselfield},
  \citenamefont {Hazumi}, \citenamefont {Heitmann}, \citenamefont {Henderson},
  \citenamefont {Henning}, \citenamefont {Hill}, \citenamefont {Hlozek},
  \citenamefont {Holder}, \citenamefont {Holzapfel}, \citenamefont {Hubmayr},
  \citenamefont {Huffenberger}, \citenamefont {Huffer}, \citenamefont {Hui},
  \citenamefont {Irwin}, \citenamefont {Johnson}, \citenamefont {Johnstone},
  \citenamefont {Jones}, \citenamefont {Karkare}, \citenamefont {Katayama},
  \citenamefont {Kerby}, \citenamefont {Kernovsky}, \citenamefont {Keskitalo},
  \citenamefont {Kisner}, \citenamefont {Knox}, \citenamefont {Kosowsky},
  \citenamefont {Kovac}, \citenamefont {Kovetz}, \citenamefont {Kuhlmann},
  \citenamefont {Kuo}, \citenamefont {Kurita}, \citenamefont {Kusaka},
  \citenamefont {Lahteenmaki}, \citenamefont {Lawrence}, \citenamefont {Lee},
  \citenamefont {Lewis}, \citenamefont {Li}, \citenamefont {Linder},
  \citenamefont {Loverde}, \citenamefont {Lowitz}, \citenamefont
  {Madhavacheril}, \citenamefont {Mantz}, \citenamefont {Matsuda},
  \citenamefont {Mauskopf}, \citenamefont {McMahon}, \citenamefont {McQuinn},
  \citenamefont {Meerburg}, \citenamefont {Melin}, \citenamefont {Meyers},
  \citenamefont {Millea}, \citenamefont {Mohr}, \citenamefont {Moncelsi},
  \citenamefont {Mroczkowski}, \citenamefont {Mukherjee}, \citenamefont
  {M{\"u}nchmeyer}, \citenamefont {Nagai}, \citenamefont {Nagy}, \citenamefont
  {Namikawa}, \citenamefont {Nati}, \citenamefont {Natoli}, \citenamefont
  {Negrello}, \citenamefont {Newburgh}, \citenamefont {Niemack}, \citenamefont
  {Nishino}, \citenamefont {Nordby}, \citenamefont {Novosad}, \citenamefont
  {O'Connor}, \citenamefont {Obied}, \citenamefont {Padin}, \citenamefont
  {Pandey}, \citenamefont {Partridge}, \citenamefont {Pierpaoli}, \citenamefont
  {Pogosian}, \citenamefont {Pryke}, \citenamefont {Puglisi}, \citenamefont
  {Racine}, \citenamefont {Raghunathan}, \citenamefont {Rahlin}, \citenamefont
  {Rajagopalan}, \citenamefont {Raveri}, \citenamefont {Reichanadter},
  \citenamefont {Reichardt}, \citenamefont {Remazeilles}, \citenamefont
  {Rocha}, \citenamefont {Roe}, \citenamefont {Roy}, \citenamefont {Ruhl},
  \citenamefont {Salatino}, \citenamefont {Saliwanchik}, \citenamefont
  {Schaan}, \citenamefont {Schillaci}, \citenamefont {Schmittfull},
  \citenamefont {Scott}, \citenamefont {Sehgal}, \citenamefont {Shandera},
  \citenamefont {Sheehy}, \citenamefont {Sherwin}, \citenamefont {Shirokoff},
  \citenamefont {Simon}, \citenamefont {Slosar}, \citenamefont {Somerville},
  \citenamefont {Spergel}, \citenamefont {Staggs}, \citenamefont {Stark},
  \citenamefont {Stompor}, \citenamefont {Story}, \citenamefont {Stoughton},
  \citenamefont {Suzuki}, \citenamefont {Tajima}, \citenamefont {Teply},
  \citenamefont {Thompson}, \citenamefont {Timbie}, \citenamefont {Tomasi},
  \citenamefont {Treu}, \citenamefont {Tristram}, \citenamefont {Tucker},
  \citenamefont {Umilt{\`a}}, \citenamefont {{van Engelen}}, \citenamefont
  {Vieira}, \citenamefont {Vieregg}, \citenamefont {Vogelsberger},
  \citenamefont {Wang}, \citenamefont {Watson}, \citenamefont {White},
  \citenamefont {Whitehorn}, \citenamefont {Wollack}, \citenamefont {Wu},
  \citenamefont {Xu}, \citenamefont {Yasini}, \citenamefont {Yeck},
  \citenamefont {Yoon}, \citenamefont {Young},\ and\ \citenamefont
  {Zonca}}]{abazajian2019b}%
  \BibitemOpen
  \bibfield  {author} {\bibinfo {author} {\bibfnamefont {K.}~\bibnamefont
  {Abazajian}}, \bibinfo {author} {\bibfnamefont {G.}~\bibnamefont {Addison}},
  \bibinfo {author} {\bibfnamefont {P.}~\bibnamefont {Adshead}}, \bibinfo
  {author} {\bibfnamefont {Z.}~\bibnamefont {Ahmed}}, \bibinfo {author}
  {\bibfnamefont {S.~W.}\ \bibnamefont {Allen}}, \bibinfo {author}
  {\bibfnamefont {D.}~\bibnamefont {Alonso}}, \bibinfo {author} {\bibfnamefont
  {M.}~\bibnamefont {Alvarez}}, \bibinfo {author} {\bibfnamefont
  {A.}~\bibnamefont {Anderson}}, \bibinfo {author} {\bibfnamefont {K.~S.}\
  \bibnamefont {Arnold}}, \bibinfo {author} {\bibfnamefont {C.}~\bibnamefont
  {Baccigalupi}}, \bibinfo {author} {\bibfnamefont {K.}~\bibnamefont {Bailey}},
  \bibinfo {author} {\bibfnamefont {D.}~\bibnamefont {Barkats}}, \bibinfo
  {author} {\bibfnamefont {D.}~\bibnamefont {Barron}}, \bibinfo {author}
  {\bibfnamefont {P.~S.}\ \bibnamefont {Barry}}, \bibinfo {author}
  {\bibfnamefont {J.~G.}\ \bibnamefont {Bartlett}}, \bibinfo {author}
  {\bibfnamefont {R.~B.}\ \bibnamefont {Thakur}}, \bibinfo {author}
  {\bibfnamefont {N.}~\bibnamefont {Battaglia}}, \bibinfo {author}
  {\bibfnamefont {E.}~\bibnamefont {Baxter}}, \bibinfo {author} {\bibfnamefont
  {R.}~\bibnamefont {Bean}}, \bibinfo {author} {\bibfnamefont {C.}~\bibnamefont
  {Bebek}}, \bibinfo {author} {\bibfnamefont {A.~N.}\ \bibnamefont {Bender}},
  \bibinfo {author} {\bibfnamefont {B.~A.}\ \bibnamefont {Benson}}, \bibinfo
  {author} {\bibfnamefont {E.}~\bibnamefont {Berger}}, \bibinfo {author}
  {\bibfnamefont {S.}~\bibnamefont {Bhimani}}, \bibinfo {author} {\bibfnamefont
  {C.~A.}\ \bibnamefont {Bischoff}}, \bibinfo {author} {\bibfnamefont
  {L.}~\bibnamefont {Bleem}}, \bibinfo {author} {\bibfnamefont
  {S.}~\bibnamefont {Bocquet}}, \bibinfo {author} {\bibfnamefont
  {K.}~\bibnamefont {Boddy}}, \bibinfo {author} {\bibfnamefont
  {M.}~\bibnamefont {Bonato}}, \bibinfo {author} {\bibfnamefont {J.~R.}\
  \bibnamefont {Bond}}, \bibinfo {author} {\bibfnamefont {J.}~\bibnamefont
  {Borrill}}, \bibinfo {author} {\bibfnamefont {F.~R.}\ \bibnamefont
  {Bouchet}}, \bibinfo {author} {\bibfnamefont {M.~L.}\ \bibnamefont {Brown}},
  \bibinfo {author} {\bibfnamefont {S.}~\bibnamefont {Bryan}}, \bibinfo
  {author} {\bibfnamefont {B.}~\bibnamefont {Burkhart}}, \bibinfo {author}
  {\bibfnamefont {V.}~\bibnamefont {Buza}}, \bibinfo {author} {\bibfnamefont
  {K.}~\bibnamefont {Byrum}}, \bibinfo {author} {\bibfnamefont
  {E.}~\bibnamefont {Calabrese}}, \bibinfo {author} {\bibfnamefont
  {V.}~\bibnamefont {Calafut}}, \bibinfo {author} {\bibfnamefont
  {R.}~\bibnamefont {Caldwell}}, \bibinfo {author} {\bibfnamefont {J.~E.}\
  \bibnamefont {Carlstrom}}, \bibinfo {author} {\bibfnamefont {J.}~\bibnamefont
  {Carron}}, \bibinfo {author} {\bibfnamefont {T.}~\bibnamefont {Cecil}},
  \bibinfo {author} {\bibfnamefont {A.}~\bibnamefont {Challinor}}, \bibinfo
  {author} {\bibfnamefont {C.~L.}\ \bibnamefont {Chang}}, \bibinfo {author}
  {\bibfnamefont {Y.}~\bibnamefont {Chinone}}, \bibinfo {author} {\bibfnamefont
  {H.-M.~S.}\ \bibnamefont {Cho}}, \bibinfo {author} {\bibfnamefont
  {A.}~\bibnamefont {Cooray}}, \bibinfo {author} {\bibfnamefont {T.~M.}\
  \bibnamefont {Crawford}}, \bibinfo {author} {\bibfnamefont {A.}~\bibnamefont
  {Crites}}, \bibinfo {author} {\bibfnamefont {A.}~\bibnamefont {Cukierman}},
  \bibinfo {author} {\bibfnamefont {F.-Y.}\ \bibnamefont {{Cyr-Racine}}},
  \bibinfo {author} {\bibfnamefont {T.}~\bibnamefont {{de Haan}}}, \bibinfo
  {author} {\bibfnamefont {G.}~\bibnamefont {{de Zotti}}}, \bibinfo {author}
  {\bibfnamefont {J.}~\bibnamefont {Delabrouille}}, \bibinfo {author}
  {\bibfnamefont {M.}~\bibnamefont {Demarteau}}, \bibinfo {author}
  {\bibfnamefont {M.}~\bibnamefont {Devlin}}, \bibinfo {author} {\bibfnamefont
  {E.}~\bibnamefont {Di~Valentino}}, \bibinfo {author} {\bibfnamefont
  {M.}~\bibnamefont {Dobbs}}, \bibinfo {author} {\bibfnamefont
  {S.}~\bibnamefont {Duff}}, \bibinfo {author} {\bibfnamefont {A.}~\bibnamefont
  {Duivenvoorden}}, \bibinfo {author} {\bibfnamefont {C.}~\bibnamefont
  {Dvorkin}}, \bibinfo {author} {\bibfnamefont {W.}~\bibnamefont {Edwards}},
  \bibinfo {author} {\bibfnamefont {J.}~\bibnamefont {Eimer}}, \bibinfo
  {author} {\bibfnamefont {J.}~\bibnamefont {Errard}}, \bibinfo {author}
  {\bibfnamefont {T.}~\bibnamefont {{Essinger-Hileman}}}, \bibinfo {author}
  {\bibfnamefont {G.}~\bibnamefont {Fabbian}}, \bibinfo {author} {\bibfnamefont
  {C.}~\bibnamefont {Feng}}, \bibinfo {author} {\bibfnamefont {S.}~\bibnamefont
  {Ferraro}}, \bibinfo {author} {\bibfnamefont {J.~P.}\ \bibnamefont
  {Filippini}}, \bibinfo {author} {\bibfnamefont {R.}~\bibnamefont {Flauger}},
  \bibinfo {author} {\bibfnamefont {B.}~\bibnamefont {Flaugher}}, \bibinfo
  {author} {\bibfnamefont {A.~A.}\ \bibnamefont {Fraisse}}, \bibinfo {author}
  {\bibfnamefont {A.}~\bibnamefont {Frolov}}, \bibinfo {author} {\bibfnamefont
  {N.}~\bibnamefont {Galitzki}}, \bibinfo {author} {\bibfnamefont
  {S.}~\bibnamefont {Galli}}, \bibinfo {author} {\bibfnamefont
  {K.}~\bibnamefont {Ganga}}, \bibinfo {author} {\bibfnamefont
  {M.}~\bibnamefont {Gerbino}}, \bibinfo {author} {\bibfnamefont
  {M.}~\bibnamefont {Gilchriese}}, \bibinfo {author} {\bibfnamefont
  {V.}~\bibnamefont {Gluscevic}}, \bibinfo {author} {\bibfnamefont
  {D.}~\bibnamefont {Green}}, \bibinfo {author} {\bibfnamefont
  {D.}~\bibnamefont {Grin}}, \bibinfo {author} {\bibfnamefont {E.}~\bibnamefont
  {Grohs}}, \bibinfo {author} {\bibfnamefont {R.}~\bibnamefont {Gualtieri}},
  \bibinfo {author} {\bibfnamefont {V.}~\bibnamefont {Guarino}}, \bibinfo
  {author} {\bibfnamefont {J.~E.}\ \bibnamefont {Gudmundsson}}, \bibinfo
  {author} {\bibfnamefont {S.}~\bibnamefont {Habib}}, \bibinfo {author}
  {\bibfnamefont {G.}~\bibnamefont {Haller}}, \bibinfo {author} {\bibfnamefont
  {M.}~\bibnamefont {Halpern}}, \bibinfo {author} {\bibfnamefont {N.~W.}\
  \bibnamefont {Halverson}}, \bibinfo {author} {\bibfnamefont {S.}~\bibnamefont
  {Hanany}}, \bibinfo {author} {\bibfnamefont {K.}~\bibnamefont {Harrington}},
  \bibinfo {author} {\bibfnamefont {M.}~\bibnamefont {Hasegawa}}, \bibinfo
  {author} {\bibfnamefont {M.}~\bibnamefont {Hasselfield}}, \bibinfo {author}
  {\bibfnamefont {M.}~\bibnamefont {Hazumi}}, \bibinfo {author} {\bibfnamefont
  {K.}~\bibnamefont {Heitmann}}, \bibinfo {author} {\bibfnamefont
  {S.}~\bibnamefont {Henderson}}, \bibinfo {author} {\bibfnamefont {J.~W.}\
  \bibnamefont {Henning}}, \bibinfo {author} {\bibfnamefont {J.~C.}\
  \bibnamefont {Hill}}, \bibinfo {author} {\bibfnamefont {R.}~\bibnamefont
  {Hlozek}}, \bibinfo {author} {\bibfnamefont {G.}~\bibnamefont {Holder}},
  \bibinfo {author} {\bibfnamefont {W.}~\bibnamefont {Holzapfel}}, \bibinfo
  {author} {\bibfnamefont {J.}~\bibnamefont {Hubmayr}}, \bibinfo {author}
  {\bibfnamefont {K.~M.}\ \bibnamefont {Huffenberger}}, \bibinfo {author}
  {\bibfnamefont {M.}~\bibnamefont {Huffer}}, \bibinfo {author} {\bibfnamefont
  {H.}~\bibnamefont {Hui}}, \bibinfo {author} {\bibfnamefont {K.}~\bibnamefont
  {Irwin}}, \bibinfo {author} {\bibfnamefont {B.~R.}\ \bibnamefont {Johnson}},
  \bibinfo {author} {\bibfnamefont {D.}~\bibnamefont {Johnstone}}, \bibinfo
  {author} {\bibfnamefont {W.~C.}\ \bibnamefont {Jones}}, \bibinfo {author}
  {\bibfnamefont {K.}~\bibnamefont {Karkare}}, \bibinfo {author} {\bibfnamefont
  {N.}~\bibnamefont {Katayama}}, \bibinfo {author} {\bibfnamefont
  {J.}~\bibnamefont {Kerby}}, \bibinfo {author} {\bibfnamefont
  {S.}~\bibnamefont {Kernovsky}}, \bibinfo {author} {\bibfnamefont
  {R.}~\bibnamefont {Keskitalo}}, \bibinfo {author} {\bibfnamefont
  {T.}~\bibnamefont {Kisner}}, \bibinfo {author} {\bibfnamefont
  {L.}~\bibnamefont {Knox}}, \bibinfo {author} {\bibfnamefont {A.}~\bibnamefont
  {Kosowsky}}, \bibinfo {author} {\bibfnamefont {J.}~\bibnamefont {Kovac}},
  \bibinfo {author} {\bibfnamefont {E.~D.}\ \bibnamefont {Kovetz}}, \bibinfo
  {author} {\bibfnamefont {S.}~\bibnamefont {Kuhlmann}}, \bibinfo {author}
  {\bibfnamefont {C.-l.}\ \bibnamefont {Kuo}}, \bibinfo {author} {\bibfnamefont
  {N.}~\bibnamefont {Kurita}}, \bibinfo {author} {\bibfnamefont
  {A.}~\bibnamefont {Kusaka}}, \bibinfo {author} {\bibfnamefont
  {A.}~\bibnamefont {Lahteenmaki}}, \bibinfo {author} {\bibfnamefont {C.~R.}\
  \bibnamefont {Lawrence}}, \bibinfo {author} {\bibfnamefont {A.~T.}\
  \bibnamefont {Lee}}, \bibinfo {author} {\bibfnamefont {A.}~\bibnamefont
  {Lewis}}, \bibinfo {author} {\bibfnamefont {D.}~\bibnamefont {Li}}, \bibinfo
  {author} {\bibfnamefont {E.}~\bibnamefont {Linder}}, \bibinfo {author}
  {\bibfnamefont {M.}~\bibnamefont {Loverde}}, \bibinfo {author} {\bibfnamefont
  {A.}~\bibnamefont {Lowitz}}, \bibinfo {author} {\bibfnamefont {M.~S.}\
  \bibnamefont {Madhavacheril}}, \bibinfo {author} {\bibfnamefont
  {A.}~\bibnamefont {Mantz}}, \bibinfo {author} {\bibfnamefont
  {F.}~\bibnamefont {Matsuda}}, \bibinfo {author} {\bibfnamefont
  {P.}~\bibnamefont {Mauskopf}}, \bibinfo {author} {\bibfnamefont
  {J.}~\bibnamefont {McMahon}}, \bibinfo {author} {\bibfnamefont
  {M.}~\bibnamefont {McQuinn}}, \bibinfo {author} {\bibfnamefont {P.~D.}\
  \bibnamefont {Meerburg}}, \bibinfo {author} {\bibfnamefont {J.-B.}\
  \bibnamefont {Melin}}, \bibinfo {author} {\bibfnamefont {J.}~\bibnamefont
  {Meyers}}, \bibinfo {author} {\bibfnamefont {M.}~\bibnamefont {Millea}},
  \bibinfo {author} {\bibfnamefont {J.}~\bibnamefont {Mohr}}, \bibinfo {author}
  {\bibfnamefont {L.}~\bibnamefont {Moncelsi}}, \bibinfo {author}
  {\bibfnamefont {T.}~\bibnamefont {Mroczkowski}}, \bibinfo {author}
  {\bibfnamefont {S.}~\bibnamefont {Mukherjee}}, \bibinfo {author}
  {\bibfnamefont {M.}~\bibnamefont {M{\"u}nchmeyer}}, \bibinfo {author}
  {\bibfnamefont {D.}~\bibnamefont {Nagai}}, \bibinfo {author} {\bibfnamefont
  {J.}~\bibnamefont {Nagy}}, \bibinfo {author} {\bibfnamefont {T.}~\bibnamefont
  {Namikawa}}, \bibinfo {author} {\bibfnamefont {F.}~\bibnamefont {Nati}},
  \bibinfo {author} {\bibfnamefont {T.}~\bibnamefont {Natoli}}, \bibinfo
  {author} {\bibfnamefont {M.}~\bibnamefont {Negrello}}, \bibinfo {author}
  {\bibfnamefont {L.}~\bibnamefont {Newburgh}}, \bibinfo {author}
  {\bibfnamefont {M.~D.}\ \bibnamefont {Niemack}}, \bibinfo {author}
  {\bibfnamefont {H.}~\bibnamefont {Nishino}}, \bibinfo {author} {\bibfnamefont
  {M.}~\bibnamefont {Nordby}}, \bibinfo {author} {\bibfnamefont
  {V.}~\bibnamefont {Novosad}}, \bibinfo {author} {\bibfnamefont
  {P.}~\bibnamefont {O'Connor}}, \bibinfo {author} {\bibfnamefont
  {G.}~\bibnamefont {Obied}}, \bibinfo {author} {\bibfnamefont
  {S.}~\bibnamefont {Padin}}, \bibinfo {author} {\bibfnamefont
  {S.}~\bibnamefont {Pandey}}, \bibinfo {author} {\bibfnamefont
  {B.}~\bibnamefont {Partridge}}, \bibinfo {author} {\bibfnamefont
  {E.}~\bibnamefont {Pierpaoli}}, \bibinfo {author} {\bibfnamefont
  {L.}~\bibnamefont {Pogosian}}, \bibinfo {author} {\bibfnamefont
  {C.}~\bibnamefont {Pryke}}, \bibinfo {author} {\bibfnamefont
  {G.}~\bibnamefont {Puglisi}}, \bibinfo {author} {\bibfnamefont
  {B.}~\bibnamefont {Racine}}, \bibinfo {author} {\bibfnamefont
  {S.}~\bibnamefont {Raghunathan}}, \bibinfo {author} {\bibfnamefont
  {A.}~\bibnamefont {Rahlin}}, \bibinfo {author} {\bibfnamefont
  {S.}~\bibnamefont {Rajagopalan}}, \bibinfo {author} {\bibfnamefont
  {M.}~\bibnamefont {Raveri}}, \bibinfo {author} {\bibfnamefont
  {M.}~\bibnamefont {Reichanadter}}, \bibinfo {author} {\bibfnamefont {C.~L.}\
  \bibnamefont {Reichardt}}, \bibinfo {author} {\bibfnamefont {M.}~\bibnamefont
  {Remazeilles}}, \bibinfo {author} {\bibfnamefont {G.}~\bibnamefont {Rocha}},
  \bibinfo {author} {\bibfnamefont {N.~A.}\ \bibnamefont {Roe}}, \bibinfo
  {author} {\bibfnamefont {A.}~\bibnamefont {Roy}}, \bibinfo {author}
  {\bibfnamefont {J.}~\bibnamefont {Ruhl}}, \bibinfo {author} {\bibfnamefont
  {M.}~\bibnamefont {Salatino}}, \bibinfo {author} {\bibfnamefont
  {B.}~\bibnamefont {Saliwanchik}}, \bibinfo {author} {\bibfnamefont
  {E.}~\bibnamefont {Schaan}}, \bibinfo {author} {\bibfnamefont
  {A.}~\bibnamefont {Schillaci}}, \bibinfo {author} {\bibfnamefont {M.~M.}\
  \bibnamefont {Schmittfull}}, \bibinfo {author} {\bibfnamefont
  {D.}~\bibnamefont {Scott}}, \bibinfo {author} {\bibfnamefont
  {N.}~\bibnamefont {Sehgal}}, \bibinfo {author} {\bibfnamefont
  {S.}~\bibnamefont {Shandera}}, \bibinfo {author} {\bibfnamefont
  {C.}~\bibnamefont {Sheehy}}, \bibinfo {author} {\bibfnamefont {B.~D.}\
  \bibnamefont {Sherwin}}, \bibinfo {author} {\bibfnamefont {E.}~\bibnamefont
  {Shirokoff}}, \bibinfo {author} {\bibfnamefont {S.~M.}\ \bibnamefont
  {Simon}}, \bibinfo {author} {\bibfnamefont {A.}~\bibnamefont {Slosar}},
  \bibinfo {author} {\bibfnamefont {R.}~\bibnamefont {Somerville}}, \bibinfo
  {author} {\bibfnamefont {D.}~\bibnamefont {Spergel}}, \bibinfo {author}
  {\bibfnamefont {S.~T.}\ \bibnamefont {Staggs}}, \bibinfo {author}
  {\bibfnamefont {A.}~\bibnamefont {Stark}}, \bibinfo {author} {\bibfnamefont
  {R.}~\bibnamefont {Stompor}}, \bibinfo {author} {\bibfnamefont {K.~T.}\
  \bibnamefont {Story}}, \bibinfo {author} {\bibfnamefont {C.}~\bibnamefont
  {Stoughton}}, \bibinfo {author} {\bibfnamefont {A.}~\bibnamefont {Suzuki}},
  \bibinfo {author} {\bibfnamefont {O.}~\bibnamefont {Tajima}}, \bibinfo
  {author} {\bibfnamefont {G.~P.}\ \bibnamefont {Teply}}, \bibinfo {author}
  {\bibfnamefont {K.}~\bibnamefont {Thompson}}, \bibinfo {author}
  {\bibfnamefont {P.}~\bibnamefont {Timbie}}, \bibinfo {author} {\bibfnamefont
  {M.}~\bibnamefont {Tomasi}}, \bibinfo {author} {\bibfnamefont {J.~I.}\
  \bibnamefont {Treu}}, \bibinfo {author} {\bibfnamefont {M.}~\bibnamefont
  {Tristram}}, \bibinfo {author} {\bibfnamefont {G.}~\bibnamefont {Tucker}},
  \bibinfo {author} {\bibfnamefont {C.}~\bibnamefont {Umilt{\`a}}}, \bibinfo
  {author} {\bibfnamefont {A.}~\bibnamefont {{van Engelen}}}, \bibinfo {author}
  {\bibfnamefont {J.~D.}\ \bibnamefont {Vieira}}, \bibinfo {author}
  {\bibfnamefont {A.~G.}\ \bibnamefont {Vieregg}}, \bibinfo {author}
  {\bibfnamefont {M.}~\bibnamefont {Vogelsberger}}, \bibinfo {author}
  {\bibfnamefont {G.}~\bibnamefont {Wang}}, \bibinfo {author} {\bibfnamefont
  {S.}~\bibnamefont {Watson}}, \bibinfo {author} {\bibfnamefont
  {M.}~\bibnamefont {White}}, \bibinfo {author} {\bibfnamefont
  {N.}~\bibnamefont {Whitehorn}}, \bibinfo {author} {\bibfnamefont {E.~J.}\
  \bibnamefont {Wollack}}, \bibinfo {author} {\bibfnamefont {W.~L.~K.}\
  \bibnamefont {Wu}}, \bibinfo {author} {\bibfnamefont {Z.}~\bibnamefont {Xu}},
  \bibinfo {author} {\bibfnamefont {S.}~\bibnamefont {Yasini}}, \bibinfo
  {author} {\bibfnamefont {J.}~\bibnamefont {Yeck}}, \bibinfo {author}
  {\bibfnamefont {K.~W.}\ \bibnamefont {Yoon}}, \bibinfo {author}
  {\bibfnamefont {E.}~\bibnamefont {Young}}, \ and\ \bibinfo {author}
  {\bibfnamefont {A.}~\bibnamefont {Zonca}},\ }\href
  {http://arxiv.org/abs/1907.04473} {\bibfield  {journal} {\bibinfo  {journal}
  {arXiv:1907.04473 [astro-ph, physics:hep-ex]}\ } (\bibinfo {year}
  {2019}{\natexlab{b}})},\ \Eprint {http://arxiv.org/abs/1907.04473}
  {arXiv:1907.04473 [astro-ph, physics:hep-ex]} \BibitemShut {NoStop}%
\bibitem [{\citenamefont {{South Pole Observatory
  Collaboration}}(2020)}]{southpoleobservatorycollaboration2020}%
  \BibitemOpen
  \bibfield  {author} {\bibinfo {author} {\bibnamefont {{South Pole Observatory
  Collaboration}}},\ }\href@noop {} {\bibfield  {journal} {\bibinfo  {journal}
  {in prep}\ } (\bibinfo {year} {2020})}\BibitemShut {NoStop}%
\bibitem [{\citenamefont {Beck}\ \emph {et~al.}(2018)\citenamefont {Beck},
  \citenamefont {Fabbian},\ and\ \citenamefont {Errard}}]{beck2018}%
  \BibitemOpen
  \bibfield  {author} {\bibinfo {author} {\bibfnamefont {D.}~\bibnamefont
  {Beck}}, \bibinfo {author} {\bibfnamefont {G.}~\bibnamefont {Fabbian}}, \
  and\ \bibinfo {author} {\bibfnamefont {J.}~\bibnamefont {Errard}},\ }\href
  {http://arxiv.org/abs/1806.01216} {\bibfield  {journal} {\bibinfo  {journal}
  {arXiv:1806.01216 [astro-ph]}\ } (\bibinfo {year} {2018})},\ \Eprint
  {http://arxiv.org/abs/1806.01216} {arXiv:1806.01216 [astro-ph]} \BibitemShut
  {NoStop}%
\bibitem [{\citenamefont {B{\"o}hm}\ \emph {et~al.}(2018)\citenamefont
  {B{\"o}hm}, \citenamefont {Sherwin}, \citenamefont {Liu}, \citenamefont
  {Hill}, \citenamefont {Schmittfull},\ and\ \citenamefont
  {Namikawa}}]{bohm2018}%
  \BibitemOpen
  \bibfield  {author} {\bibinfo {author} {\bibfnamefont {V.}~\bibnamefont
  {B{\"o}hm}}, \bibinfo {author} {\bibfnamefont {B.~D.}\ \bibnamefont
  {Sherwin}}, \bibinfo {author} {\bibfnamefont {J.}~\bibnamefont {Liu}},
  \bibinfo {author} {\bibfnamefont {J.~C.}\ \bibnamefont {Hill}}, \bibinfo
  {author} {\bibfnamefont {M.}~\bibnamefont {Schmittfull}}, \ and\ \bibinfo
  {author} {\bibfnamefont {T.}~\bibnamefont {Namikawa}},\ }\href
  {http://arxiv.org/abs/1806.01157} {\bibfield  {journal} {\bibinfo  {journal}
  {arXiv:1806.01157 [astro-ph]}\ } (\bibinfo {year} {2018})},\ \Eprint
  {http://arxiv.org/abs/1806.01157} {arXiv:1806.01157 [astro-ph]} \BibitemShut
  {NoStop}%
\bibitem [{\citenamefont {Lewis}\ \emph {et~al.}(2017)\citenamefont {Lewis},
  \citenamefont {Hall},\ and\ \citenamefont {Challinor}}]{lewis2017}%
  \BibitemOpen
  \bibfield  {author} {\bibinfo {author} {\bibfnamefont {A.}~\bibnamefont
  {Lewis}}, \bibinfo {author} {\bibfnamefont {A.}~\bibnamefont {Hall}}, \ and\
  \bibinfo {author} {\bibfnamefont {A.}~\bibnamefont {Challinor}},\ }\href
  {http://adsabs.harvard.edu/abs/2017arXiv170602673L} {\bibfield  {journal}
  {\bibinfo  {journal} {ArXiv e-prints}\ }\textbf {\bibinfo {volume} {1706}},\
  \bibinfo {pages} {arXiv:1706.02673} (\bibinfo {year} {2017})}\BibitemShut
  {NoStop}%
\bibitem [{\citenamefont {Pratten}\ and\ \citenamefont
  {Lewis}(2016)}]{pratten2016}%
  \BibitemOpen
  \bibfield  {author} {\bibinfo {author} {\bibfnamefont {G.}~\bibnamefont
  {Pratten}}\ and\ \bibinfo {author} {\bibfnamefont {A.}~\bibnamefont
  {Lewis}},\ }\href {\doibase 10.1088/1475-7516/2016/08/047} {\bibfield
  {journal} {\bibinfo  {journal} {Journal of Cosmology and Astroparticle
  Physics}\ }\textbf {\bibinfo {volume} {2016}},\ \bibinfo {pages} {047}
  (\bibinfo {year} {2016})},\ \Eprint {http://arxiv.org/abs/1605.05662}
  {arXiv:1605.05662} \BibitemShut {NoStop}%
\bibitem [{\citenamefont {Gelman}(2006)}]{gelman2006}%
  \BibitemOpen
  \bibfield  {author} {\bibinfo {author} {\bibfnamefont {A.}~\bibnamefont
  {Gelman}},\ }\href {\doibase 10.1214/06-BA117A} {\bibfield  {journal}
  {\bibinfo  {journal} {Bayesian Anal.}\ }\textbf {\bibinfo {volume} {1}},\
  \bibinfo {pages} {515} (\bibinfo {year} {2006})}\BibitemShut {NoStop}%
\bibitem [{\citenamefont {{Wandelt}}\ \emph {et~al.}(2004)\citenamefont
  {{Wandelt}}, \citenamefont {{Larson}},\ and\ \citenamefont
  {{Lakshminarayanan}}}]{WandeltLarsonLakshminarayanan2004}%
  \BibitemOpen
  \bibfield  {author} {\bibinfo {author} {\bibfnamefont {B.~D.}\ \bibnamefont
  {{Wandelt}}}, \bibinfo {author} {\bibfnamefont {D.~L.}\ \bibnamefont
  {{Larson}}}, \ and\ \bibinfo {author} {\bibfnamefont {A.}~\bibnamefont
  {{Lakshminarayanan}}},\ }\href {\doibase 10.1103/PhysRevD.70.083511}
  {\bibfield  {journal} {\bibinfo  {journal} {\prd}\ }\textbf {\bibinfo
  {volume} {70}},\ \bibinfo {eid} {083511} (\bibinfo {year} {2004})},\ \Eprint
  {http://arxiv.org/abs/astro-ph/0310080} {arXiv:astro-ph/0310080 [astro-ph]}
  \BibitemShut {NoStop}%
\bibitem [{\citenamefont {{Jasche}}\ and\ \citenamefont
  {{Wandelt}}(2013)}]{JascheWandelt2013}%
  \BibitemOpen
  \bibfield  {author} {\bibinfo {author} {\bibfnamefont {J.}~\bibnamefont
  {{Jasche}}}\ and\ \bibinfo {author} {\bibfnamefont {B.~D.}\ \bibnamefont
  {{Wandelt}}},\ }\href {\doibase 10.1093/mnras/stt449} {\bibfield  {journal}
  {\bibinfo  {journal} {\mnras}\ }\textbf {\bibinfo {volume} {432}},\ \bibinfo
  {pages} {894} (\bibinfo {year} {2013})},\ \Eprint
  {http://arxiv.org/abs/1203.3639} {arXiv:1203.3639 [astro-ph.CO]} \BibitemShut
  {NoStop}%
\bibitem [{\citenamefont {{Lavaux}}\ \emph {et~al.}(2019)\citenamefont
  {{Lavaux}}, \citenamefont {{Jasche}},\ and\ \citenamefont
  {{Leclercq}}}]{LavauxJascheLeclercq2019}%
  \BibitemOpen
  \bibfield  {author} {\bibinfo {author} {\bibfnamefont {G.}~\bibnamefont
  {{Lavaux}}}, \bibinfo {author} {\bibfnamefont {J.}~\bibnamefont {{Jasche}}},
  \ and\ \bibinfo {author} {\bibfnamefont {F.}~\bibnamefont {{Leclercq}}},\
  }\href@noop {} {\bibfield  {journal} {\bibinfo  {journal} {arXiv e-prints}\
  ,\ \bibinfo {eid} {arXiv:1909.06396}} (\bibinfo {year} {2019})},\ \Eprint
  {http://arxiv.org/abs/1909.06396} {arXiv:1909.06396 [astro-ph.CO]}
  \BibitemShut {NoStop}%
\bibitem [{\citenamefont {{Ramanah}}\ \emph {et~al.}(2019)\citenamefont
  {{Ramanah}}, \citenamefont {{Lavaux}}, \citenamefont {{Jasche}},\ and\
  \citenamefont {{Wand elt}}}]{DoogeshEtAl2019}%
  \BibitemOpen
  \bibfield  {author} {\bibinfo {author} {\bibfnamefont {D.~K.}\ \bibnamefont
  {{Ramanah}}}, \bibinfo {author} {\bibfnamefont {G.}~\bibnamefont {{Lavaux}}},
  \bibinfo {author} {\bibfnamefont {J.}~\bibnamefont {{Jasche}}}, \ and\
  \bibinfo {author} {\bibfnamefont {B.~D.}\ \bibnamefont {{Wand elt}}},\ }\href
  {\doibase 10.1051/0004-6361/201834117} {\bibfield  {journal} {\bibinfo
  {journal} {\aap}\ }\textbf {\bibinfo {volume} {621}},\ \bibinfo {eid} {A69}
  (\bibinfo {year} {2019})},\ \Eprint {http://arxiv.org/abs/1808.07496}
  {arXiv:1808.07496 [astro-ph.CO]} \BibitemShut {NoStop}%
\bibitem [{\citenamefont {Anderes}\ \emph {et~al.}(2015)\citenamefont
  {Anderes}, \citenamefont {Wandelt},\ and\ \citenamefont
  {Lavaux}}]{anderes2015}%
  \BibitemOpen
  \bibfield  {author} {\bibinfo {author} {\bibfnamefont {E.}~\bibnamefont
  {Anderes}}, \bibinfo {author} {\bibfnamefont {B.~D.}\ \bibnamefont
  {Wandelt}}, \ and\ \bibinfo {author} {\bibfnamefont {G.}~\bibnamefont
  {Lavaux}},\ }\href {\doibase 10.1088/0004-637X/808/2/152} {\bibfield
  {journal} {\bibinfo  {journal} {The Astrophysical Journal}\ }\textbf
  {\bibinfo {volume} {808}},\ \bibinfo {pages} {152} (\bibinfo {year}
  {2015})}\BibitemShut {NoStop}%
\bibitem [{\citenamefont {Jewell}\ \emph {et~al.}(2009)\citenamefont {Jewell},
  \citenamefont {Eriksen}, \citenamefont {Wandelt}, \citenamefont {O'Dwyer},
  \citenamefont {Huey},\ and\ \citenamefont {G{\'o}rski}}]{jewell2009}%
  \BibitemOpen
  \bibfield  {author} {\bibinfo {author} {\bibfnamefont {J.~B.}\ \bibnamefont
  {Jewell}}, \bibinfo {author} {\bibfnamefont {H.~K.}\ \bibnamefont {Eriksen}},
  \bibinfo {author} {\bibfnamefont {B.~D.}\ \bibnamefont {Wandelt}}, \bibinfo
  {author} {\bibfnamefont {I.~J.}\ \bibnamefont {O'Dwyer}}, \bibinfo {author}
  {\bibfnamefont {G.}~\bibnamefont {Huey}}, \ and\ \bibinfo {author}
  {\bibfnamefont {K.~M.}\ \bibnamefont {G{\'o}rski}},\ }\href {\doibase
  10.1088/0004-637X/697/1/258} {\bibfield  {journal} {\bibinfo  {journal} {The
  Astrophysical Journal}\ }\textbf {\bibinfo {volume} {697}},\ \bibinfo {pages}
  {258} (\bibinfo {year} {2009})}\BibitemShut {NoStop}%
\bibitem [{\citenamefont {Racine}\ \emph {et~al.}(2016)\citenamefont {Racine},
  \citenamefont {Jewell}, \citenamefont {Eriksen},\ and\ \citenamefont
  {Wehus}}]{racine2016}%
  \BibitemOpen
  \bibfield  {author} {\bibinfo {author} {\bibfnamefont {B.}~\bibnamefont
  {Racine}}, \bibinfo {author} {\bibfnamefont {J.~B.}\ \bibnamefont {Jewell}},
  \bibinfo {author} {\bibfnamefont {H.~K.}\ \bibnamefont {Eriksen}}, \ and\
  \bibinfo {author} {\bibfnamefont {I.~K.}\ \bibnamefont {Wehus}},\ }\href
  {\doibase 10.3847/0004-637X/820/1/31} {\bibfield  {journal} {\bibinfo
  {journal} {The Astrophysical Journal}\ }\textbf {\bibinfo {volume} {820}},\
  \bibinfo {pages} {31} (\bibinfo {year} {2016})}\BibitemShut {NoStop}%
\bibitem [{\citenamefont {M{\"u}nchmeyer}\ and\ \citenamefont
  {Smith}(2019)}]{munchmeyer2019}%
  \BibitemOpen
  \bibfield  {author} {\bibinfo {author} {\bibfnamefont {M.}~\bibnamefont
  {M{\"u}nchmeyer}}\ and\ \bibinfo {author} {\bibfnamefont {K.~M.}\
  \bibnamefont {Smith}},\ }\href {http://arxiv.org/abs/1905.05846} {\bibfield
  {journal} {\bibinfo  {journal} {arXiv:1905.05846 [astro-ph]}\ } (\bibinfo
  {year} {2019})},\ \Eprint {http://arxiv.org/abs/1905.05846} {arXiv:1905.05846
  [astro-ph]} \BibitemShut {NoStop}%
\bibitem [{\citenamefont {{Elsner}}\ and\ \citenamefont
  {{Wandelt}}(2013)}]{elsner2013efficient}%
  \BibitemOpen
  \bibfield  {author} {\bibinfo {author} {\bibfnamefont {F.}~\bibnamefont
  {{Elsner}}}\ and\ \bibinfo {author} {\bibfnamefont {B.~D.}\ \bibnamefont
  {{Wandelt}}},\ }\href {\doibase 10.1051/0004-6361/201220586} {\bibfield
  {journal} {\bibinfo  {journal} {\aap}\ }\textbf {\bibinfo {volume} {549}},\
  \bibinfo {eid} {A111} (\bibinfo {year} {2013})},\ \Eprint
  {http://arxiv.org/abs/1210.4931} {arXiv:1210.4931 [astro-ph.CO]} \BibitemShut
  {NoStop}%
\bibitem [{\citenamefont {{Kodi Ramanah}}\ \emph {et~al.}(2019)\citenamefont
  {{Kodi Ramanah}}, \citenamefont {{Lavaux}},\ and\ \citenamefont
  {{Wandelt}}}]{2019MNRAS.490..947K}%
  \BibitemOpen
  \bibfield  {author} {\bibinfo {author} {\bibfnamefont {D.}~\bibnamefont
  {{Kodi Ramanah}}}, \bibinfo {author} {\bibfnamefont {G.}~\bibnamefont
  {{Lavaux}}}, \ and\ \bibinfo {author} {\bibfnamefont {B.~D.}\ \bibnamefont
  {{Wandelt}}},\ }\href {\doibase 10.1093/mnras/stz2608} {\bibfield  {journal}
  {\bibinfo  {journal} {\mnras}\ }\textbf {\bibinfo {volume} {490}},\ \bibinfo
  {pages} {947} (\bibinfo {year} {2019})},\ \Eprint
  {http://arxiv.org/abs/1906.10704} {arXiv:1906.10704 [astro-ph.CO]}
  \BibitemShut {NoStop}%
\bibitem [{\citenamefont {Kahan}(1965)}]{kahan1965}%
  \BibitemOpen
  \bibfield  {author} {\bibinfo {author} {\bibfnamefont {W.}~\bibnamefont
  {Kahan}},\ }\href {https://dl.acm.org/doi/abs/10.1145/363707.363723}
  {\bibfield  {journal} {\bibinfo  {journal} {Communications of the ACM}\ }
  (\bibinfo {year} {1965})}\BibitemShut {NoStop}%
\bibitem [{\citenamefont {Hoffman}\ and\ \citenamefont
  {Gelman}(2014)}]{hoffman2014}%
  \BibitemOpen
  \bibfield  {author} {\bibinfo {author} {\bibfnamefont {M.~D.}\ \bibnamefont
  {Hoffman}}\ and\ \bibinfo {author} {\bibfnamefont {A.}~\bibnamefont
  {Gelman}},\ }\href {http://jmlr.org/papers/v15/hoffman14a.html} {\bibfield
  {journal} {\bibinfo  {journal} {Journal of Machine Learning Research}\
  }\textbf {\bibinfo {volume} {15}},\ \bibinfo {pages} {1593} (\bibinfo {year}
  {2014})}\BibitemShut {NoStop}%
\bibitem [{\citenamefont {Neal}(2003)}]{neal2003}%
  \BibitemOpen
  \bibfield  {author} {\bibinfo {author} {\bibfnamefont {R.~M.}\ \bibnamefont
  {Neal}},\ }\href {https://www.jstor.org/stable/3448413} {\bibfield  {journal}
  {\bibinfo  {journal} {The Annals of Statistics}\ }\textbf {\bibinfo {volume}
  {31}},\ \bibinfo {pages} {705} (\bibinfo {year} {2003})}\BibitemShut
  {NoStop}%
\bibitem [{\citenamefont {Neal}(1995)}]{neal1995}%
  \BibitemOpen
  \bibfield  {author} {\bibinfo {author} {\bibfnamefont {R.~M.}\ \bibnamefont
  {Neal}},\ }\href {http://adsabs.harvard.edu/abs/1995bayes.an..6004N} {\ ,\
  \bibinfo {pages} {arXiv:bayes} (\bibinfo {year} {1995})}\BibitemShut
  {NoStop}%
\bibitem [{\citenamefont {Lewis}(2019)}]{lewis2019}%
  \BibitemOpen
  \bibfield  {author} {\bibinfo {author} {\bibfnamefont {A.}~\bibnamefont
  {Lewis}},\ }\href {http://adsabs.harvard.edu/abs/2019arXiv191013970L}
  {\bibfield  {journal} {\bibinfo  {journal} {arXiv e-prints}\ }\textbf
  {\bibinfo {volume} {1910}},\ \bibinfo {pages} {arXiv:1910.13970} (\bibinfo
  {year} {2019})}\BibitemShut {NoStop}%
\bibitem [{\citenamefont {Barron}\ \emph {et~al.}(2017)\citenamefont {Barron},
  \citenamefont {Chinone}, \citenamefont {Kusaka}, \citenamefont {Borril},
  \citenamefont {Errard}, \citenamefont {Feeney}, \citenamefont {Ferraro},
  \citenamefont {Keskitalo}, \citenamefont {Lee}, \citenamefont {Roe},
  \citenamefont {Sherwin},\ and\ \citenamefont {Suzuki}}]{barron2017}%
  \BibitemOpen
  \bibfield  {author} {\bibinfo {author} {\bibfnamefont {D.}~\bibnamefont
  {Barron}}, \bibinfo {author} {\bibfnamefont {Y.}~\bibnamefont {Chinone}},
  \bibinfo {author} {\bibfnamefont {A.}~\bibnamefont {Kusaka}}, \bibinfo
  {author} {\bibfnamefont {J.}~\bibnamefont {Borril}}, \bibinfo {author}
  {\bibfnamefont {J.}~\bibnamefont {Errard}}, \bibinfo {author} {\bibfnamefont
  {S.}~\bibnamefont {Feeney}}, \bibinfo {author} {\bibfnamefont
  {S.}~\bibnamefont {Ferraro}}, \bibinfo {author} {\bibfnamefont
  {R.}~\bibnamefont {Keskitalo}}, \bibinfo {author} {\bibfnamefont {A.~T.}\
  \bibnamefont {Lee}}, \bibinfo {author} {\bibfnamefont {N.~A.}\ \bibnamefont
  {Roe}}, \bibinfo {author} {\bibfnamefont {B.~D.}\ \bibnamefont {Sherwin}}, \
  and\ \bibinfo {author} {\bibfnamefont {A.}~\bibnamefont {Suzuki}},\ }\href
  {http://arxiv.org/abs/1702.07467} {\bibfield  {journal} {\bibinfo  {journal}
  {arXiv:1702.07467 [astro-ph]}\ } (\bibinfo {year} {2017})},\ \Eprint
  {http://arxiv.org/abs/1702.07467} {arXiv:1702.07467 [astro-ph]} \BibitemShut
  {NoStop}%
\bibitem [{\citenamefont {Coulton}\ \emph {et~al.}(2019)\citenamefont
  {Coulton}, \citenamefont {Meerburg}, \citenamefont {Baker}, \citenamefont
  {Hotinli}, \citenamefont {Duivenvoorden},\ and\ \citenamefont {{van
  Engelen}}}]{coulton2019}%
  \BibitemOpen
  \bibfield  {author} {\bibinfo {author} {\bibfnamefont {W.~R.}\ \bibnamefont
  {Coulton}}, \bibinfo {author} {\bibfnamefont {P.~D.}\ \bibnamefont
  {Meerburg}}, \bibinfo {author} {\bibfnamefont {D.~G.}\ \bibnamefont {Baker}},
  \bibinfo {author} {\bibfnamefont {S.}~\bibnamefont {Hotinli}}, \bibinfo
  {author} {\bibfnamefont {A.~J.}\ \bibnamefont {Duivenvoorden}}, \ and\
  \bibinfo {author} {\bibfnamefont {A.}~\bibnamefont {{van Engelen}}},\ }\href
  {http://adsabs.harvard.edu/abs/2019arXiv191207619C} {\bibfield  {journal}
  {\bibinfo  {journal} {arXiv e-prints}\ }\textbf {\bibinfo {volume} {1912}},\
  \bibinfo {pages} {arXiv:1912.07619} (\bibinfo {year} {2019})}\BibitemShut
  {NoStop}%
\bibitem [{\citenamefont {Meerburg}\ \emph {et~al.}(2016)\citenamefont
  {Meerburg}, \citenamefont {Meyers}, \citenamefont {{van Engelen}},\ and\
  \citenamefont {{Ali-Ha{\"i}moud}}}]{meerburg2016}%
  \BibitemOpen
  \bibfield  {author} {\bibinfo {author} {\bibfnamefont {P.~D.}\ \bibnamefont
  {Meerburg}}, \bibinfo {author} {\bibfnamefont {J.}~\bibnamefont {Meyers}},
  \bibinfo {author} {\bibfnamefont {A.}~\bibnamefont {{van Engelen}}}, \ and\
  \bibinfo {author} {\bibfnamefont {Y.}~\bibnamefont {{Ali-Ha{\"i}moud}}},\
  }\href {\doibase 10.1103/PhysRevD.93.123511} {\bibfield  {journal} {\bibinfo
  {journal} {Physical Review D}\ }\textbf {\bibinfo {volume} {93}},\ \bibinfo
  {pages} {123511} (\bibinfo {year} {2016})}\BibitemShut {NoStop}%
\bibitem [{\citenamefont {Box}(1979)}]{BOX1979201}%
  \BibitemOpen
  \bibfield  {author} {\bibinfo {author} {\bibfnamefont {G.}~\bibnamefont
  {Box}},\ }in\ \href {\doibase
  https://doi.org/10.1016/B978-0-12-438150-6.50018-2} {\emph {\bibinfo
  {booktitle} {Robustness in Statistics}}},\ \bibinfo {editor} {edited by\
  \bibinfo {editor} {\bibfnamefont {R.~L.}\ \bibnamefont {LAUNER}}\ and\
  \bibinfo {editor} {\bibfnamefont {G.~N.}\ \bibnamefont {WILKINSON}}}\
  (\bibinfo  {publisher} {Academic Press},\ \bibinfo {year} {1979})\ pp.\
  \bibinfo {pages} {201 -- 236}\BibitemShut {NoStop}%
\bibitem [{\citenamefont {Rubin}(1984)}]{rubin1984}%
  \BibitemOpen
  \bibfield  {author} {\bibinfo {author} {\bibfnamefont {D.~B.}\ \bibnamefont
  {Rubin}},\ }\href {\doibase 10.1214/aos/1176346785} {\bibfield  {journal}
  {\bibinfo  {journal} {Ann. Statist.}\ }\textbf {\bibinfo {volume} {12}},\
  \bibinfo {pages} {1151} (\bibinfo {year} {1984})}\BibitemShut {NoStop}%
\bibitem [{\citenamefont {{Feeney}}\ \emph {et~al.}(2019)\citenamefont
  {{Feeney}}, \citenamefont {{Peiris}}, \citenamefont {{Williamson}},
  \citenamefont {{Nissanke}}, \citenamefont {{Mortlock}}, \citenamefont
  {{Alsing}},\ and\ \citenamefont {{Scolnic}}}]{2019PhRvL.122f1105F}%
  \BibitemOpen
  \bibfield  {author} {\bibinfo {author} {\bibfnamefont {S.~M.}\ \bibnamefont
  {{Feeney}}}, \bibinfo {author} {\bibfnamefont {H.~V.}\ \bibnamefont
  {{Peiris}}}, \bibinfo {author} {\bibfnamefont {A.~R.}\ \bibnamefont
  {{Williamson}}}, \bibinfo {author} {\bibfnamefont {S.~M.}\ \bibnamefont
  {{Nissanke}}}, \bibinfo {author} {\bibfnamefont {D.~J.}\ \bibnamefont
  {{Mortlock}}}, \bibinfo {author} {\bibfnamefont {J.}~\bibnamefont
  {{Alsing}}}, \ and\ \bibinfo {author} {\bibfnamefont {D.}~\bibnamefont
  {{Scolnic}}},\ }\href {\doibase 10.1103/PhysRevLett.122.061105} {\bibfield
  {journal} {\bibinfo  {journal} {\prl}\ }\textbf {\bibinfo {volume} {122}},\
  \bibinfo {eid} {061105} (\bibinfo {year} {2019})},\ \Eprint
  {http://arxiv.org/abs/1802.03404} {arXiv:1802.03404 [astro-ph.CO]}
  \BibitemShut {NoStop}%
\bibitem [{\citenamefont {Goodman}\ and\ \citenamefont
  {Weare}(2010)}]{goodman2010}%
  \BibitemOpen
  \bibfield  {author} {\bibinfo {author} {\bibfnamefont {J.}~\bibnamefont
  {Goodman}}\ and\ \bibinfo {author} {\bibfnamefont {J.}~\bibnamefont
  {Weare}},\ }\href {\doibase 10.2140/camcos.2010.5.65} {\bibfield  {journal}
  {\bibinfo  {journal} {Communications in Applied Mathematics and Computational
  Science}\ }\textbf {\bibinfo {volume} {5}},\ \bibinfo {pages} {65} (\bibinfo
  {year} {2010})}\BibitemShut {NoStop}%
\bibitem [{\citenamefont {Bezanson}\ \emph {et~al.}(2017)\citenamefont
  {Bezanson}, \citenamefont {Edelman}, \citenamefont {Karpinski},\ and\
  \citenamefont {Shah}}]{bezanson2017}%
  \BibitemOpen
  \bibfield  {author} {\bibinfo {author} {\bibfnamefont {J.}~\bibnamefont
  {Bezanson}}, \bibinfo {author} {\bibfnamefont {A.}~\bibnamefont {Edelman}},
  \bibinfo {author} {\bibfnamefont {S.}~\bibnamefont {Karpinski}}, \ and\
  \bibinfo {author} {\bibfnamefont {V.}~\bibnamefont {Shah}},\ }\href {\doibase
  10.1137/141000671} {\bibfield  {journal} {\bibinfo  {journal} {SIAM Review}\
  }\textbf {\bibinfo {volume} {59}},\ \bibinfo {pages} {65} (\bibinfo {year}
  {2017})}\BibitemShut {NoStop}%
\bibitem [{\citenamefont {Innes}(2018)}]{innes2018}%
  \BibitemOpen
  \bibfield  {author} {\bibinfo {author} {\bibfnamefont {M.}~\bibnamefont
  {Innes}},\ }\href {http://adsabs.harvard.edu/abs/2018arXiv181007951I}
  {\bibfield  {journal} {\bibinfo  {journal} {arXiv e-prints}\ }\textbf
  {\bibinfo {volume} {1810}},\ \bibinfo {pages} {arXiv:1810.07951} (\bibinfo
  {year} {2018})}\BibitemShut {NoStop}%
\bibitem [{\citenamefont {Crites}\ \emph {et~al.}(2015)\citenamefont {Crites},
  \citenamefont {Henning}, \citenamefont {Ade}, \citenamefont {Aird},
  \citenamefont {Austermann}, \citenamefont {Beall}, \citenamefont {Bender},
  \citenamefont {Benson}, \citenamefont {Bleem}, \citenamefont {Carlstrom},
  \citenamefont {Chang}, \citenamefont {Chiang}, \citenamefont {Cho},
  \citenamefont {Citron}, \citenamefont {Crawford}, \citenamefont {{de Haan}},
  \citenamefont {Dobbs}, \citenamefont {Everett}, \citenamefont {Gallicchio},
  \citenamefont {Gao}, \citenamefont {George}, \citenamefont {Gilbert},
  \citenamefont {Halverson}, \citenamefont {Hanson}, \citenamefont
  {Harrington}, \citenamefont {Hilton}, \citenamefont {Holder}, \citenamefont
  {Holzapfel}, \citenamefont {Hoover}, \citenamefont {Hou}, \citenamefont
  {Hrubes}, \citenamefont {Huang}, \citenamefont {Hubmayr}, \citenamefont
  {Irwin}, \citenamefont {Keisler}, \citenamefont {Knox}, \citenamefont {Lee},
  \citenamefont {Leitch}, \citenamefont {Li}, \citenamefont {Liang},
  \citenamefont {{Luong-Van}}, \citenamefont {McMahon}, \citenamefont {Mehl},
  \citenamefont {Meyer}, \citenamefont {Mocanu}, \citenamefont {Montroy},
  \citenamefont {Natoli}, \citenamefont {Nibarger}, \citenamefont {Novosad},
  \citenamefont {Padin}, \citenamefont {Pryke}, \citenamefont {Reichardt},
  \citenamefont {Ruhl}, \citenamefont {Saliwanchik}, \citenamefont {Sayre},
  \citenamefont {Schaffer}, \citenamefont {Smecher}, \citenamefont {Stark},
  \citenamefont {Story}, \citenamefont {Tucker}, \citenamefont {Vanderlinde},
  \citenamefont {Vieira}, \citenamefont {Wang}, \citenamefont {Whitehorn},
  \citenamefont {Yefremenko},\ and\ \citenamefont {Zahn}}]{crites2015}%
  \BibitemOpen
  \bibfield  {author} {\bibinfo {author} {\bibfnamefont {A.~T.}\ \bibnamefont
  {Crites}}, \bibinfo {author} {\bibfnamefont {J.~W.}\ \bibnamefont {Henning}},
  \bibinfo {author} {\bibfnamefont {P.~A.~R.}\ \bibnamefont {Ade}}, \bibinfo
  {author} {\bibfnamefont {K.~A.}\ \bibnamefont {Aird}}, \bibinfo {author}
  {\bibfnamefont {J.~E.}\ \bibnamefont {Austermann}}, \bibinfo {author}
  {\bibfnamefont {J.~A.}\ \bibnamefont {Beall}}, \bibinfo {author}
  {\bibfnamefont {A.~N.}\ \bibnamefont {Bender}}, \bibinfo {author}
  {\bibfnamefont {B.~A.}\ \bibnamefont {Benson}}, \bibinfo {author}
  {\bibfnamefont {L.~E.}\ \bibnamefont {Bleem}}, \bibinfo {author}
  {\bibfnamefont {J.~E.}\ \bibnamefont {Carlstrom}}, \bibinfo {author}
  {\bibfnamefont {C.~L.}\ \bibnamefont {Chang}}, \bibinfo {author}
  {\bibfnamefont {H.~C.}\ \bibnamefont {Chiang}}, \bibinfo {author}
  {\bibfnamefont {H.-M.}\ \bibnamefont {Cho}}, \bibinfo {author} {\bibfnamefont
  {R.}~\bibnamefont {Citron}}, \bibinfo {author} {\bibfnamefont {T.~M.}\
  \bibnamefont {Crawford}}, \bibinfo {author} {\bibfnamefont {T.}~\bibnamefont
  {{de Haan}}}, \bibinfo {author} {\bibfnamefont {M.~A.}\ \bibnamefont
  {Dobbs}}, \bibinfo {author} {\bibfnamefont {W.}~\bibnamefont {Everett}},
  \bibinfo {author} {\bibfnamefont {J.}~\bibnamefont {Gallicchio}}, \bibinfo
  {author} {\bibfnamefont {J.}~\bibnamefont {Gao}}, \bibinfo {author}
  {\bibfnamefont {E.~M.}\ \bibnamefont {George}}, \bibinfo {author}
  {\bibfnamefont {A.}~\bibnamefont {Gilbert}}, \bibinfo {author} {\bibfnamefont
  {N.~W.}\ \bibnamefont {Halverson}}, \bibinfo {author} {\bibfnamefont
  {D.}~\bibnamefont {Hanson}}, \bibinfo {author} {\bibfnamefont
  {N.}~\bibnamefont {Harrington}}, \bibinfo {author} {\bibfnamefont {G.~C.}\
  \bibnamefont {Hilton}}, \bibinfo {author} {\bibfnamefont {G.~P.}\
  \bibnamefont {Holder}}, \bibinfo {author} {\bibfnamefont {W.~L.}\
  \bibnamefont {Holzapfel}}, \bibinfo {author} {\bibfnamefont {S.}~\bibnamefont
  {Hoover}}, \bibinfo {author} {\bibfnamefont {Z.}~\bibnamefont {Hou}},
  \bibinfo {author} {\bibfnamefont {J.~D.}\ \bibnamefont {Hrubes}}, \bibinfo
  {author} {\bibfnamefont {N.}~\bibnamefont {Huang}}, \bibinfo {author}
  {\bibfnamefont {J.}~\bibnamefont {Hubmayr}}, \bibinfo {author} {\bibfnamefont
  {K.~D.}\ \bibnamefont {Irwin}}, \bibinfo {author} {\bibfnamefont
  {R.}~\bibnamefont {Keisler}}, \bibinfo {author} {\bibfnamefont
  {L.}~\bibnamefont {Knox}}, \bibinfo {author} {\bibfnamefont {A.~T.}\
  \bibnamefont {Lee}}, \bibinfo {author} {\bibfnamefont {E.~M.}\ \bibnamefont
  {Leitch}}, \bibinfo {author} {\bibfnamefont {D.}~\bibnamefont {Li}}, \bibinfo
  {author} {\bibfnamefont {C.}~\bibnamefont {Liang}}, \bibinfo {author}
  {\bibfnamefont {D.}~\bibnamefont {{Luong-Van}}}, \bibinfo {author}
  {\bibfnamefont {J.~J.}\ \bibnamefont {McMahon}}, \bibinfo {author}
  {\bibfnamefont {J.}~\bibnamefont {Mehl}}, \bibinfo {author} {\bibfnamefont
  {S.~S.}\ \bibnamefont {Meyer}}, \bibinfo {author} {\bibfnamefont
  {L.}~\bibnamefont {Mocanu}}, \bibinfo {author} {\bibfnamefont {T.~E.}\
  \bibnamefont {Montroy}}, \bibinfo {author} {\bibfnamefont {T.}~\bibnamefont
  {Natoli}}, \bibinfo {author} {\bibfnamefont {J.~P.}\ \bibnamefont
  {Nibarger}}, \bibinfo {author} {\bibfnamefont {V.}~\bibnamefont {Novosad}},
  \bibinfo {author} {\bibfnamefont {S.}~\bibnamefont {Padin}}, \bibinfo
  {author} {\bibfnamefont {C.}~\bibnamefont {Pryke}}, \bibinfo {author}
  {\bibfnamefont {C.~L.}\ \bibnamefont {Reichardt}}, \bibinfo {author}
  {\bibfnamefont {J.~E.}\ \bibnamefont {Ruhl}}, \bibinfo {author}
  {\bibfnamefont {B.~R.}\ \bibnamefont {Saliwanchik}}, \bibinfo {author}
  {\bibfnamefont {J.~T.}\ \bibnamefont {Sayre}}, \bibinfo {author}
  {\bibfnamefont {K.~K.}\ \bibnamefont {Schaffer}}, \bibinfo {author}
  {\bibfnamefont {G.}~\bibnamefont {Smecher}}, \bibinfo {author} {\bibfnamefont
  {A.~A.}\ \bibnamefont {Stark}}, \bibinfo {author} {\bibfnamefont {K.~T.}\
  \bibnamefont {Story}}, \bibinfo {author} {\bibfnamefont {C.}~\bibnamefont
  {Tucker}}, \bibinfo {author} {\bibfnamefont {K.}~\bibnamefont {Vanderlinde}},
  \bibinfo {author} {\bibfnamefont {J.~D.}\ \bibnamefont {Vieira}}, \bibinfo
  {author} {\bibfnamefont {G.}~\bibnamefont {Wang}}, \bibinfo {author}
  {\bibfnamefont {N.}~\bibnamefont {Whitehorn}}, \bibinfo {author}
  {\bibfnamefont {V.}~\bibnamefont {Yefremenko}}, \ and\ \bibinfo {author}
  {\bibfnamefont {O.}~\bibnamefont {Zahn}},\ }\href {\doibase
  10.1088/0004-637X/805/1/36} {\bibfield  {journal} {\bibinfo  {journal} {The
  Astrophysical Journal}\ }\textbf {\bibinfo {volume} {805}},\ \bibinfo {pages}
  {36} (\bibinfo {year} {2015})}\BibitemShut {NoStop}%
\bibitem [{\citenamefont {Blei}\ \emph {et~al.}(2017)\citenamefont {Blei},
  \citenamefont {Kucukelbir},\ and\ \citenamefont {McAuliffe}}]{blei2017}%
  \BibitemOpen
  \bibfield  {author} {\bibinfo {author} {\bibfnamefont {D.~M.}\ \bibnamefont
  {Blei}}, \bibinfo {author} {\bibfnamefont {A.}~\bibnamefont {Kucukelbir}}, \
  and\ \bibinfo {author} {\bibfnamefont {J.~D.}\ \bibnamefont {McAuliffe}},\
  }\href {\doibase 10.1080/01621459.2017.1285773} {\bibfield  {journal}
  {\bibinfo  {journal} {Journal of the American Statistical Association}\
  }\textbf {\bibinfo {volume} {112}},\ \bibinfo {pages} {859} (\bibinfo {year}
  {2017})},\ \Eprint {http://arxiv.org/abs/1601.00670} {arXiv:1601.00670}
  \BibitemShut {NoStop}%
\bibitem [{\citenamefont {Seljak}\ and\ \citenamefont {Yu}(2019)}]{seljak2019}%
  \BibitemOpen
  \bibfield  {author} {\bibinfo {author} {\bibfnamefont {U.}~\bibnamefont
  {Seljak}}\ and\ \bibinfo {author} {\bibfnamefont {B.}~\bibnamefont {Yu}},\
  }\href {http://adsabs.harvard.edu/abs/2019arXiv190104454S} {\bibfield
  {journal} {\bibinfo  {journal} {arXiv e-prints}\ }\textbf {\bibinfo {volume}
  {1901}},\ \bibinfo {pages} {arXiv:1901.04454} (\bibinfo {year} {2019})},\
  \Eprint {http://arxiv.org/abs/1901.04454} {arXiv:1901.04454} \BibitemShut
  {NoStop}%
\bibitem [{\citenamefont {Knollm{\"u}ller}\ and\ \citenamefont
  {En{\ss}lin}(2019)}]{knollmuller2019}%
  \BibitemOpen
  \bibfield  {author} {\bibinfo {author} {\bibfnamefont {J.}~\bibnamefont
  {Knollm{\"u}ller}}\ and\ \bibinfo {author} {\bibfnamefont {T.~A.}\
  \bibnamefont {En{\ss}lin}},\ }\href
  {http://adsabs.harvard.edu/abs/2019arXiv190111033K} {\bibfield  {journal}
  {\bibinfo  {journal} {arXiv e-prints}\ }\textbf {\bibinfo {volume} {1901}},\
  \bibinfo {pages} {arXiv:1901.11033} (\bibinfo {year} {2019})},\ \Eprint
  {http://arxiv.org/abs/1901.11033} {arXiv:1901.11033} \BibitemShut {NoStop}%
\bibitem [{\citenamefont {Seljak}\ \emph {et~al.}(2017)\citenamefont {Seljak},
  \citenamefont {Aslanyan}, \citenamefont {Feng},\ and\ \citenamefont
  {Modi}}]{seljak2017}%
  \BibitemOpen
  \bibfield  {author} {\bibinfo {author} {\bibfnamefont {U.}~\bibnamefont
  {Seljak}}, \bibinfo {author} {\bibfnamefont {G.}~\bibnamefont {Aslanyan}},
  \bibinfo {author} {\bibfnamefont {Y.}~\bibnamefont {Feng}}, \ and\ \bibinfo
  {author} {\bibfnamefont {C.}~\bibnamefont {Modi}},\ }\href
  {http://adsabs.harvard.edu/abs/2017arXiv170606645S} {\bibfield  {journal}
  {\bibinfo  {journal} {ArXiv e-prints}\ }\textbf {\bibinfo {volume} {1706}},\
  \bibinfo {pages} {arXiv:1706.06645} (\bibinfo {year} {2017})}\BibitemShut
  {NoStop}%
\bibitem [{\citenamefont {Marin}\ \emph {et~al.}(2011)\citenamefont {Marin},
  \citenamefont {Pudlo}, \citenamefont {Robert},\ and\ \citenamefont
  {Ryder}}]{marin2011}%
  \BibitemOpen
  \bibfield  {author} {\bibinfo {author} {\bibfnamefont {J.-M.}\ \bibnamefont
  {Marin}}, \bibinfo {author} {\bibfnamefont {P.}~\bibnamefont {Pudlo}},
  \bibinfo {author} {\bibfnamefont {C.~P.}\ \bibnamefont {Robert}}, \ and\
  \bibinfo {author} {\bibfnamefont {R.}~\bibnamefont {Ryder}},\ }\href
  {http://adsabs.harvard.edu/abs/2011arXiv1101.0955M} {\bibfield  {journal}
  {\bibinfo  {journal} {arXiv e-prints}\ }\textbf {\bibinfo {volume} {1101}},\
  \bibinfo {pages} {arXiv:1101.0955} (\bibinfo {year} {2011})}\BibitemShut
  {NoStop}%
\bibitem [{\citenamefont {Price}\ \emph {et~al.}(2018)\citenamefont {Price},
  \citenamefont {Drovandi}, \citenamefont {Lee},\ and\ \citenamefont
  {Nott}}]{price2018}%
  \BibitemOpen
  \bibfield  {author} {\bibinfo {author} {\bibfnamefont {L.~F.}\ \bibnamefont
  {Price}}, \bibinfo {author} {\bibfnamefont {C.~C.}\ \bibnamefont {Drovandi}},
  \bibinfo {author} {\bibfnamefont {A.}~\bibnamefont {Lee}}, \ and\ \bibinfo
  {author} {\bibfnamefont {D.~J.}\ \bibnamefont {Nott}},\ }\href {\doibase
  10.1080/10618600.2017.1302882} {\bibfield  {journal} {\bibinfo  {journal}
  {Journal of Computational and Graphical Statistics}\ }\textbf {\bibinfo
  {volume} {27}},\ \bibinfo {pages} {1} (\bibinfo {year} {2018})}\BibitemShut
  {NoStop}%
\end{thebibliography}%

\end{document}